\documentclass{aa}

\usepackage[version=3]{mhchem}
\usepackage{graphicx}
\usepackage{subfigure}
\usepackage{amsmath}
\usepackage[justification=centering]{caption}

\usepackage{txfonts}
\usepackage{natbib}
\bibpunct{(}{)}{;}{a}{}{,} 

\newcommand{\bi}{\begin{itemize}}
\newcommand{\ei}{\end{itemize}}
\newcommand{\be}{\begin{enumerate}}
\newcommand{\ee}{\end{enumerate}}

\newcommand{\beq}{\begin{equation}}
\newcommand{\eeq}{\end{equation}}

\begin{document}

\title{Formation of cometary \ce{O2} ice and related ice species on grain surfaces in the midplane of the pre-Solar nebula}

\author{Christian Eistrup \inst{1}
\and Catherine Walsh\inst{2}}

\institute{Leiden Observatory, Leiden University, P.O. Box 9513, 2300 RA Leiden, the Netherlands\\
\email{eistrup@strw.leidenuniv.nl}
\and School of Physics and Astronomy, University of Leeds, Leeds LS2 9JT, UK\\
\email{c.walsh1@leeds.ac.uk}}

\date{Received $\cdots$ / Accepted $\cdots$}

\titlerunning{Formation of cometary \ce{O2} ice...}
\authorrunning{Eistrup \& Walsh}
\abstract
{The detection of abundant \ce{O2} at 1-10\% relative to \ce{H2O} ice in the comae of comets 1P/Halley and 67P/Churyumov-Gerasimenko, motivated attempts to explain the origin of the high \ce{O2} ice abundance. Recent chemical modelling of the outer, colder regions of a protoplanetary disk midplane has shown production of \ce{O2} ice at the same abundance as that measured in the comet.}
{A thorough investigation is carried out to constrain the conditions under which \ce{O2} ice could have been produced through kinetic chemistry in the pre-Solar nebula midplane.}
{An updated chemical kinetics code is utilised to evolve chemistry under pre-Solar nebula midplane conditions. Four different chemical starting conditions, and the effects of various chemical parameters are tested.}
{Using the fiducial network, and for either reset conditions (atomic initial abundances) or atomic oxygen only conditions, the abundance level of \ce{O2} ice measured in the comets can be reproduced at an intermediate time, after 0.1-2 Myr of evolution, depending on ionisation level. When including \ce{O3} chemistry, the abundance of \ce{O2} ice is much lower than the cometary abundance (by several orders of magnitude). \ce{H2O2} and \ce{O3} ices are abundantly produced (at around the level of \ce{O2} ice) in disagreement with their respective abundances or upper limits from observations of comet 67P. Upon closer investigation of the parameter space, and varying parameters for grain-surface chemistry, it is found that for temperatures 15-25 K, densities of $10^{9}-10^{10}$ cm$^{-3}$, and a barrier for quantum tunnelling set to 2 {\AA}, the measured level of \ce{O2} ice can be reproduced with the new chemical network, including an updated binding energy for atomic oxygen (1660K). However, the abundances of \ce{H2O2} and \ce{O3} ices still disagree with the observations. A larger activation energy for the \ce{O + O2 -> O3} reaction ($E_{\rm{act}}$>1000 K) helps to reproduce the non-detection of \ce{O3} ice in the comet, as well as reproducing the observed abundances of \ce{H2O2} and \ce{O2} ices. The only other case where the \ce{O2} ice matches the observed abundance, and \ce{O3} and \ce{H2O2} ice are lower, is the case when starting with an appreciable amount of oxygen locked in \ce{O2}.}
{The parameter space investigation revealed a sweet spot for production of \ce{O2} ice at an abundance matching those in 67P and 1P, and \ce{O3} and \ce{H2O2} ices abundances matching those in 67P. This means that there is a radial region in the pre-Solar nebula from 120-150 AU, within which \ce{O2} could have been produced in-situ via ice chemistry on grain surfaces. However, it is apparent that there is a high degree of sensitivity of the chemistry to the assumed chemical parameters (e.g. binding energy, activation barrier width, and quantum tunnelling barrier). Hence, because the more likely scenario starting with a percentage of elemental oxygen locked in \ce{O2} also reproduces the \ce{O2} ice abundance in 67P at early stages, this supports previous suggestions that the cometary \ce{O2} ice could have a primordial origin.}

\keywords{protoplanetary disks -- astrochemistry -- comets: composition -- molecular processes}

\maketitle


\section{Introduction}
\label{intro}

The detection of abundant molecular oxygen at 1-10\% (average 3.8${\pm 0.85}$\%) relative to \ce{H2O} ice in the coma of comet 67P/Churyumov-Gerasimenko \citep[herinafter 67P:][]{bieler15} came as a surprise, as it was the first detection of \ce{O2} in a comet. This was not expected because \ce{O2} ice has been found to be efficiently converted to \ce{H2O} ice in laboratory studies under interstellar conditions \citep[e.g.][]{ioppolo2008}. Subsequent to this detection, a re-analysis of Comet 1P/Halley data from the \emph{Giotto} mission \citep{rubin2015} indicated a similar \ce{O2}/\ce{H2O} ice ratio (3.8$\pm1.7$\%), suggesting that indeed \ce{O2} ice may be a common ice species in Solar System comets. These detections thus prompted speculation as to the chemical origin of the \ce{O2} ice. \citet{taquet2016} modelled the chemical evolution of material from the pre-stellar core stage to the midplane of the formed protoplanetary disk, and found that \ce{O2} ice can be produced at the early stages and survive the transport to the disk midplane. \citet{mousis2016} found that, if \ce{O2} ice is formed from radiolysis of \ce{H2O} ice through the reaction \ce{2\rm{i}H2O ->[$\gamma$] 2\rm{i}H2 + \rm{i}O2} (where ``i'' denotes a molecule in the ice form), then this likely did not happen during the pre-Solar nebula (PSN) disk phase, but rather in the parent cloud, thus supporting the findings of \citet{taquet2016}. \citet{dulieu2017} performed laboratory experiments to investigate if dismutation of \ce{H2O2} ice (\ce{2iH2O2 -> 2iH2O + iO2}) on the cometary surface could be the origin of the \ce{O2} detection. However, this explanation requires a high initial abundance of \ce{H2O2} ice relative to \ce{H2O} ice (twice the detected abundance of \ce{O2}, or $\sim$7\%), and a high efficiency for the conversion of \ce{H2O2} to \ce{O2} in order to match the low detected level of \ce{H2O2} relative to \ce{O2} of $\sim 6\times 10^{-4}$. \ce{O3} ice, a molecule chemically related to \ce{O2}, \ce{H2O2} and \ce{H2O}, was not detected in the coma of comet 67P, and has an upper limit of $10^{-6}$ with respect to \ce{H2O} ice. It is worth to note here that molecular oxygen is also produced when \ce{CO2} ice is exposed to far-UV radiation \citep[see e.g.][]{martindomenech2015}. However, although this may be viable chemical route to \ce{O2} ice, it does not explain the strong association between the production rates for \ce{H2O} and \ce{O2} seen for comet 67P \citep{bieler15}. On the other hand, experiments investigating chemistry in \ce{CO2} ice irradiated with 5 keV ions (\ce{H+} and \ce{He+}) and electrons \citep{ennis2011,jones2014} favour the production of ozone (\ce{O3}) over molecular oxygen (\ce{O2}). These experiments mimic the conditions that ices are exposed to in the outer Solar System, and upon cosmic ray impact.

In \citet{eistrup2016} (hereafter Paper 1), it was found through chemical kinetic modelling of protoplanetary disk midplanes that \ce{O2} ice could be produced to match the measured cometary abundance by 1 Myr, if the chemical starting conditions were purely atomised and the ionisation level was low ($\sim10^{-19}$ s$^{-1}$). Purely atomised starting conditions reflect the assumption that an energetic stellar outburst or accretion heating close to the star could have fully dissociated all volatiles in the midplane, and low ionisation means that the only ionisation source in the midplane is the decay of short-lived radionuclides. This latter scenario assumes that a magnetic field could have shielded the midplane from cosmic rays \citep[as proposed by][]{cleeves13crex}. This finding sparked interest into whether or not the chemical origin of the \ce{O2} ice could be chemical processing of the icy material in the pre-Solar nebular (PSN) disk midplane. Most recently \citet{mousis2018} explored the possibility that turbulent transport of icy grains between the disk midplane and the upper layers of the disk exposed the grains to a stronger cosmic-ray flux thus chemically processing \ce{H2O} ice to produce \ce{O2} ice via radiolysis. They find that on a 10 Myr timescale \ce{O2} ice in the midplane remains underproduced by up to two orders of magnitude relative to the abundances observed in the comets. \citet{eistrup2018} also found that this abundance for \ce{O2} ice be reached on similarly long timescales.

Based on the promising results in Paper 1, this work investigates under which conditions in the PSN midplane \ce{O2} ice could have been chemically produced in-situ to reach the observed abundance level in the comets. This work differs from that in \citet{mousis2018} in that a full chemical kinetics network is used that follows the chemical connection between \ce{H2O}, \ce{O2} and related ice species. A disk model more suitable for the PSN is used, and a more thorough investigation of the \ce{O2} kinetic ice chemistry is conducted. Initial chemical abundances, physical conditions, parameters for grain-surface chemistry, as well as the inclusion of \ce{O3} in the chemical network are all tested to see the effect on the \ce{O2} ice production and abundance.

\section{Methods}
\label{methods}

\begin{figure*}[h]
\subfigure{\includegraphics[width=0.512\textwidth]{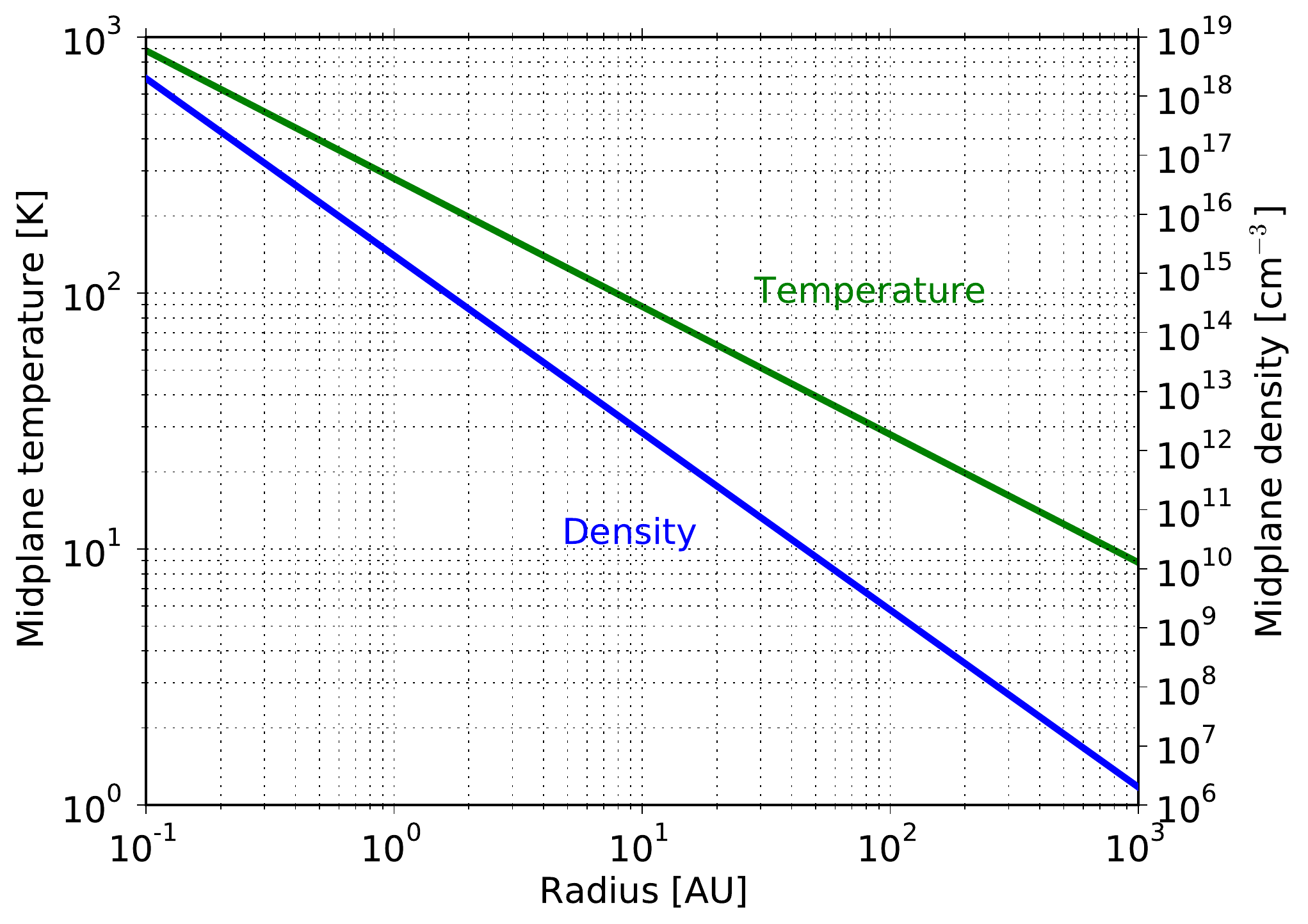}}
\subfigure{\includegraphics[width=0.488\textwidth]{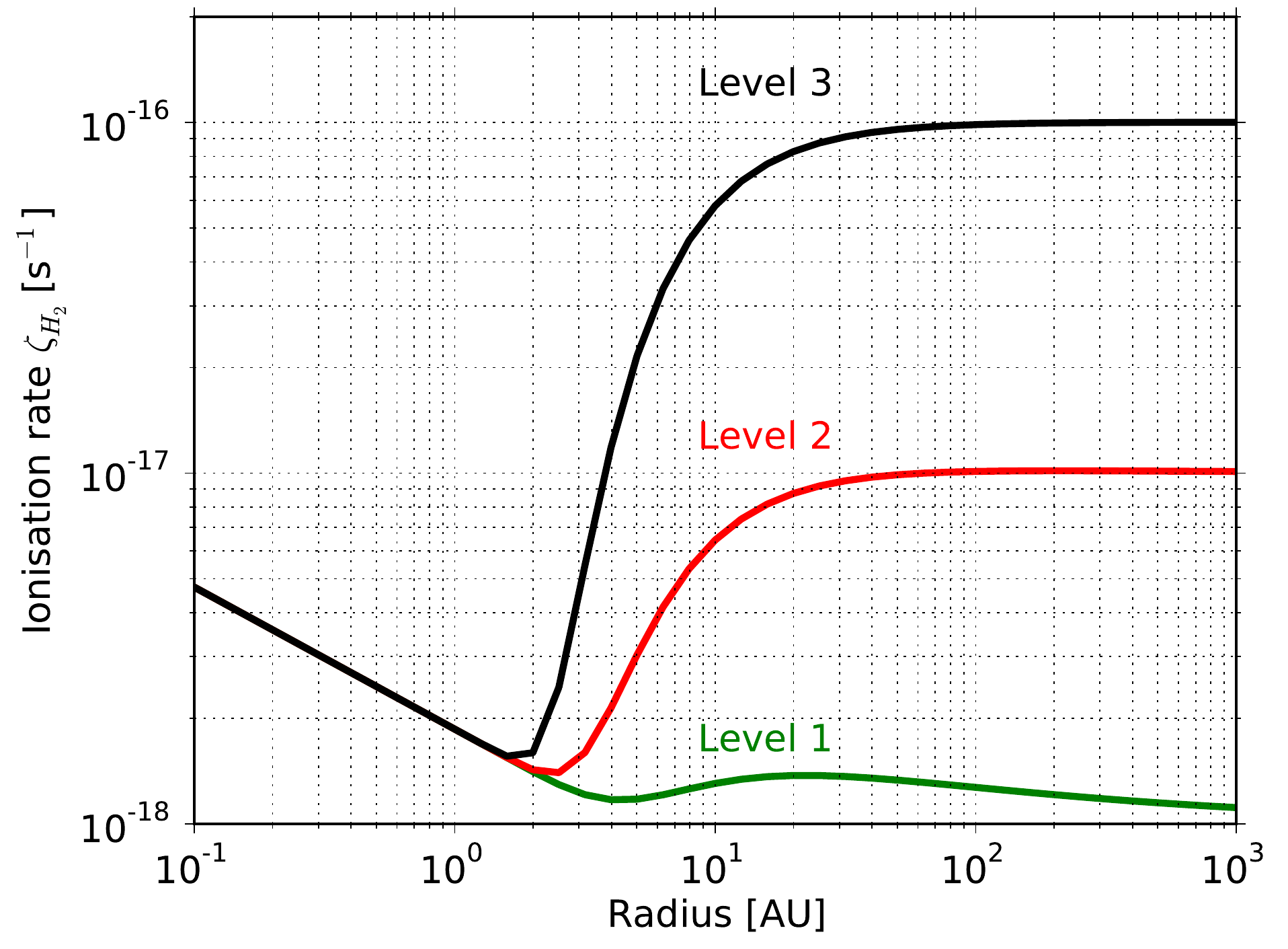}}\\
\caption{a): midplane temperature and density profiles for the pre-Solar nebula. b): ionisation levels utilised in the models. Level 1: SLRs only. Level 2: SLRs + fiducial CRs. Level 3: SLRs + enhanced CRs.}
\label{high_mass_phys}
\end{figure*}

In order to investigate the chemical evolution in the PSN midplane, the physical disk model for the PSN from \citet{hayashi1981} is utilised. This disk model estimates the structure and mass of the PSN, assuming the total current mass of the planets in the Solar System to be distributed as dust and gas in a protoplanetary disk in equilibrium. The mass of this nebula is $M_{\rm{MMSN}}$=0.08 $M_{\odot}$, based on integrating the PSN surface density structure from \citet{aikawa1997}\footnote{$\Sigma(R)=54(R/10\mathrm{AU})^{-3/2}$g cm$^{-2}$} from 0.1-1000 AU. This provides temperature and density profiles for the disk midplane, as well as the surface density profile. The latter is used to calculate the attenuation of cosmic rays (herefrom CRs) impinging on the disk, thereby estimating the contribution of CRs to the disk midplane ionisation, as was discussed in detail in Paper 1. The disk structure is static in time, which was shown in \citet{eistrup2018} to not cause a significantly different chemical evolution from an evolving disk structure, apart from the inward shifting of molecular icelines in response to decreasing temperature with time. 

In Figure \ref{high_mass_phys}a this PSN midplane temperature and density structure is plotted from 0.1-1000 AU. The temperature decreases by two orders of magnitude from $\sim$900 K to 9 K from the inner to the outer disk, and the density drops from $10^{18}$ cm$^{-3}$ to $10^{6}$ cm$^{-3}$. The dynamic ranges, especially in density, are thus larger than what was used in Paper 1. However, this disk is also extending out to 1000 AU instead of 30 AU in Paper 1. This is because this PSN disk midplane is significantly warmer and more massive than the disk structure in Paper 1 (0.08 $M_{\odot}$ versus 0.01 $M_{\odot}$), thus the relevant temperature regime for \ce{O2} ice chemistry (10-30 K) is found outside 100 AU here.

\subsection{Ionisation levels}
\label{ion_levels}

In Paper 1 it was confirmed that ionisation is an important driver of chemistry in disk midplanes \citep[ see e.g.][]{aikawa1997,walsh15}. It was also found that abundant \ce{O2} ice was produced at low ionisation (only radionuclide decay, and no contribution from cosmic rays). 
Therefore, three different levels of ionisation are explored here: a low level (Level 1, SLRs only), a high level (Level 2, the fiducial level which includes galactic cosmic rays), and an extra high level (Level 3, which includes enhanced cosmic rays), see Figure \ref{high_mass_phys}b. The two former levels assume the same contributions to the midplane ionisation as was the case for the low and high level in Paper 1: ionisation Level 1 includes only the contribution from short-lived radionuclides (SLRs) in the midplane. Ionisation Level 2 includes in addition CRs impinging on the disk, using the canonical cosmic ray ionisation rate for the local ISM ($\zeta_{\rm{H_{2}}}=10^{-17}$ s$^{-1}$). Ionisation Level 3 assumes a CR ionisation contribution ten times higher than for ionisation Level 2, based on estimates from e.g. \citet{dishoeck1986}, \citet{dalgarno2006} and \citet{indriolo2015} that the galactic cosmic ray rate $\zeta_{GCR}$ could be in the range $\zeta_{\rm{H}}=10^{-17}-10^{-16}$ s$^{-1}$ in diffuse clouds. We explore this enhanced CR scenario because \ce{O2} ice has been found to be efficiently synthesised in experiments studying \ce{H2O} ice radiolysis in relation to Solar System icy bodies \citep[see, e.g., lab work by][and others]{teolis2017}. For Levels 1 and 2, the ionisation ranges between $10^{-18}-10^{-17}$ s$^{-1}$ throughout the disk whereas for Level 3, $\sim10^{-16}$ s$^{-1}$ is reached at radii larger than 10 AU. In the inner disk inside $\sim$2 AU, all three ionisation levels are the same, because the surface densities here ($>$600 g cm$^{-2}$) attenuate the impinging CRs, so that only the SLRs contribute. In the outer, more diffuse disk, the contribution from CRs impinging on the disk becomes more dominant, thus leading to three markedly different ionisation levels.

\subsection{Chemical network}
\label{chem_network}
The chemical model used here includes gas-phase chemistry, gas-grain interactions
and grain-surface chemistry.  The gas-phase chemistry is from the
latest release of the UMIST Database for Astrochemistry
\citep[][]{mcelroy13} termed {\sc Rate}12.  
The rates for gas-grain interactions and grain-surface chemistry are calculated 
as described in \citet[][and references therein]{walsh15}.  
A gas/dust mass ratio of 100 is adopted. This setup is similar to the ``Full chemistry'' setup from Paper 1. The adopted grain size is 0.1~$\mu$m. 

Four scenarios of initial chemical abundances are explored: the first two are cloud inheritance (hereafter ``Inheritance'' scenario) and chemical reset (hereafter ``Reset'' scenario). In the inheritance scenario the initial abundances are molecular (but no \ce{O2} initially), and assumed inherited from the parent molecular cloud. In the reset scenario, the initial abundances are atomic, because all molecules are assumed dissociated prior to arrival in the disk midplane, due to exposure to a sufficiently strong accretion shock en route into the disk, or accretion heating very close to the star leading to high temperatures. The ``reset'' scenario is of particular interest for this paper because it was for this scenario in Paper 1 that the \ce{O2} ice to \ce{H2O} ice ratio was similar to that found in comets 67P and 1P \citep{bieler15,rubin2015}. Table \ref{old_abund} lists the initial atomic and molecular abundances used in these two scenarios. Tables \ref{bind_e} and \ref{act_e} list the relevant binding energies for species, and activation barriers for reactions applicable to the \ce{O2} chemistry, respectively. Radiolysis of \ce{H2O} ice through the reaction \ce{2\rm{i}H2O ->[$\gamma$] 2\rm{i}H2 + \rm{i}O2} and dismutation of \ce{H2O2} ice through the reaction \ce{2iH2O2 -> 2iH2O + iO2} are not explicitly included in the network. For the inheritance scenario, in this chemical network, atomic oxygen can be produced in-situ in the ice mantles via the cosmic-ray induced photodissociation of \ce{H2O} ice and other abundant oxygen-bearing molecules therein (e.g., \ce{CO2} ice).

The third and fourth scenarios for the initial chemical abundances are more simple: first, the ``Water''-scenario, which assumes initial abundances of gas-phase H, He, \ce{H2} and \ce{H2O}. This scenario is intended to investigate if \ce{O2} ice can be produced through processing of \ce{H2O} ice. Second, the ``Atomic oxygen''-scenario, which assumes gas-phase H, He, \ce{H2} and O as initial abundances. Here it is investigated whether or not \ce{O2} ice production from atoms depends on the presence of elements other than oxygen. \citet{eistrup2018} showed that \ce{H2S} ice on the grain surfaces acts as a catalyst for the conversion of adsorbed oxygen atoms into \ce{O2} ice. Whether or not the sulphur-bearing species are essential for the production of \ce{O2} ice can be tested in these two scenarios, because they exclude sulphur, and only include a source of oxygen in the form of oxygen atoms or \ce{H2O}. 

Lastly, it is explored if the observed cometary \ce{O2} ice abundance can be maintained if the PSN started out with a percentage (5\%) of elemental oxygen locked up in primordial \ce{O2}, as was suggested by \citet{taquet2016}.

\section{Results}
\label{results}

In this section the chemical evolution of various species in the PSN are presented in a number of figures. Due to the focus on \ce{O2} ice in this paper, the attention will be on species that are chemically related to \ce{O2} ice \citep[a schematic overview of these species can be found in][]{cuppen2010}.
\subsection{PSN abundance evolution}
\subsubsection{Inheritance scenario}

\begin{figure*}[h]
\subfigure{\includegraphics[width=0.33\textwidth]{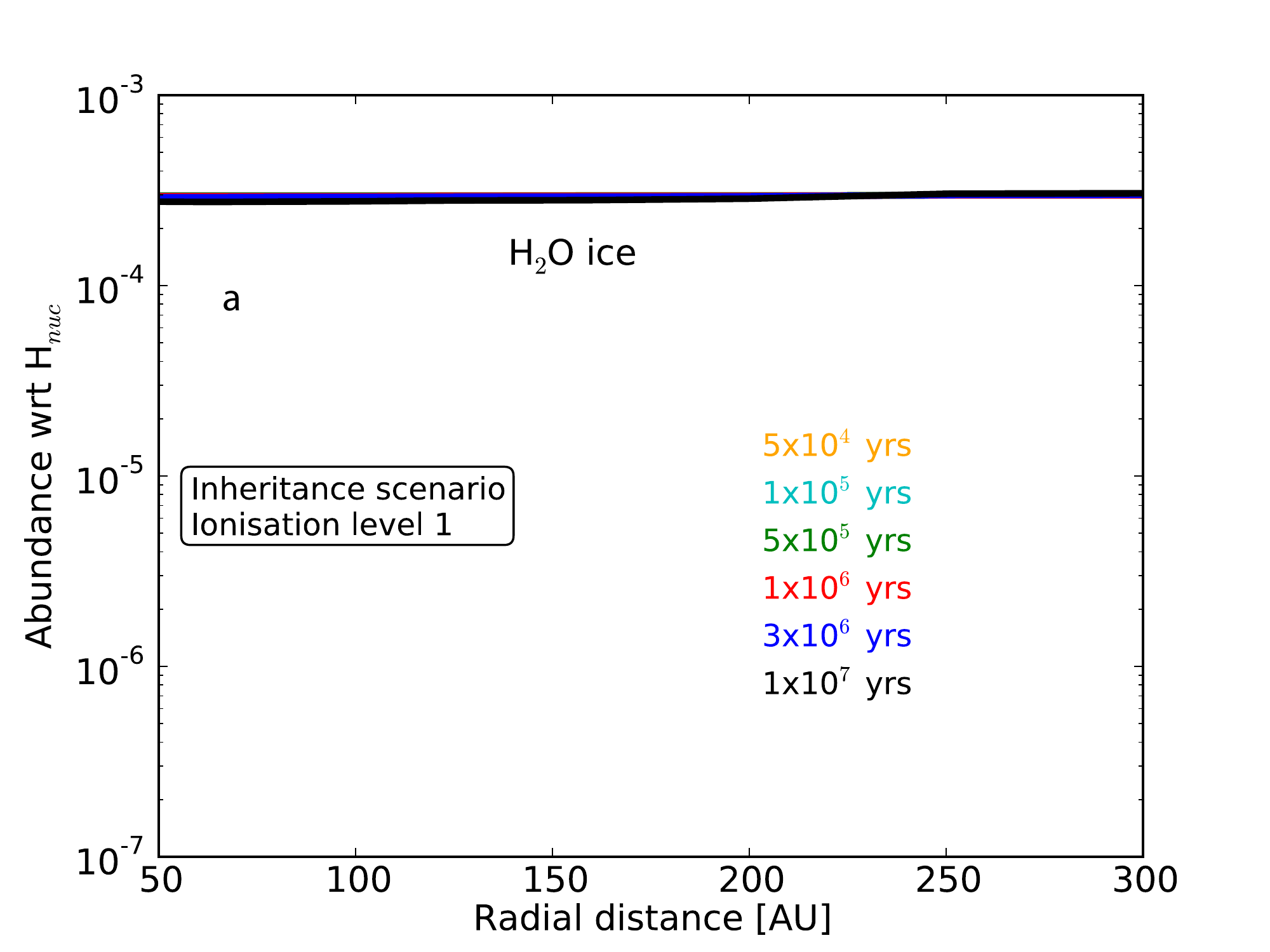}}
\subfigure{\includegraphics[width=0.33\textwidth]{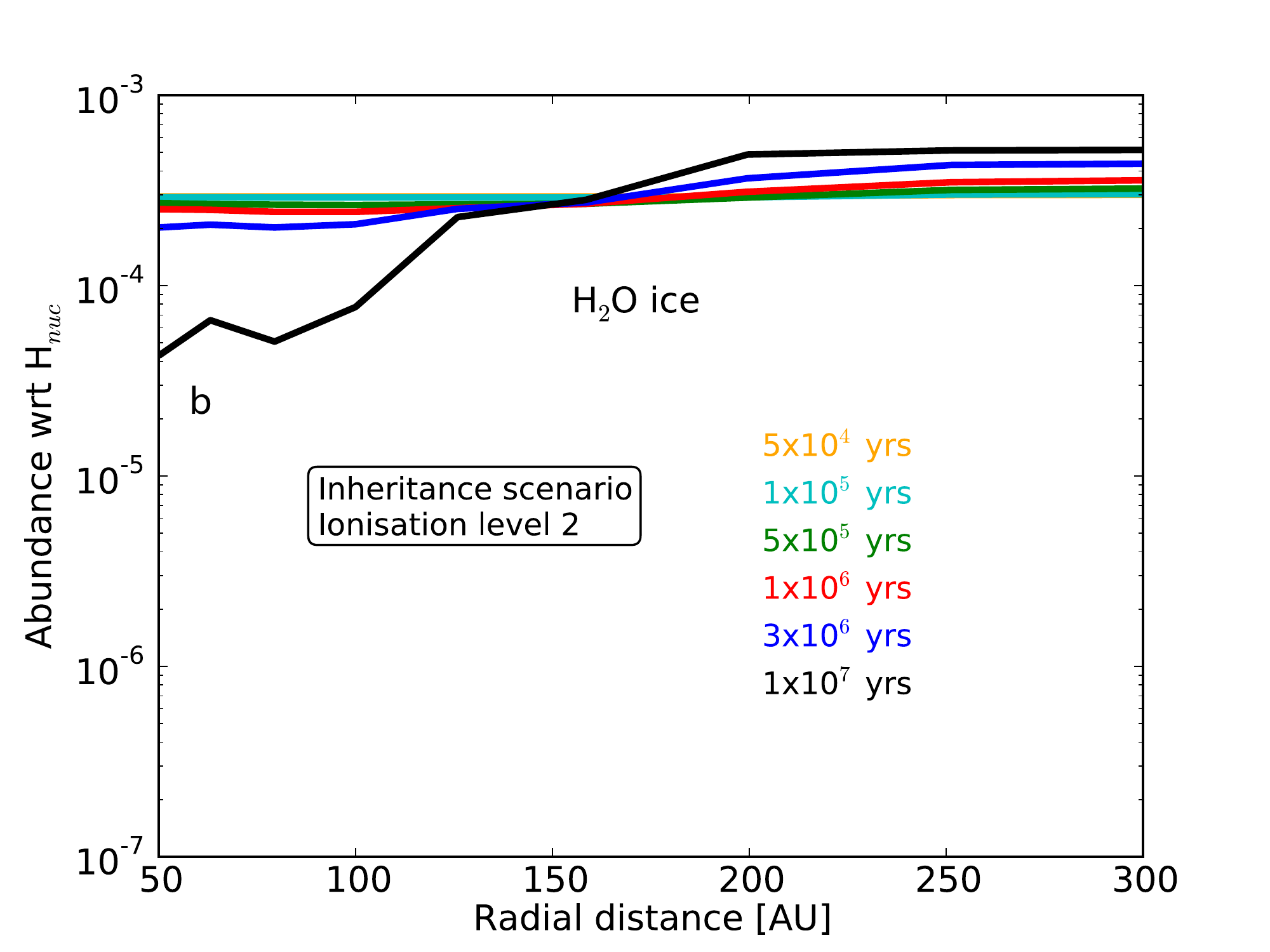}}
\subfigure{\includegraphics[width=0.33\textwidth]{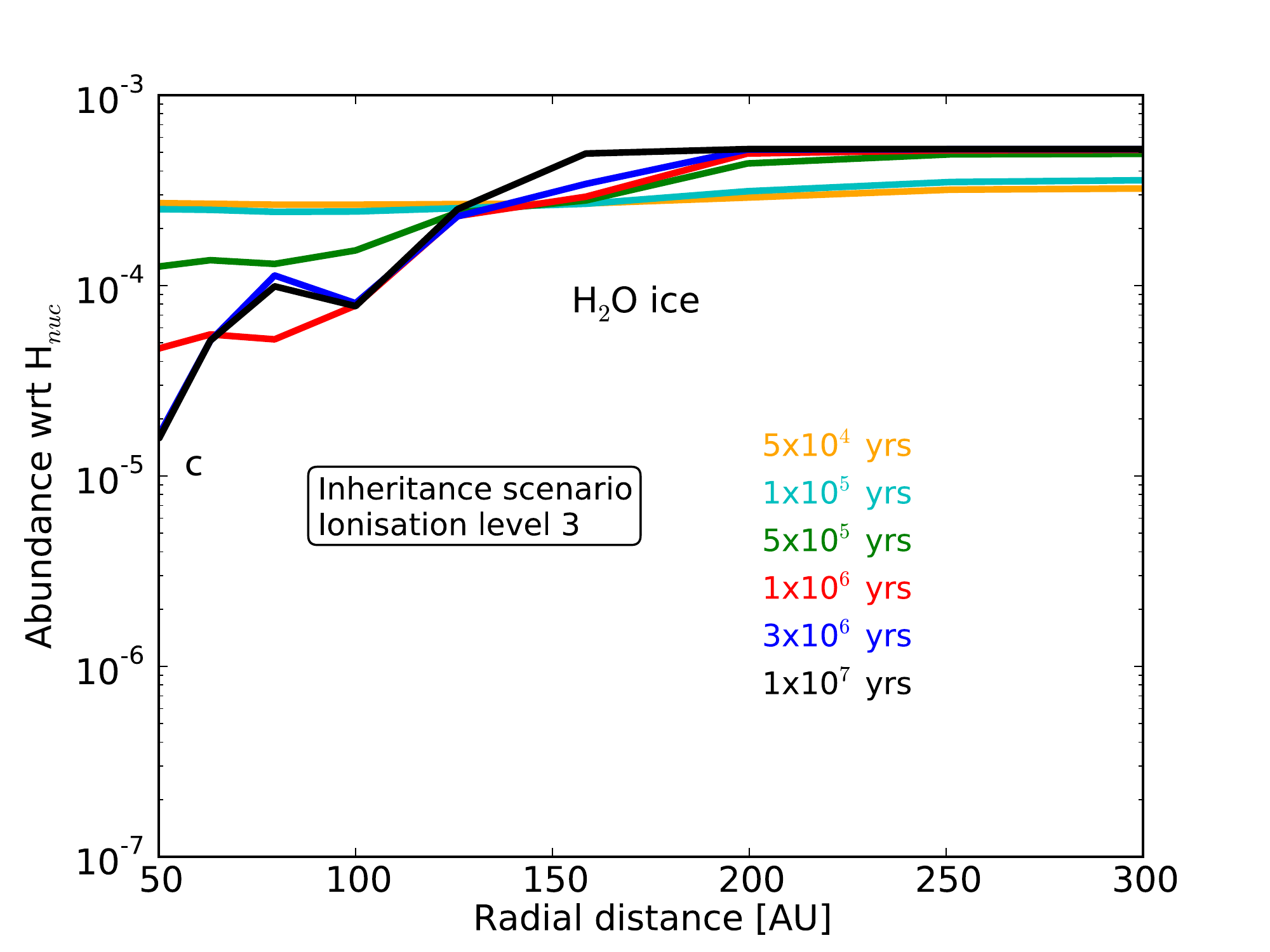}}\\
\subfigure{\includegraphics[width=0.33\textwidth]{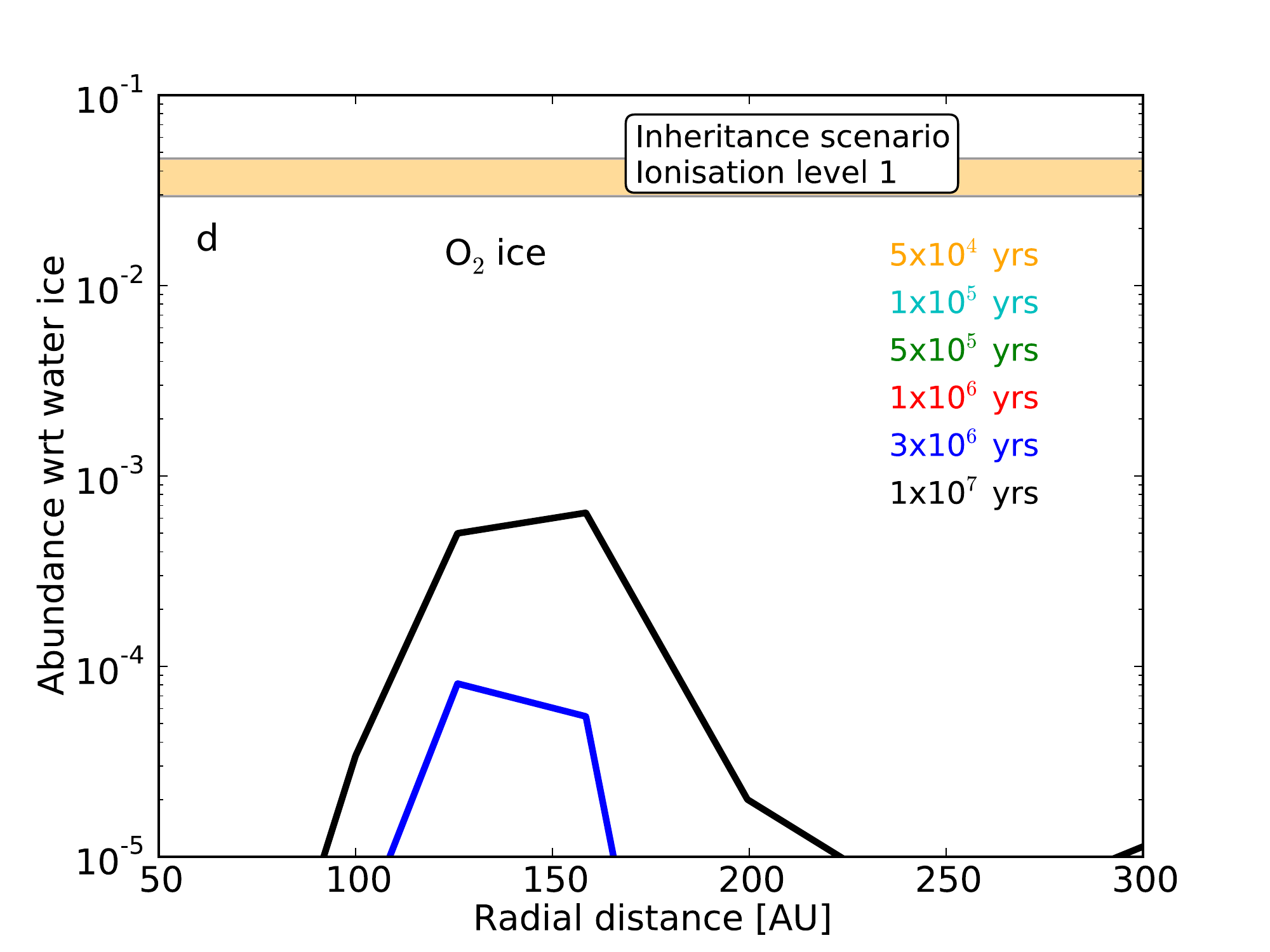}}
\subfigure{\includegraphics[width=0.33\textwidth]{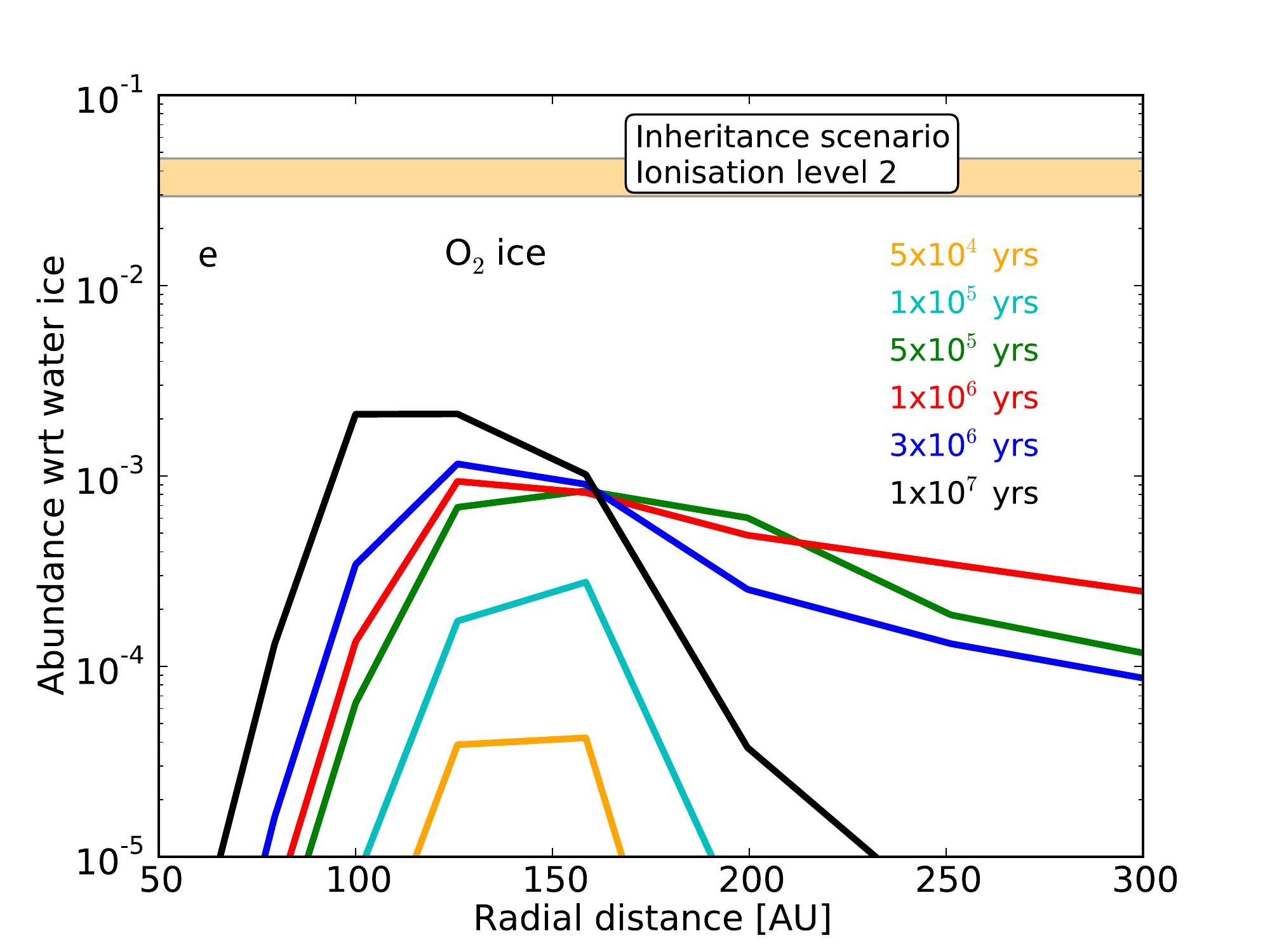}}
\subfigure{\includegraphics[width=0.33\textwidth]{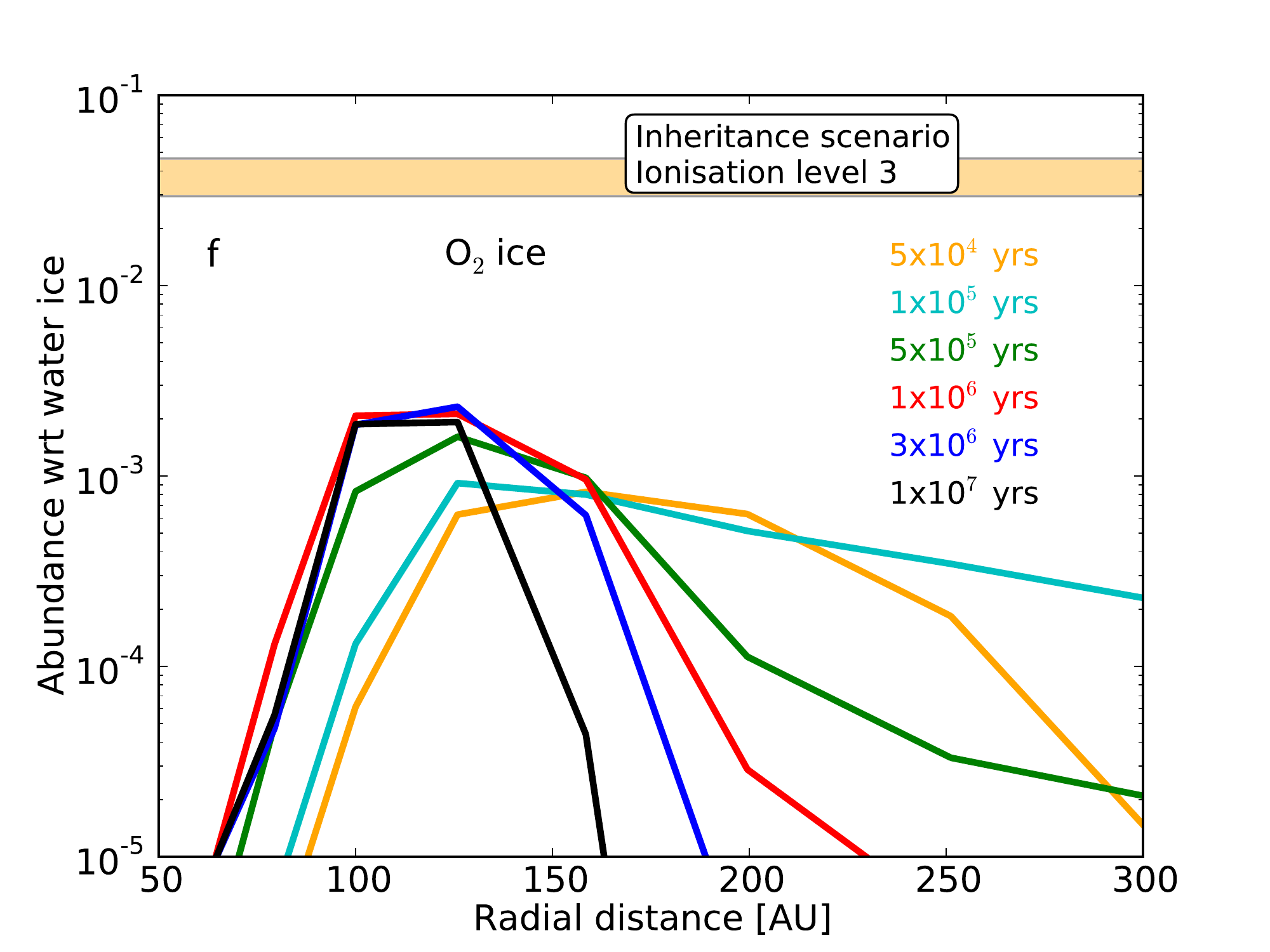}}\\
\subfigure{\includegraphics[width=0.33\textwidth]{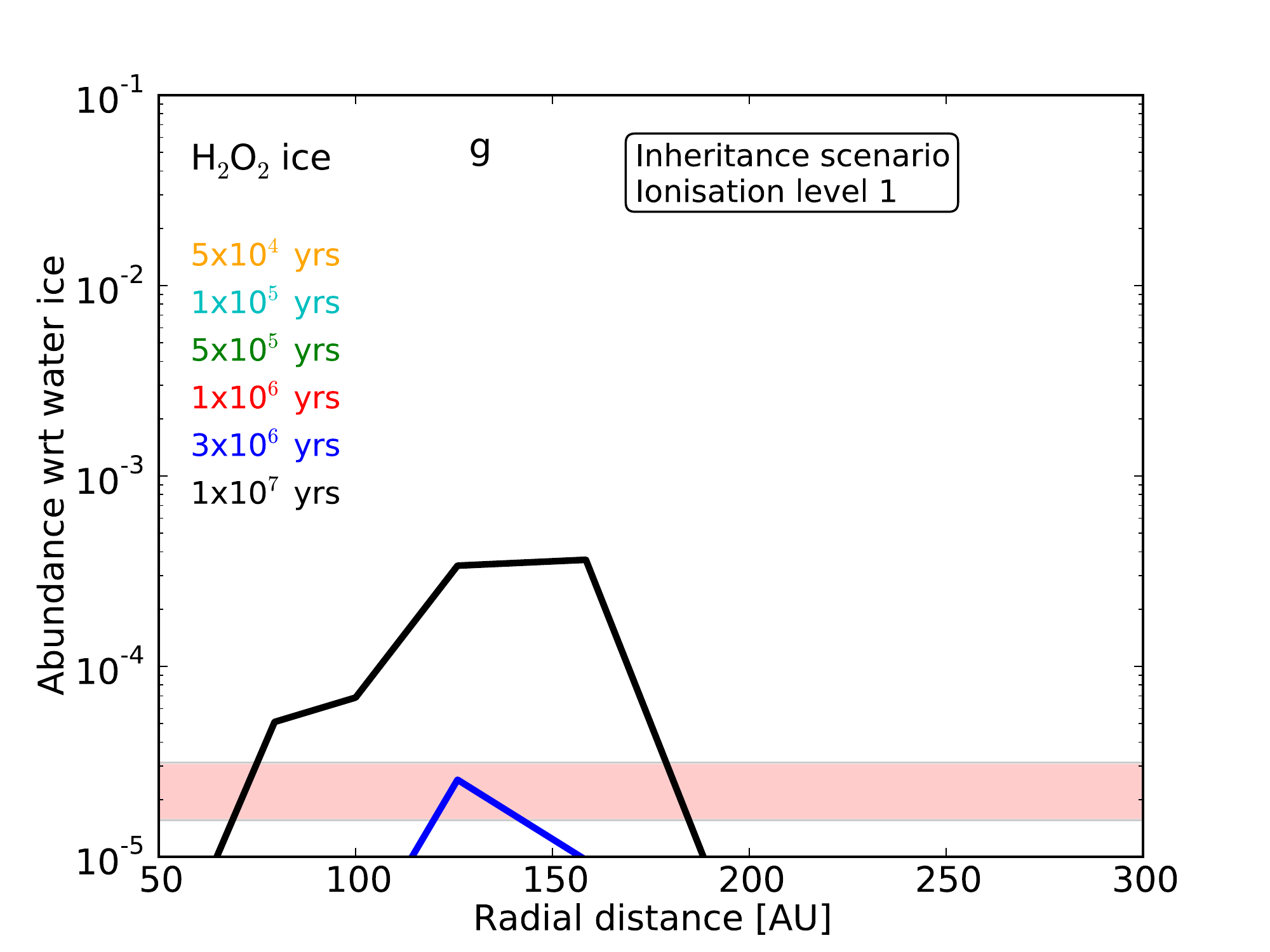}}
\subfigure{\includegraphics[width=0.33\textwidth]{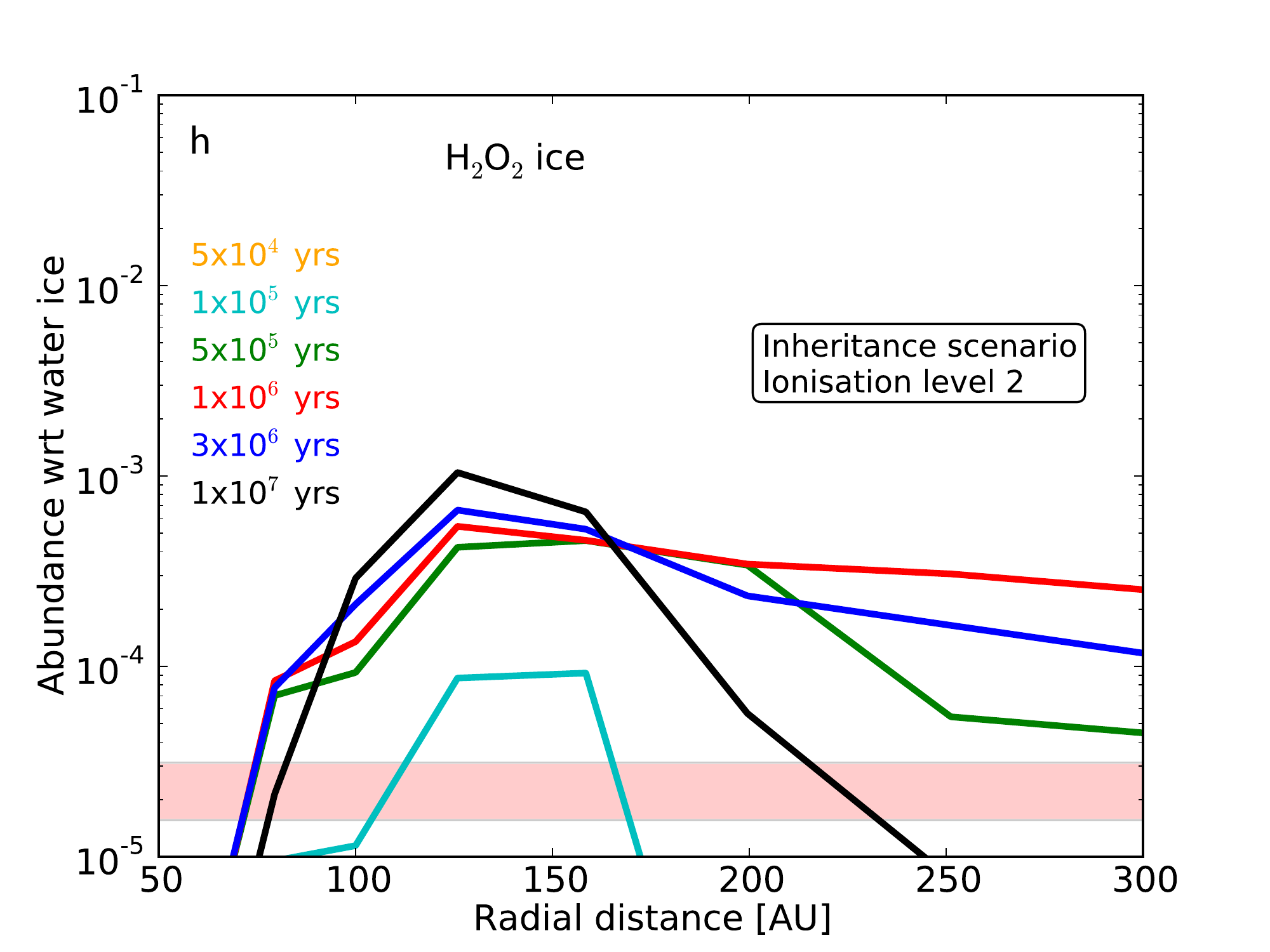}}
\subfigure{\includegraphics[width=0.33\textwidth]{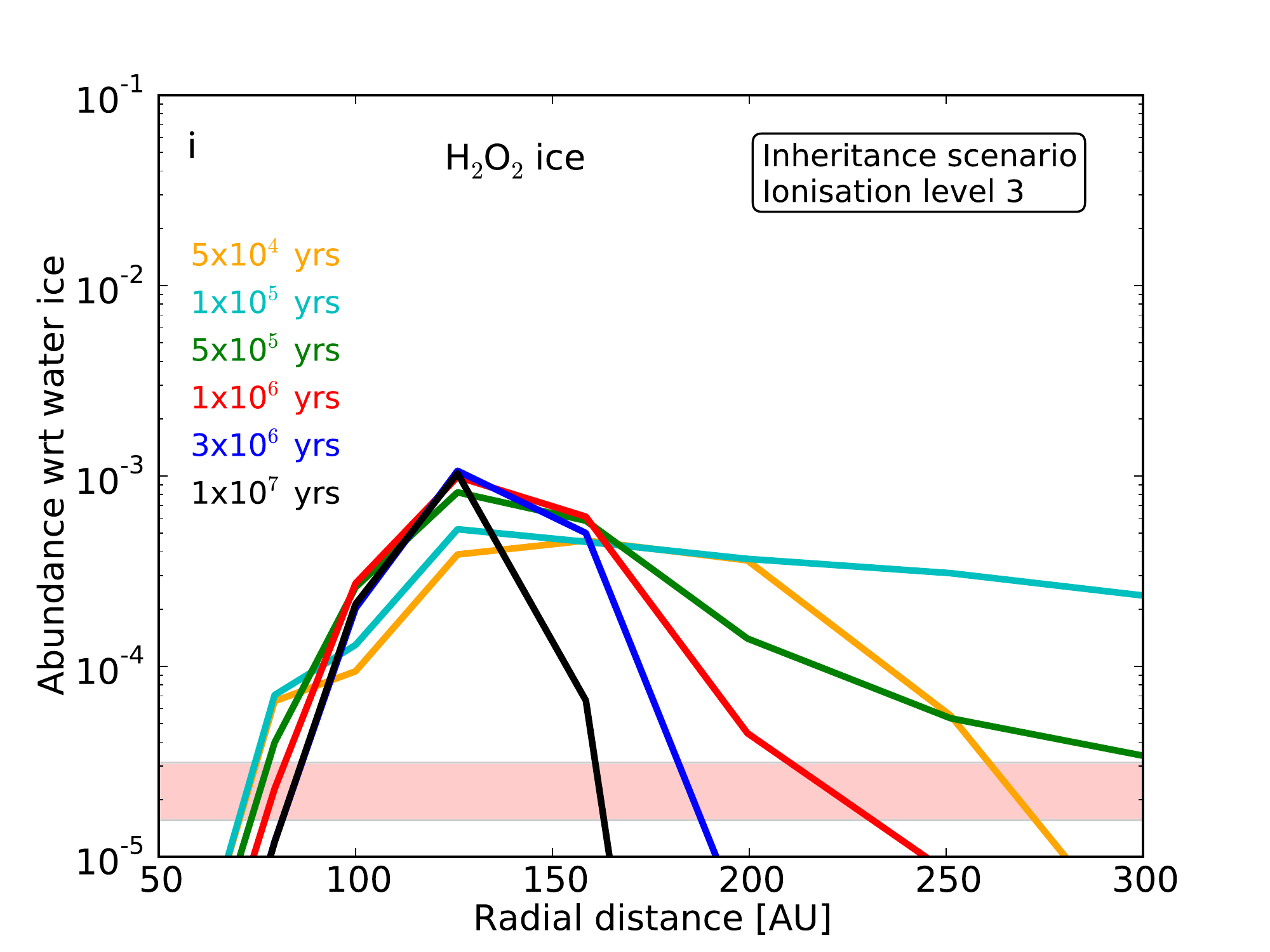}}\\
\caption{Radial abundance profiles for the inheritance scenario at multiple evolutionary time steps. Top to bottom are \ce{H2O}, \ce{O2} and \ce{H2O2}. Left to right are increasing ionisation levels. For \ce{O2} and \ce{H2O2} the limits detected in comet 67P are indicated as yellow and red shaded areas, respectively. The chemical network utilised does not include \ce{O3} chemistry.}
\label{abun_evol_inh_old_water}
\end{figure*}

Figure \ref{abun_evol_inh_old_water} features abundances as a function of midplane radius for different evolutionary times up to 10 Myr for the inheritance scenario. Left to right are increasing ionisation levels, and top to bottom are \ce{H2O}, \ce{O2} and \ce{H2O2} ices, respectively. \ce{H2O} ice is plotted with respect to H$_{nuc}$ abundance, whereas \ce{O2} and \ce{H2O2} ice are plotted with respect to \ce{H2O} ice. The is done to match the convention used in the reporting of cometary abundances. The model details are listed in each panel, along with color coding indicating evolution times. In the middle and lower panels the respective observed limits for \ce{O2} and \ce{H2O2} ices are marked with orange and red shaded regions, respectively. The orange shading for \ce{O2} ice marks the limits to the mean \ce{O2} ice abundance observed in comet 67P (3.8${\pm 0.85}\times 10^{-2}$ with respect to \ce{H2O} ice), and the red shading marks the limits for \ce{H2O2} ice ($6\pm 0.7 \times 10^{-4}$ wrt \ce{O2} ice, thus $2.34\pm 0.8\times 10^{-5}$ wrt \ce{H2O} ice \citep{bieler15}). Note that in 67P variations in \ce{O2} over \ce{H2O} ice were seen spanning 1-10\%.

An evolution time of 10 Myr is chosen because this is assumed to be the maximum lifetime of a gaseous protoplanetary disk, and based on the results from \citet{eistrup2018}, the outer icy disk midplane should have reached steady state by this time. Since the focus in this work is the outer, icy PSN disk midplane, the radial range starts at 50 AU, which lies inside the \ce{O2} iceline ($\sim120$ AU at $\sim$29 K).

For ionisation Level 1 in Fig. \ref{abun_evol_inh_old_water}a, \ce{H2O} ice stays abundant throughout the evolution. For higher ionisations, the \ce{H2O} ice abundance decreases with time inside, and increases with time outside $\sim120$ AU. This radius marks the \ce{O2} iceline. Panels d to f show the evolution of \ce{O2} ice for the different ionisation levels. The profiles for \ce{O2} ice for ionisation Levels 1 and 2 peak at similar abundance levels of 2-3$\times 10^{-3}$ with respect to \ce{H2O} ice, which is an order of magnitude lower than the observed abundance. It is seen that for higher ionisation, this abundance is reached faster, with ionisation Level 1 producing of order $10^{-3}$ with respect to \ce{H2O} ice only by 10 Myr. 

It is shown that for \ce{H2O2} ice abundance profiles featured in panels g to i they evolve in a similar way to \ce{O2} (note the same $y$-axis range as for \ce{O2} ice), and reach peak abundances at $\sim$120 AU by 10 Myr, that are a factor of 2-3 lower than those for \ce{O2} ice. This abundance level is 30-100 times higher than that observed for \ce{H2O2} ice in comet 67P. For all ionisation levels there are narrow radial regions where the \ce{H2O2} ice abundance matches the observed levels over a time range defined by the ionisation level (e.g. between 70-80 AU ionisation Level 3 at all times). However, at these points the \ce{O2} ice abundance does not match that observed. 

For the inheritance scenario it is thus seen that \ce{O2} ice is underproduced by at least an order of magnitude, and \ce{H2O2} is overproduced by up to two orders of magnitude.

\subsubsection{Reset scenario}

\begin{figure*}[h]
\subfigure{\includegraphics[width=0.33\textwidth]{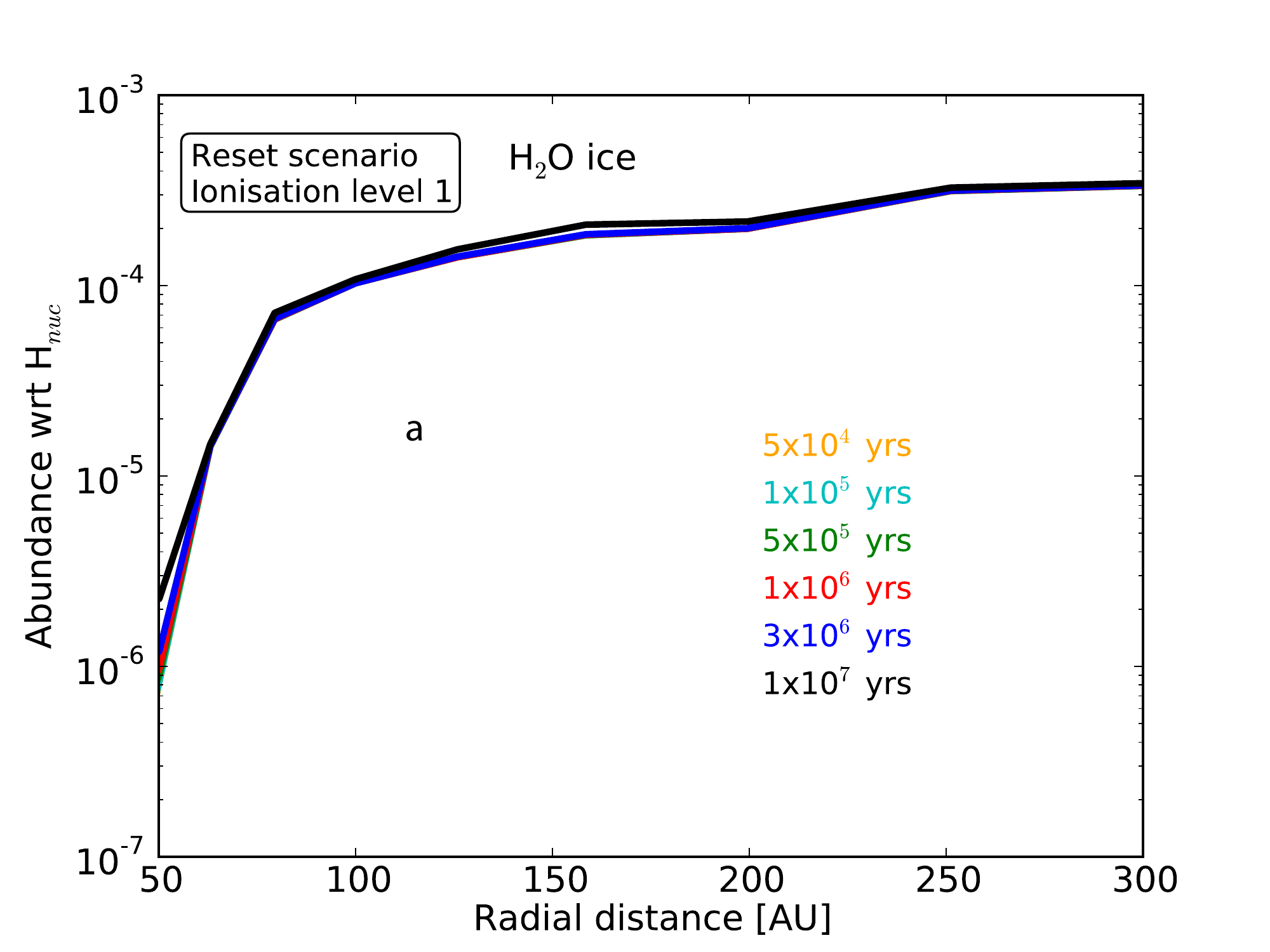}}
\subfigure{\includegraphics[width=0.33\textwidth]{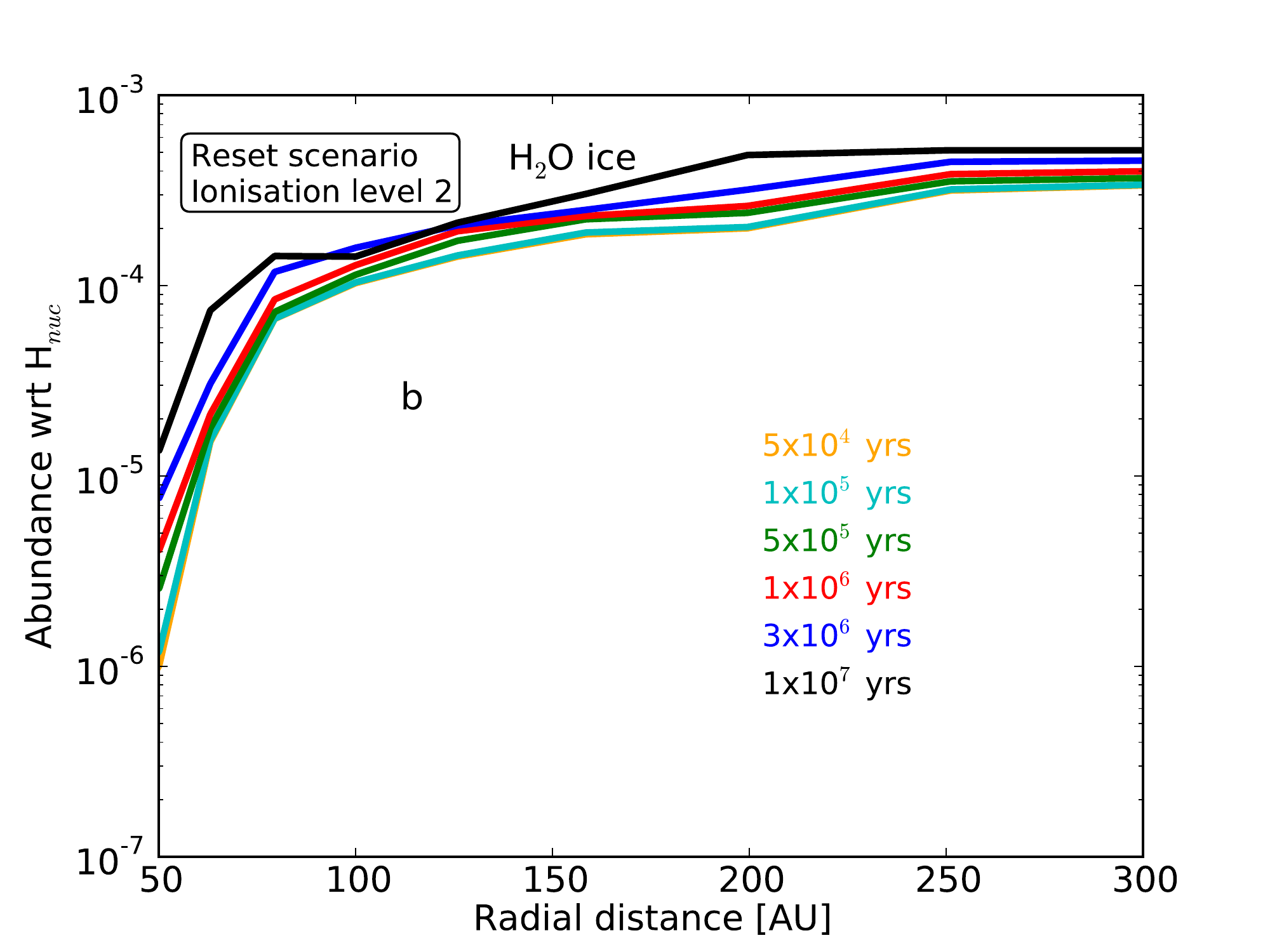}}
\subfigure{\includegraphics[width=0.33\textwidth]{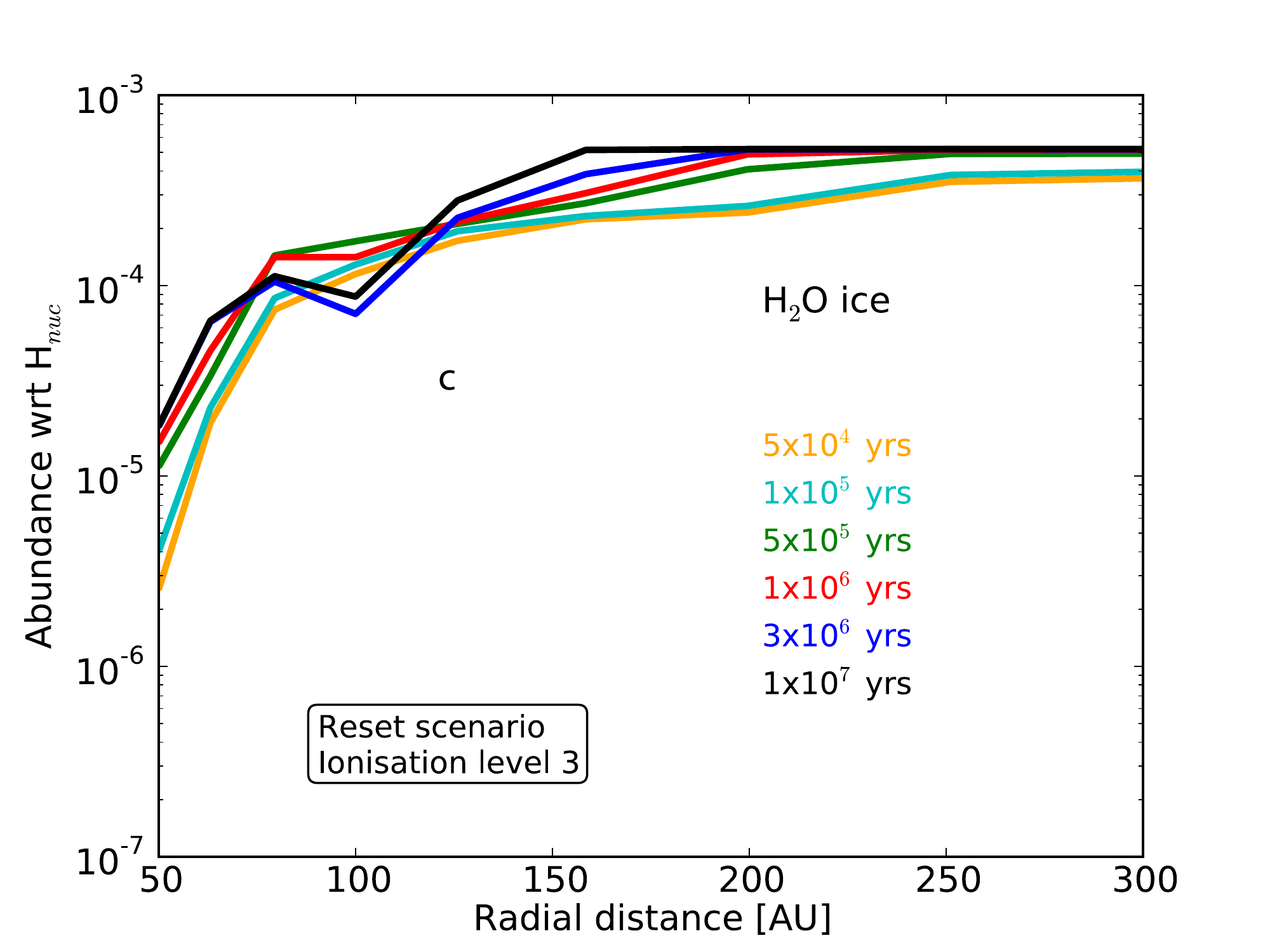}}\\
\subfigure{\includegraphics[width=0.33\textwidth]{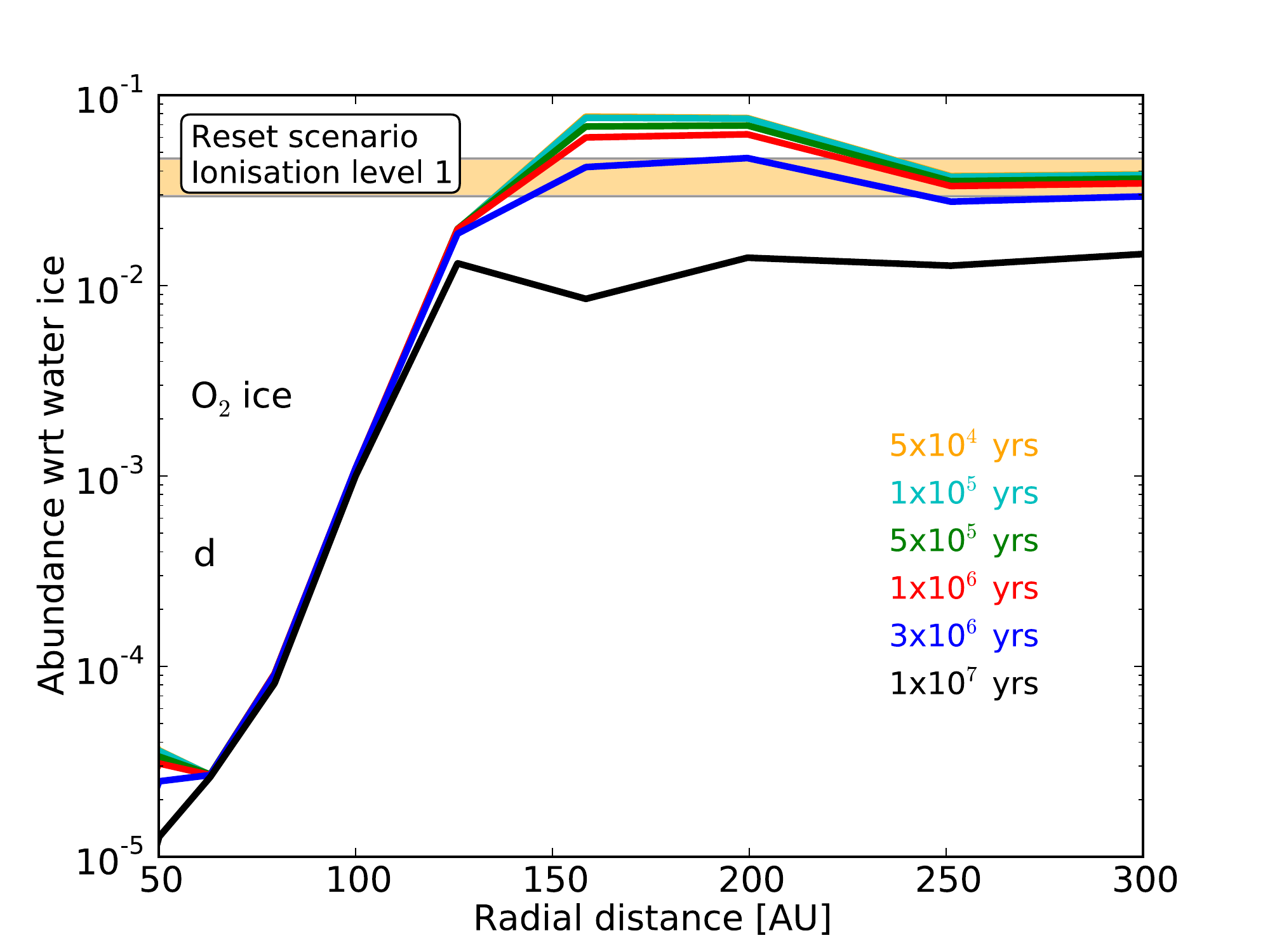}}
\subfigure{\includegraphics[width=0.33\textwidth]{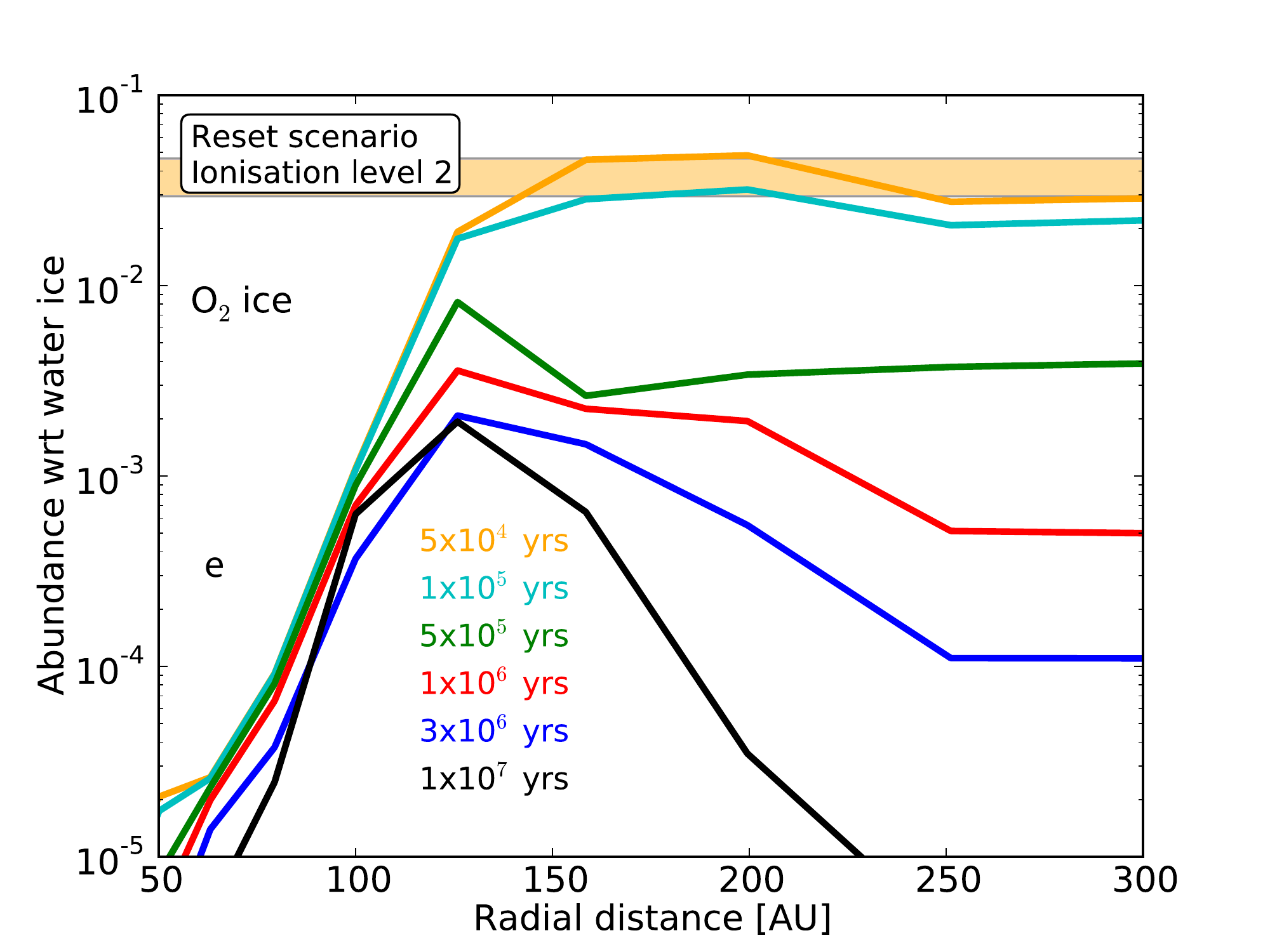}}
\subfigure{\includegraphics[width=0.33\textwidth]{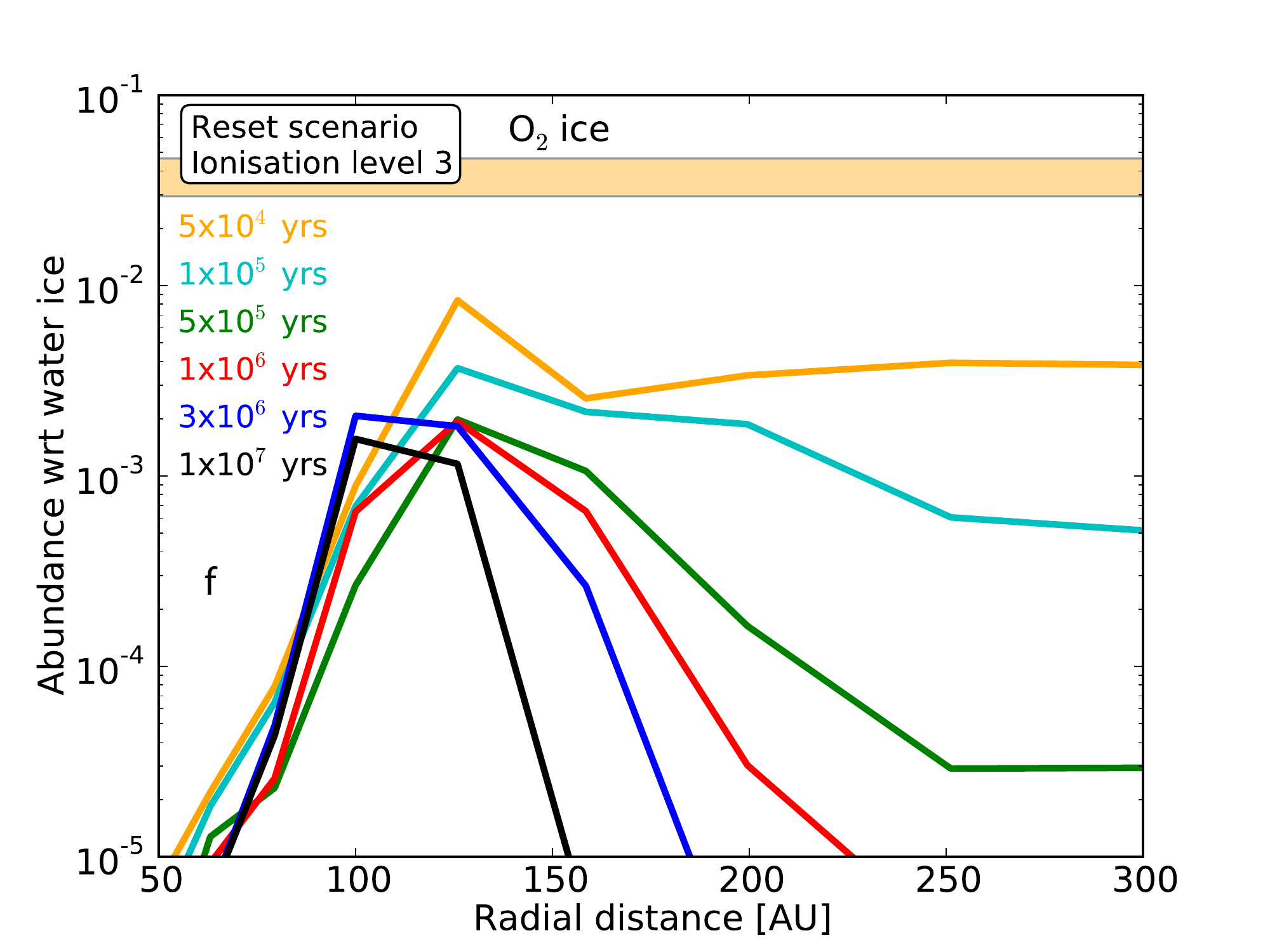}}\\
\subfigure{\includegraphics[width=0.33\textwidth]{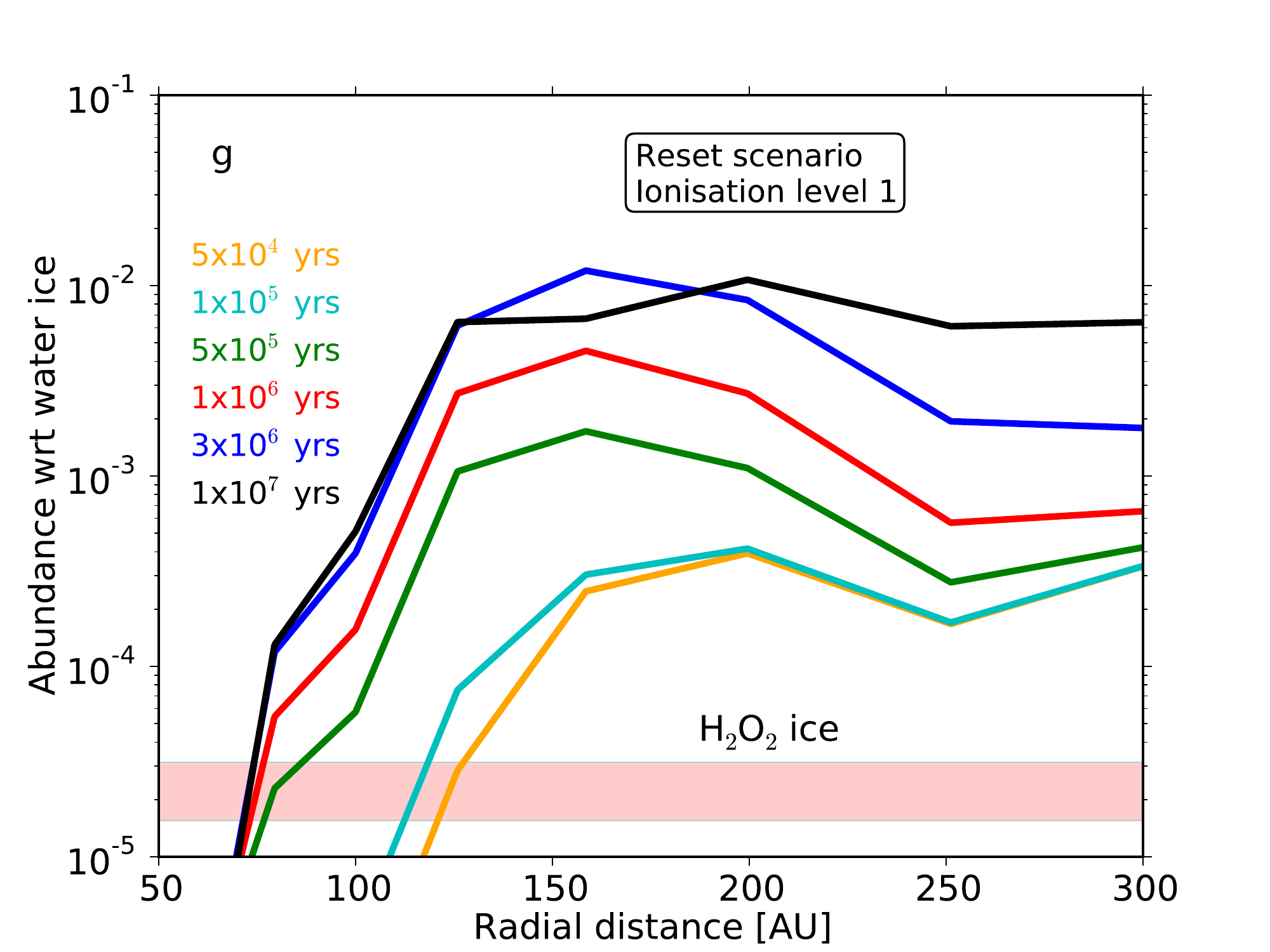}}
\subfigure{\includegraphics[width=0.33\textwidth]{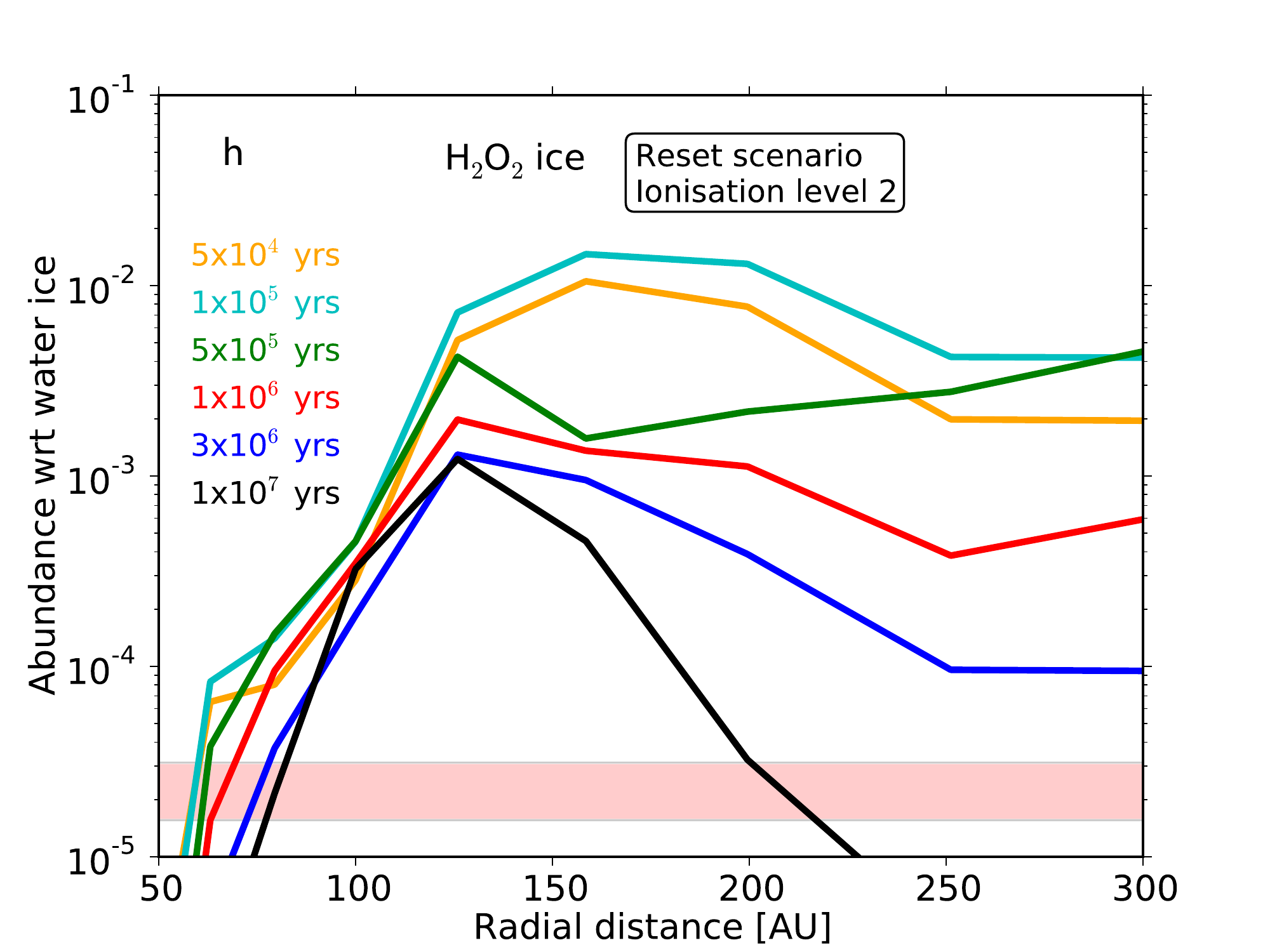}}
\subfigure{\includegraphics[width=0.33\textwidth]{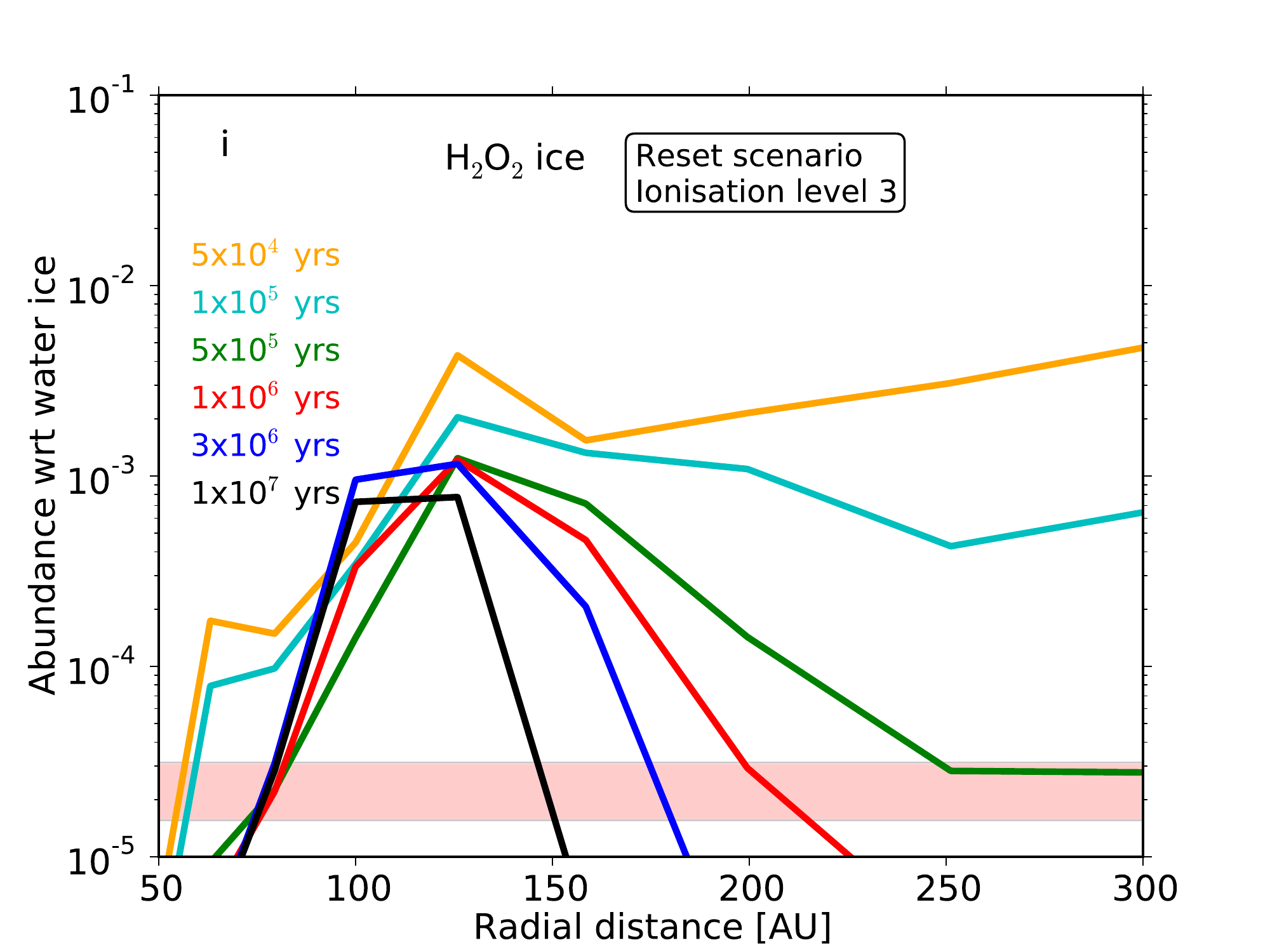}}\\
\caption{Radial abundance profiles for the reset scenario at multiple evolutionary time steps. Top to bottom are \ce{H2O}, \ce{O2} and \ce{H2O2} ices. Left to right are increasing ionisation levels. For \ce{O2} and \ce{H2O2} the limits detected in comet 67P are indicated as yellow and red shaded areas, respectively. The chemical network utilised does not include \ce{O3} chemistry.}
\label{abun_evol_res_old_water}
\end{figure*}

The abundance evolution profiles for the reset scenario are shown in Fig. \ref{abun_evol_res_old_water}. For \ce{O2} ice in panels d to f it is seen that a high abundance ($10^{-2}-10^{-1}$ with respect to \ce{H2O} ice) is reached for both ionisation Levels 1 and 2, with each reproducing the observed abundance for a large radial range covering 150-300 AU. For ionisation Level 1 the \ce{O2} ice abundance matching the observation is maintained up to 2 Myr. Until 1 Myr between 150-220 AU, the modelled abundance lies above the observations (>5$\times10^{-2}$ with respect to \ce{H2O} ice). For Level 2 the abundance level matching the observed values happens between $5\times10^{4}-10^{5}$ yrs in the range 150-200 AU, but subsequently drops to reach 2-3$\times 10^{-3}$ with respect to \ce{H2O} ice by 10 Myr. This was also the abundance level reached for \ce{O2} ice by 10 Myr for the inheritance scenario. For ionisation Level 3, the abundance is below the observed level already by 5$\times 10^{4}$ yrs of evolution, and by 10 Myr it also has reached the level of 2-3$\times 10^{-3}$ with respect to \ce{H2O} ice. For ionisation Level 1, the abundance remains higher than 2-3$\times 10^{-3}$ with respect to \ce{H2O} ice throughout the evolution, indicating that, for both inheritance and reset, at this ionisation level, 10 Myr of chemical evolution is not enough to bring the \ce{O2} ice abundance to what appears to be its steady state level according to the results from the inheritance scenario in Fig. \ref{abun_evol_inh_old_water} \citep[for a description of disk midplane chemical steady state, see][]{eistrup2018}.

\ce{H2O2} ice is seen in Fig. \ref{abun_evol_res_old_water}, panels g to i, to evolve significantly over time. 
For all ionisation levels, \ce{H2O2} ice is produced over time with the timescale dependent on ionisation level. The fastest production is seen for Level 3, where the peak abundance of >$10^{-2}$ with respect to \ce{H2O} ice is reached already by $10^{5}$ yrs. On the other hand it takes 0.5 Myr and 3 Myr to reach a similar abundance for Levels 2 and 1, respectively.  The \ce{H2O2} ice is subsequently destroyed after reaching its peak. For ionisation Levels 2 and 3 a steady state abundance of order 10$^{-3}$ with respect to \ce{H2O} ice is reached at radii $\sim$100-120 AU. Beyond this radius it has disappeared by 10 Myr.

\subsubsection{Water and atomic oxygen starting conditions}
\label{oxygen_start}

\begin{figure}
\subfigure{\includegraphics[width=0.5\textwidth]{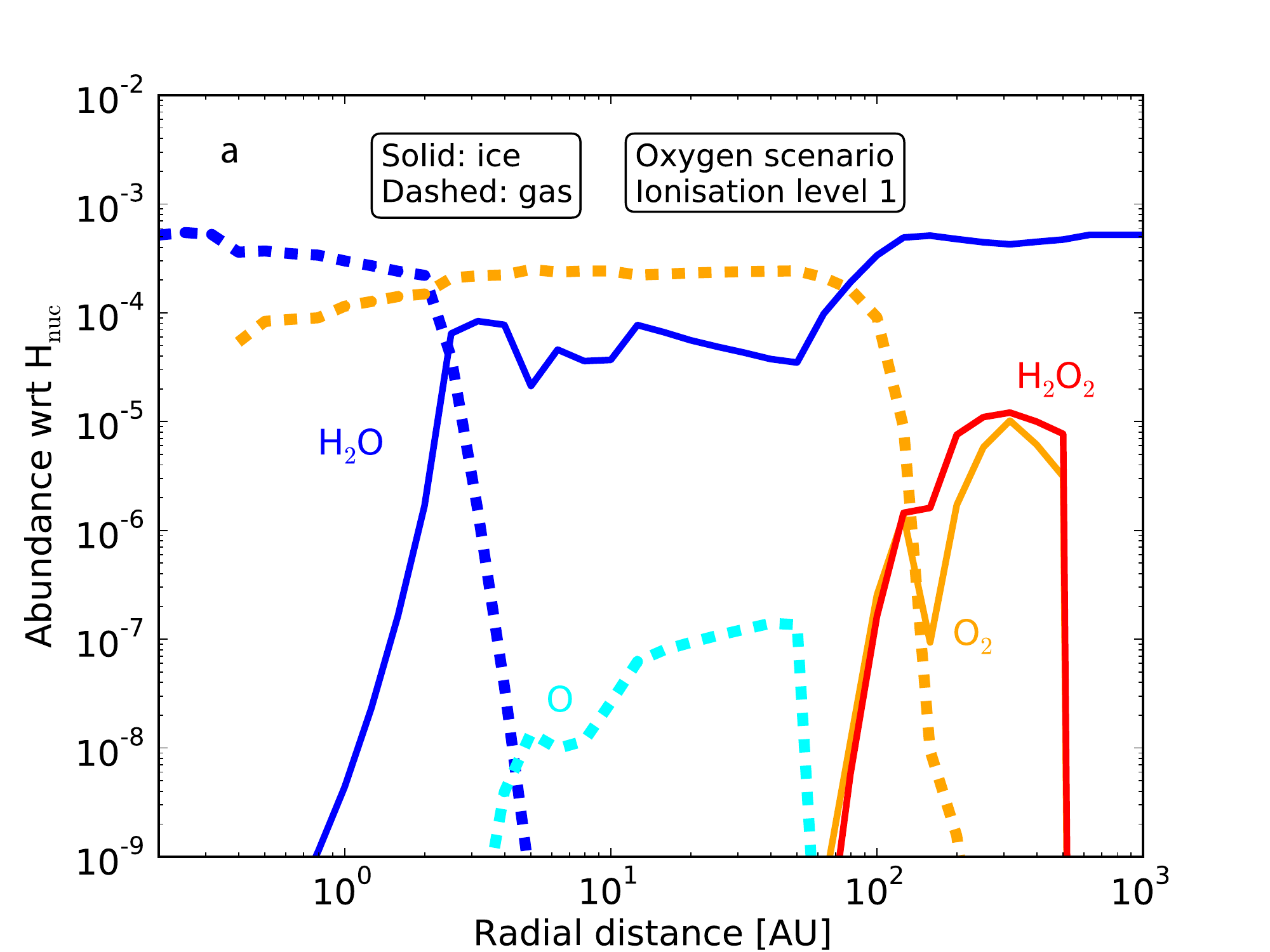}}\\
\subfigure{\includegraphics[width=0.5\textwidth]{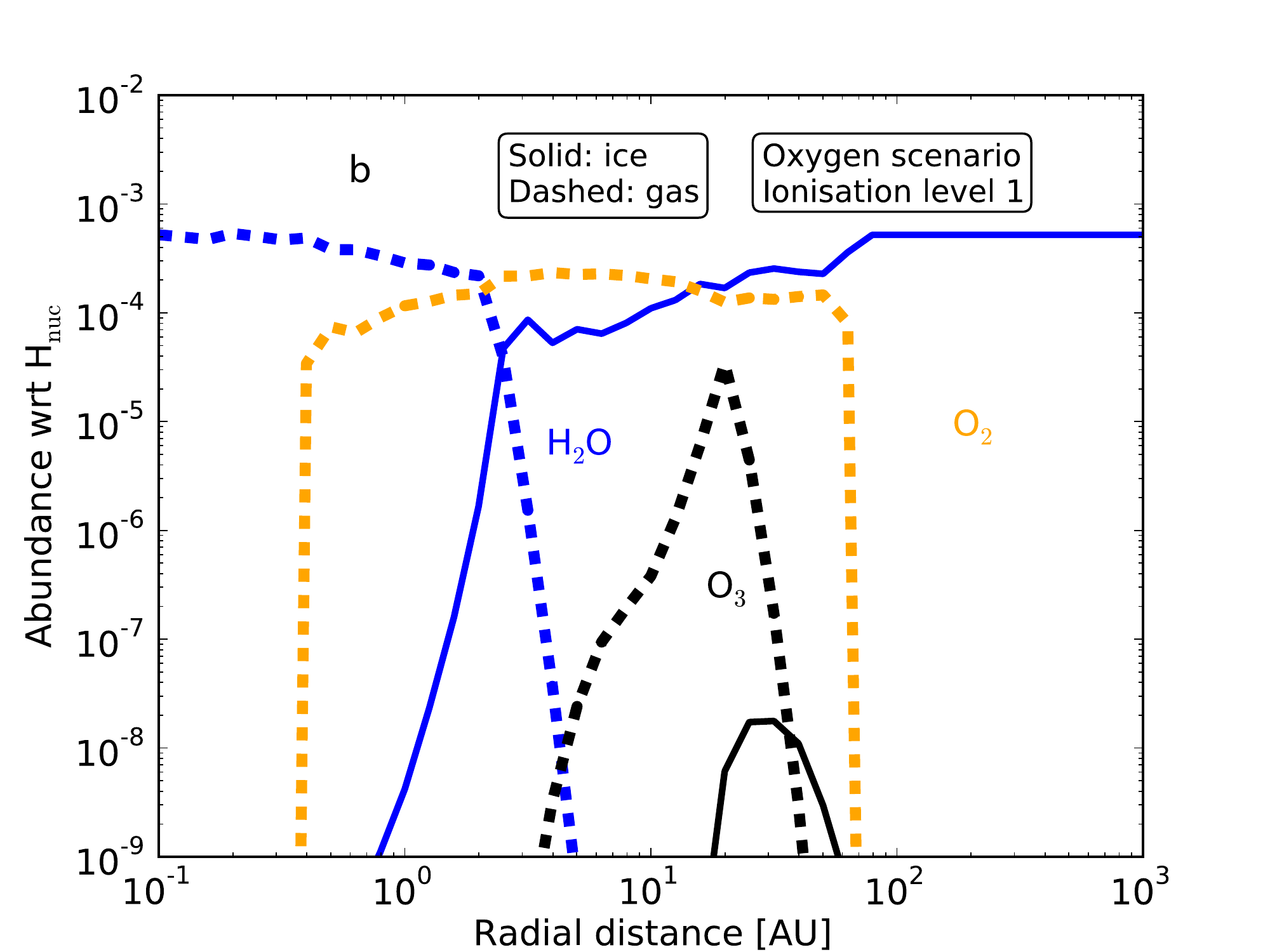}}
\caption{Abundances by 10 Myr evolution in the oxygen-only scenario, for ionisation Level 1. a) without \ce{O3} chemistry included. b) with \ce{O3} chemistry included.}
\label{oxygen_10}
\end{figure}

Models were also run starting with simpler sets of initial abundances, either with all oxygen already present in \ce{H2O} (``Water scenario'') or all free (``Atomic oxygen scenario''), in addition to H and He. This was to test the formation of \ce{O2} ice solely from processing of \ce{H2O} ice, and the formation via assembly from free oxygen atoms, in the absence of other chemistry. It was found that the water scenario did not, at any point in time or radius, reproduce \ce{O2} ice to the abundance seen in the comets. The maximum abundance was <$10^{-7}$ with respect to \ce{H2O} ice, regardless of ionisation level. This is shown in Fig. \ref{abun_10}, where the gas and ice abundances of O, \ce{O2}, \ce{H2O} and \ce{H2O2} are plotted as a function of radius by 10 Myr of evolution.

The atomic oxygen scenario, on the other hand, can reproduce the observed abundance of \ce{O2} ice by 10 Myr, as is seen in panel a of Fig. \ref{oxygen_10}. This panel of Fig. \ref{oxygen_10} shows the abundances of \ce{H2O}, O, \ce{O3} and \ce{H2O2} with respect to H$_{\rm{nuc}}$ as a function of radius, by 10 Myr of evolution. The radial range (80-300 AU) for this reproduction is similar to the range in the reset scenario, but in the atomic oxygen scenario \ce{H2O2} ice is found to be more efficiently formed than \ce{O2} ice. This scenario will be revisited later.

\subsection{Including ozone ice chemistry}

\begin{figure*}[h]
\subfigure{\includegraphics[width=0.33\textwidth]{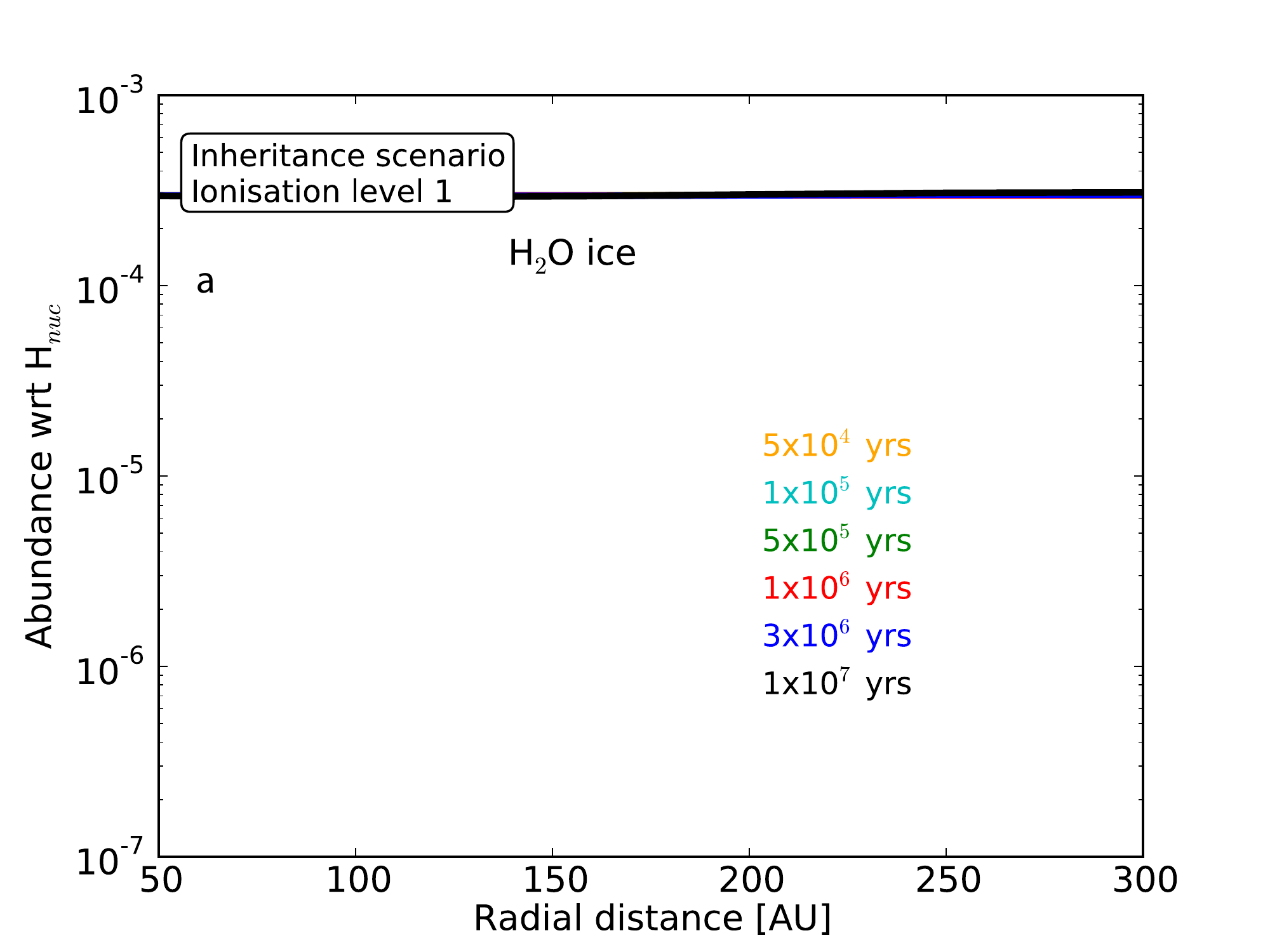}}
\subfigure{\includegraphics[width=0.33\textwidth]{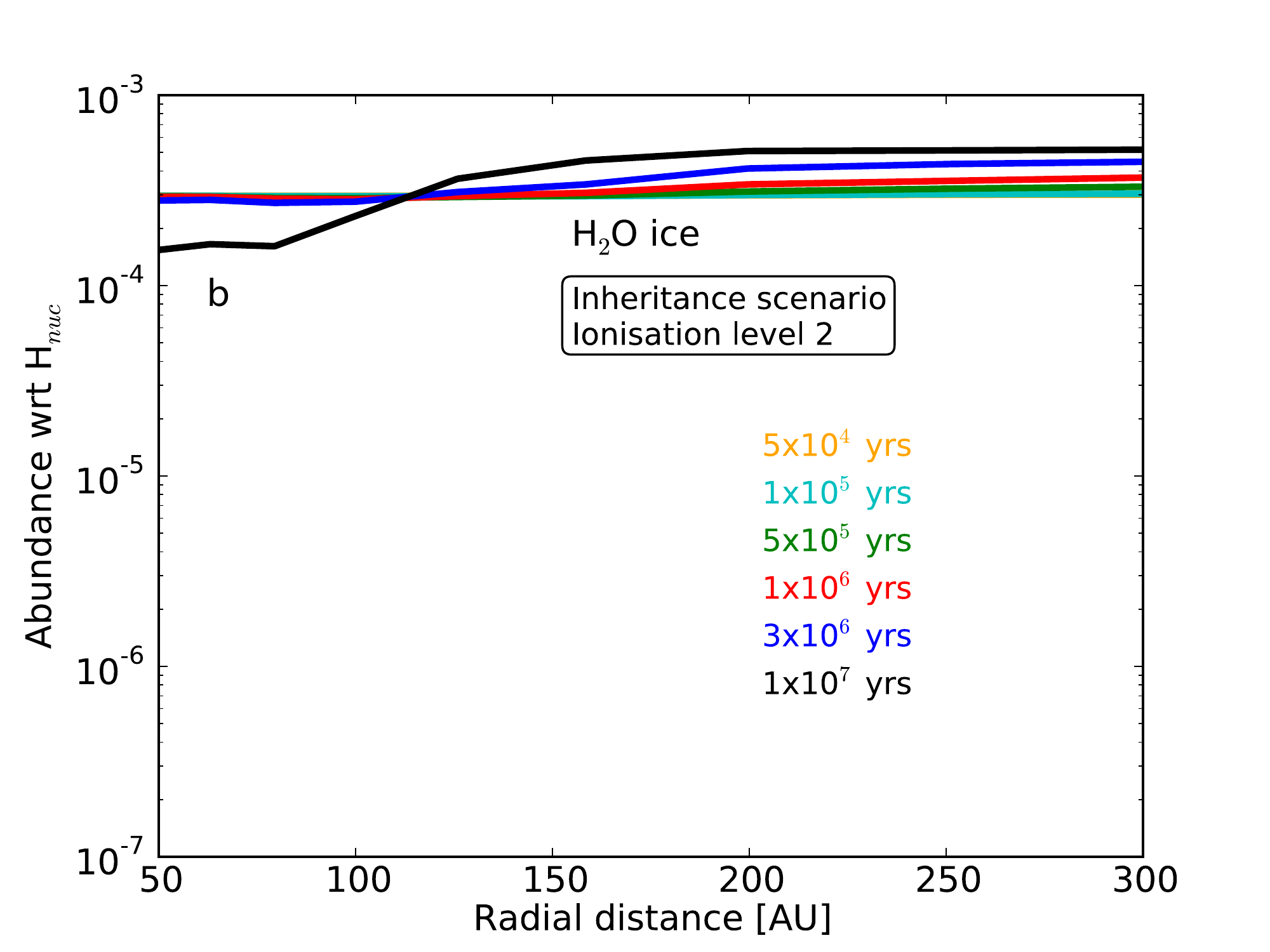}}
\subfigure{\includegraphics[width=0.33\textwidth]{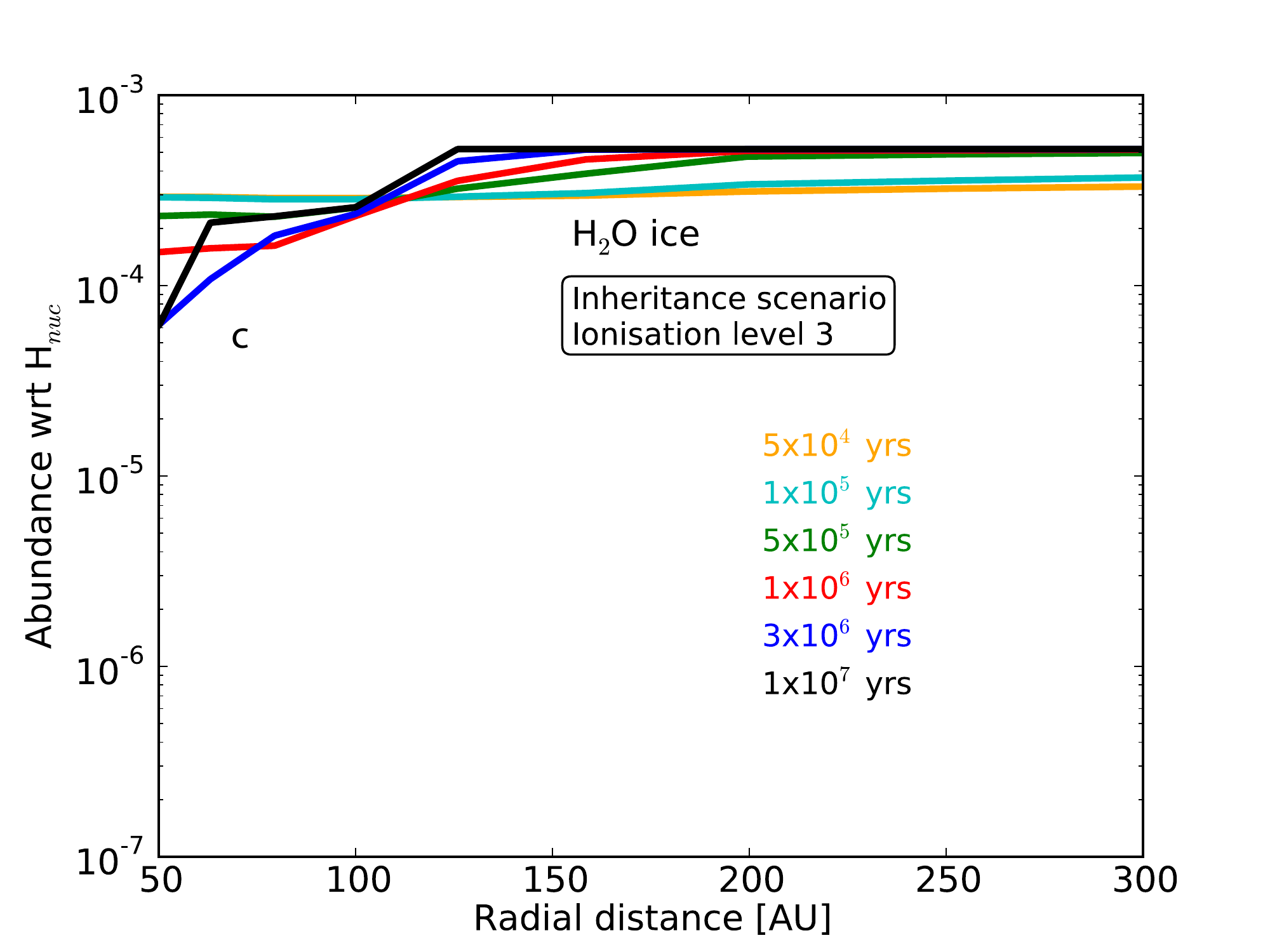}}\\
\subfigure{\includegraphics[width=0.33\textwidth]{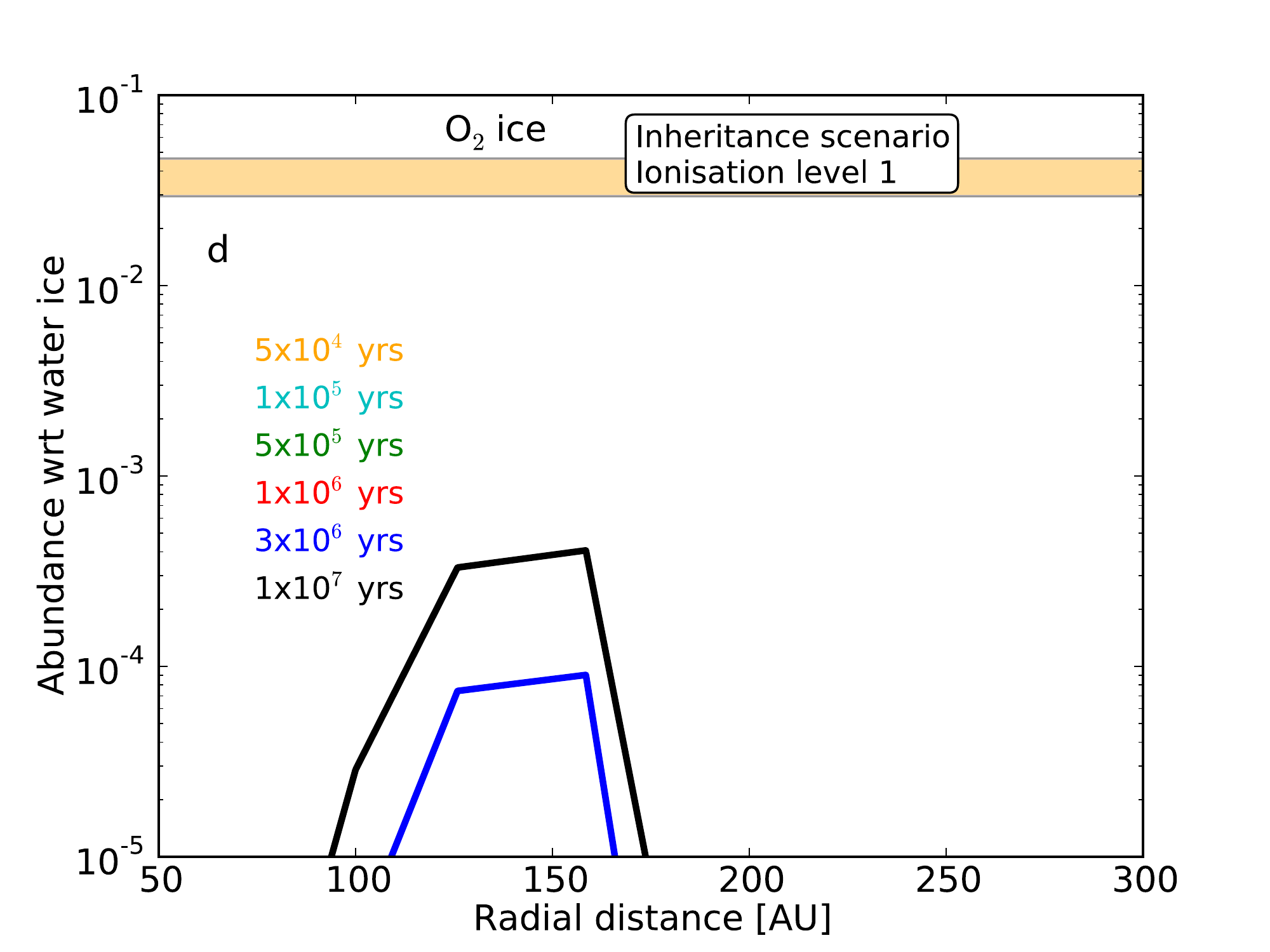}}
\subfigure{\includegraphics[width=0.33\textwidth]{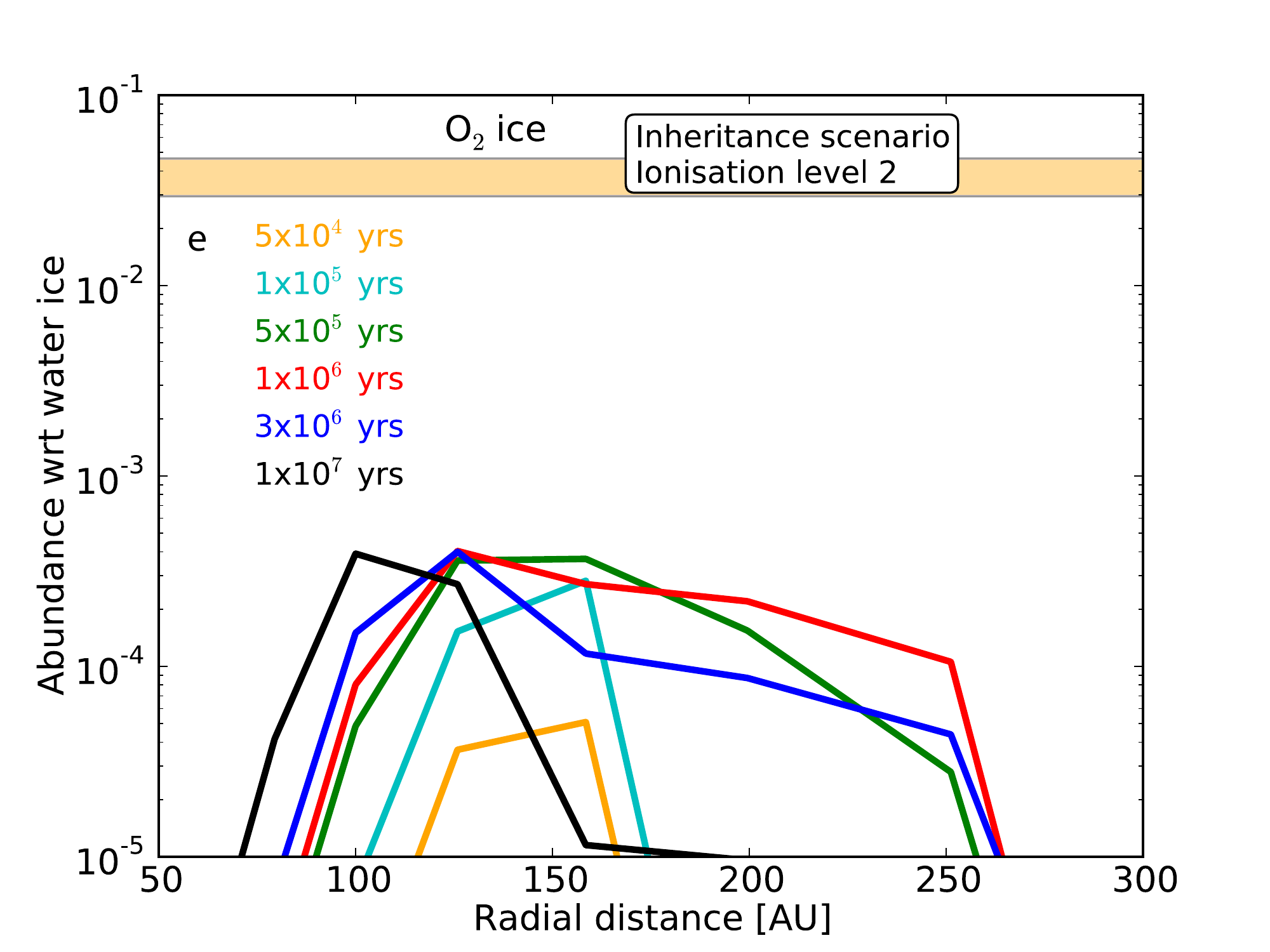}}
\subfigure{\includegraphics[width=0.33\textwidth]{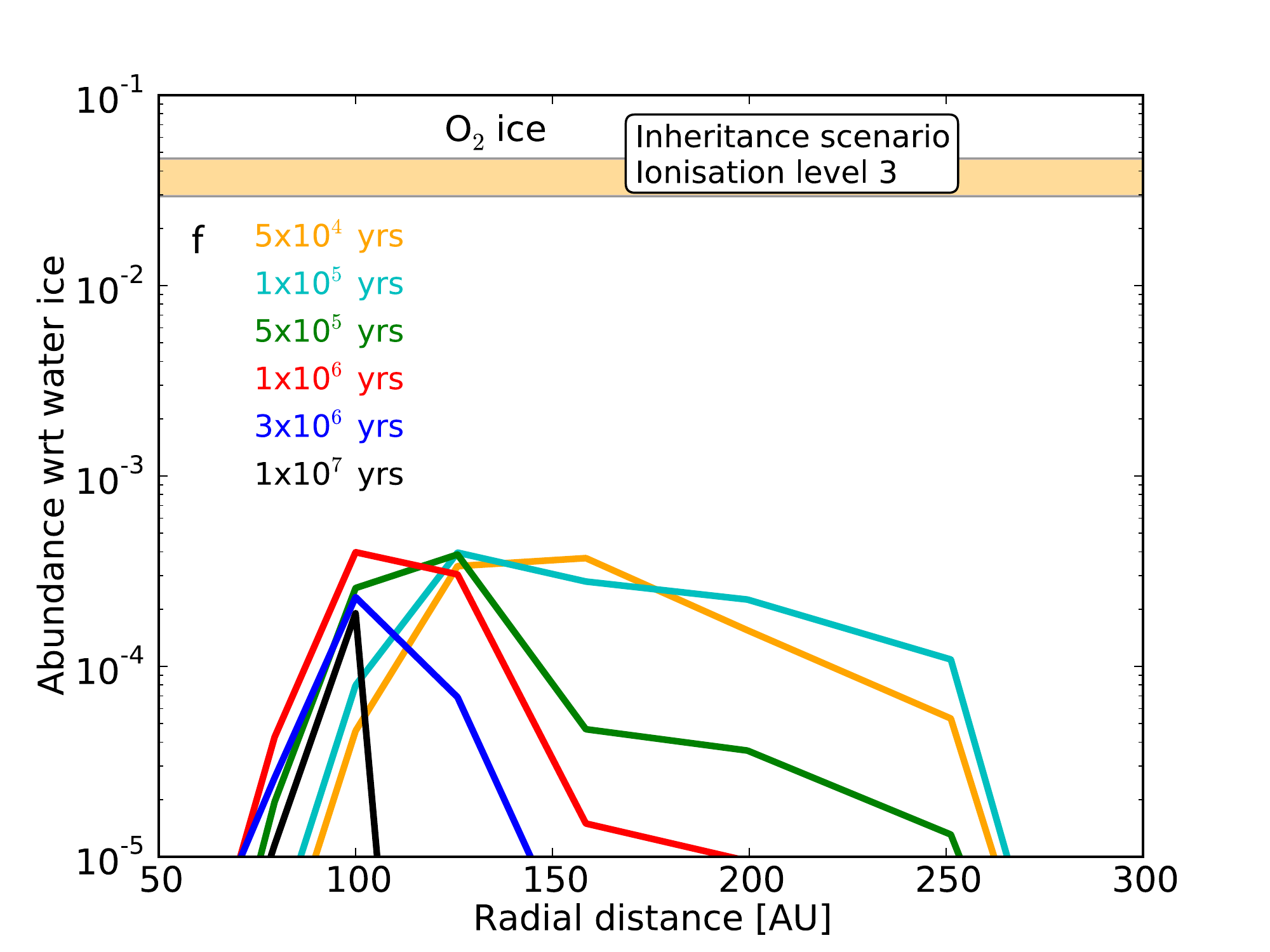}}\\
\subfigure{\includegraphics[width=0.33\textwidth]{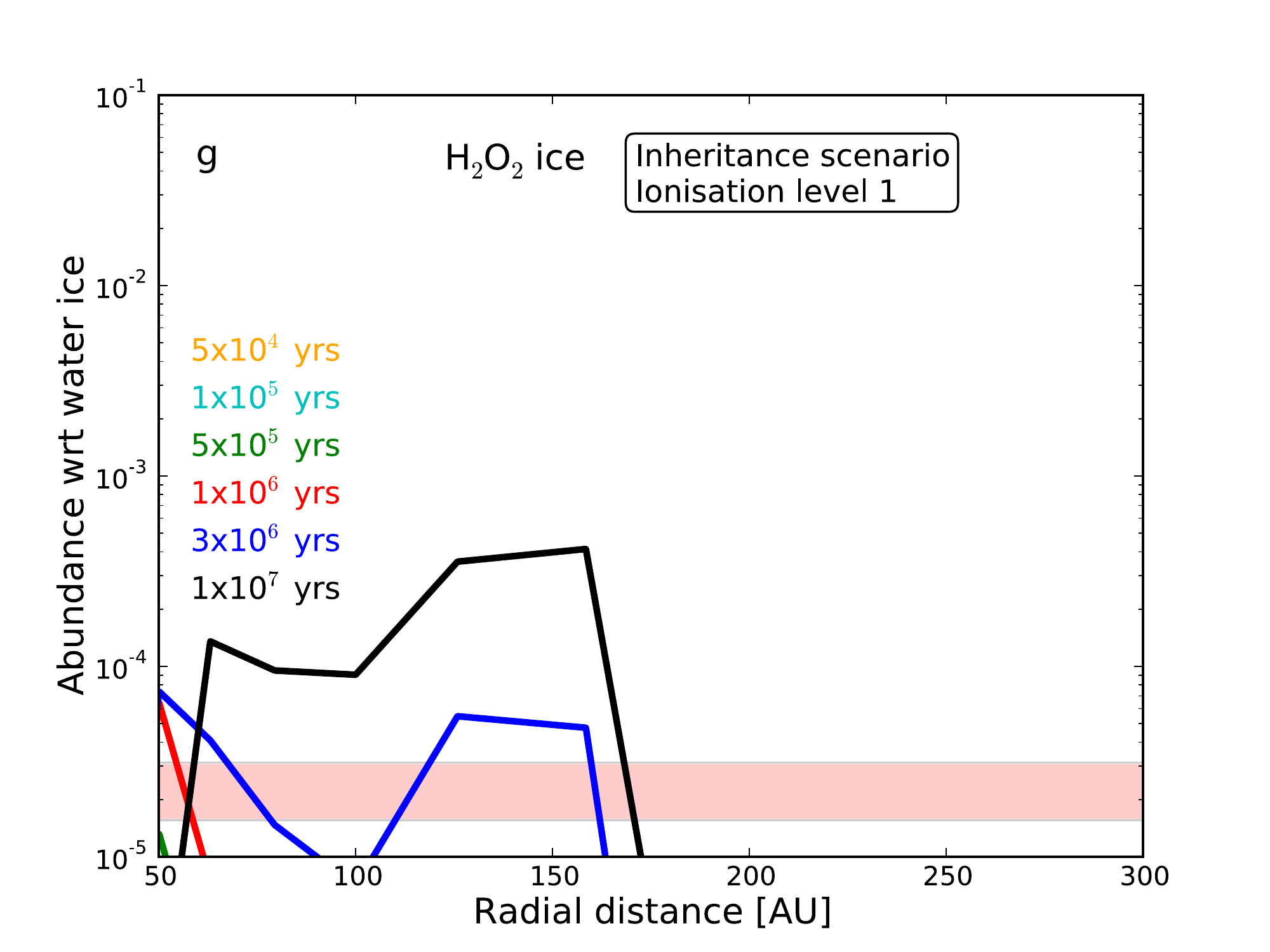}}
\subfigure{\includegraphics[width=0.33\textwidth]{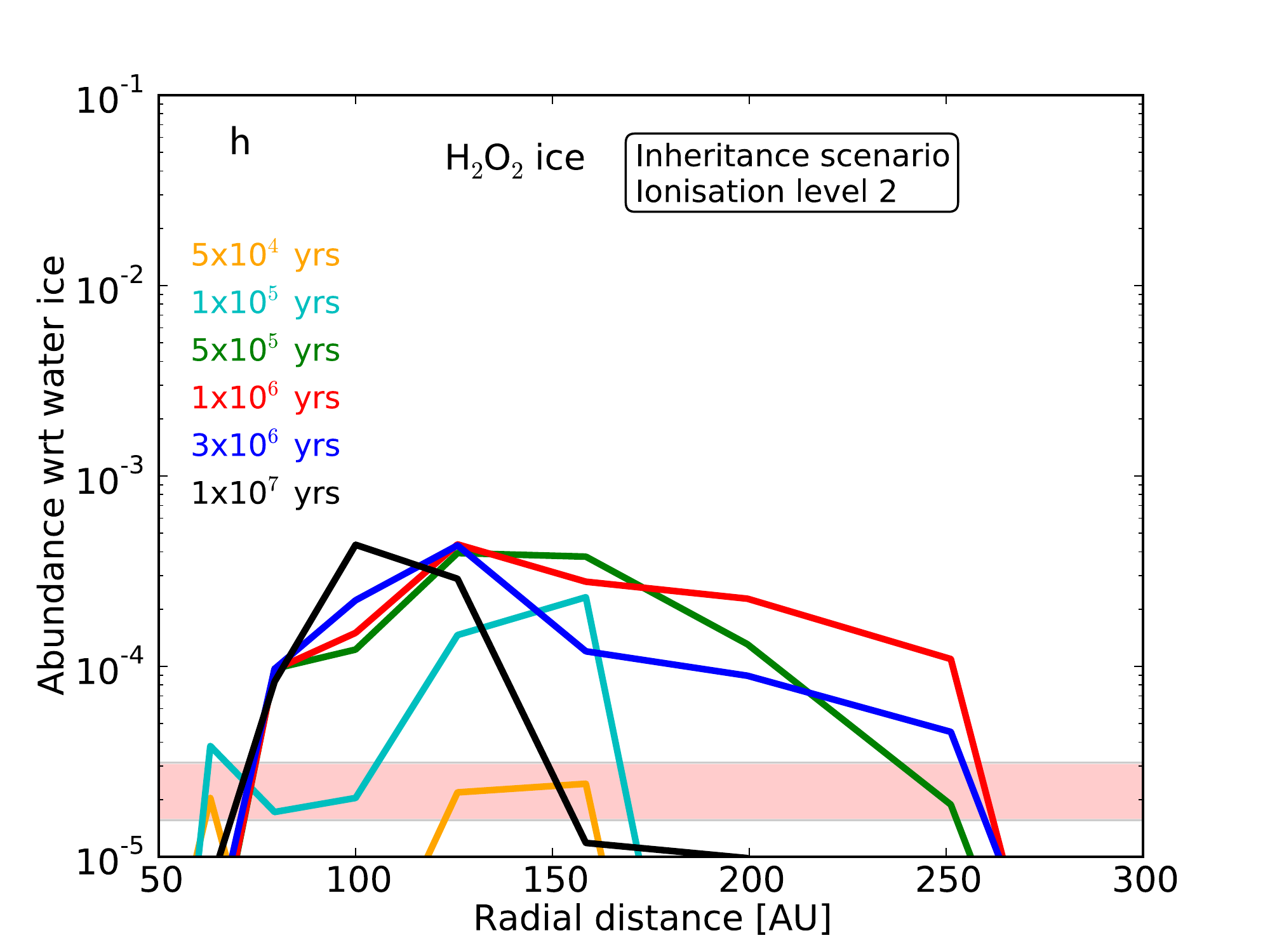}}
\subfigure{\includegraphics[width=0.33\textwidth]{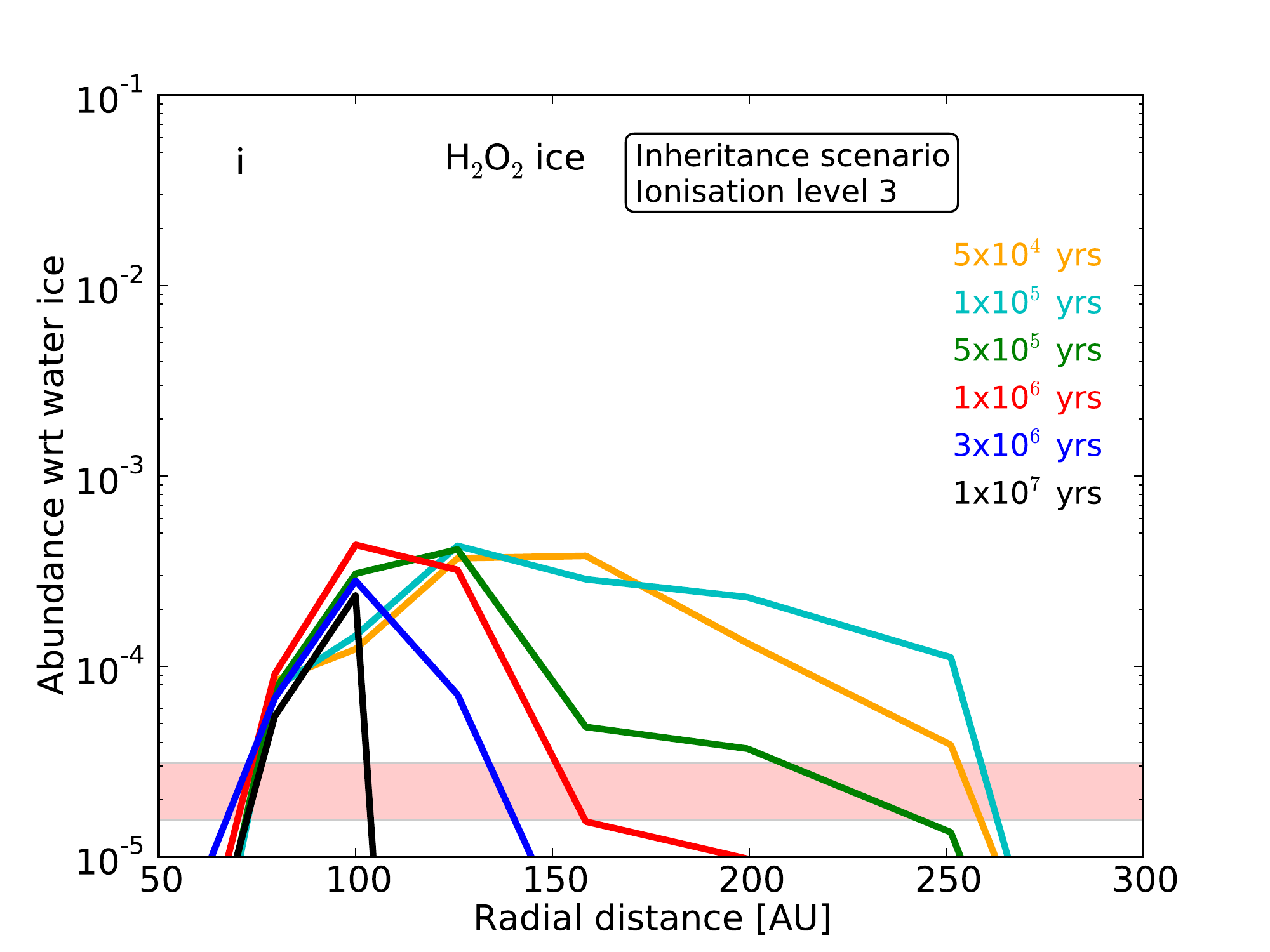}}\\
\subfigure{\includegraphics[width=0.33\textwidth]{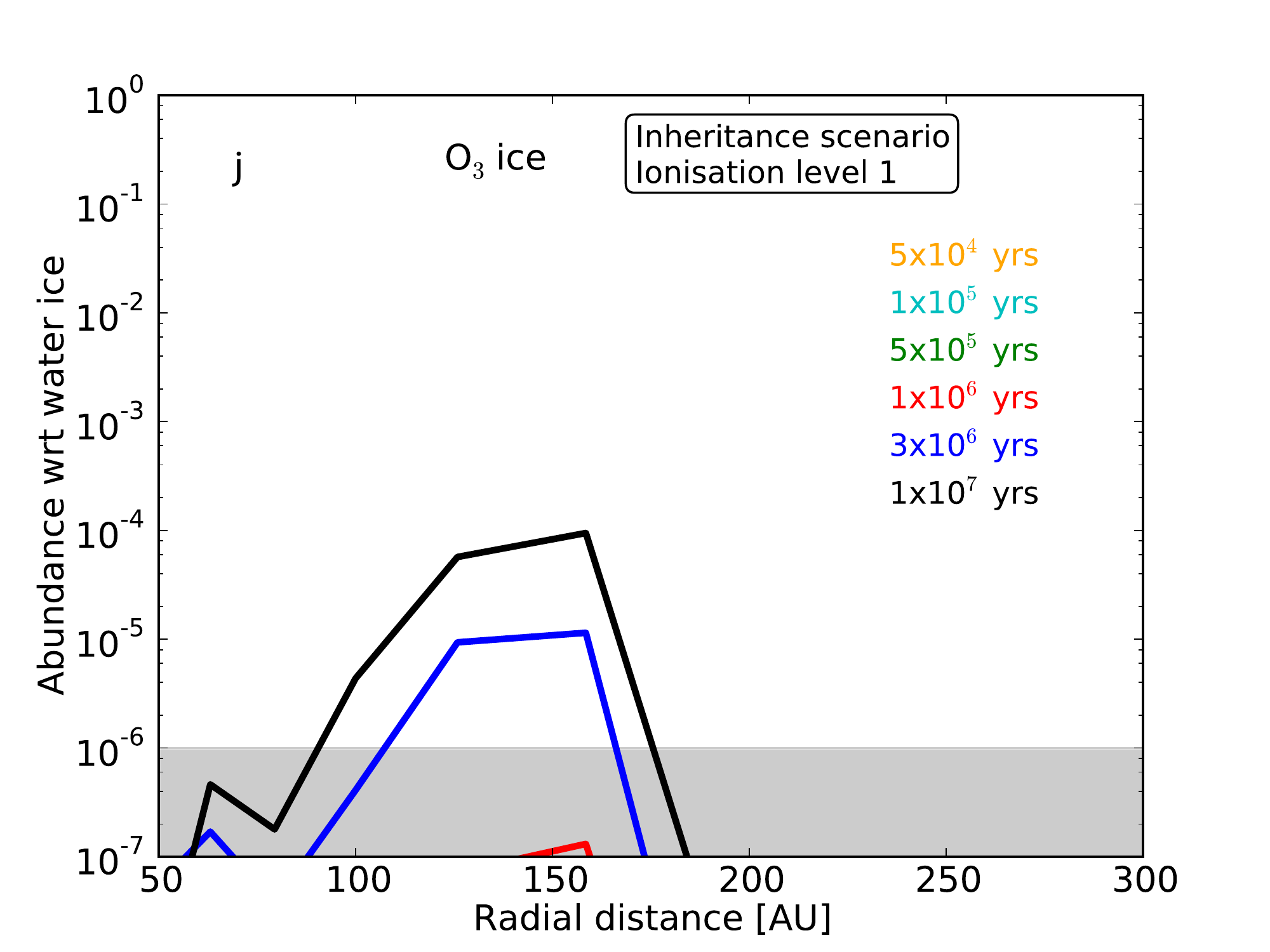}}
\subfigure{\includegraphics[width=0.33\textwidth]{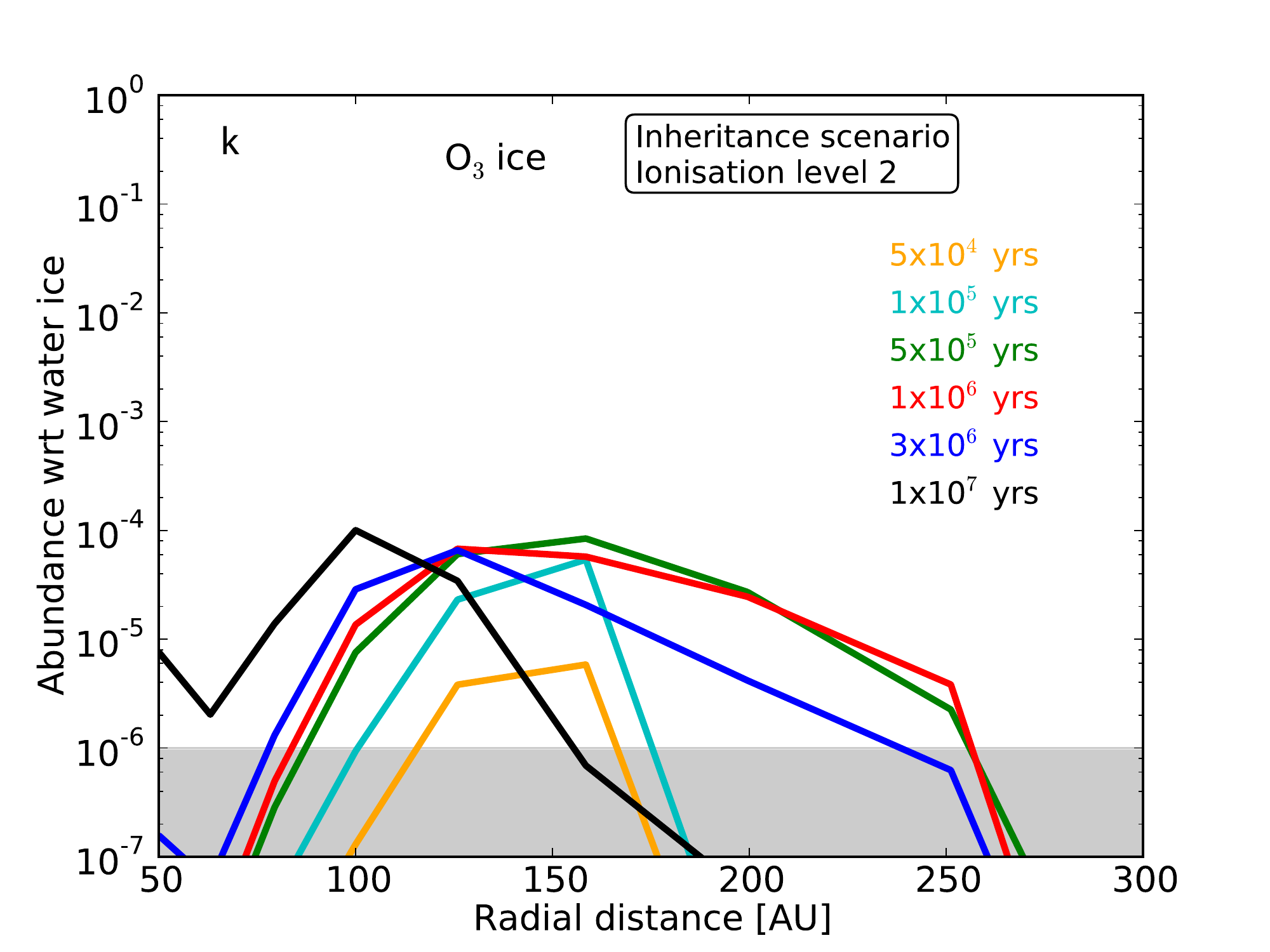}}
\subfigure{\includegraphics[width=0.33\textwidth]{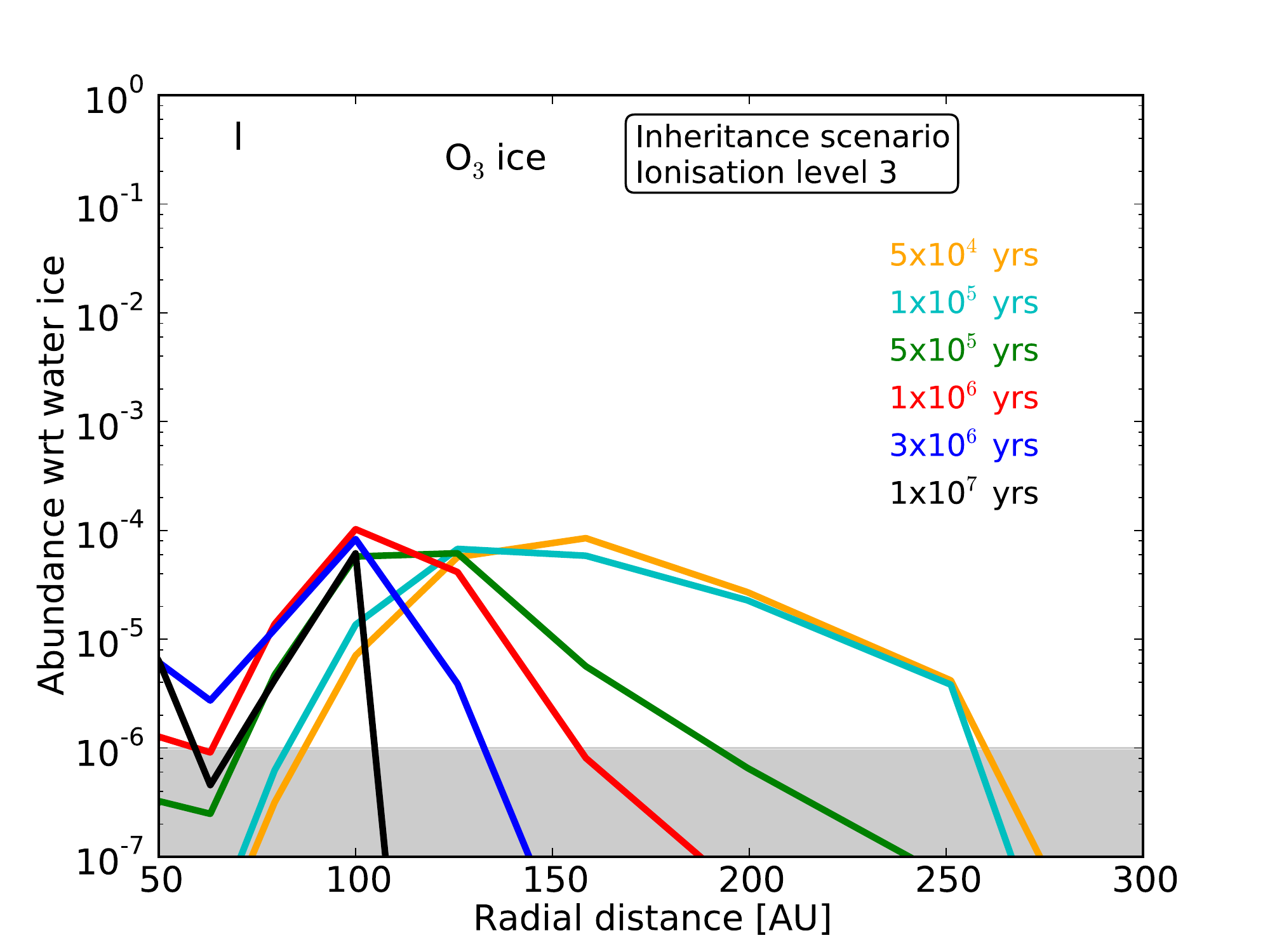}}\\
\caption{Radial abundance profiles for the inheritance scenario at multiple evolutionary time steps. Top to bottom are \ce{H2O}, \ce{O2} and \ce{H2O2} and \ce{O3} ices. Left to right are increasing ionisation levels. For \ce{O2}, \ce{H2O2}, and \ce{O3} the limits, and upper limit, detected in comet 67P, are indicated as yellow, red and grey shaded areas, respectively. The chemical network utilised includes \ce{O3} chemistry.}
\label{ozone_time_mol_water}
\end{figure*}

\begin{figure*}[h]
\subfigure{\includegraphics[width=0.33\textwidth]{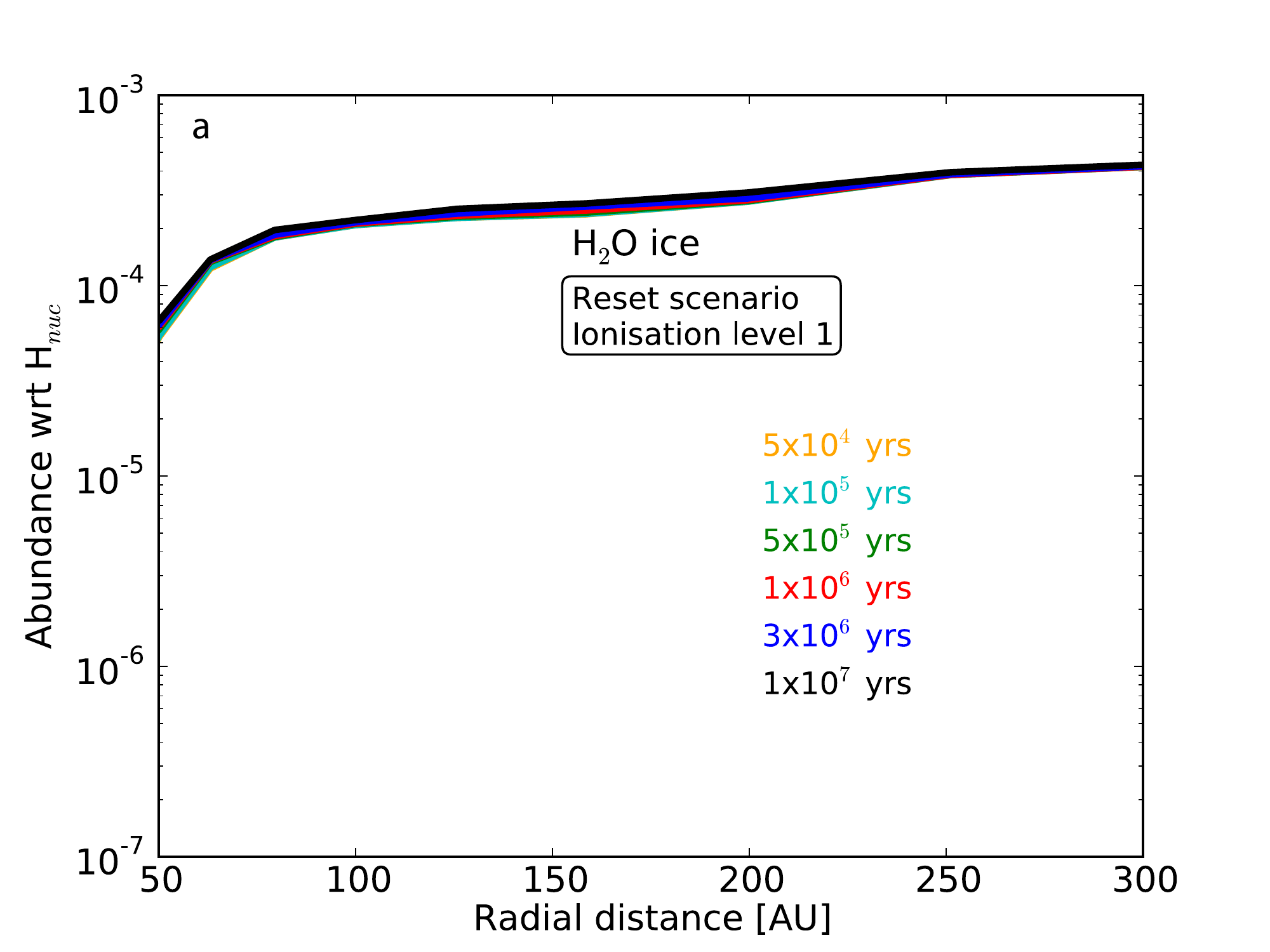}}
\subfigure{\includegraphics[width=0.33\textwidth]{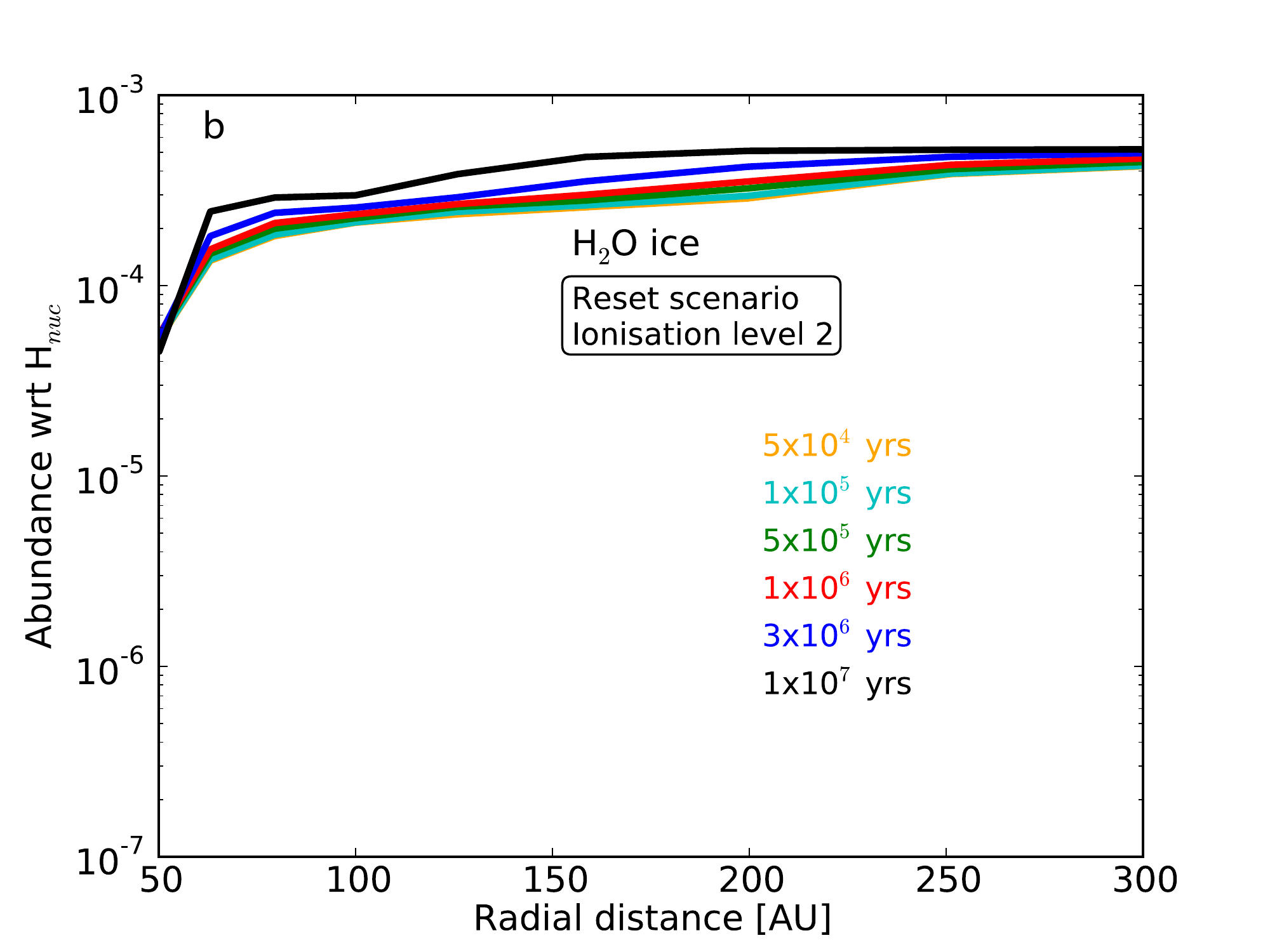}}
\subfigure{\includegraphics[width=0.33\textwidth]{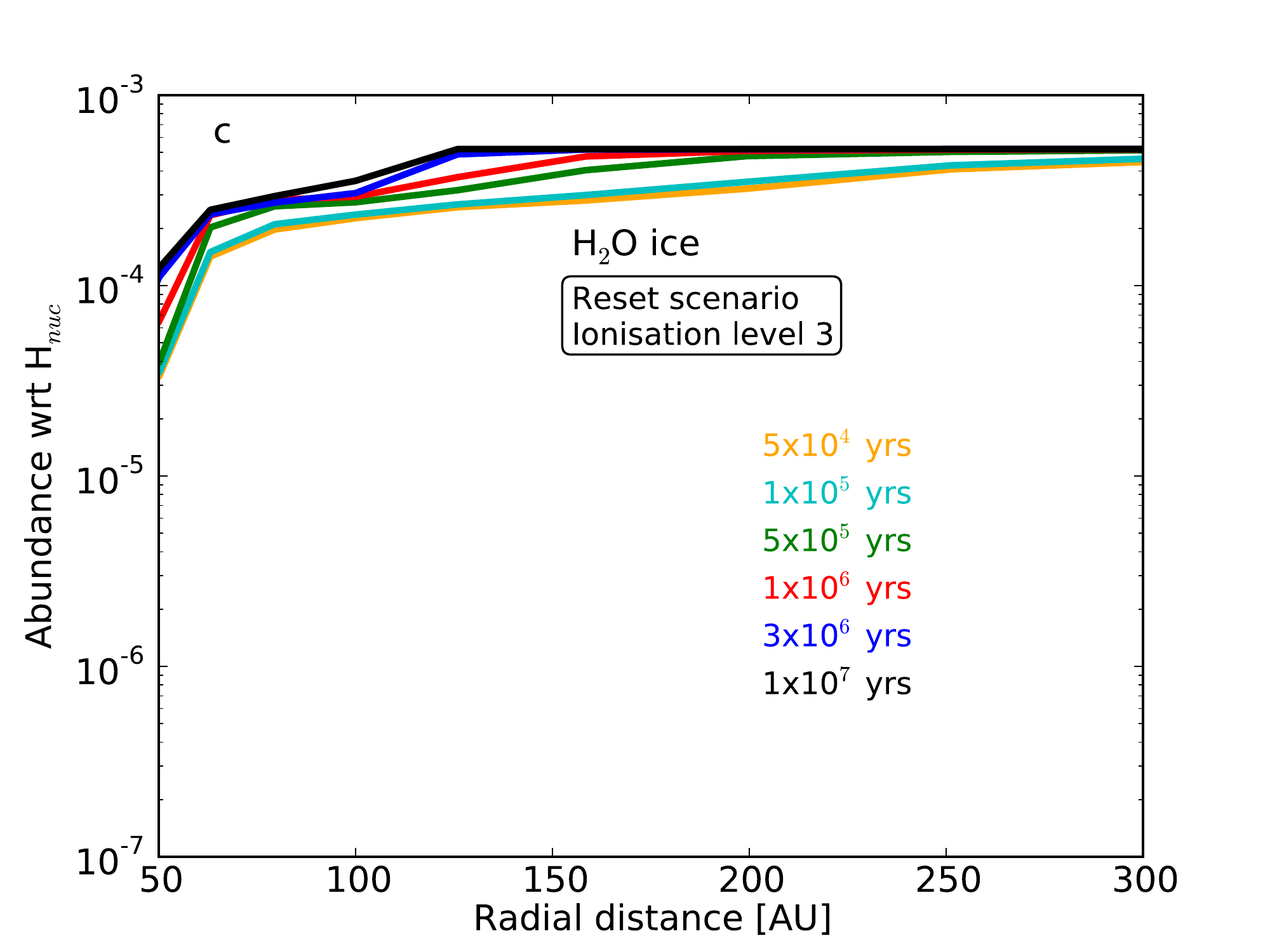}}\\
\subfigure{\includegraphics[width=0.33\textwidth]{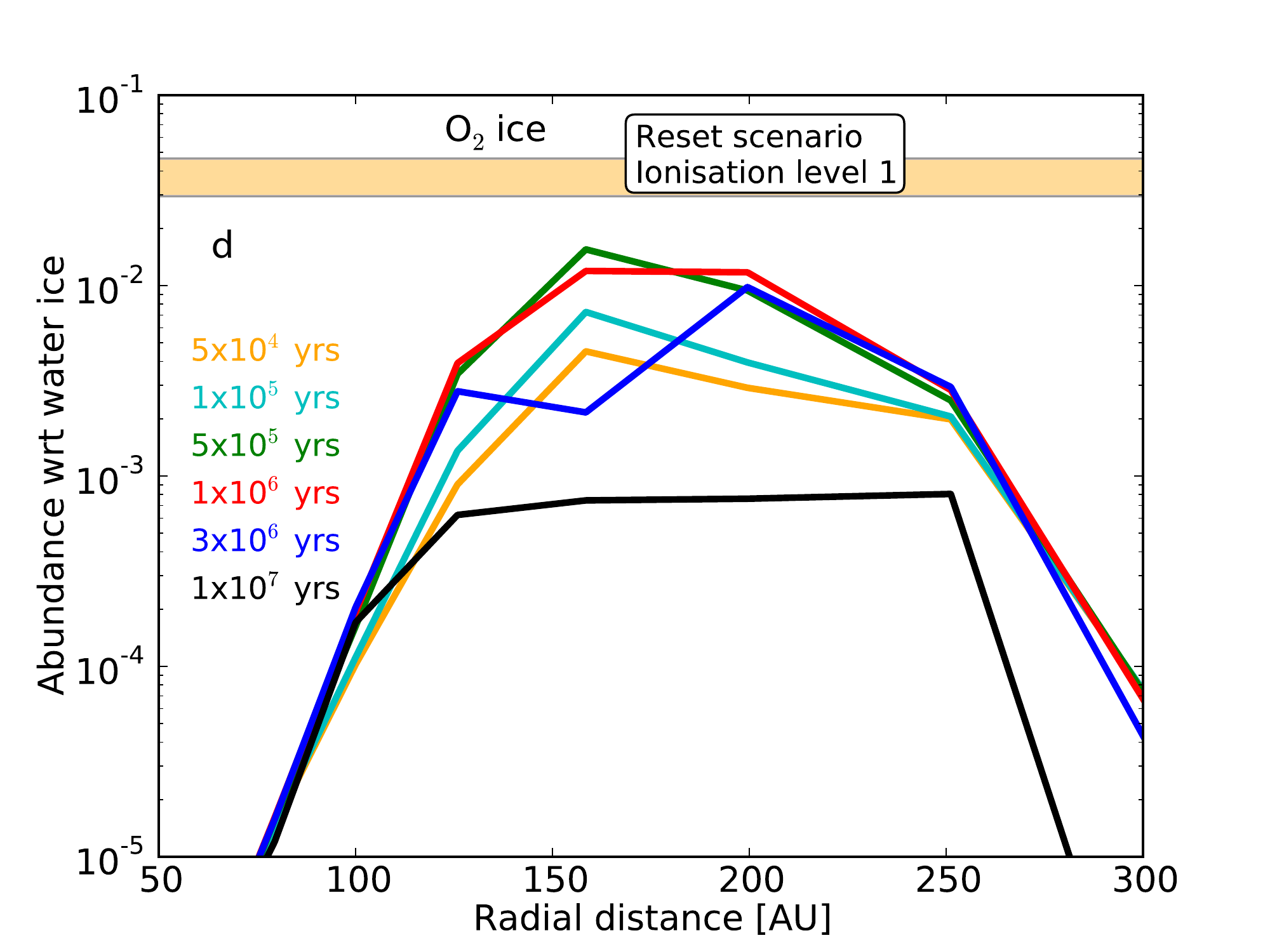}}
\subfigure{\includegraphics[width=0.33\textwidth]{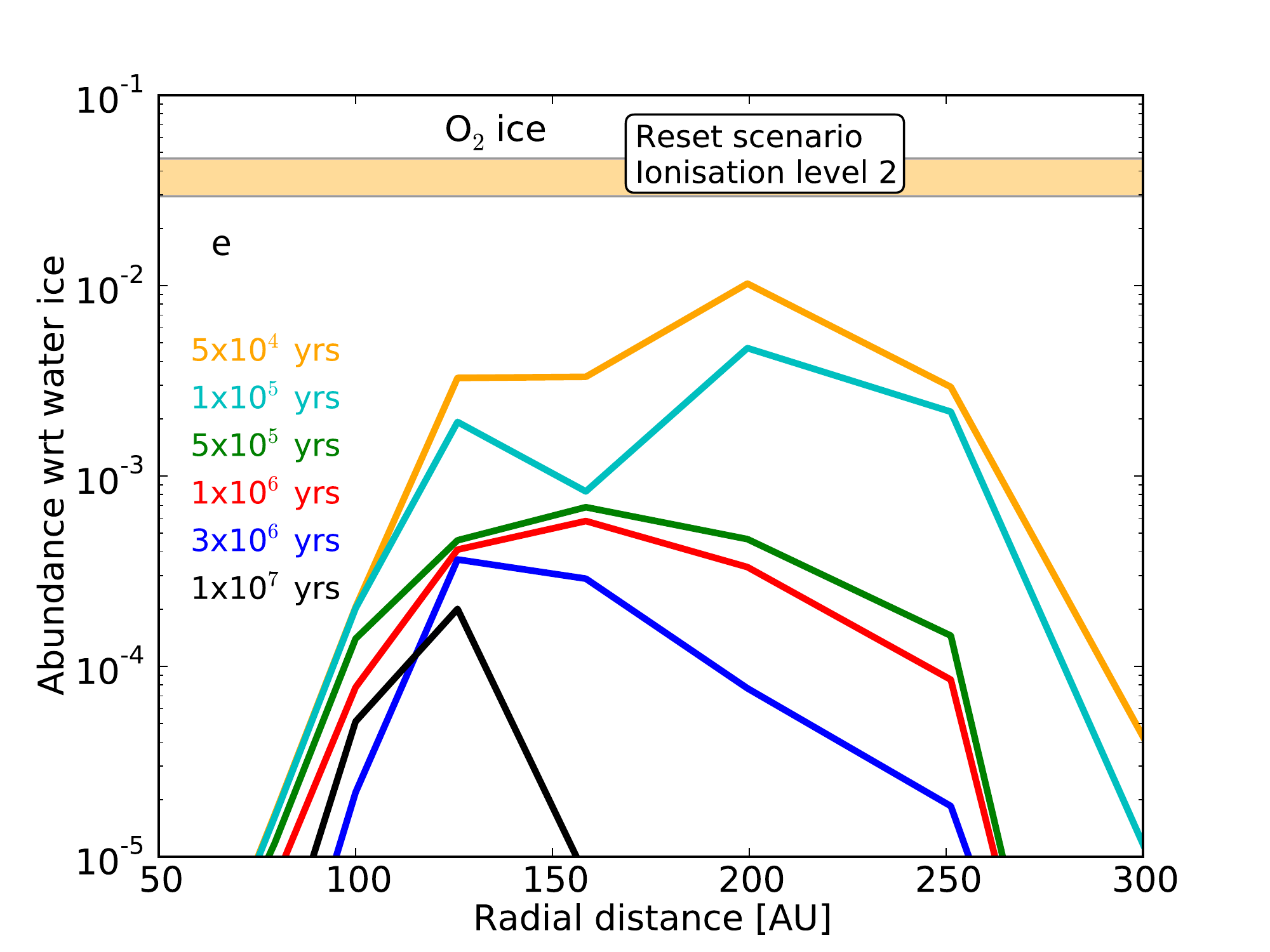}}
\subfigure{\includegraphics[width=0.33\textwidth]{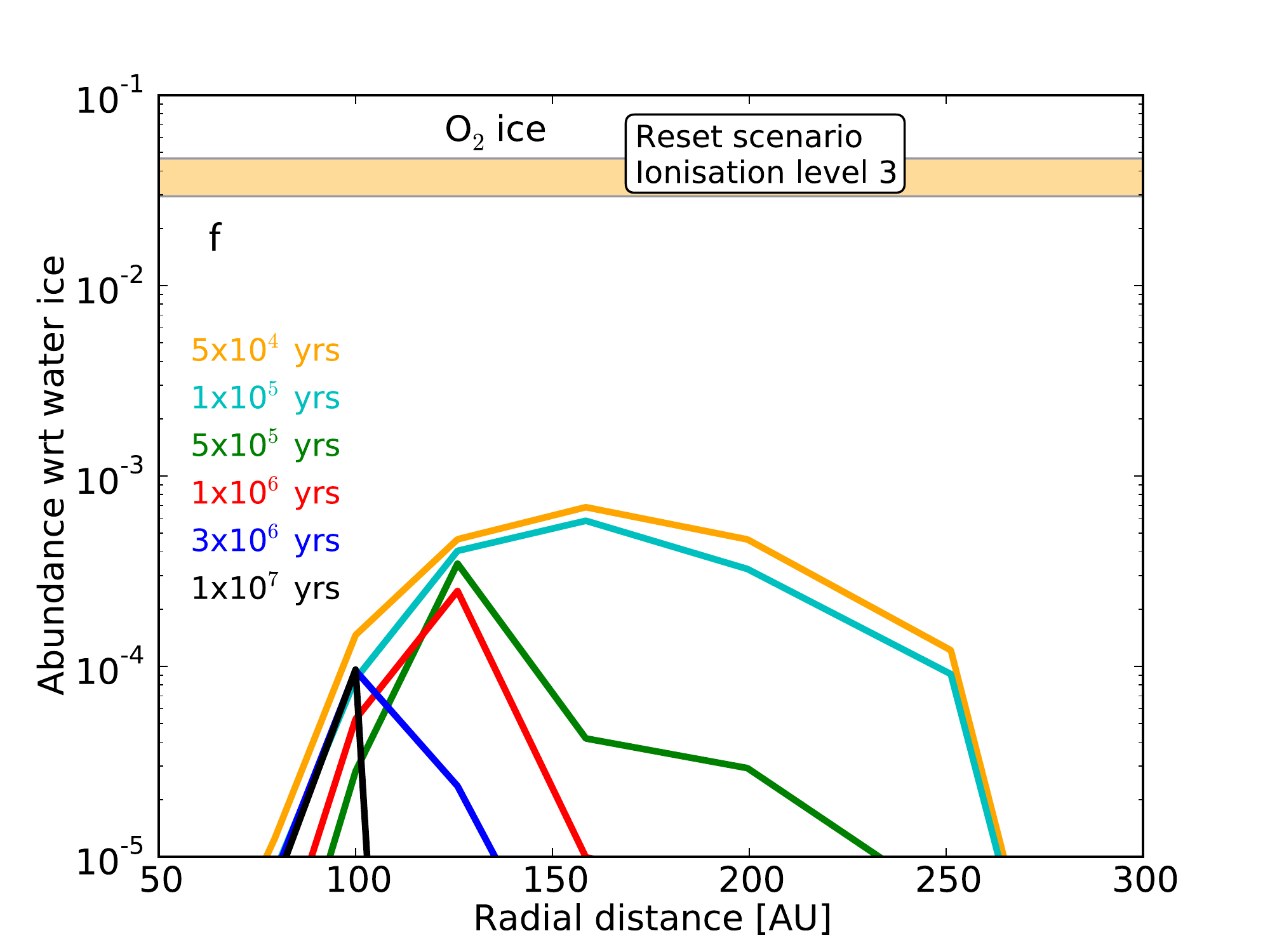}}\\
\subfigure{\includegraphics[width=0.33\textwidth]{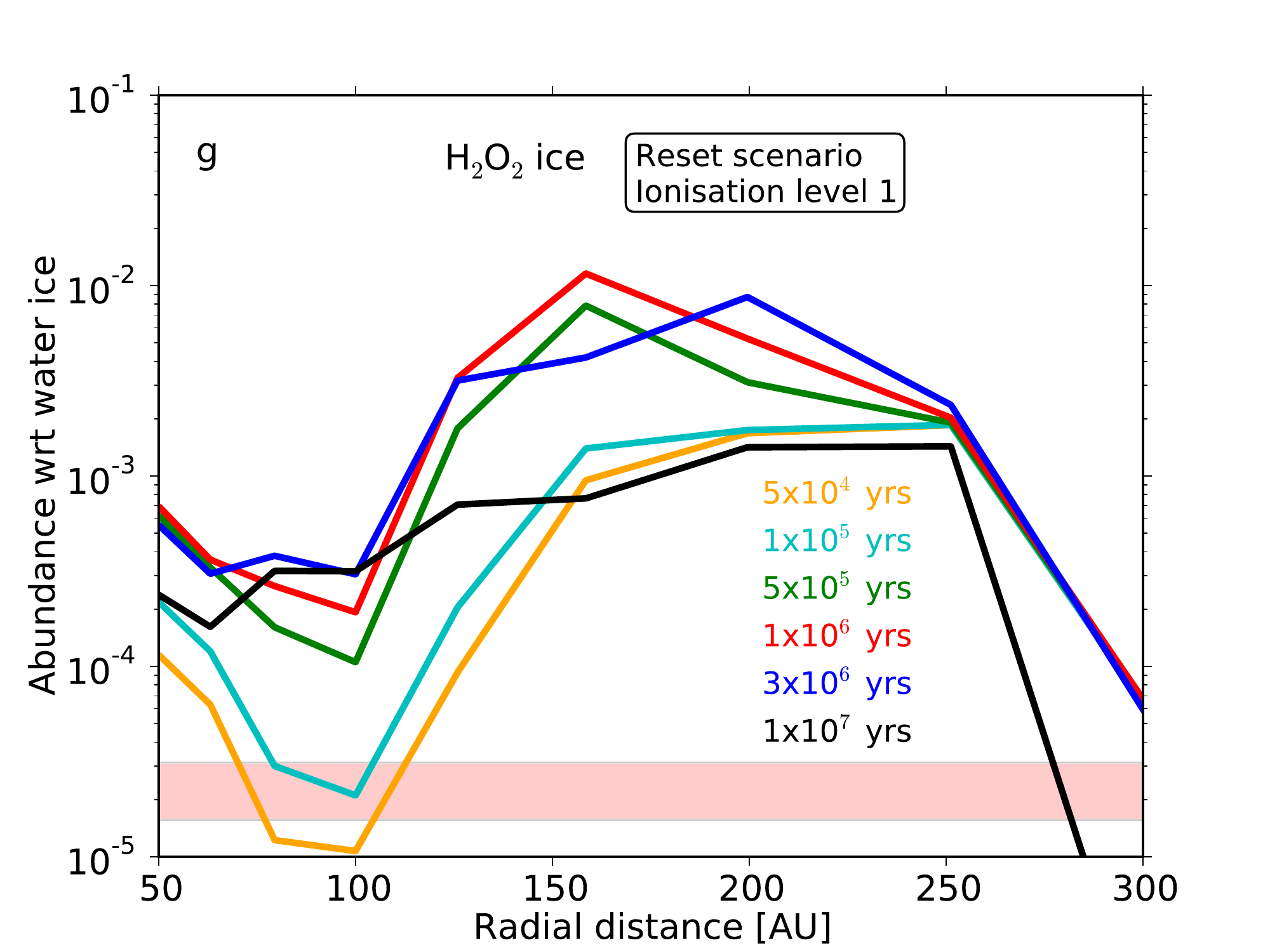}}
\subfigure{\includegraphics[width=0.33\textwidth]{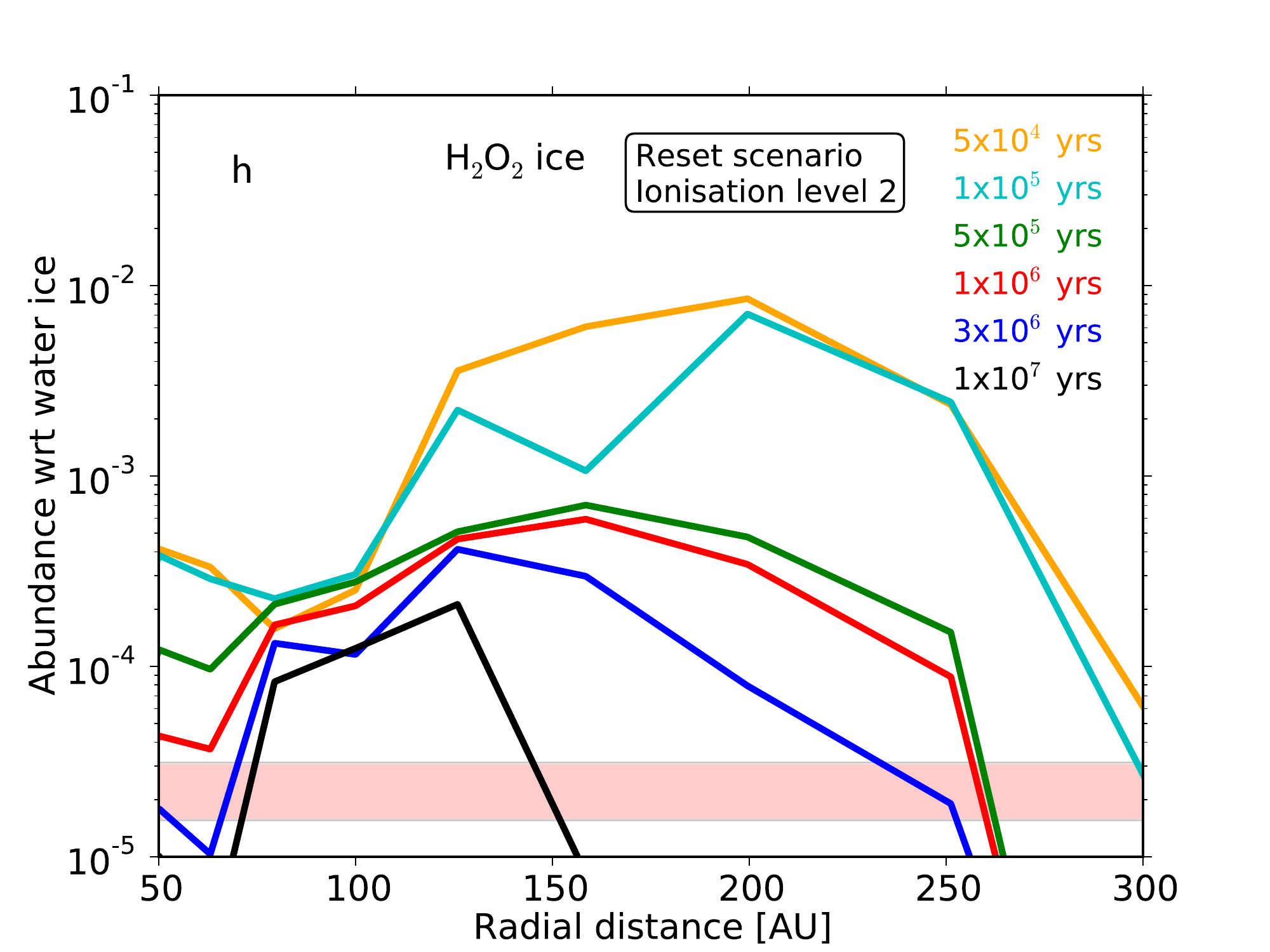}}
\subfigure{\includegraphics[width=0.33\textwidth]{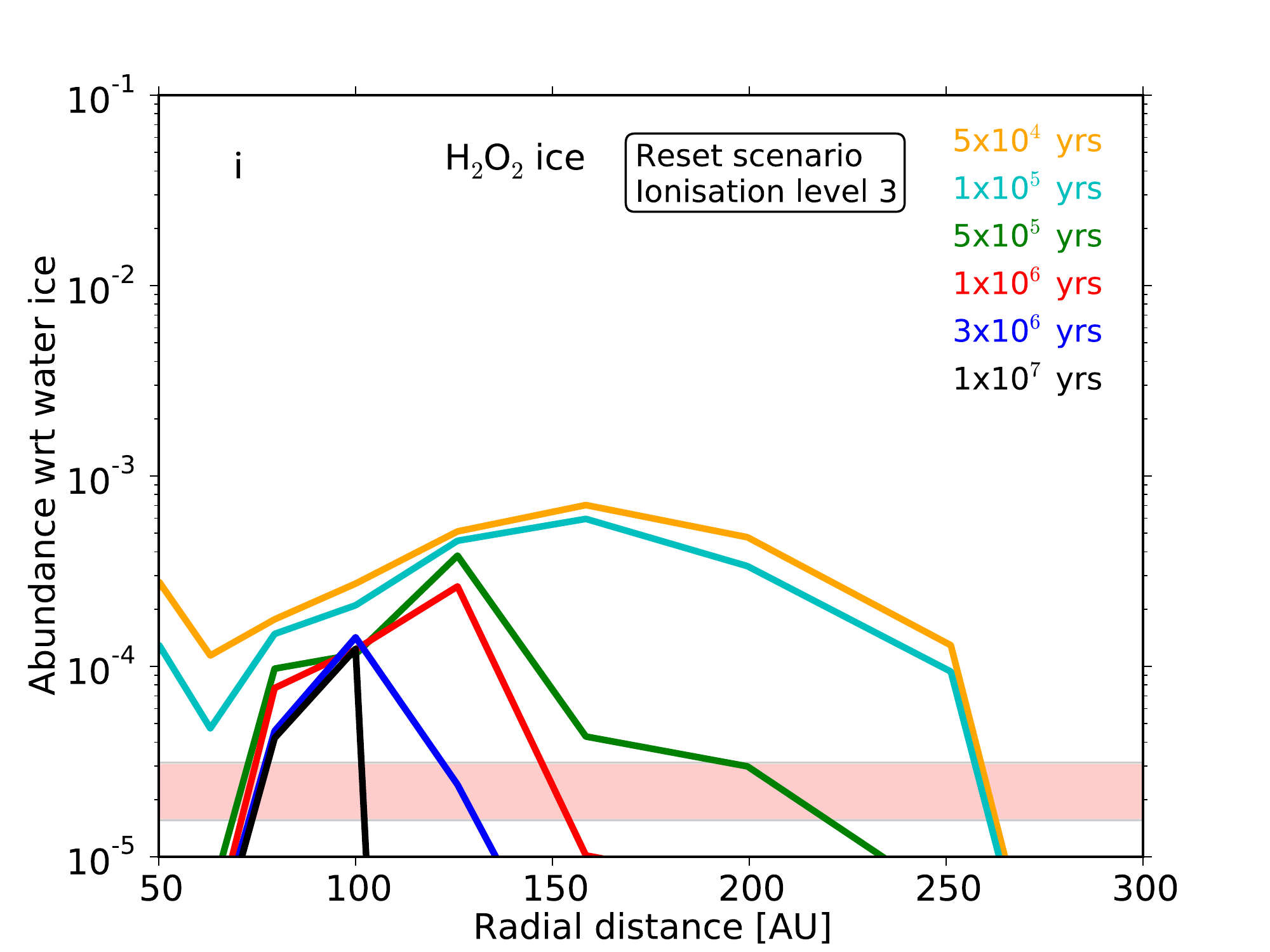}}\\
\subfigure{\includegraphics[width=0.33\textwidth]{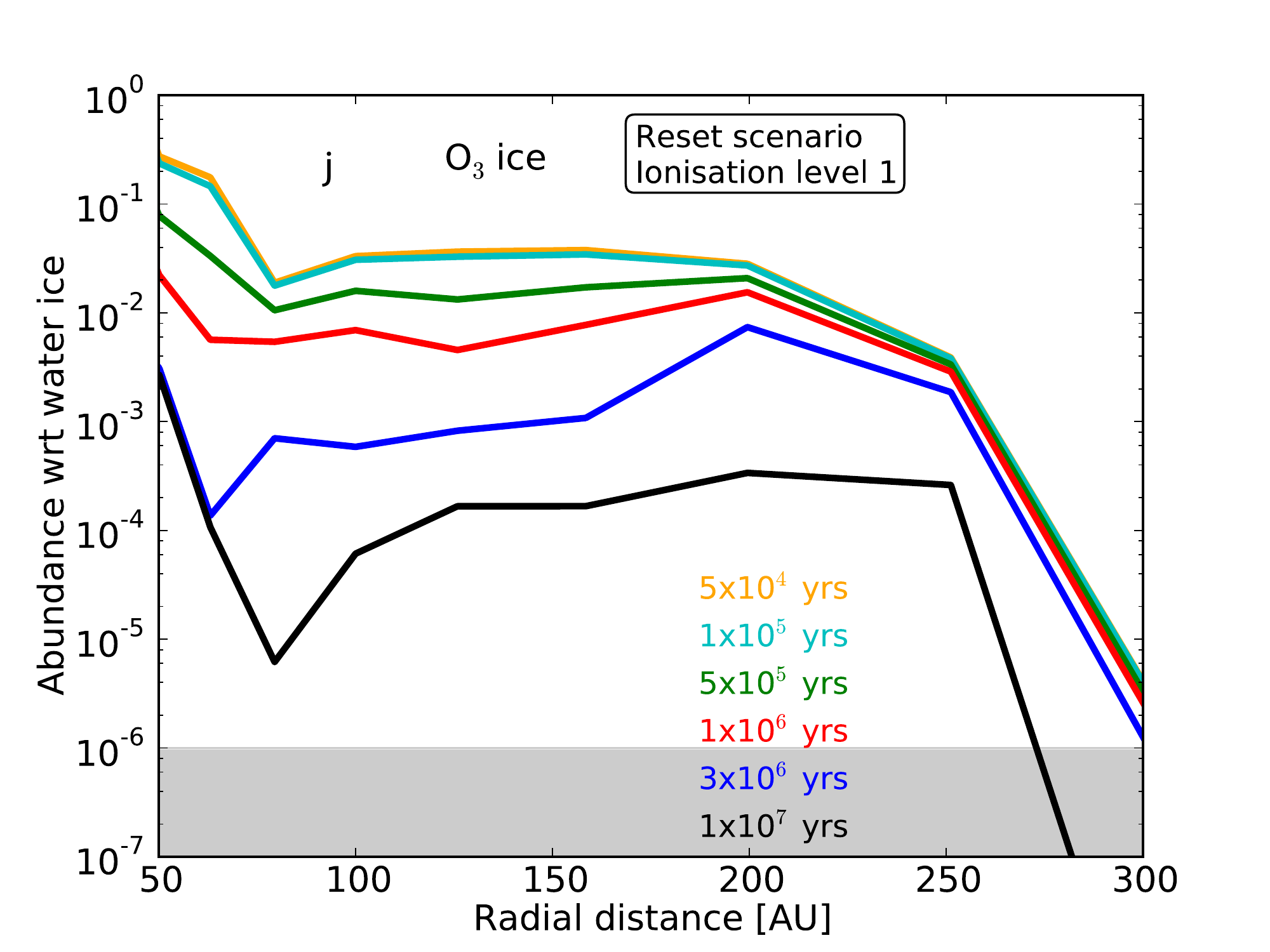}}
\subfigure{\includegraphics[width=0.33\textwidth]{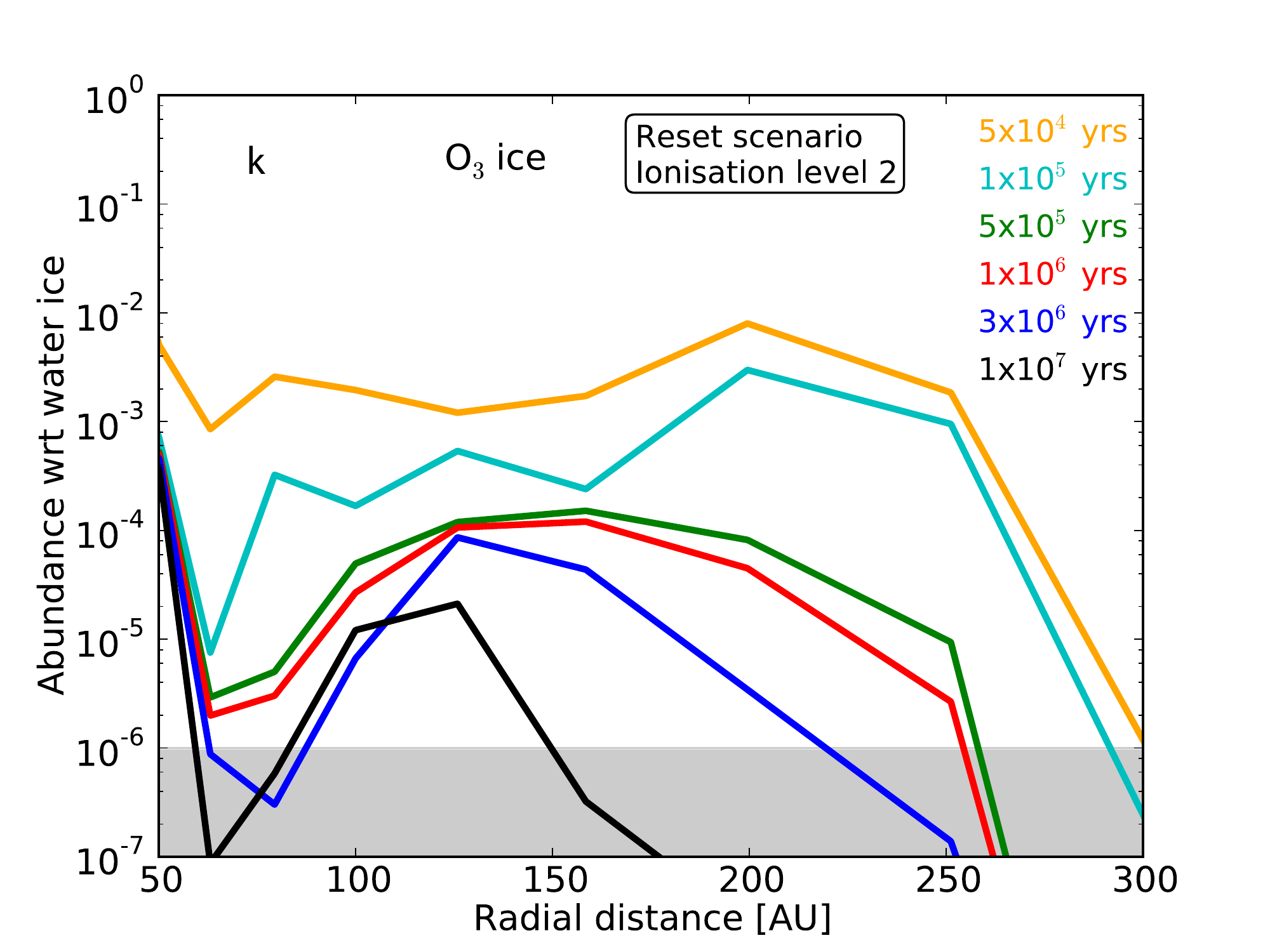}}
\subfigure{\includegraphics[width=0.33\textwidth]{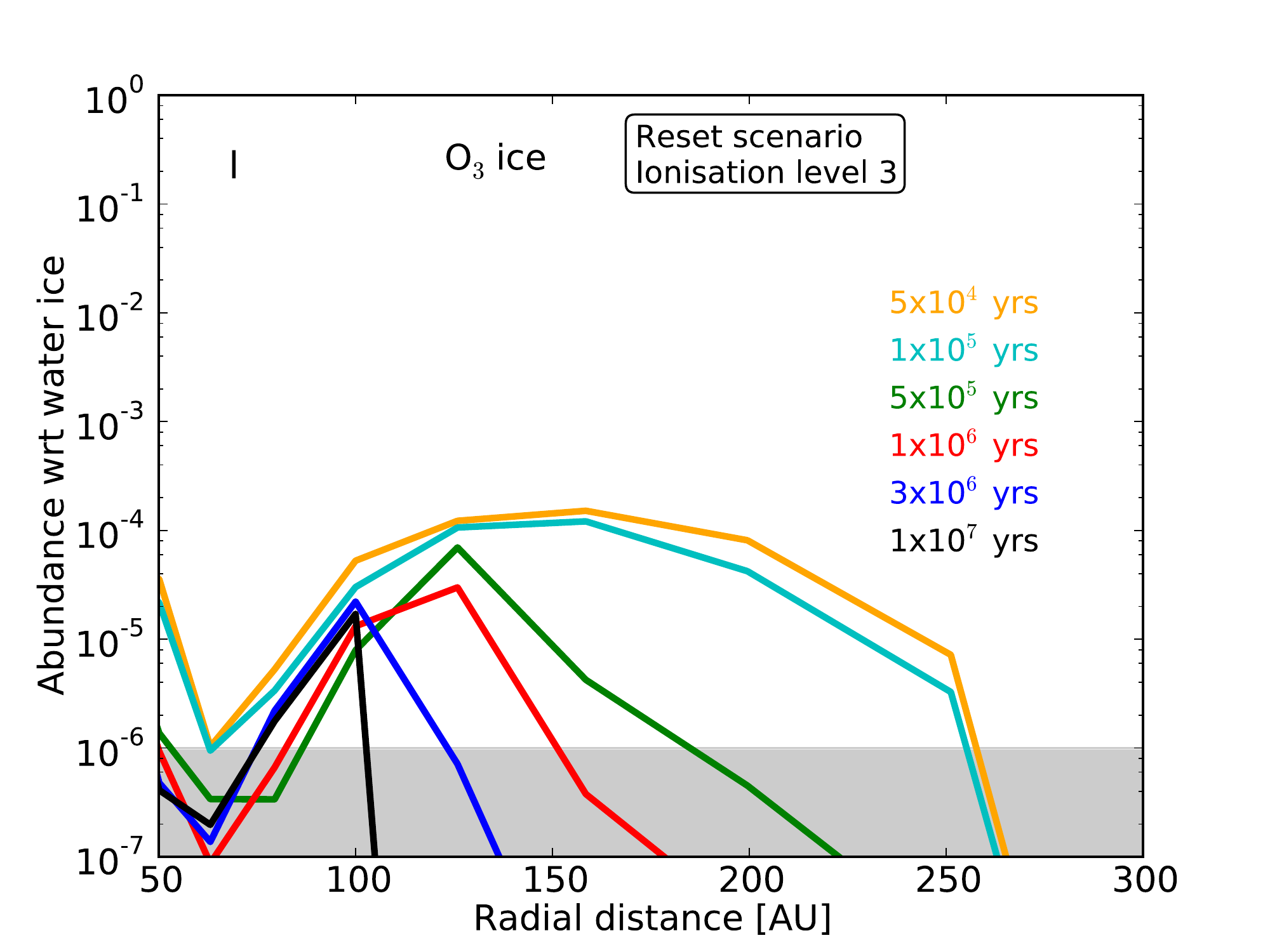}}\\
\caption{Radial abundance profiles for the reset scenario at multiple evolutionary time steps. Top to bottom are \ce{H2O}, \ce{O2} and \ce{H2O2} and \ce{O3} ices. Left to right are increasing ionisation levels. For \ce{O2}, \ce{H2O2}, and \ce{O3} the limits, and upper limit, detected in comet 67P, are indicated as yellow, red and grey shaded areas, respectively. The chemical network utilised includes \ce{O3} chemistry.}
\label{ozone_time_atom_water}
\end{figure*}

The chemical network utilised so far has not included \ce{O3}. This is because it has not been considered to be an important player in ISM chemistry, as it is formed primarily in the gas-phase via a three-body mechanism. However, data exist on \ce{O3} ice chemistry, as outlined in e.g. \citet{cuppen2010} and \citet{lamberts2013}. The production pathway to \ce{O3} ice is 
\begin{equation}
\ce{iO2 ->[\mathrm{iO}] iO3},\nonumber
\end{equation}
with a fiducial barrier of 500 K, and the destruction pathway is
\begin{equation}
\ce{iO3 ->[\mathrm{iH}] iO2 + iOH}, \nonumber
\end{equation}
which has no barrier. Hence, it is expected that the pathway to \ce{O3} ice via \ce{O2} ice may impact on the \ce{O2} ice abundance, and will be dependent on the both the availability of free oxygen atoms on the ice, and the efficiency of hydrogenation.

Figure \ref{ozone_time_mol_water} presents the evolving abundances of \ce{H2O} ice, \ce{O2} ice \ce{H2O2} ice and \ce{O3} ice as a function of radius for the inheritance scenario, thus now including \ce{O3} ice chemistry. These panels are directly comparable with those in Fig. \ref{abun_evol_inh_old_water}. The abundance evolution of \ce{O3} ice is shown in the bottom row. The abundance evolution seen for \ce{O2} ice in panels d to f and for \ce{H2O2} ice in panels g to i are similar to the trends in Fig. \ref{abun_evol_inh_old_water} for the chemistry without \ce{O3}. However, the abundances reached by 10 Myr with \ce{O3} chemistry are lower than those without: both \ce{O2} ice and \ce{H2O2} ice abundances peak at $\sim 5\times10^{-4}$ with respect to \ce{H2O} ice, thus a factor of two lower for \ce{H2O2} ice and a factor of five lower for \ce{O2} ice when compared with the scenario without \ce{O3} chemistry. The peak \ce{O2} ice abundance is almost two orders of magnitude lower than the observed value, and the peak of the \ce{H2O2} ice is about 20 times higher than the observed value.

For \ce{O3} ice in panels j to l a steady build-up over time is seen. Note that the range on the $y$-axis range is different for \ce{O3} ice than for \ce{H2O2} and \ce{O2} ices. The \ce{O3} ice reaches maximum abundance fastest at ionisation Level 3 ($\sim10^{-4}$ with respect to \ce{H2O} ice), thus a factor of five lower than that reached for \ce{H2O2} ice and \ce{O2} ice, yet two orders of magnitude higher than the observed upper limit at $10^{-6}$ with respect to \ce{H2O} ice for \ce{O3} ice in comet 67P. Only at larger radii (between 150 and 270 AU, depending on evolution time) do the \ce{O3} and \ce{H2O2} ice abundances fall within the observed limits, although \ce{O2} ice does not. All three ice species thus show similarities in their evolution, and order-of-magnitude similarities in their peak abundances. The observed large differences between their respective abundances (\ce{O3}:\ce{H2O2}:\ce{O2}$\lesssim$1:23.5:39200) are not reproduced by these models.

For the reset scenario, including \ce{O3} chemistry, Fig. \ref{ozone_time_atom_water} shows a higher abundance level of \ce{O2}, \ce{H2O2} and \ce{O3} ices than for the inheritance scenario. At ionisation Level 1, \ce{O2} and \ce{H2O2} ices peak at $\sim10^{-2}$ with respect to \ce{H2O} ice at $\sim150$ AU between 0.5-1 Myr evolution, but subsequently drop by about an order of magnitude by 10 Myr. For ionisation Levels 2 and 3, abundance levels for both species of $\sim10^{-2}$ with respect to \ce{H2O} ice are only achieved by $5\times10^{4}$ yrs, and the peak abundances by 10 Myr are at $1-2\times10^{-4}$ with respect to \ce{H2O} ice, thus neither matching the observed values of \ce{H2O2} ice nor of \ce{O2} ice. At no point in time or radius does the reset scenario including \ce{O3} chemistry reproduce the observed mean abundance of \ce{O2} ice. However, for ionisation Levels 1 and 2 the abundance between 150-200 AU reaches above 1\% of \ce{H2O} ice, which is within the observed range in comet 67P (1-10\%).

The \ce{O3} abundance for the reset scenario at ionisation Level 1 starts out high at $1-3\times10^{-2}$ with respect to \ce{H2O} ice between 100-200 AU up to 5$\times10^{5}$ yrs. s seen in Fig. \ref{ozone_time_atom_water}. Note that the higher ratios reached within 70 AU is due to generally low absolute ice abundances (see panels a to c in Fig. \ref{ozone_time_atom_water}). For ionisation Levels 2 and 3, these high abundances reduce over time leading to a peak of at a few times $10^{-5}$ with respect to \ce{H2O} ice by 10 Myr between 100-120 AU. The maximum abundance achieved by 10 Myr is an order of magnitude higher than the observed upper limit. Only in the wings of	 the abundance profiles do the models reproduce this upper limit for the entire radial range, depending on evolution time and ionisation level.

The scenarios starting with only \ce{H2O} or elemental oxygen were also investigated after the inclusion of \ce{O3} chemistry. However, neither of them reproduced a significant amount of \ce{O2} ice. The abundances by 10 Myr are shown for the oxygen-only scenario in Fig. \ref{oxygen_10}b. Upon inclusion of \ce{O3} chemistry, the peak abundance of \ce{O2} ice from panel a (without \ce{O3}) drops by at least four orders of magnitude to <$10^{-9}$ with respect to \ce{H2O} ice in panel b. Starting with atomic oxygen only therefore does not reproduce the observed abundance. For the water scenario, there is also no \ce{O2} ice. An overview of gas and ice abundances of \ce{H2O}, \ce{O2}, \ce{O3} and \ce{H2O2} as a function of radius by 10 Myr of evolution for the inheritance, reset, water and oxygen scenarios is shown in Fig. \ref{ozone_10}.

In none of the tested cases including \ce{O3} chemistry is the \ce{O2} ice abundance reproduced to match the levels observed in the comets.


\subsection{Exploring the sensitivity of the abundances to assumed grain-surface parameters}
The underproduction of \ce{O2} ice and overproduction of \ce{H2O2} and \ce{O3} ices may be a result of the adsorption and reaction parameters assumed for the network of reactions involving oxygen and hydrogen. If, for example, the reaction \ce{iO2 ->[\mathrm{iO}] iO3} is too efficient on the grain-surfaces, this could lead to over- and under-production of \ce{O3} and \ce{O2} ices, respectively. This reaction is dependent on the adsorption of atomic oxygen, for which the utilised binding energy here is $E_{\rm{bin}}$=800 K. This is based on estimates from \citet{tielens1987}. However, recent experimental work by \citet{he2015} experimentally determined a binding energy of $E_{\rm{bin}}$=1660 K, thus more than doubling the former value. This higher binding energy should act to keep oxygen atoms from desorbing at higher temperatures, thereby facilitating every step in the reaction pathway leading from 
\ce{iO ->[\mathrm{iO}] iO2 ->[\mathrm{iO}] iO3}. On the other hand, at low temperature the mobility of atomic oxygen will be reduced. This latter case could possibly lead to a decreased production of \ce{O3} ice.

In this subsection, the sensitivity of the chemistry to several parameters is explored for single-point models (i.e. single $n$, $T$) which cover the temperature and density ranges in the PSN midplane where \ce{O2} ice production is expected, based on the model results described thus far. The first parameter explored is the barrier width for quantum tunnelling $b_{\rm{qt}}$which is set to either 1 or 2 {\AA} representing the limits to the range of values usually assumed for $b_{\rm{qt}}$ when modelling grain-surface chemistry, see \citet{cuppen2017}, and references therein. The second parameter is the ratio of the diffusion energy to the molecular binding energy $E_{\rm{diff}}/E_{\rm{bin}}$, which is taken to be either 0.3 or 0.5 \citep[see][and references therein]{ruffle2000,garrod2006}. This amounts to four combinations of the two reaction parameters. The values of each parameter associated with a given model setup is given in each panel. The values for the grain-surface chemistry used in \citet{eistrup2016,eistrup2018}, are $b_{\rm{qt}}$= 1{\AA} and $E_{\rm{diff}}/E_{\rm{bin}}$ = 0.5. Only the reset scenario is studied here because this is the case for which maximal \ce{O2} ice formation is seen.

\begin{table*}
\captionsetup{justification=centering}

\caption{Binding energies $E_{\rm{bin}}$[K]	 for relevant species}             
\centering
\begin{tabular}{l c c c}              
\hline\hline                        
Species 	& UDfA (used here) 	      & UDfA reference    & Penteado et al. (2017)\\              
\hline                              
\\
   O			& 800 		&  \citet{tielens1987} & 1660$\pm60$ \\  
   OH	 		& 2850		&  \citet{garrod2006}& 3210$\pm1550$\\
   \ce{O2H} 	& 3650		& \citet{garrod2006} & 800\\
   \ce{H2O2} 	& 5700		& \citet{garrod2006}& 6000$\pm100$\\ 
   \ce{O2}		& 1000		& \citet{garrod2006} & 898$\pm30$\\
   \ce{O3}		& 1800		& \citet{garrod2006} & 2100$\pm100$\\
      \ce{H2O}	& 5700	      	& \citet{brown2007} & 4800$\pm100$\\
   H			& 600		& \citet{cazaux2002}& 650$\pm100$\\
   \ce{H2}		& 430		& \citet{garrod2006}& 500$\pm100$\\
   \\
   \hline                                   
\label{bind_e}
\end{tabular}
\end{table*}

\begin{table*}
\captionsetup{justification=centering}

\caption{Activation energies $E_{\rm{act}}$[K] for relevant grain-surface reactions}             
\centering                       
\begin{tabular}{l c}              
\hline\hline                        
Reaction 	& $E_{\rm{act}}$[K]\\              
\hline                              
\\
   \ce{OH + H -> H2O}			& 0 			 \\  
   \ce{OH + H2 -> H2O + H}	 	& 2100		\\
   \ce{O3 + H -> O2 + OH} 		& 0		 \\
   \ce{O + O -> O2}				& 0		\\
   \ce{O + O2 -> O3} 				& 500	\\ 
   \ce{OH + O -> O2H}			& 0		 \\
   \ce{O2H + H2 -> H2O2 + H}		& 5000	      	\\
   \ce{H + H2O2 -> H2O + OH}		& 2000		\\
   \\
   \hline                                   
\label{act_e}
\end{tabular}
\end{table*}

\subsubsection{Abundance mosaics}
\label{abun_mosaic}

\begin{figure*}[h]
\subfigure{\includegraphics[width=0.22\textwidth]{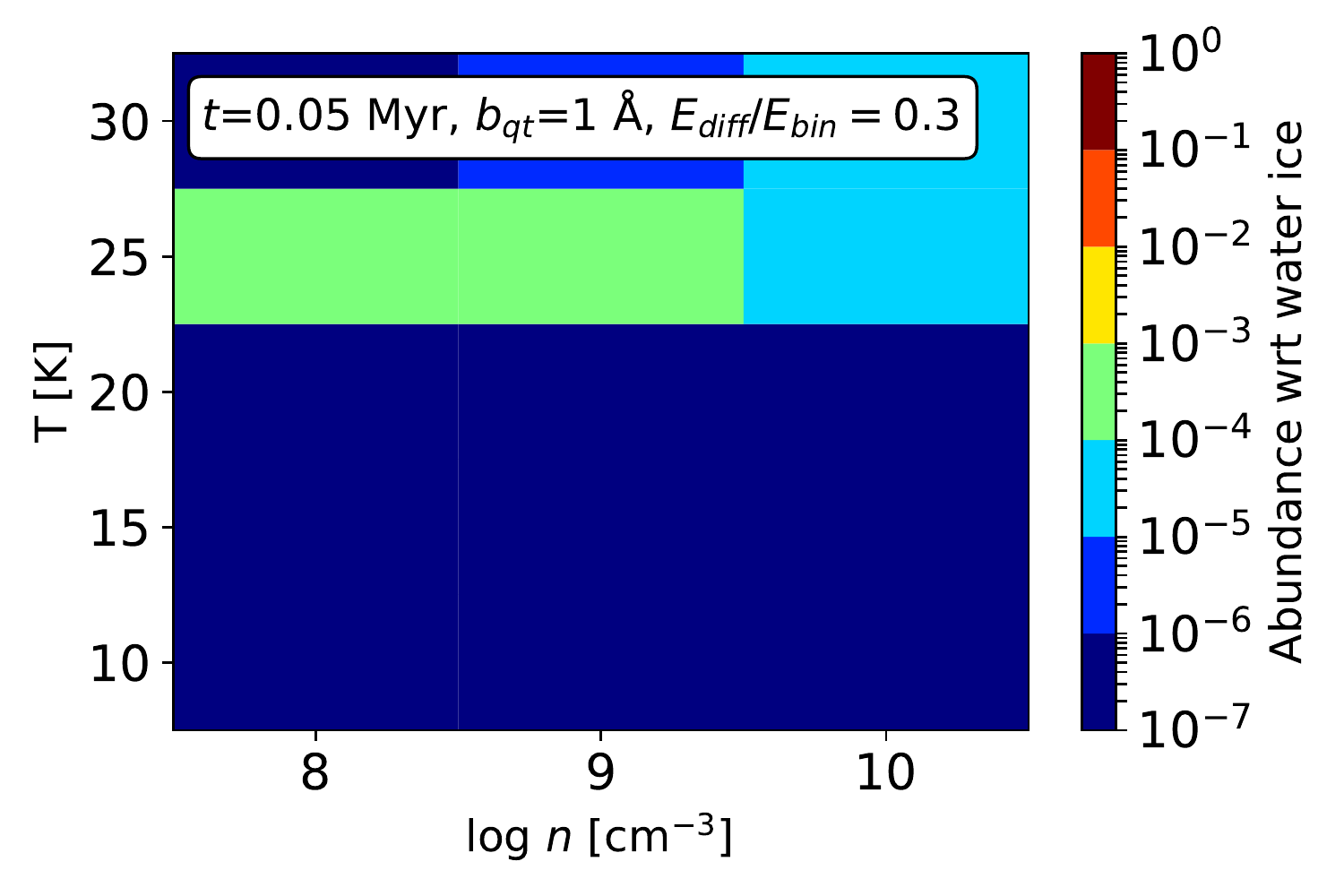}}
\subfigure{\includegraphics[width=0.22\textwidth]{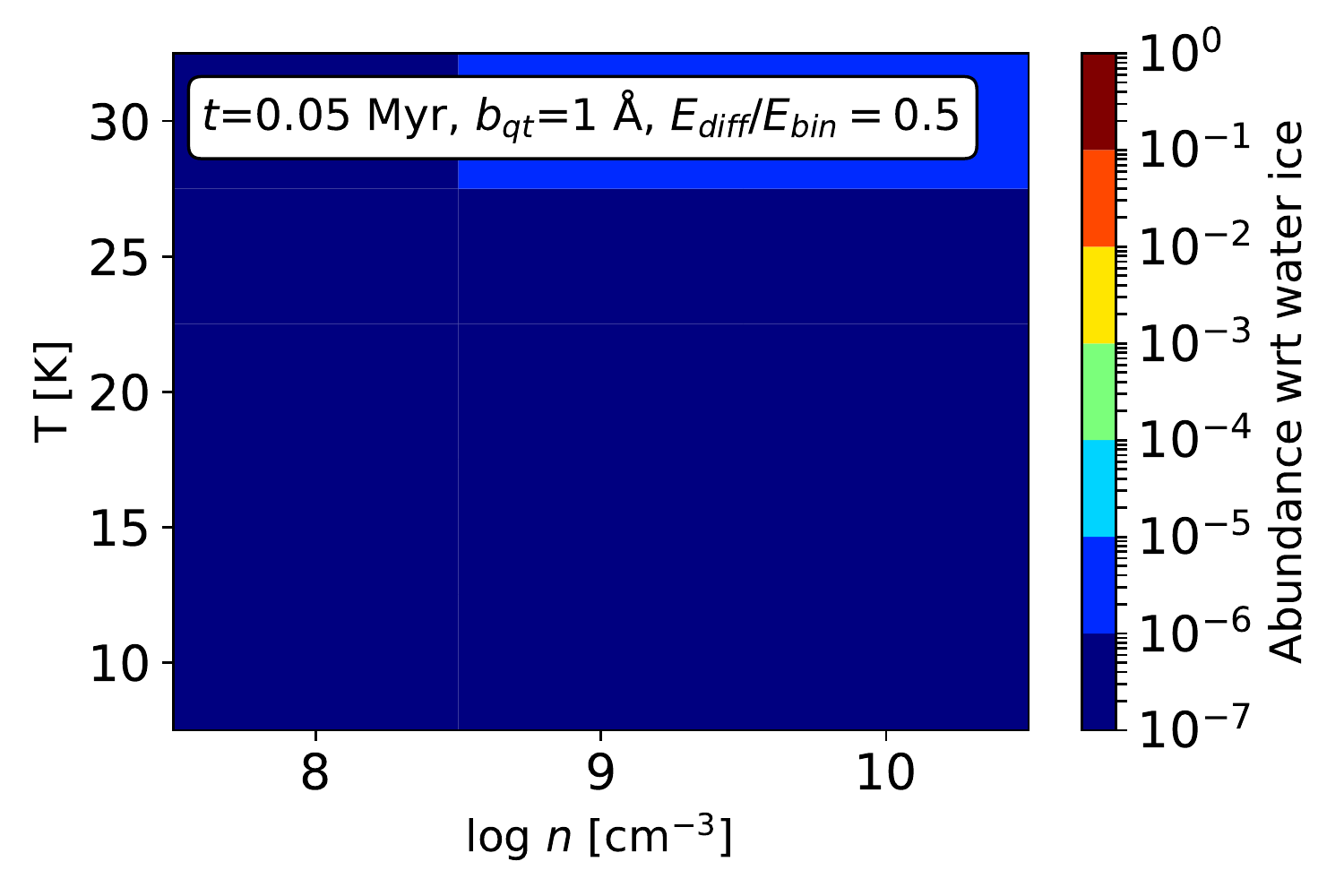}}
\subfigure{\includegraphics[width=0.22\textwidth]{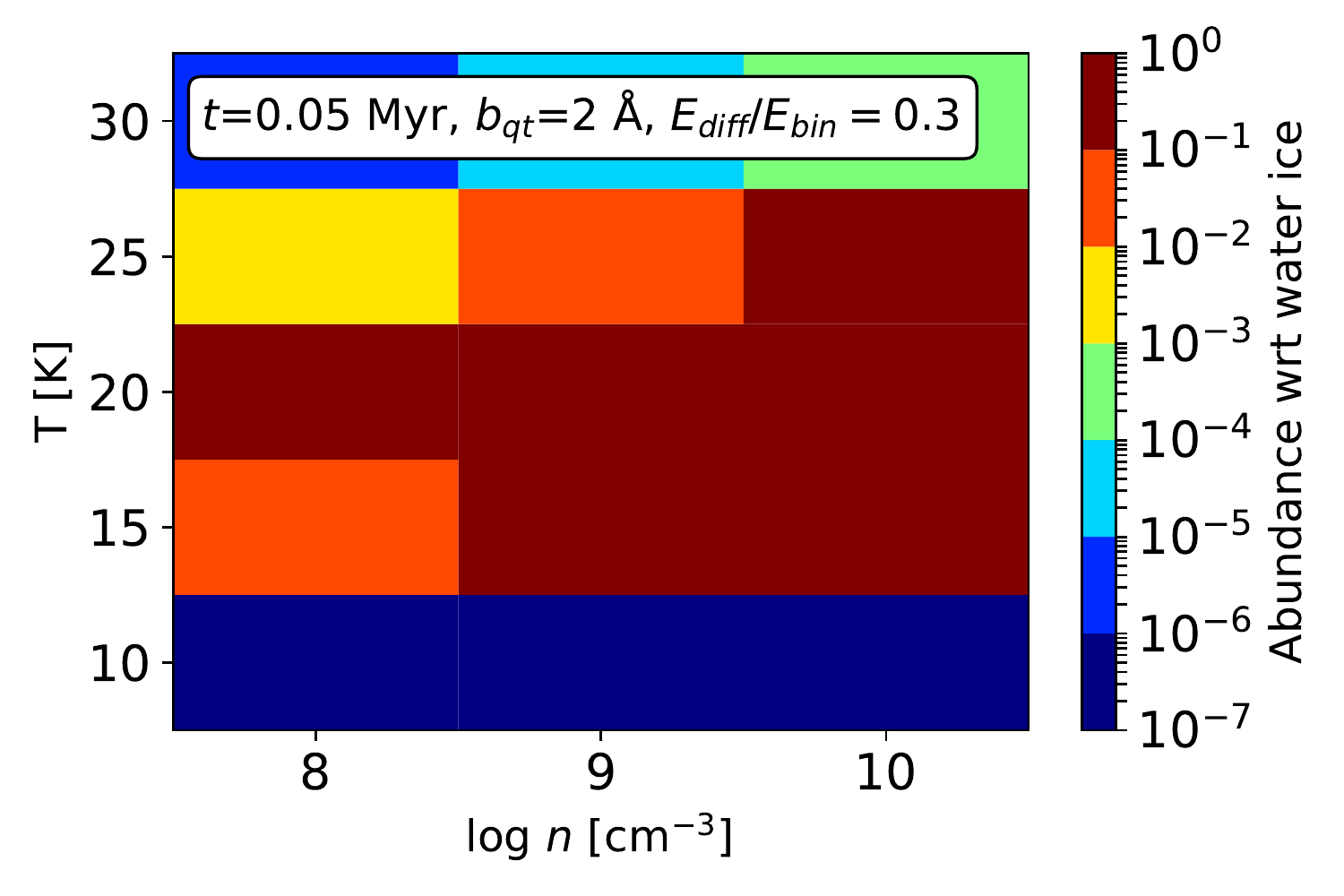}}
\subfigure{\includegraphics[width=0.22\textwidth]{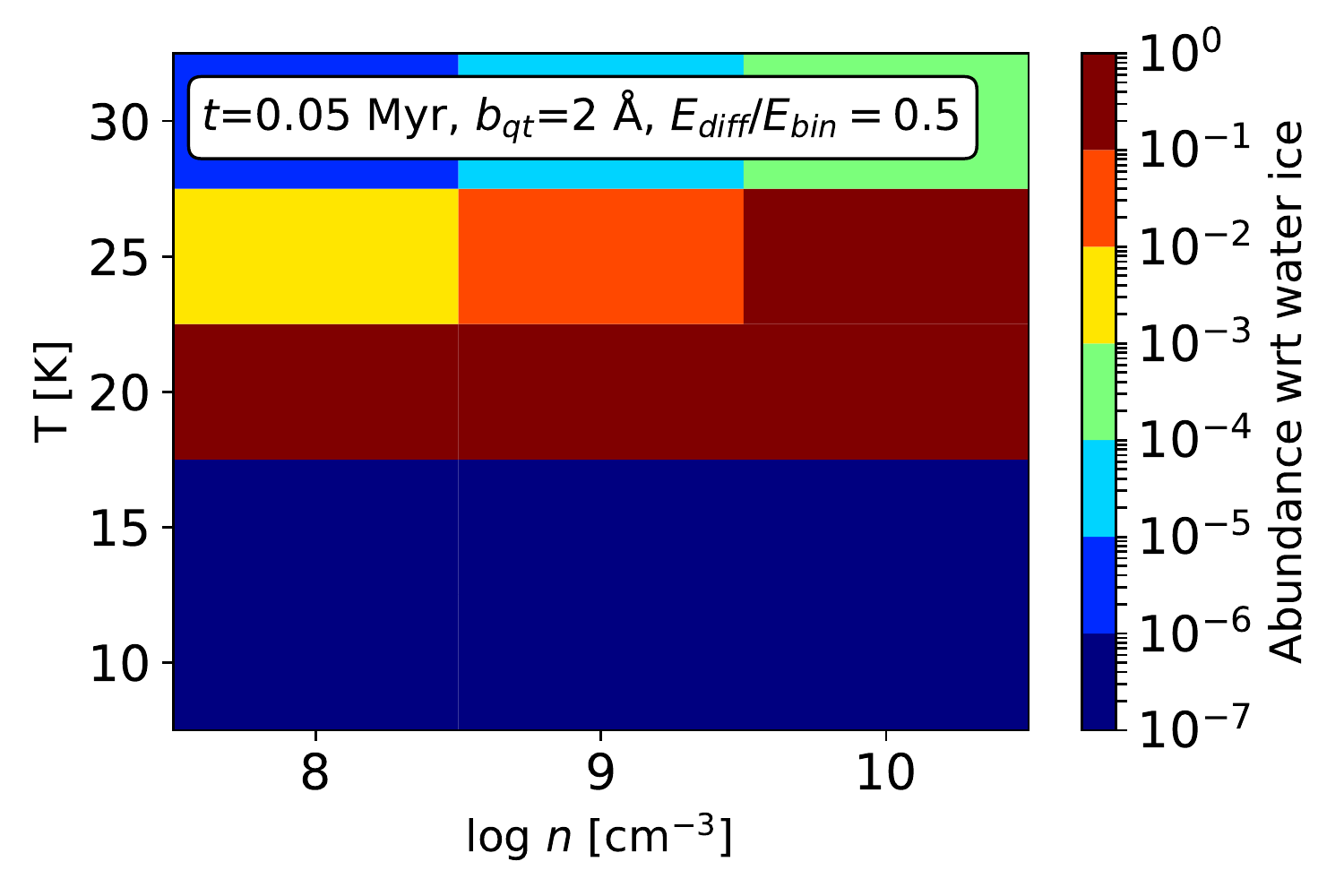}}\\
\subfigure{\includegraphics[width=0.22\textwidth]{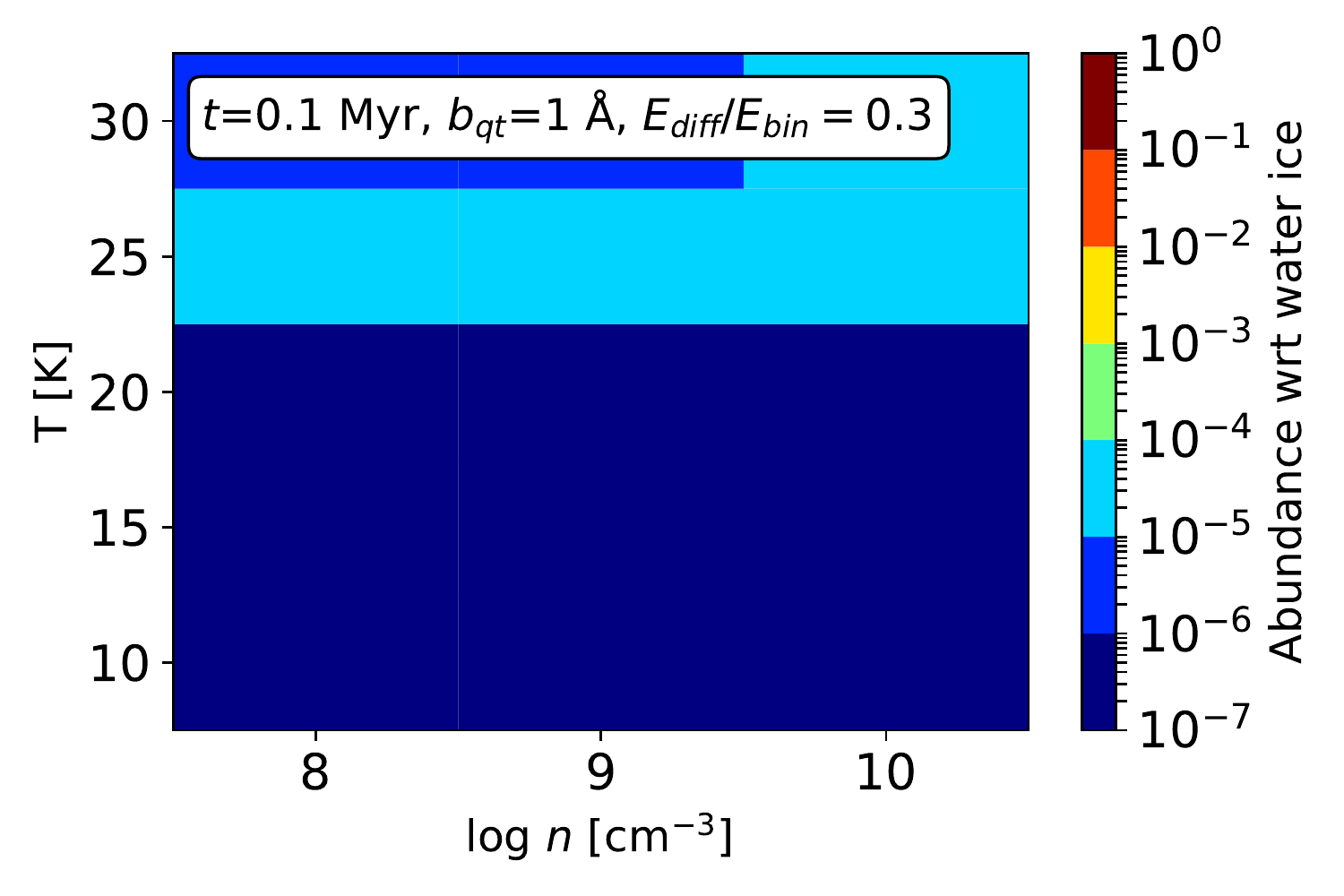}}
\subfigure{\includegraphics[width=0.22\textwidth]{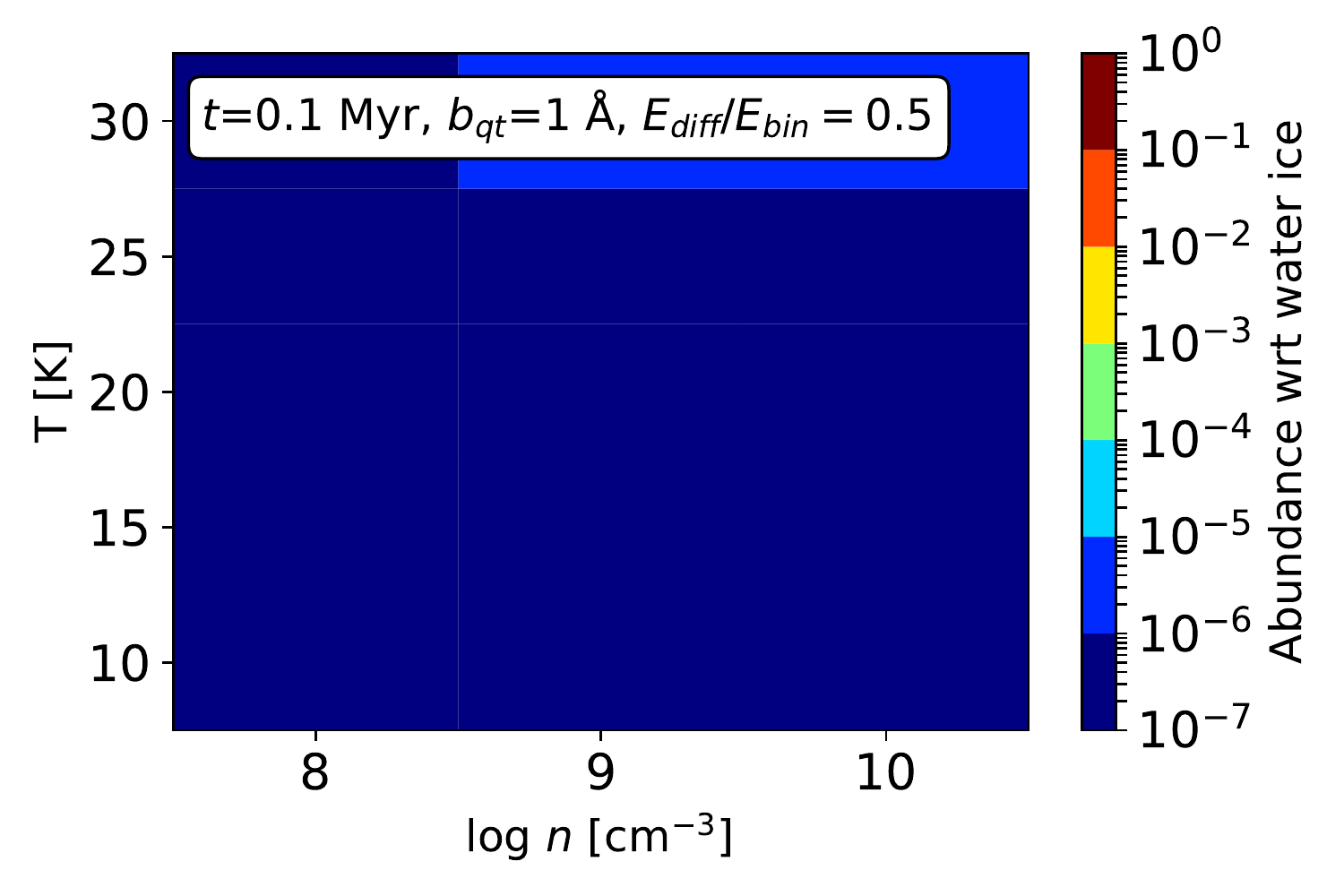}}
\subfigure{\includegraphics[width=0.22\textwidth]{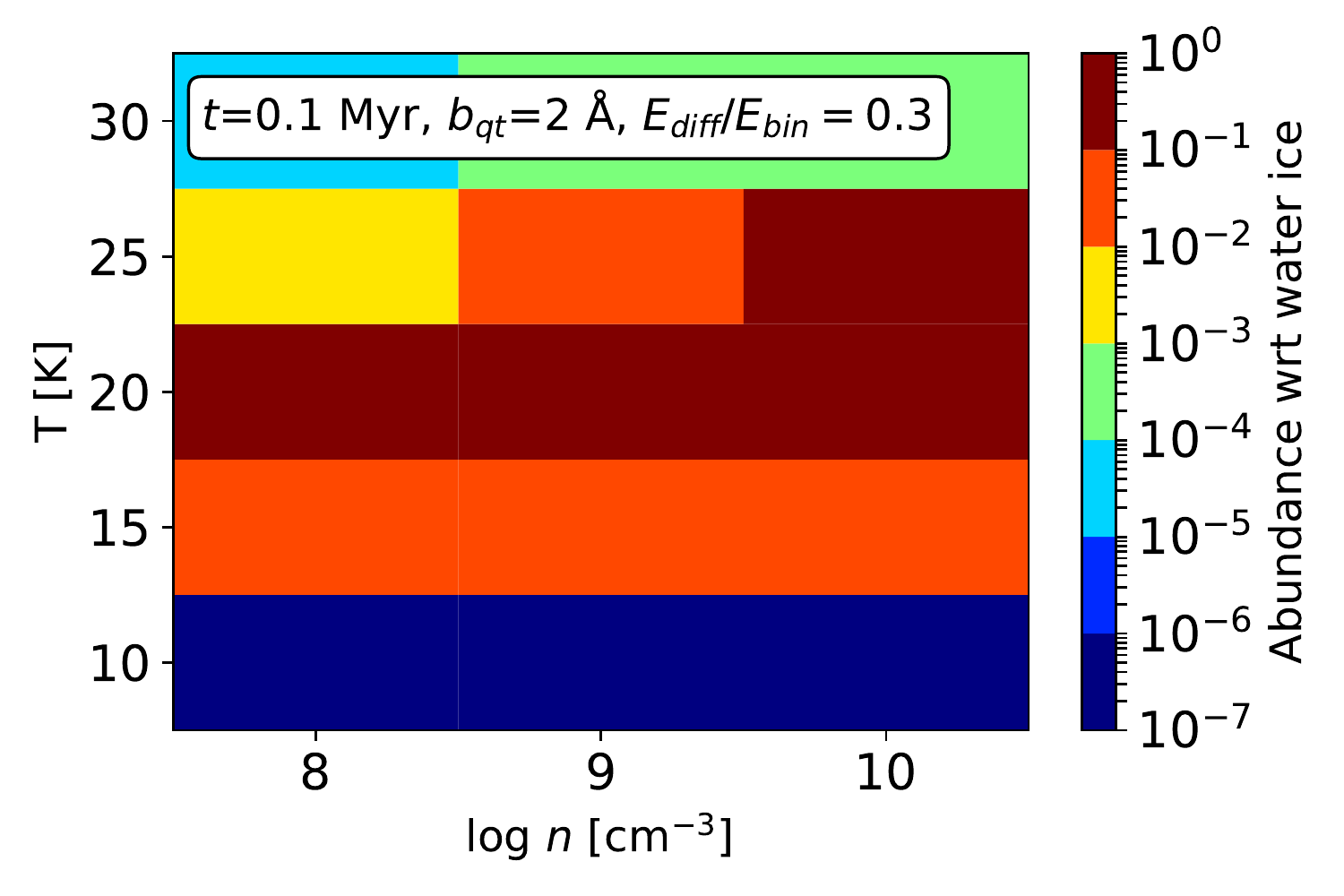}}
\subfigure{\includegraphics[width=0.22\textwidth]{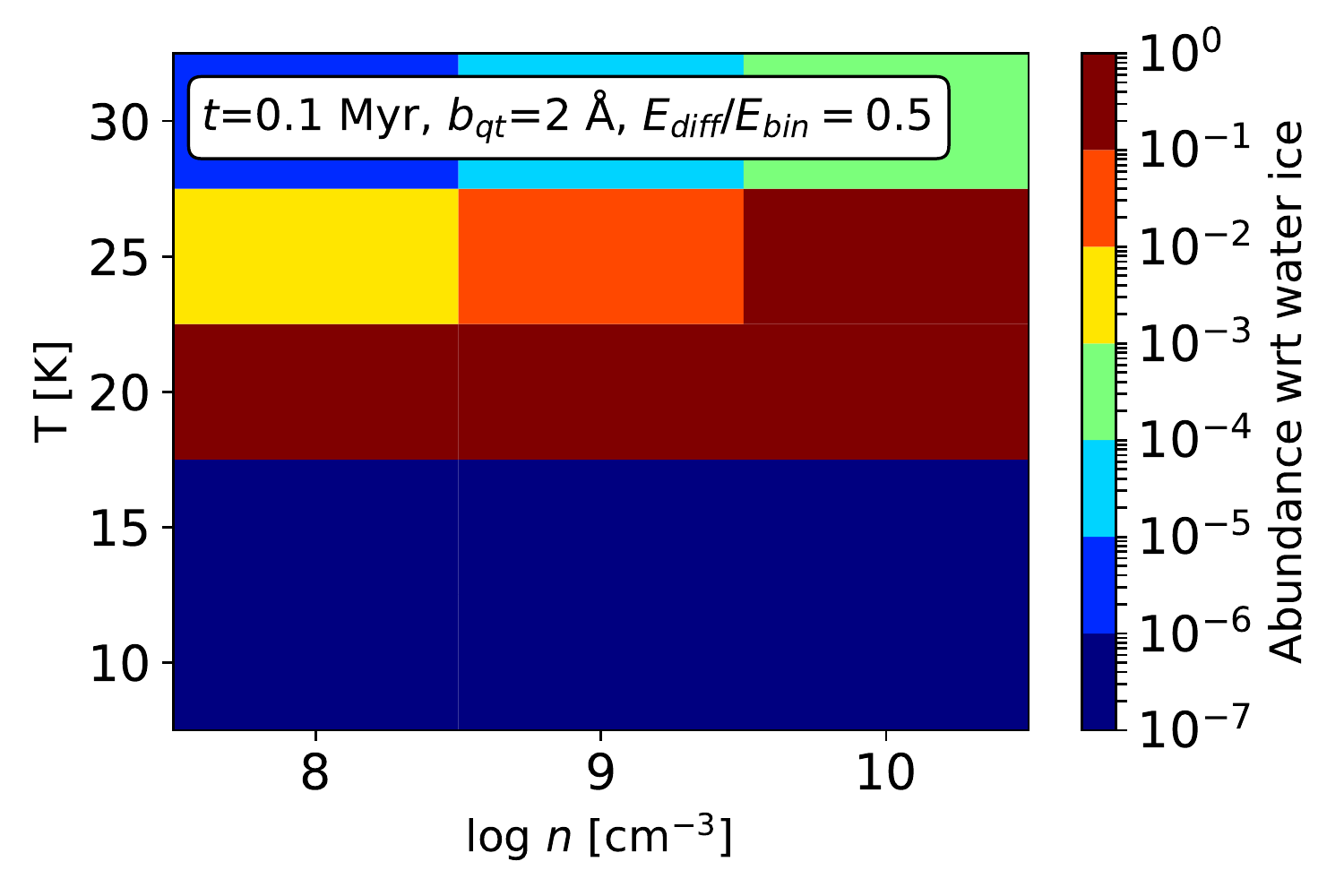}}\\
\subfigure{\includegraphics[width=0.22\textwidth]{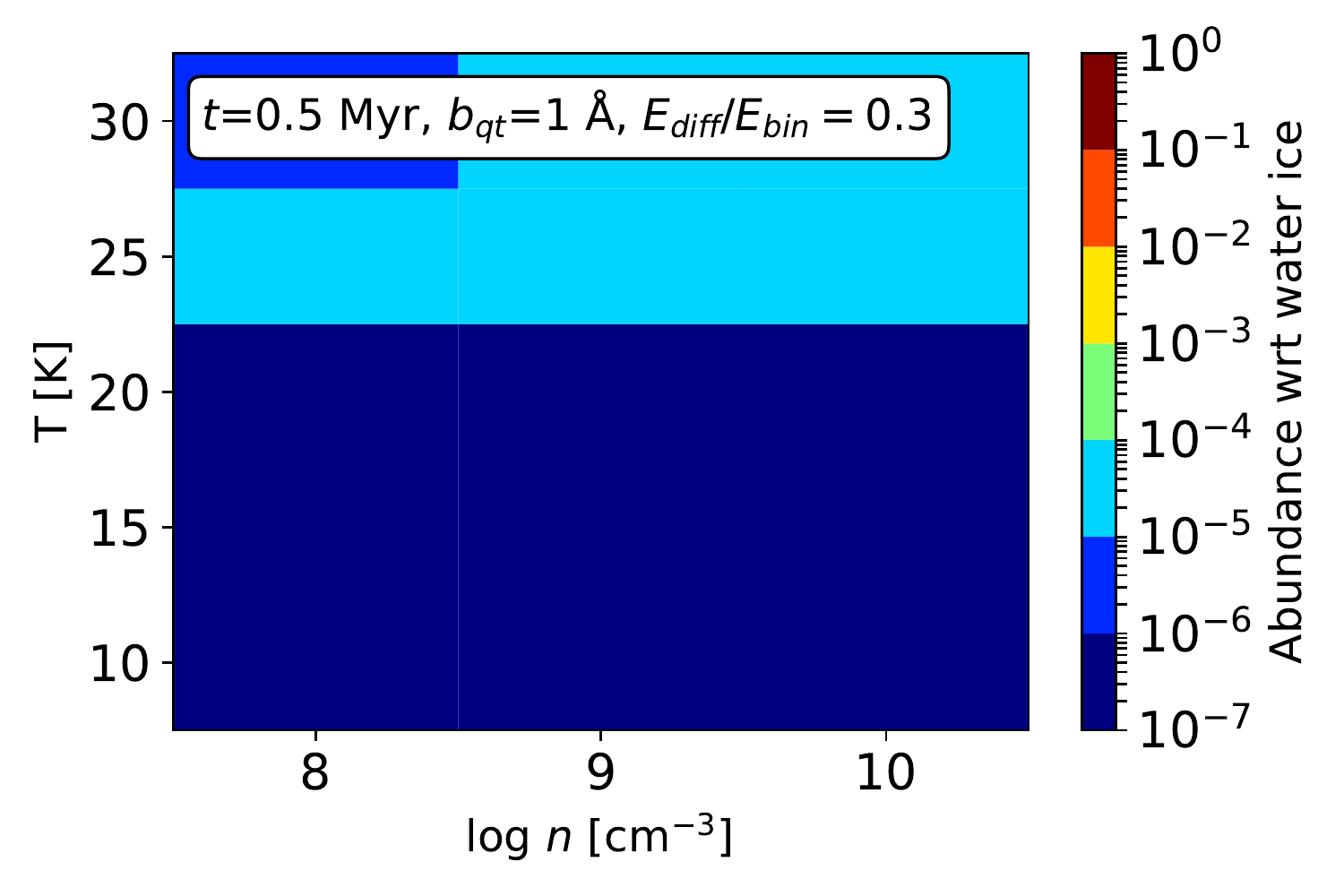}}
\subfigure{\includegraphics[width=0.22\textwidth]{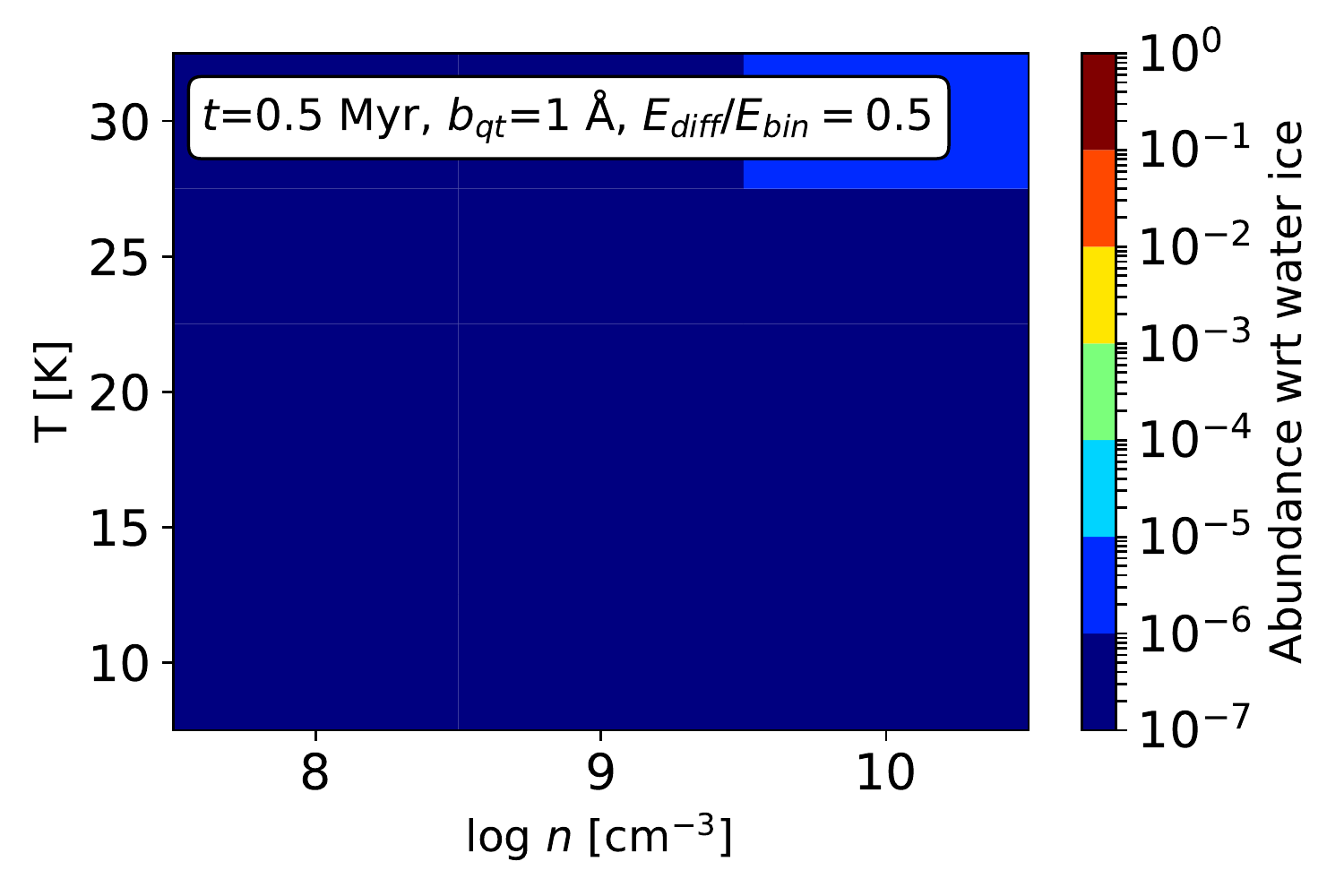}}
\subfigure{\includegraphics[width=0.22\textwidth]{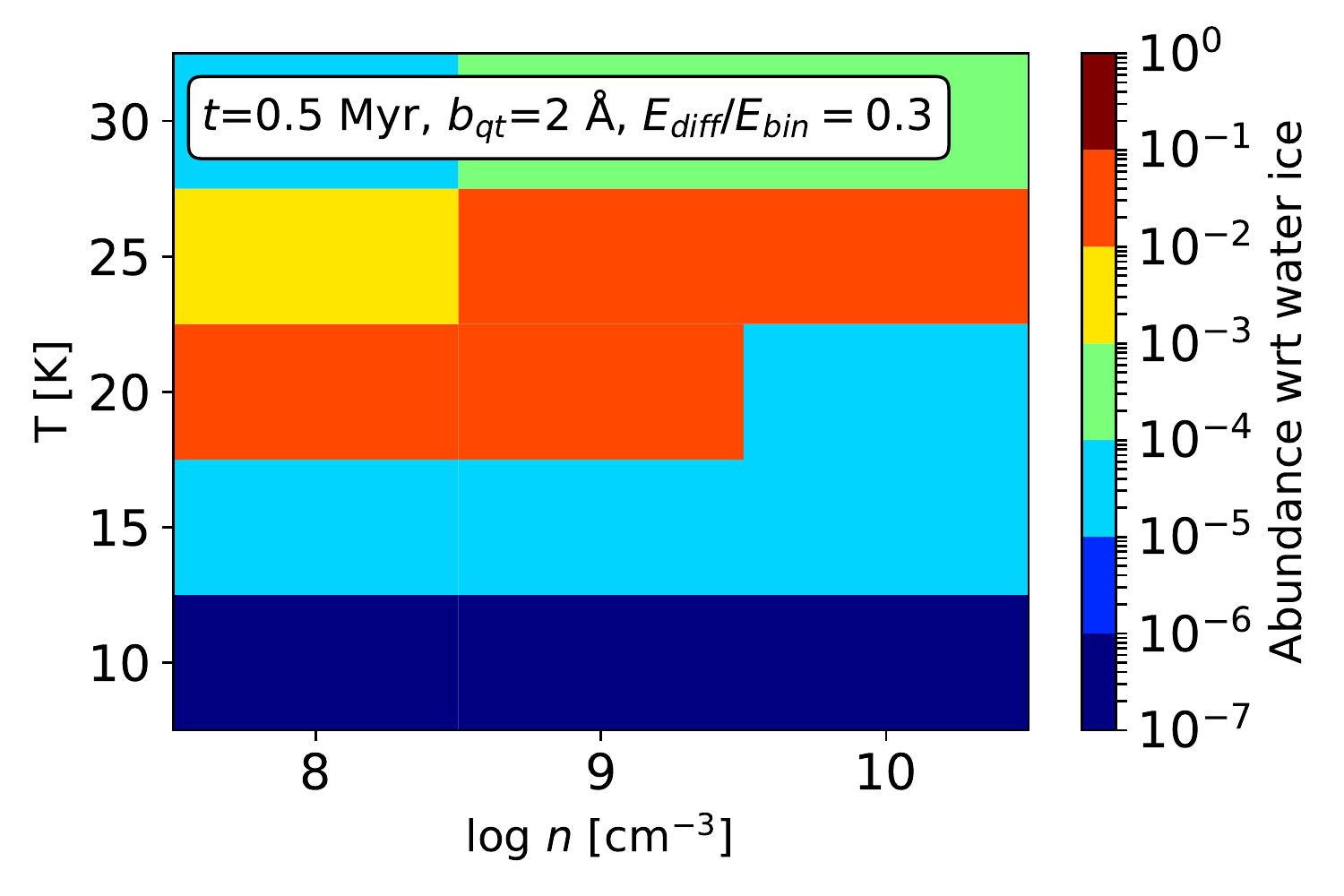}}
\subfigure{\includegraphics[width=0.22\textwidth]{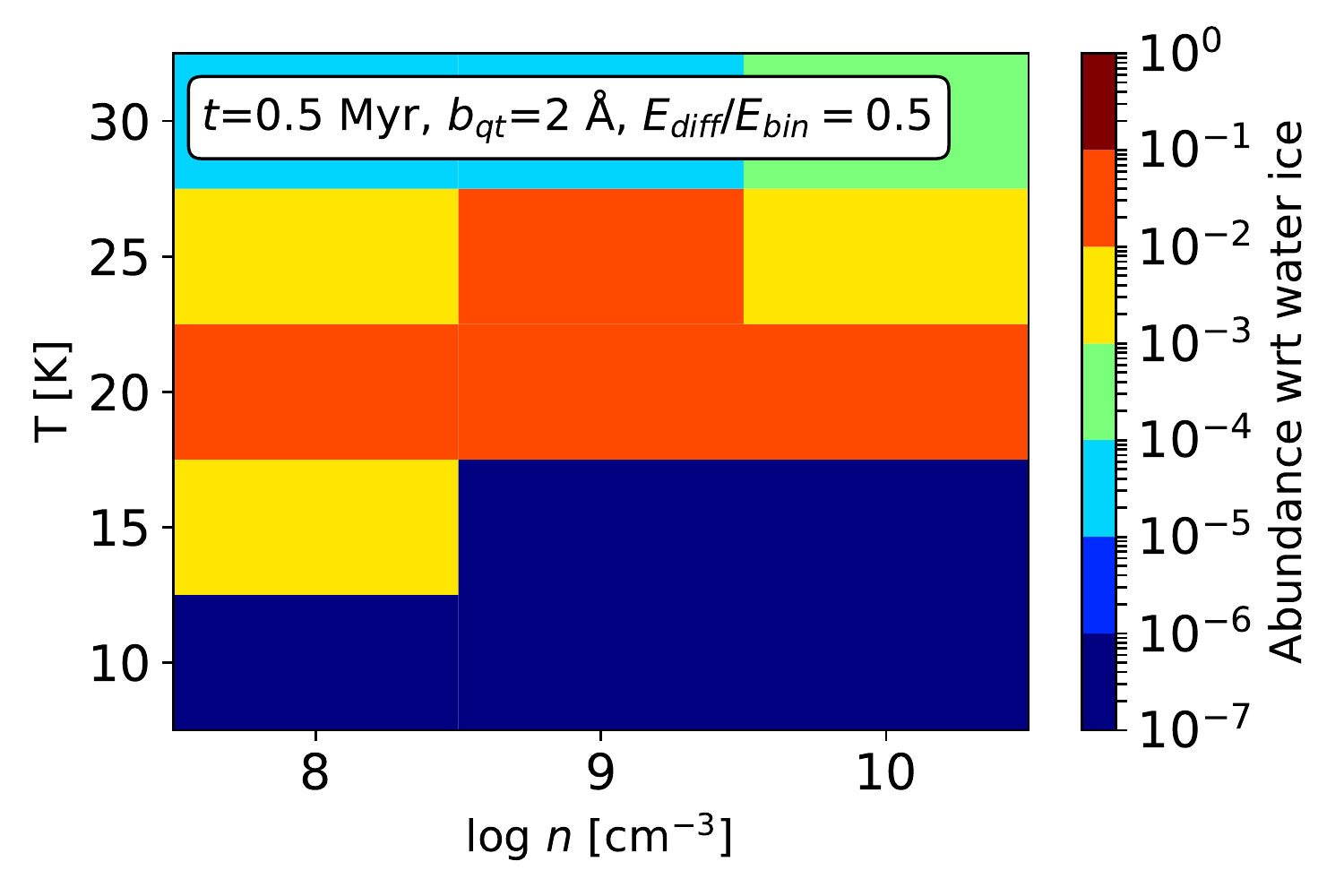}}\\
\subfigure{\includegraphics[width=0.22\textwidth]{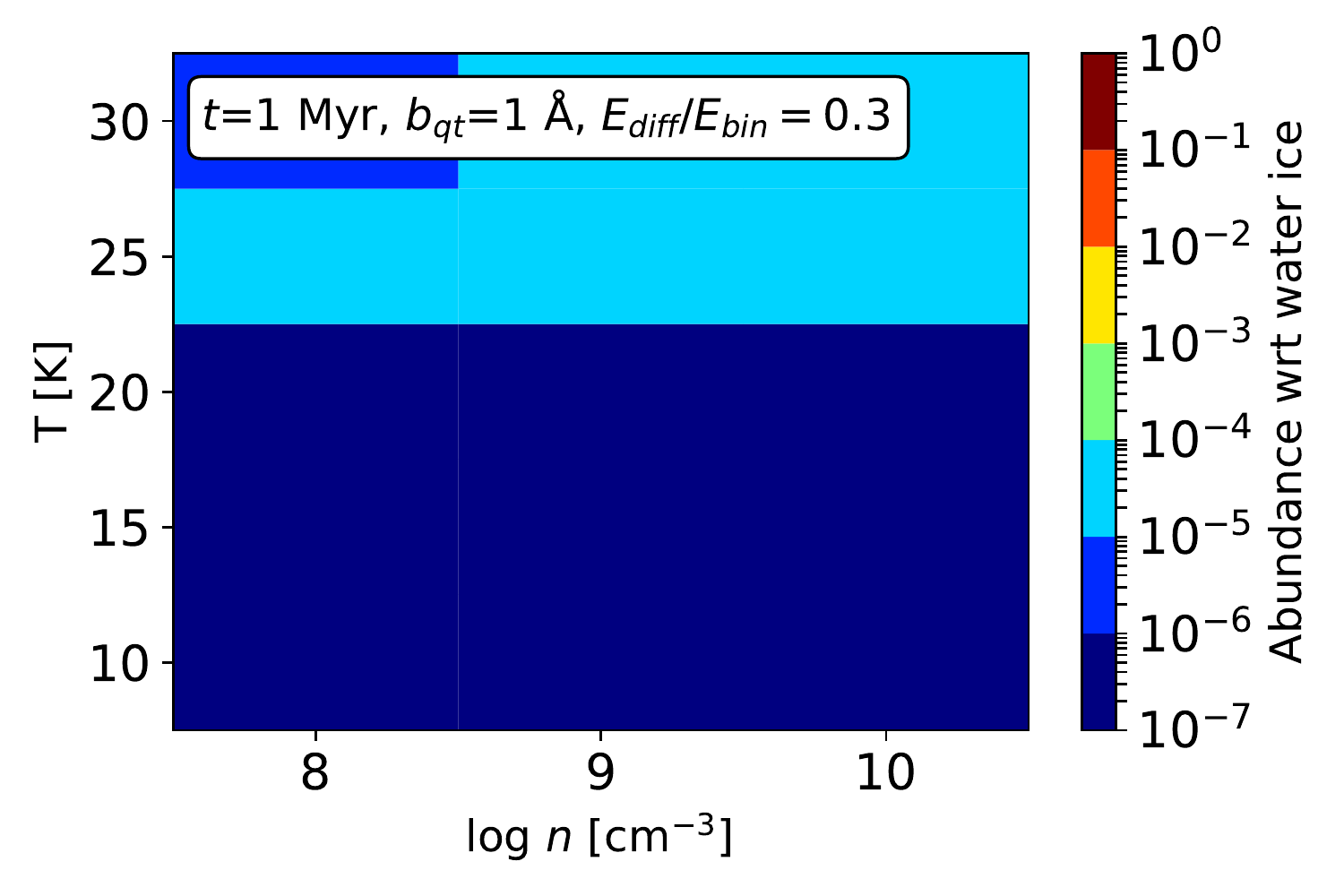}}
\subfigure{\includegraphics[width=0.22\textwidth]{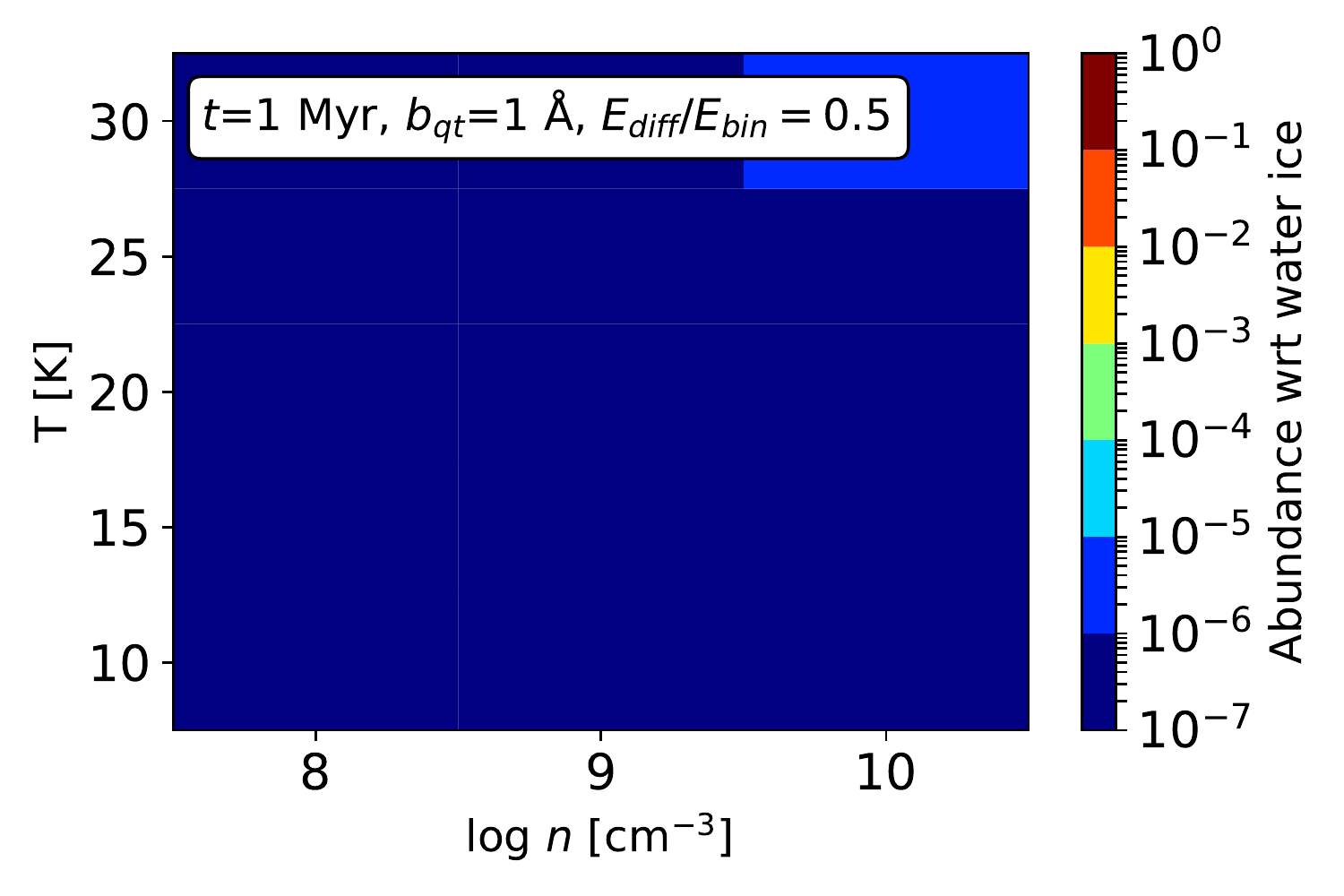}}
\subfigure{\includegraphics[width=0.22\textwidth]{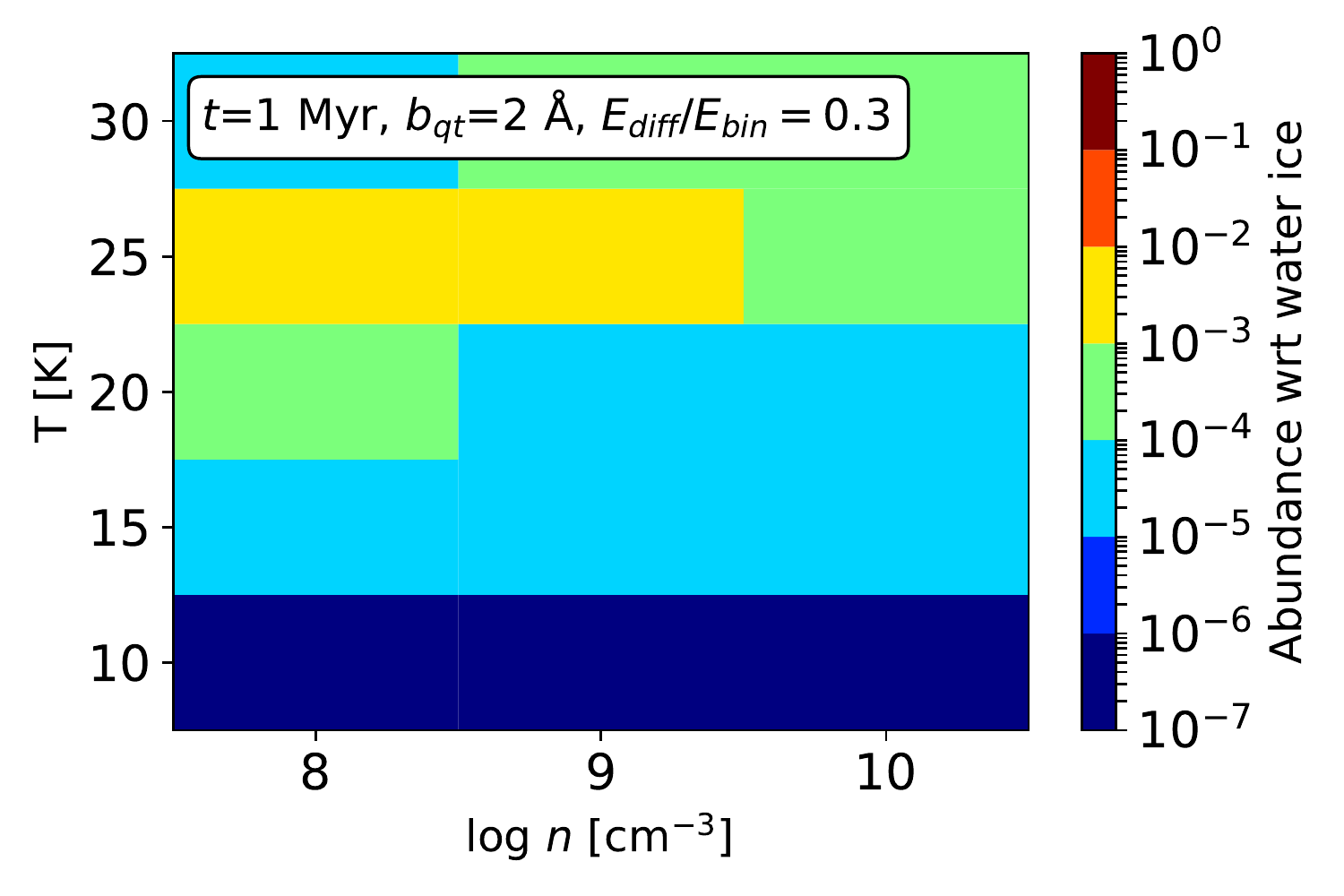}}
\subfigure{\includegraphics[width=0.22\textwidth]{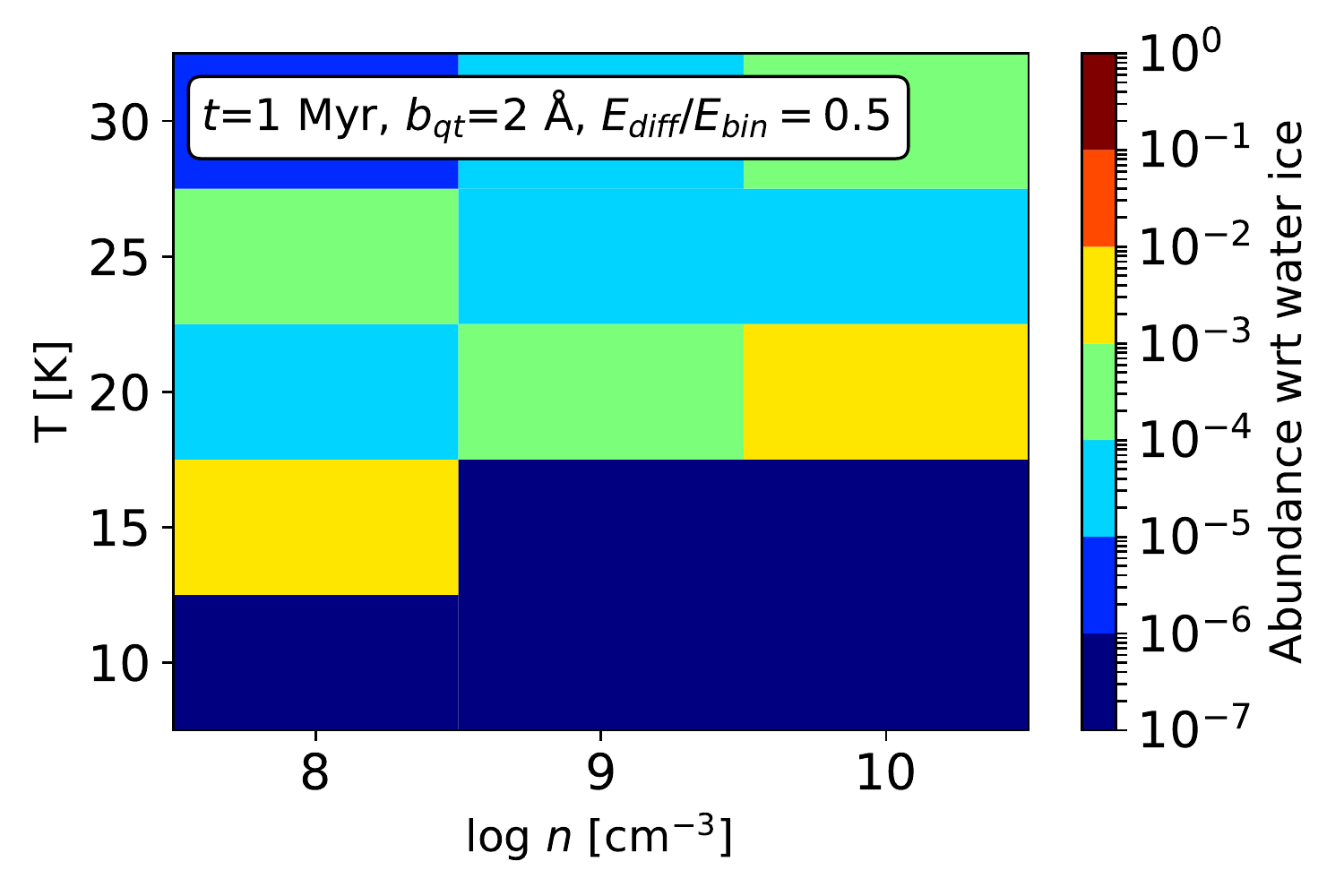}}\\
\subfigure{\includegraphics[width=0.22\textwidth]{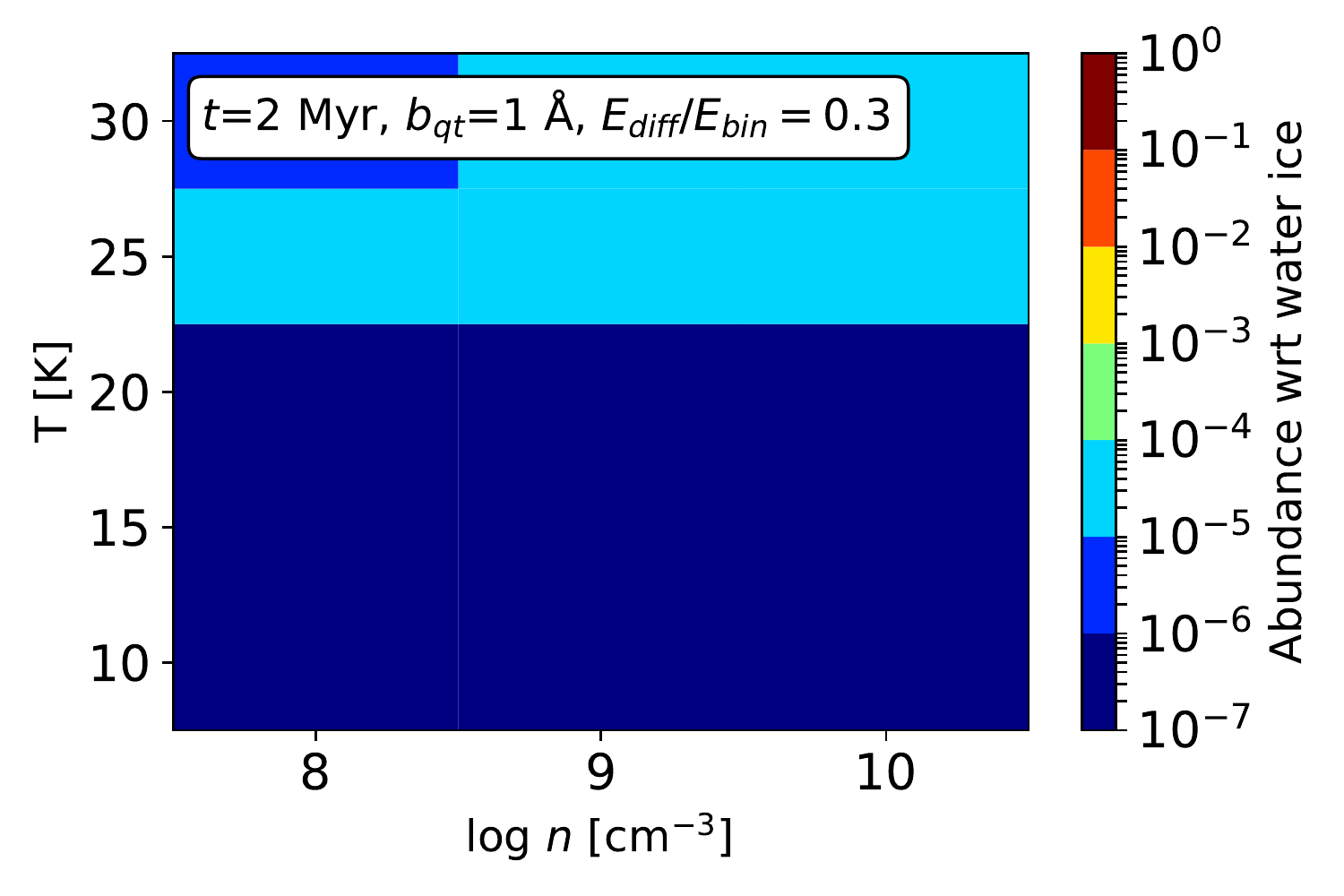}}
\subfigure{\includegraphics[width=0.22\textwidth]{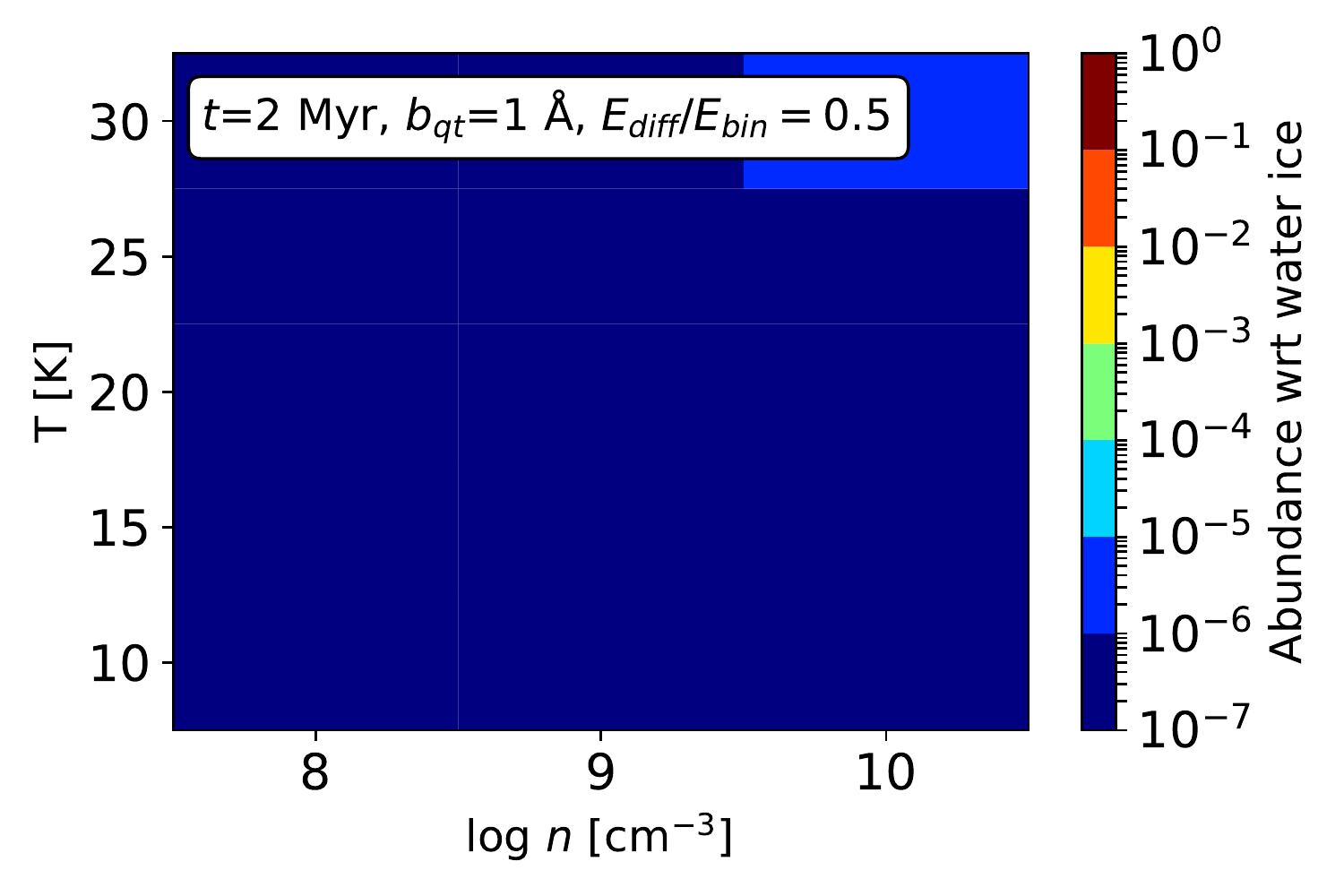}}
\subfigure{\includegraphics[width=0.22\textwidth]{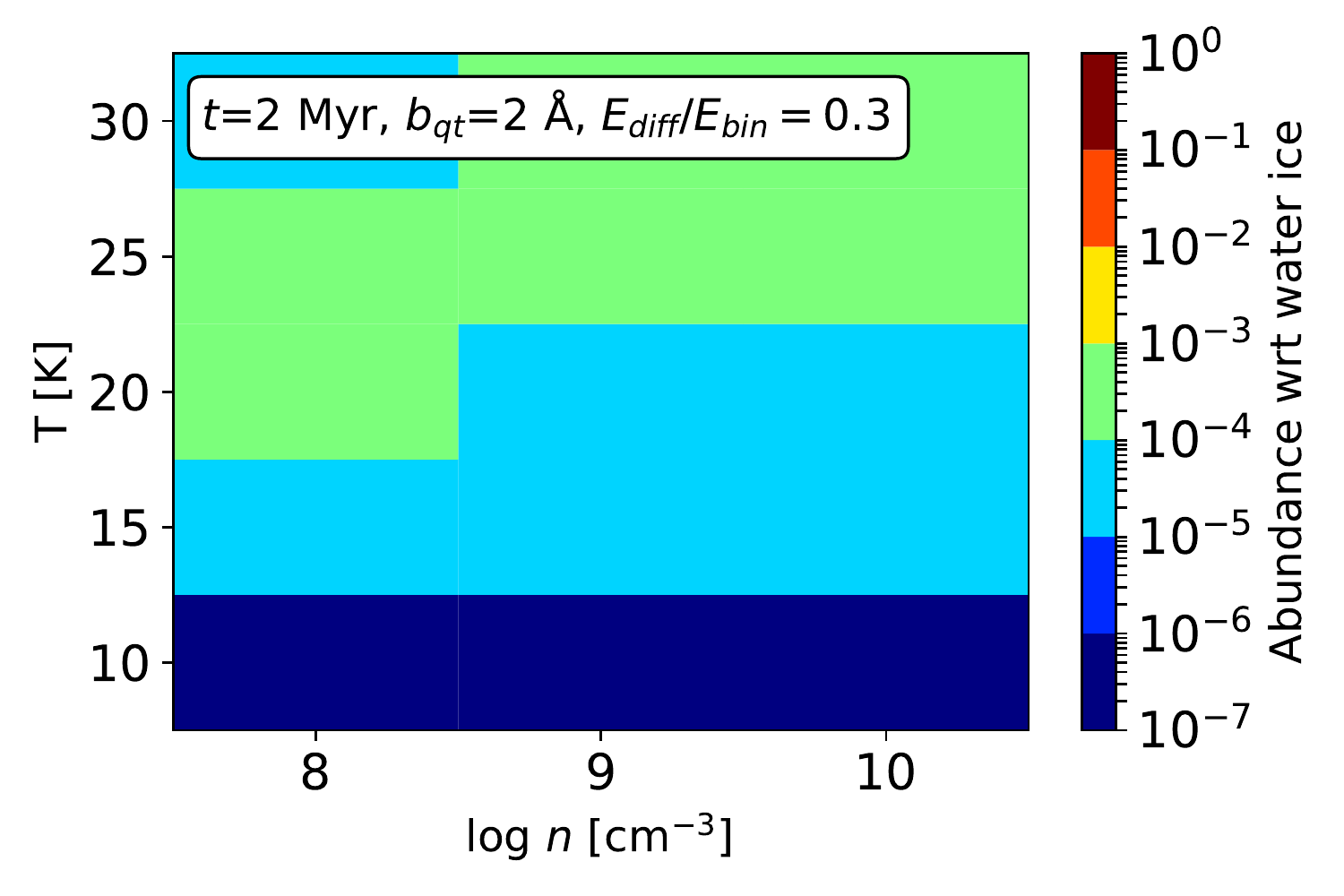}}
\subfigure{\includegraphics[width=0.22\textwidth]{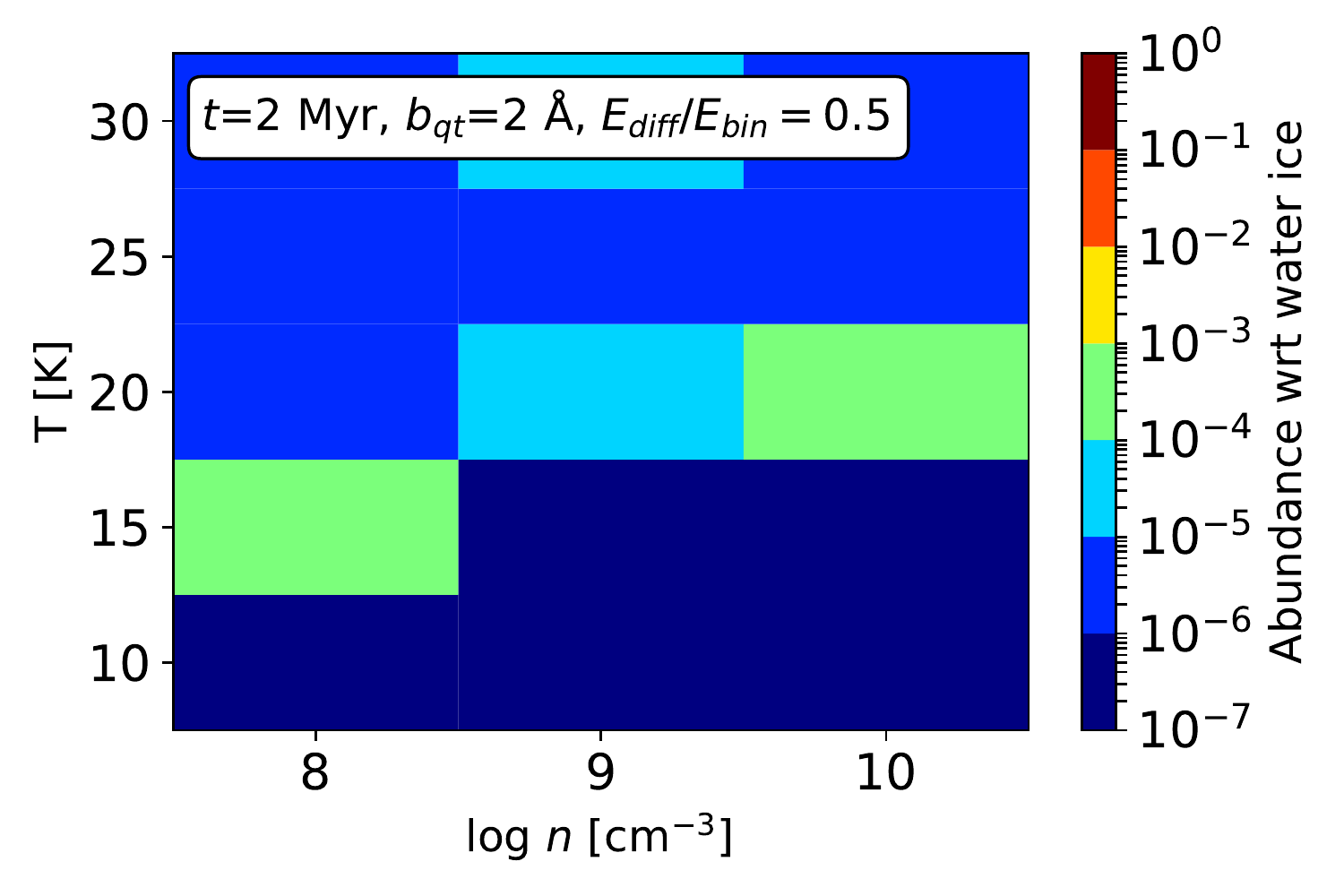}}\\
\subfigure{\includegraphics[width=0.22\textwidth]{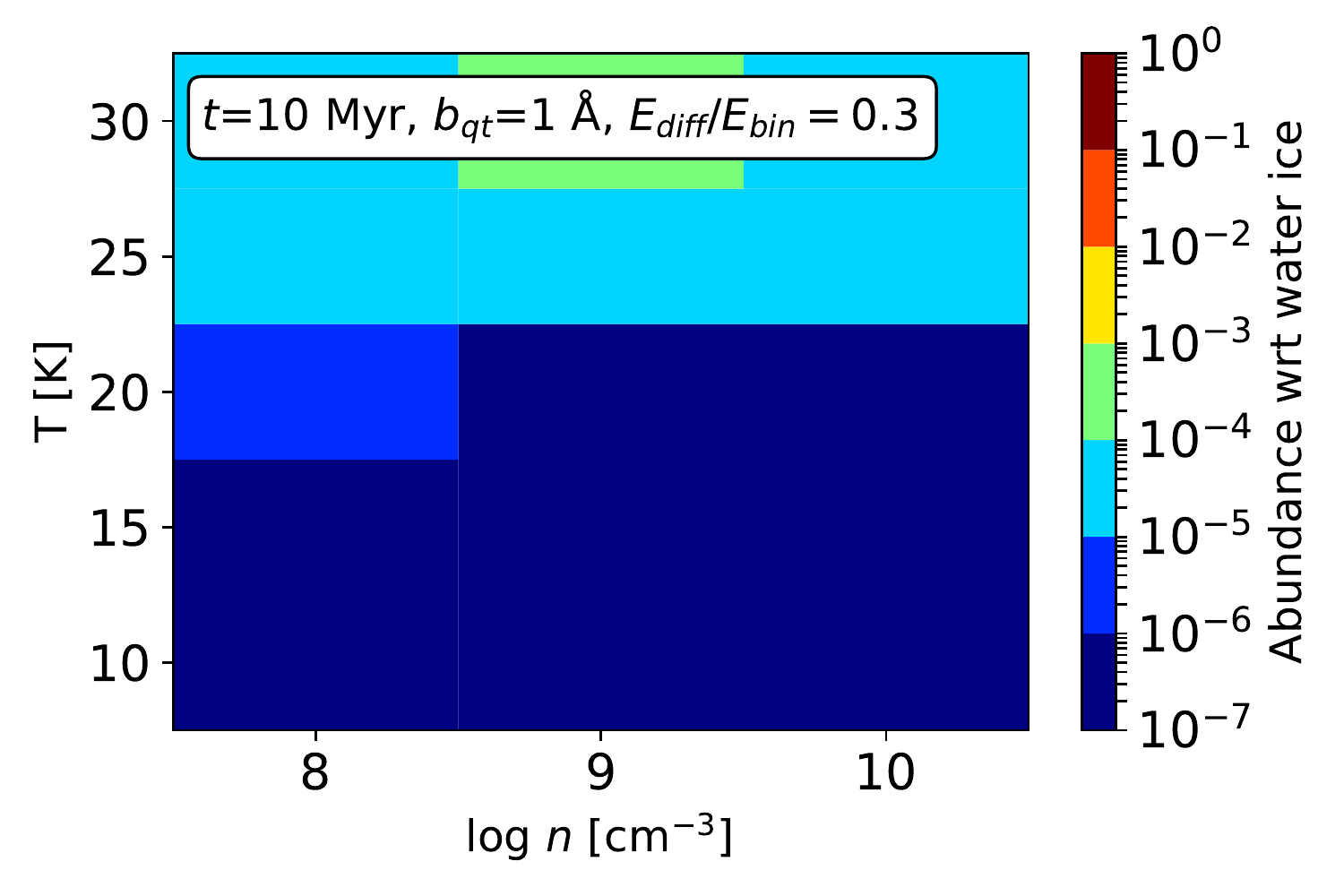}}
\subfigure{\includegraphics[width=0.22\textwidth]{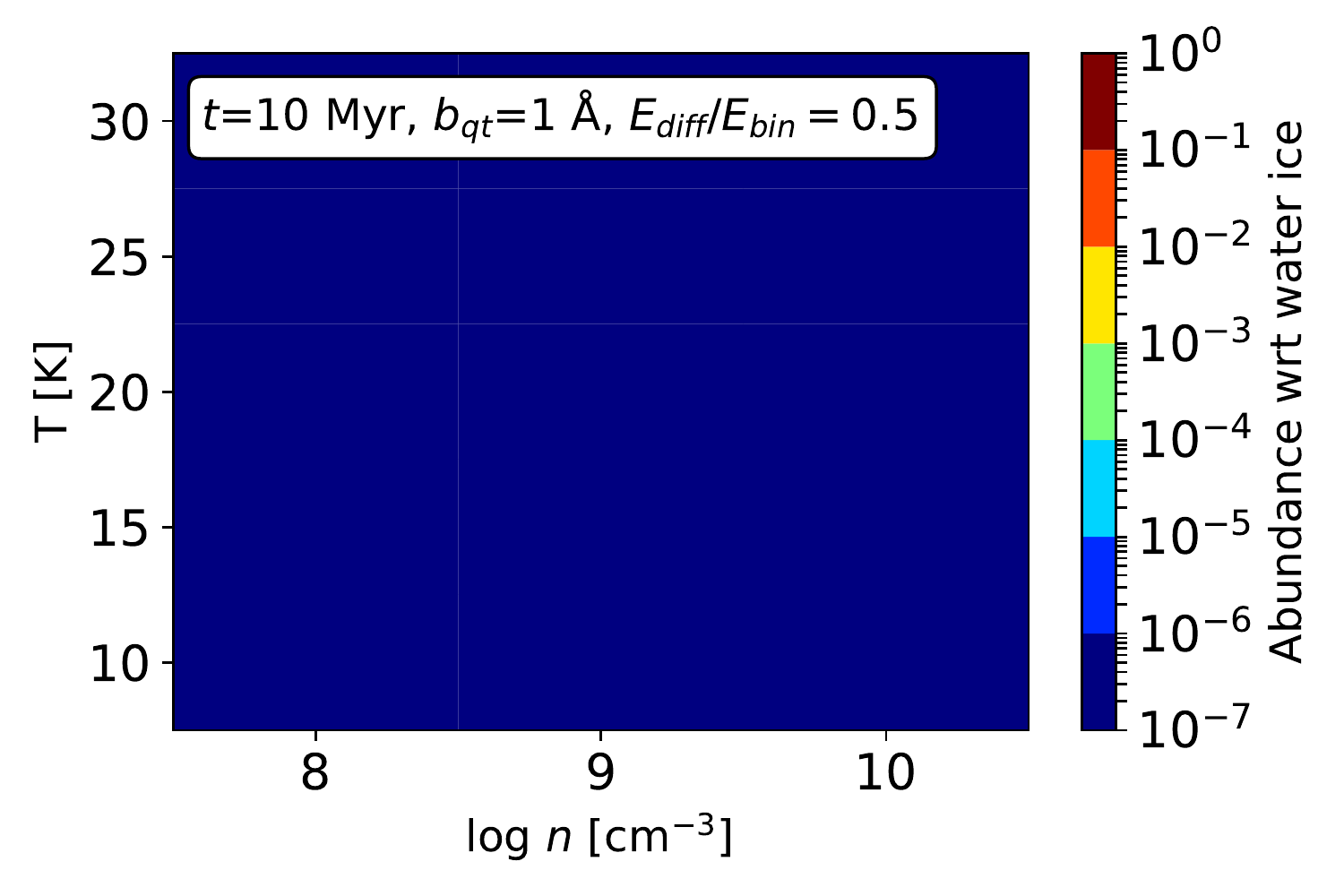}}
\subfigure{\includegraphics[width=0.22\textwidth]{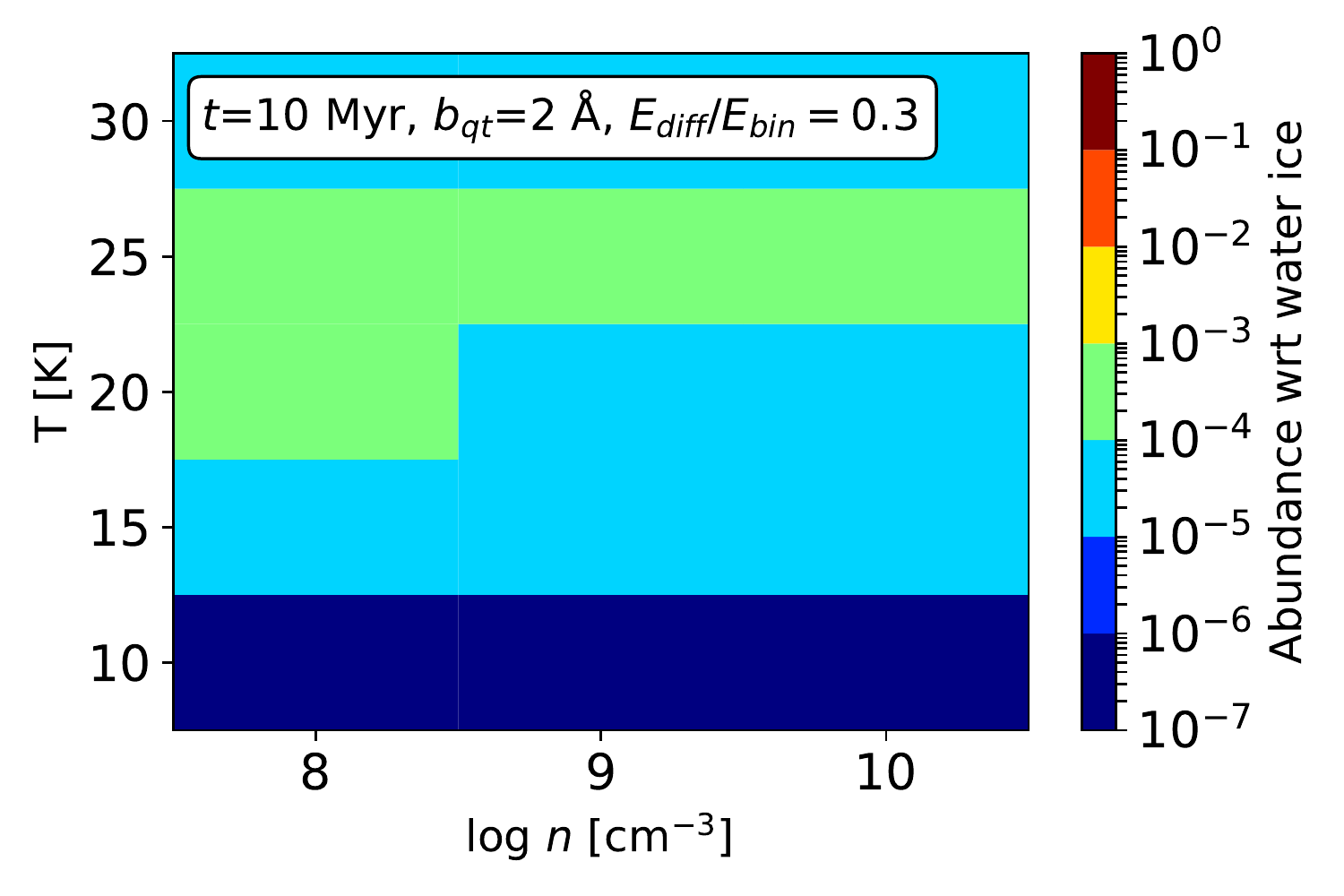}}
\subfigure{\includegraphics[width=0.22\textwidth]{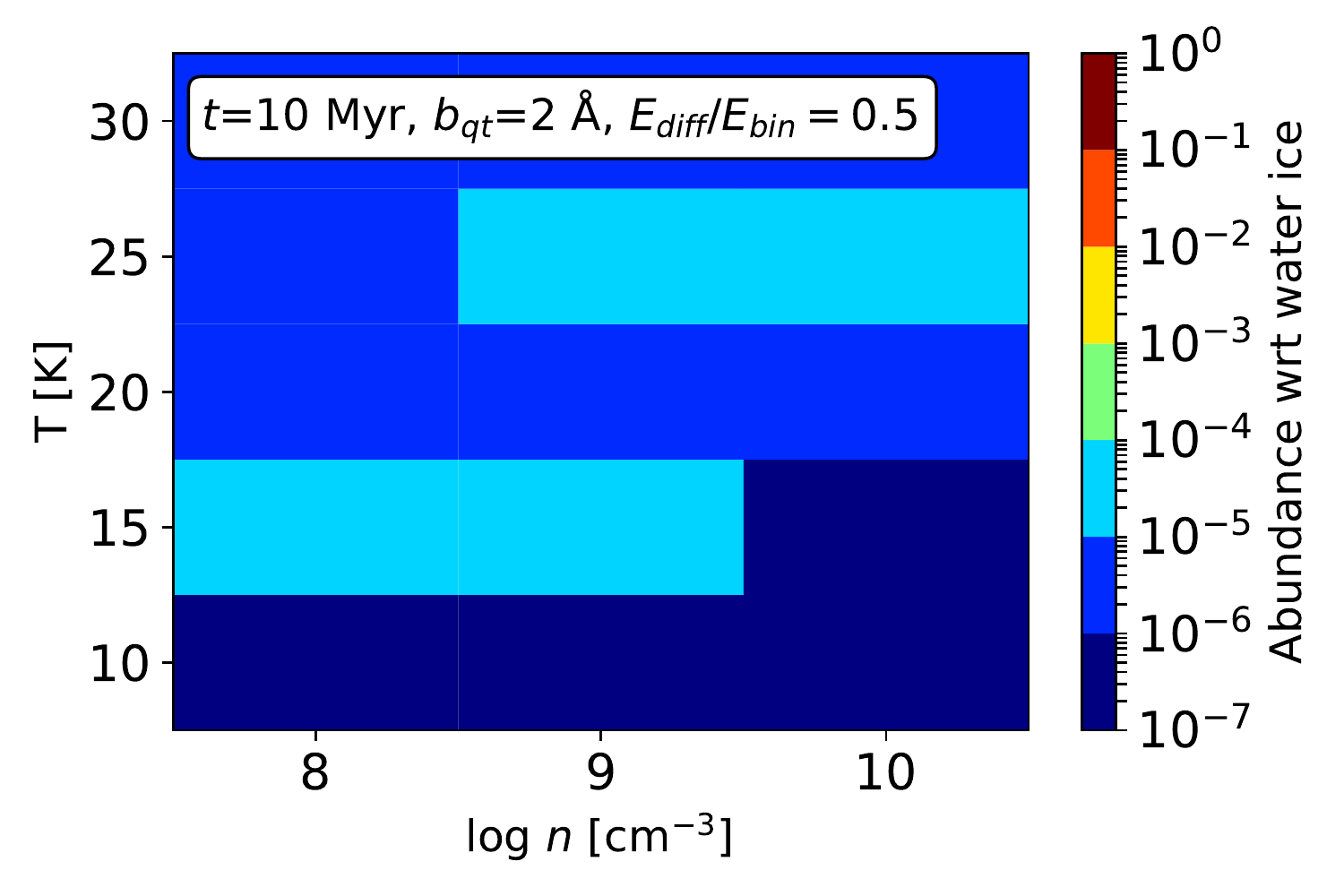}}\\
\caption{Abundances (given as colors) for \ce{O2} ice as function of midplane density ($x$-axes), and temperature ($y$-axes) at different evolutionary steps for the reset scenario. From left to right are abundances from model runs with different parameters for grain-surface reactions: columns one and two feature $b_{qt}$= 1 {\AA}, columns three and four feature $b_{qt}$= 1 {\AA}, columns one and three are with $E_{\rm{diff}}/E_{\rm{bin}}$= 0.3, and columns two and four are with $E_{\rm{diff}}/E_{\rm{bin}}$= 0.5. Top to bottom are different evolutionary times, from 0.05 Myr (top) to 10 Myr (bottom). To the right of each plot is a colorbar, indicating the abundance level with respect to \ce{H2O} ice for each color. The chemical network utilised includes \ce{O3} chemistry. Orange matches cometary \ce{O2} abundances.}
\label{chem_params_o2}
\end{figure*}

Figures \ref{chem_params_h2o}, \ref{chem_params_o2}, \ref{chem_params_o3} and \ref{chem_params_h2o2} show the abundances of \ce{H2O} ice with respect to H$_{\rm{nuc}}$, and \ce{O2} ice, \ce{O3} ice and \ce{H2O2} ice with respect to \ce{H2O} ice, again, maintaining the convention used by the comet community. From top to bottom span evolution times from 5$\times10^{4}$ yr up to 10$^{7}$ yr. In each panel in the figures a mosaic of color-indicated abundances shows the results across different combinations of midplane temperature (ranging from 10-30 K on the $y$-axis), and densities (ranging from $10{^8}-10{^{10}}$ cm$^{-3}$ on the $x$-axis). The ionisation is kept constant at $\zeta = 10^{-17}$ s$^{-1}$, approximately the same as ionisation Level 2 from Fig. \ref{high_mass_phys} in the outer disk midplane. This ionisation level was seen in Fig. \ref{abun_evol_res_old_water} (model without \ce{O3}) to match the observed \ce{O2} ice abundance early in the evolution (up to 0.1 Myr). 


From left to right in each of the Figs. \ref{chem_params_h2o} to \ref{chem_params_o3} two grain-surface reactions parameters are changed. The first one is the barrier width for quantum tunnelling $b_{\rm{qt}}$ which is either 1 {\AA} (columns one and two, in each figure) or 2 {\AA} (columns three and four, in each figure). The second parameter is the ratio of diffusion energy to molecular binding energy $E_{\rm{diff}}/E_{\rm{bin}}$, which can be either 0.3 (columns one and three, in each figure) or 0.5 (columns two and four, in each figure). This amount to four combinations of the two reaction parameters, thus the four columns of mosaics in each figure. The values of each parameter associated with a given model setup is given in each panel.

The range of evolution time steps is chosen to cover both the evolutionary stage at which \ce{O2} ice was abundant in Fig. \ref{abun_evol_res_old_water} ($5\times 10^{4}$ yr, for ionisation Level 2 $\sim10^{-17}$ s$^{-1}$), as well as the later stages of disk evolution. The colors in the mosaic are chosen to enable distinguishing between abundance levels that change per order-of-magnitude, with dark red (top color in color bar) representing the highest abundance, and dark blue (bottom color in color bar) representing the lowest. The range of abundance levels in each figure is chosen to cover the full range of abundances produced for each species in the models. By showing the abundance evolutions in this way model results matching the observed abundances of e.g. \ce{O2} ice (1-10\% of \ce{H2O} ice) will show as orange (second highest color in colorbar) in Fig. \ref{chem_params_o2}.

For \ce{H2O} ice in Fig. \ref{chem_params_h2o} it is seen that for columns one and two ($b_{\rm{qt}}$=1 {\AA}) the abundance level largely remains within the initial order of magnitude of $10^{-4}$ with respect to \ce{H2O} ice. For columns three and four ($b_{\rm{qt}}$=2 {\AA}), some change is seen: especially for the third column ($E_{\rm{diff}}/E_{\rm{bin}}$=0.3), the abundance is an order of magnitude lower for temperatures 25-30 K than for 10-20 K, and the abundance increases with time at 20 K. Turning to Fig. \ref{chem_params_o2} for \ce{O2} ice, limited production is seen for columns one and two, with an early peak abundance between $10^{-4}-10^{-3}$ with respect to \ce{H2O} ice at 25 K for $n=10^{8}-10^{9}$ cm$^{-3}$ by $5\times 10^{4}$ yr. More interesting are columns three and four: until 0.5 Myr, abundances of $10^{-2}-1$ with respect to \ce{H2O} ice are reached for temperatures 15-25 K for all densities. Thus the \ce{O2} ice abundance lies above or reproduces the observed values. For these cases, by >1 Myr, the produced \ce{O2} ice is destroyed and reaches $10^{-5}-10^{-3}$ with respect to \ce{H2O} ice by 10 Myr.

Having narrowed in on this parameter range (early evolutionary times, and $b_{\rm{qt}}$=2 {\AA}), it can now be checked if \ce{H2O2} and \ce{O3} ices can match the observed levels for the same model parameters. Fig. \ref{chem_params_o3} shows the same suite of plots for \ce{O3} ice. Results in columns three and four up to 0.5 Myr evolution show that \ce{O3} ice is abundantly produced, at a similar or higher level than \ce{O2} ice, thus not matching the observed upper limit (darkest shade of blue, <$10^{-6}$ with respect to \ce{H2O} ice). The subsequent destruction by >1 Myr does not bring the abundance level much further down, and the \ce{O3} ice abundance is generally similar to that for \ce{O2} ice. \ce{O3} ice only matches the observed upper limit for $T=10$ K.

In Fig. \ref{chem_params_h2o2}, \ce{H2O2} ice is abundantly produced for $b_{\rm{qt}}$=2 {\AA} and $E_{\rm{diff}}/E_{\rm{bin}}$=0.3. For 25 K it is consistently at a level of $10^{-1}-1$ with respect to \ce{H2O} ice during the entire evolution. For 20 K it is also abundant, although 1-2 orders of magnitude lower than that found at 25 K, and it is less abundant than \ce{O2} ice until $\sim0.5$ Myr by 1-2 orders of magnitude. At later times, for 20-25 K, \ce{H2O2} ice becomes 1-3 orders of magnitude more abundant than \ce{O2} ice. For $E_{\rm{diff}}/E_{\rm{bin}}$=0.5 it is produced at levels $10^{-6}-10^{-3}$ with respect to \ce{H2O} ice, depending on density.

\subsubsection{Abundance evolutions for selected parameter sets}
\label{select_sets_evolutions}

\begin{figure*}[h]
\subfigure{\includegraphics[width=0.33\textwidth]{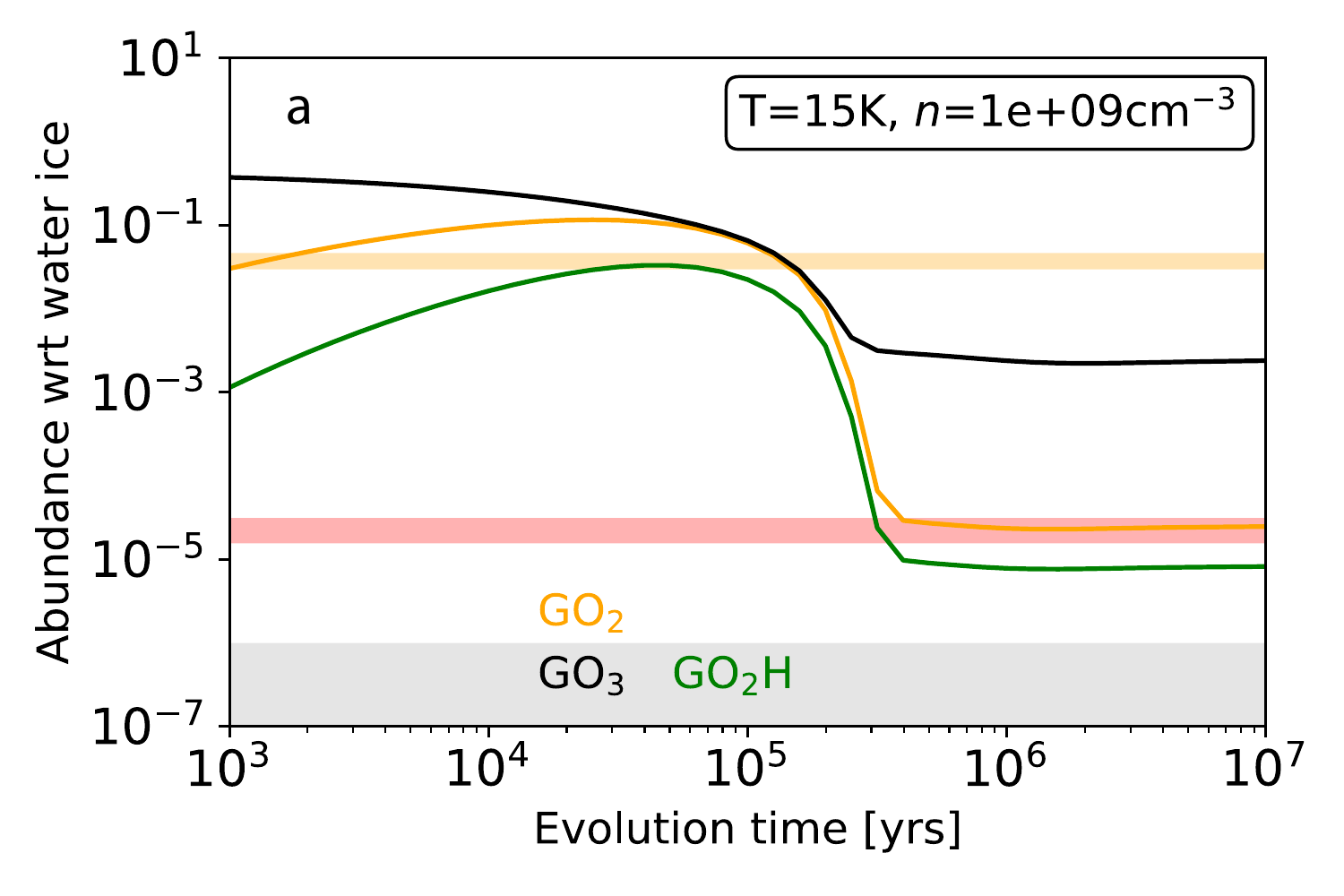}}
\subfigure{\includegraphics[width=0.33\textwidth]{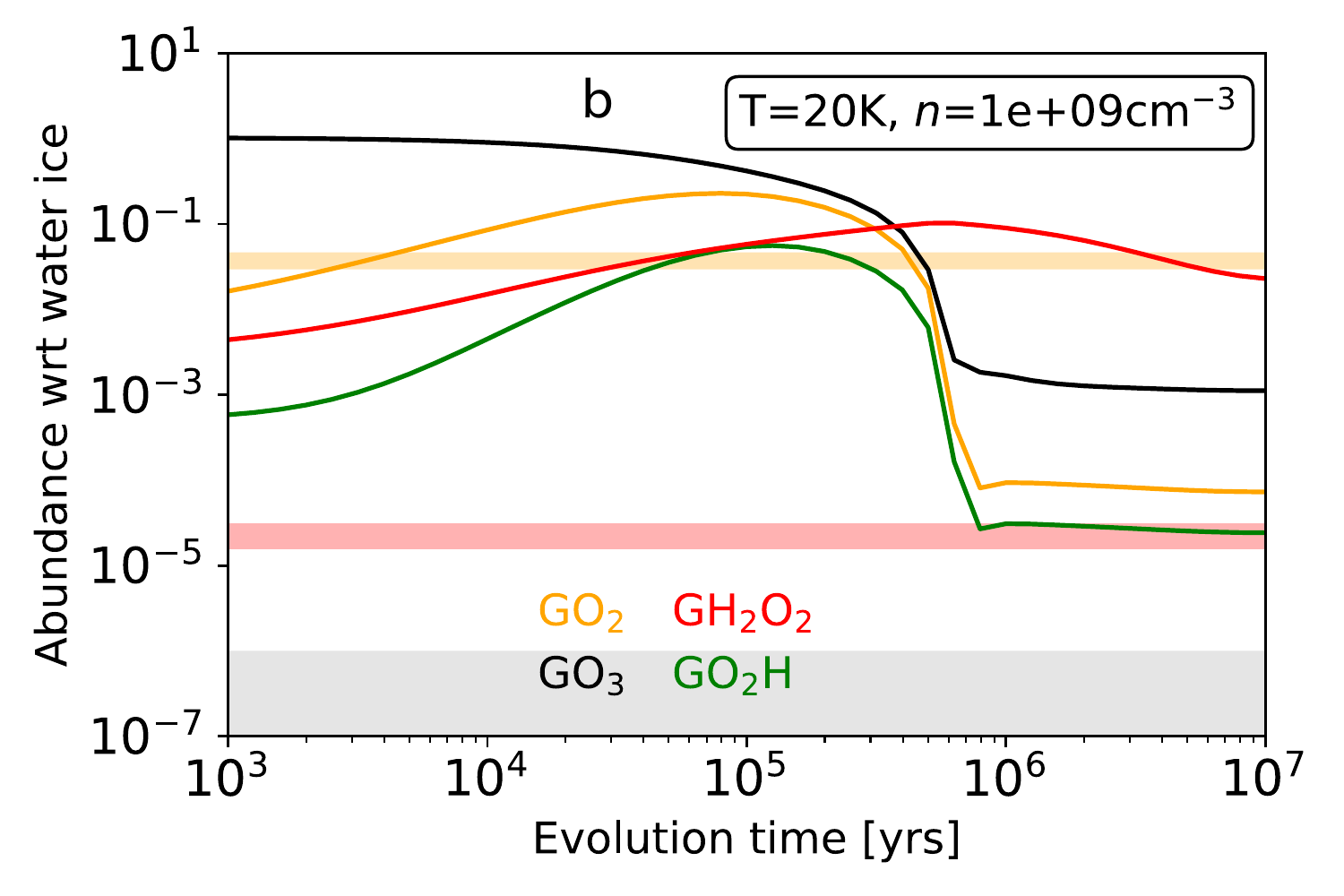}}
\subfigure{\includegraphics[width=0.33\textwidth]{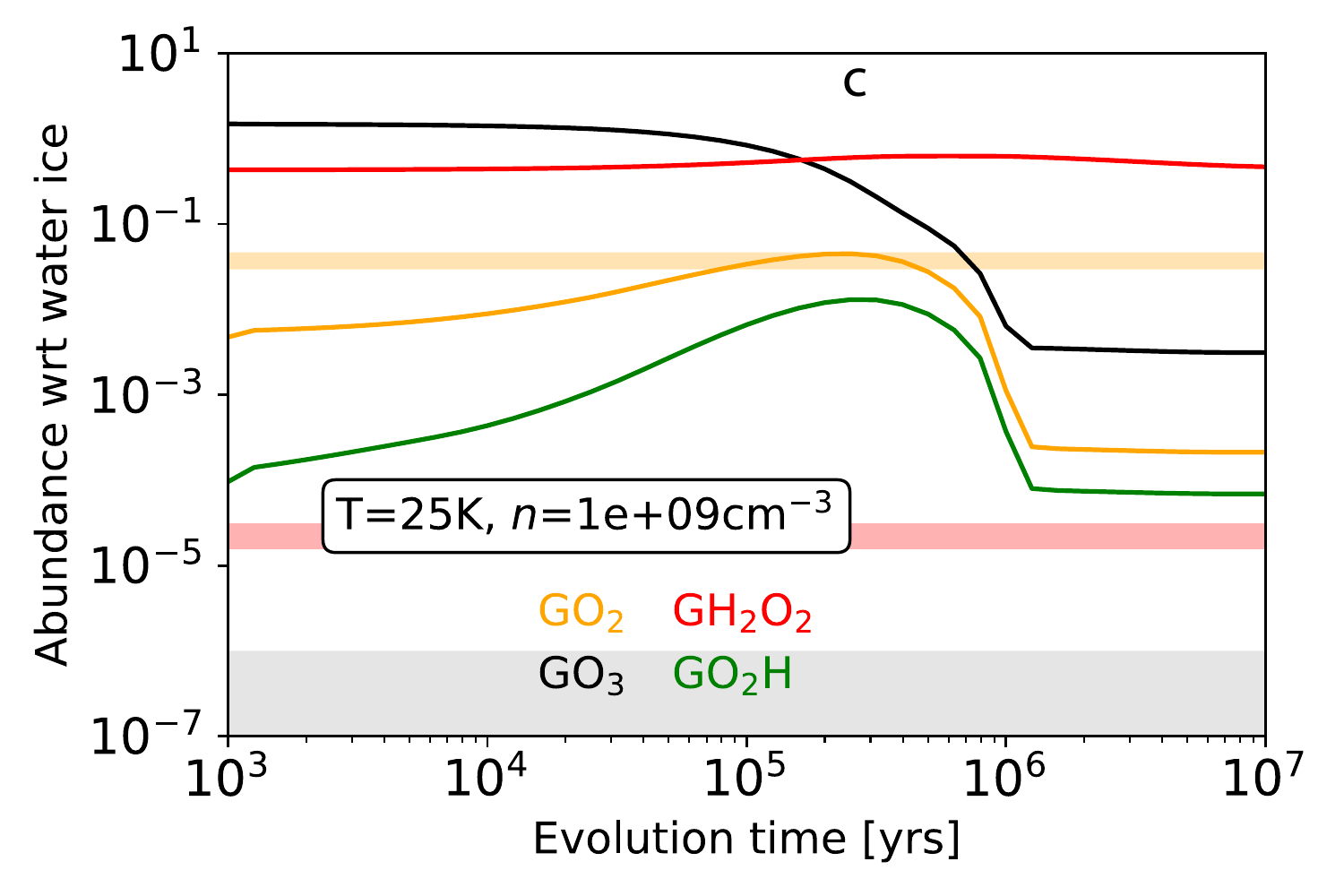}}\\
\subfigure{\includegraphics[width=0.33\textwidth]{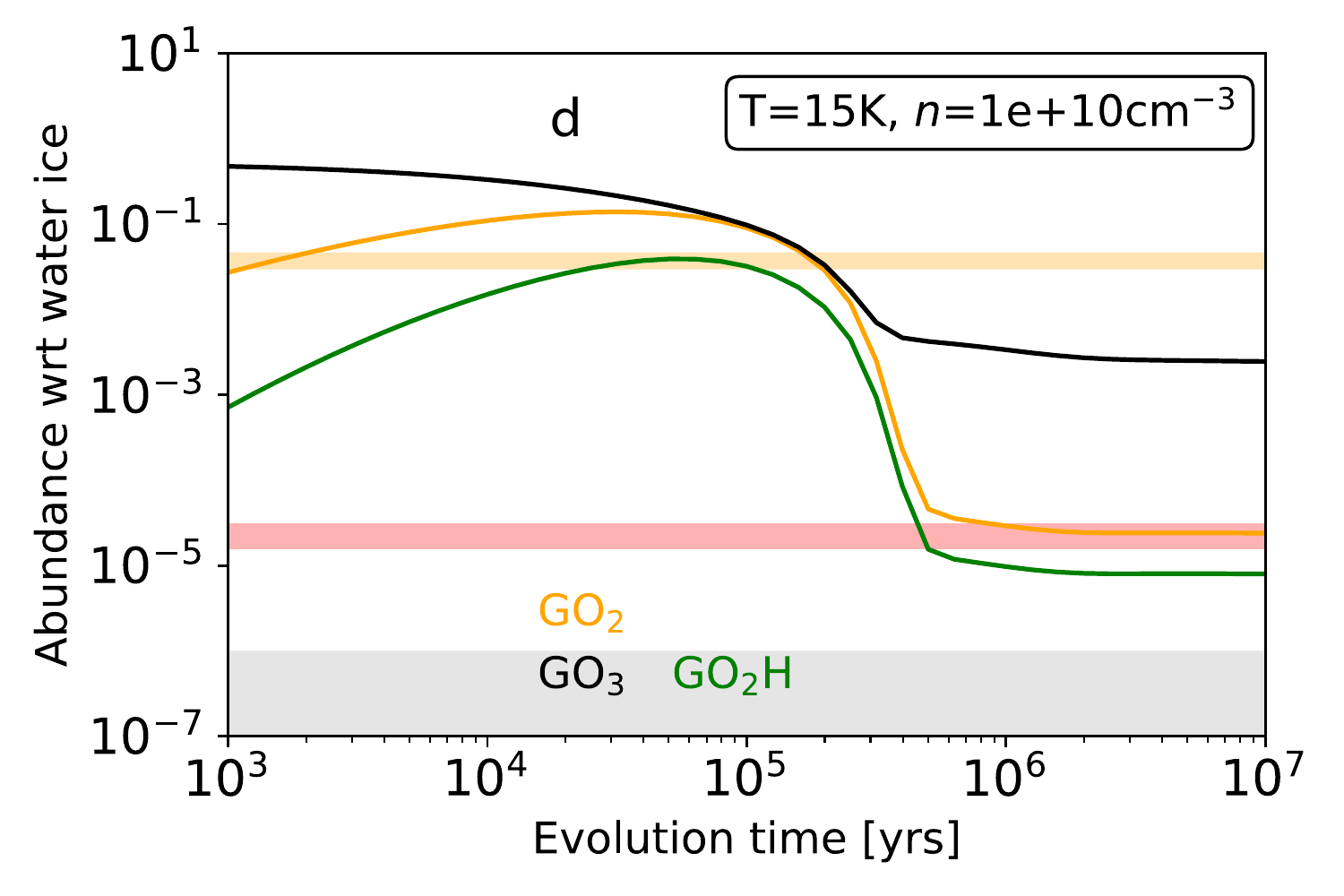}}
\subfigure{\includegraphics[width=0.33\textwidth]{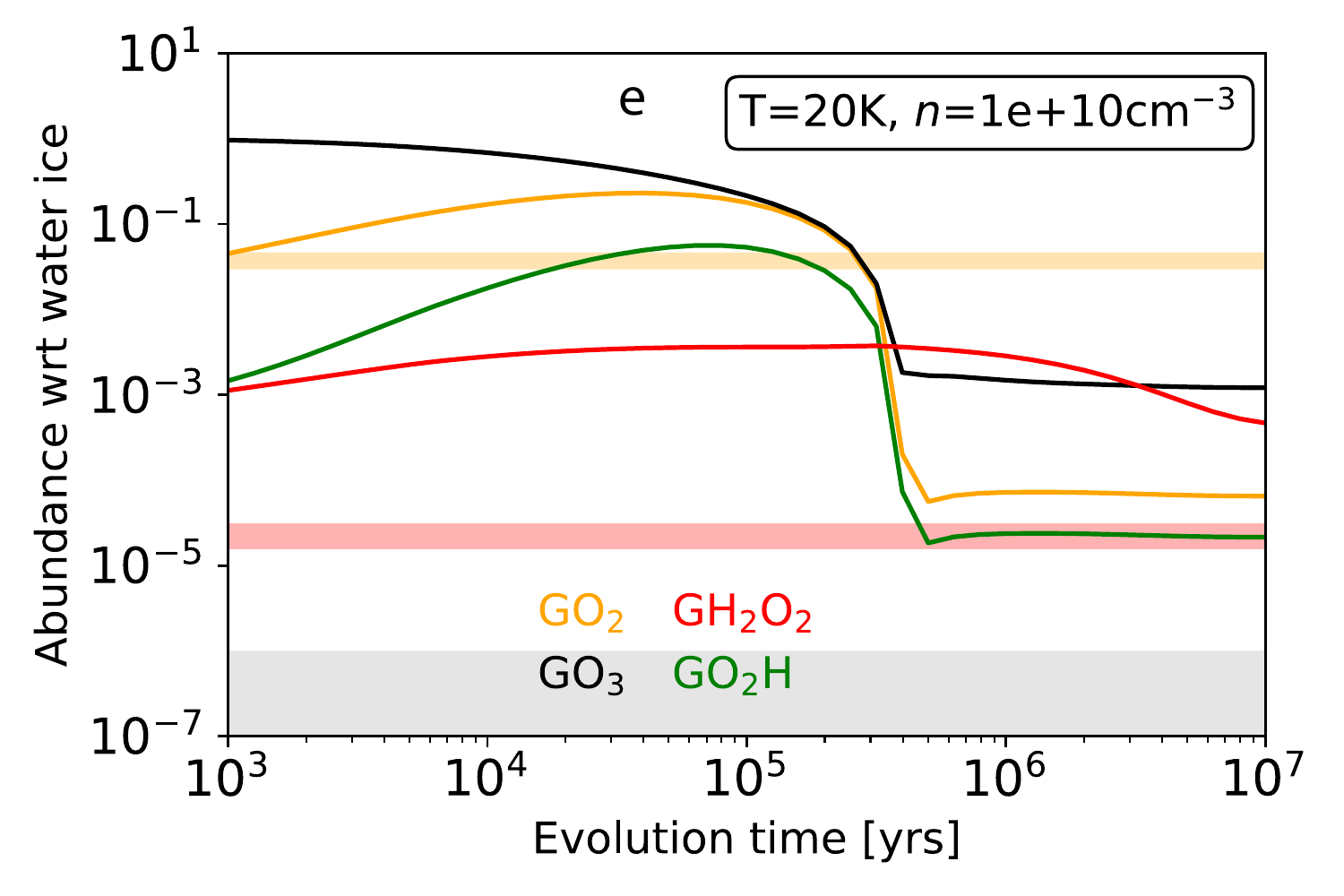}}
\subfigure{\includegraphics[width=0.33\textwidth]{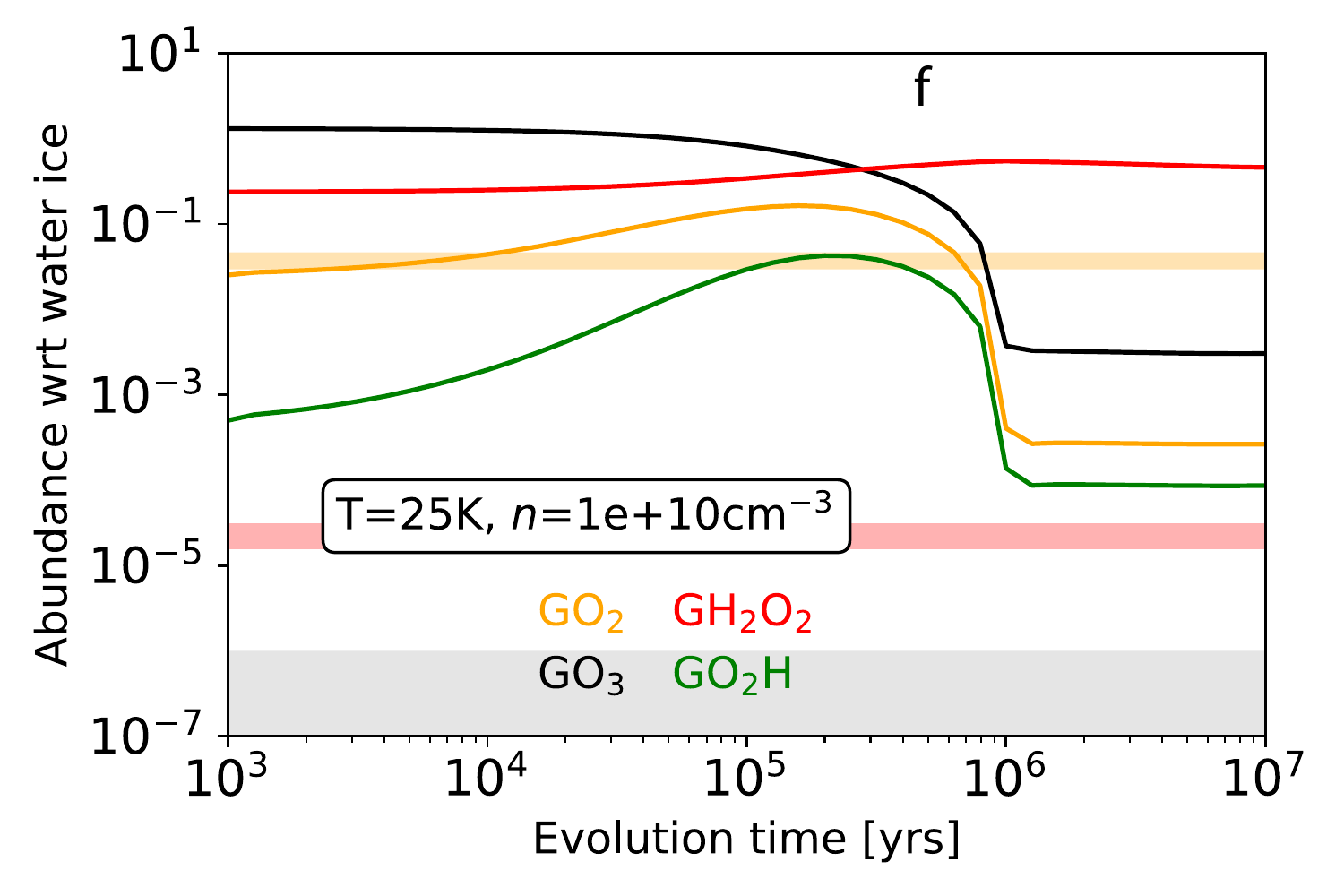}}\\
\caption{Evolving abundances as function of time for the reset scenario for six different combinations of temperature (15, 20, or 25 K), density ($10^{9}$ or $10^{10}$ cm$^{-3}$) and molecular diffusion-to-binding energy ratio of 0.3. The barrier width for quantum tunnelling is $b_{qt}=2$ {\AA}. The orange shaded regions indicates the limits to the measured abundance of \ce{O2} ice in the coma of comet 67P. The red shaded regions indicates the limits to the measured abundance of \ce{H2O2}, and the top of the grey shaded region marks the upper limit to the measured abundance of \ce{O3} ice in comet 67P. The chemical network utilised includes \ce{O3} chemistry.}
\label{red_evolutions_0_3}
\end{figure*}

\begin{figure*}[h]
\subfigure{\includegraphics[width=0.33\textwidth]{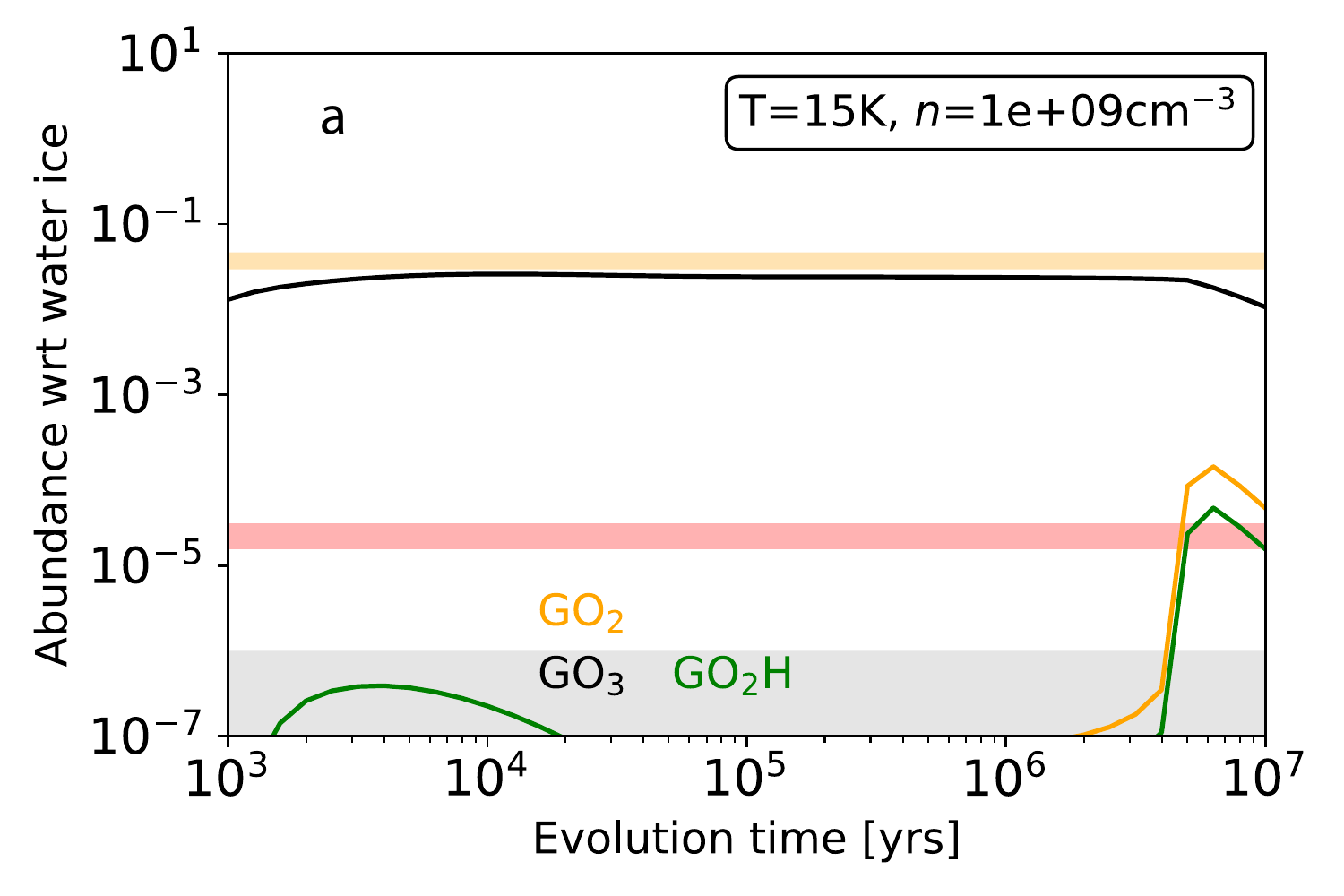}}
\subfigure{\includegraphics[width=0.33\textwidth]{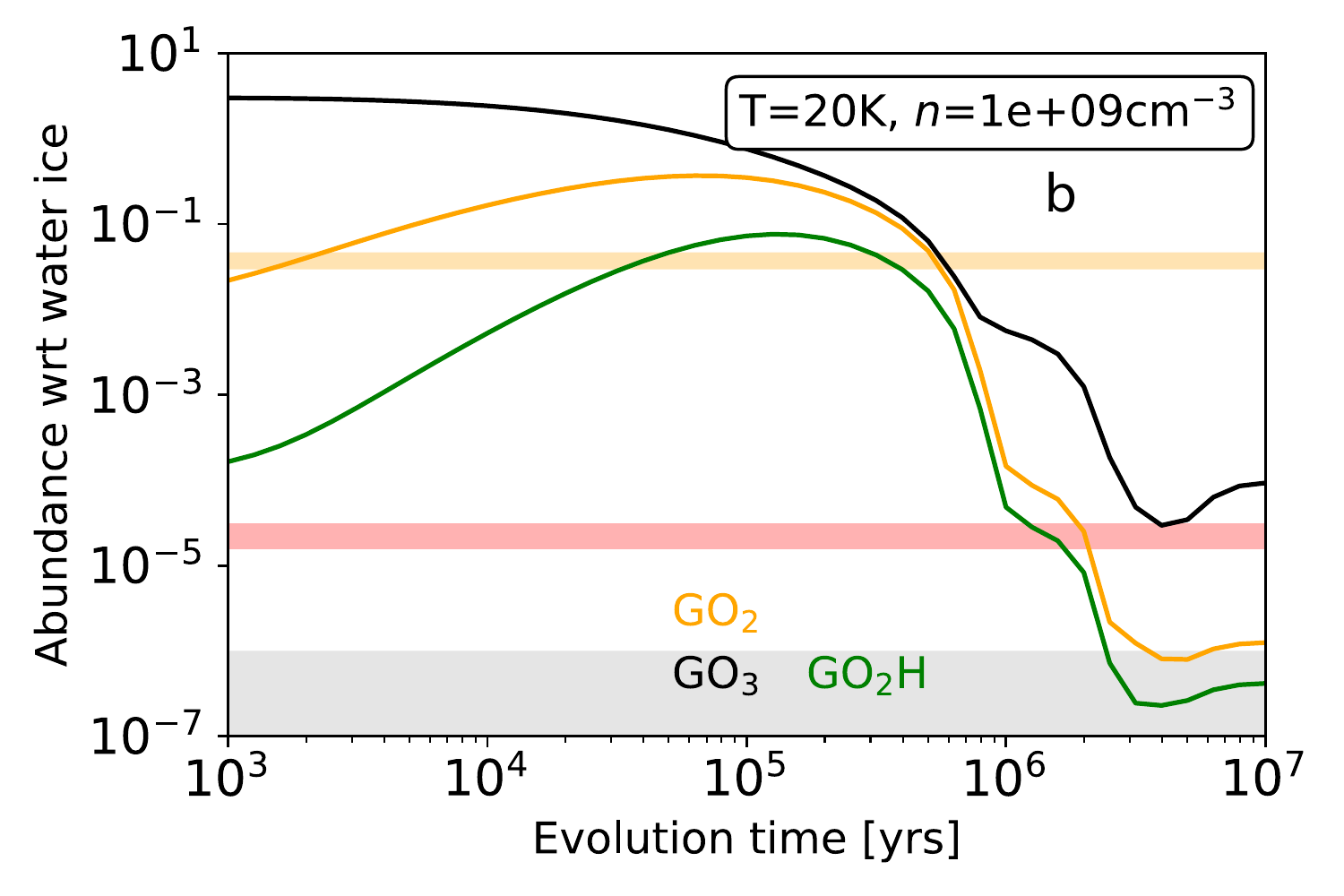}}
\subfigure{\includegraphics[width=0.33\textwidth]{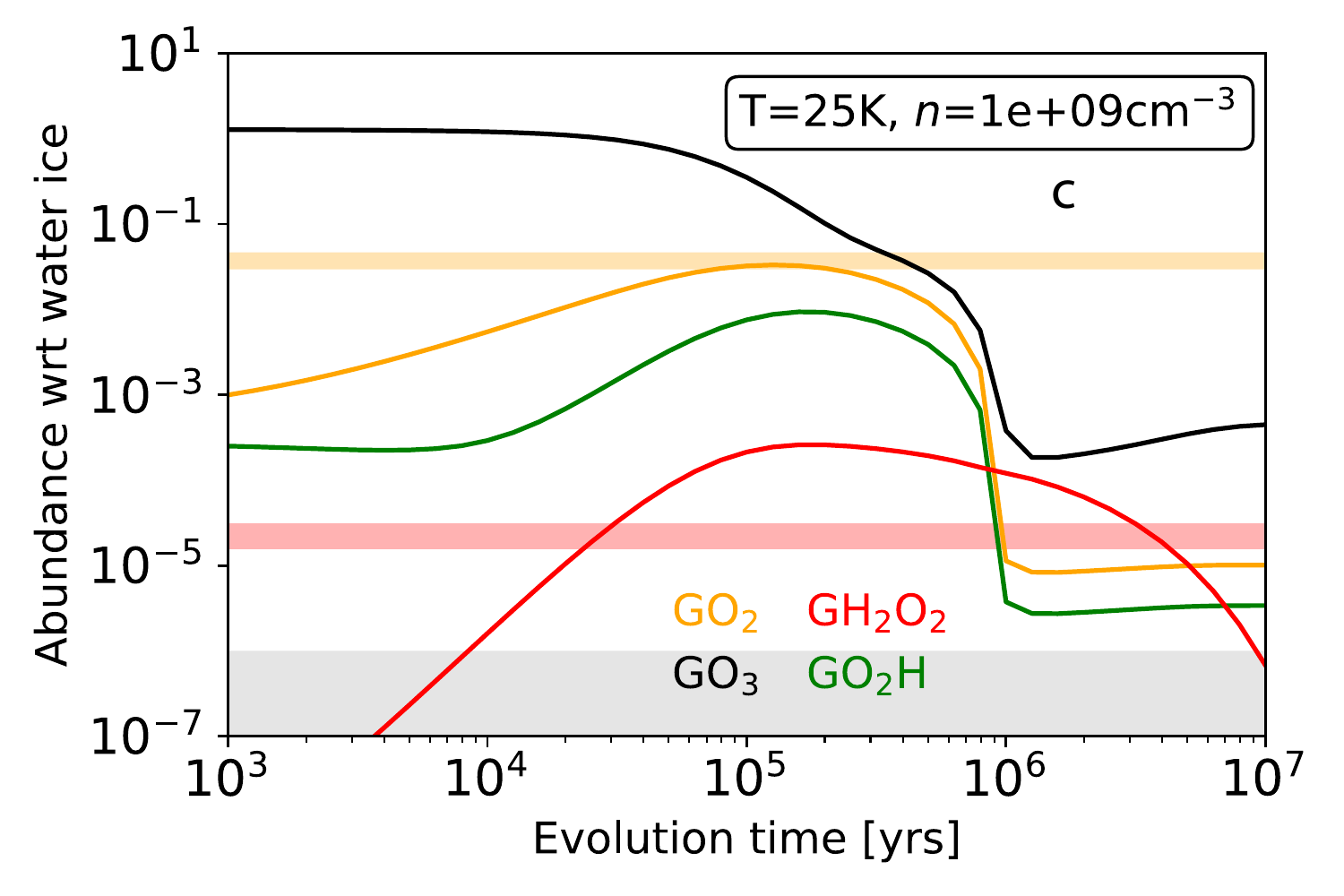}}\\
\subfigure{\includegraphics[width=0.33\textwidth]{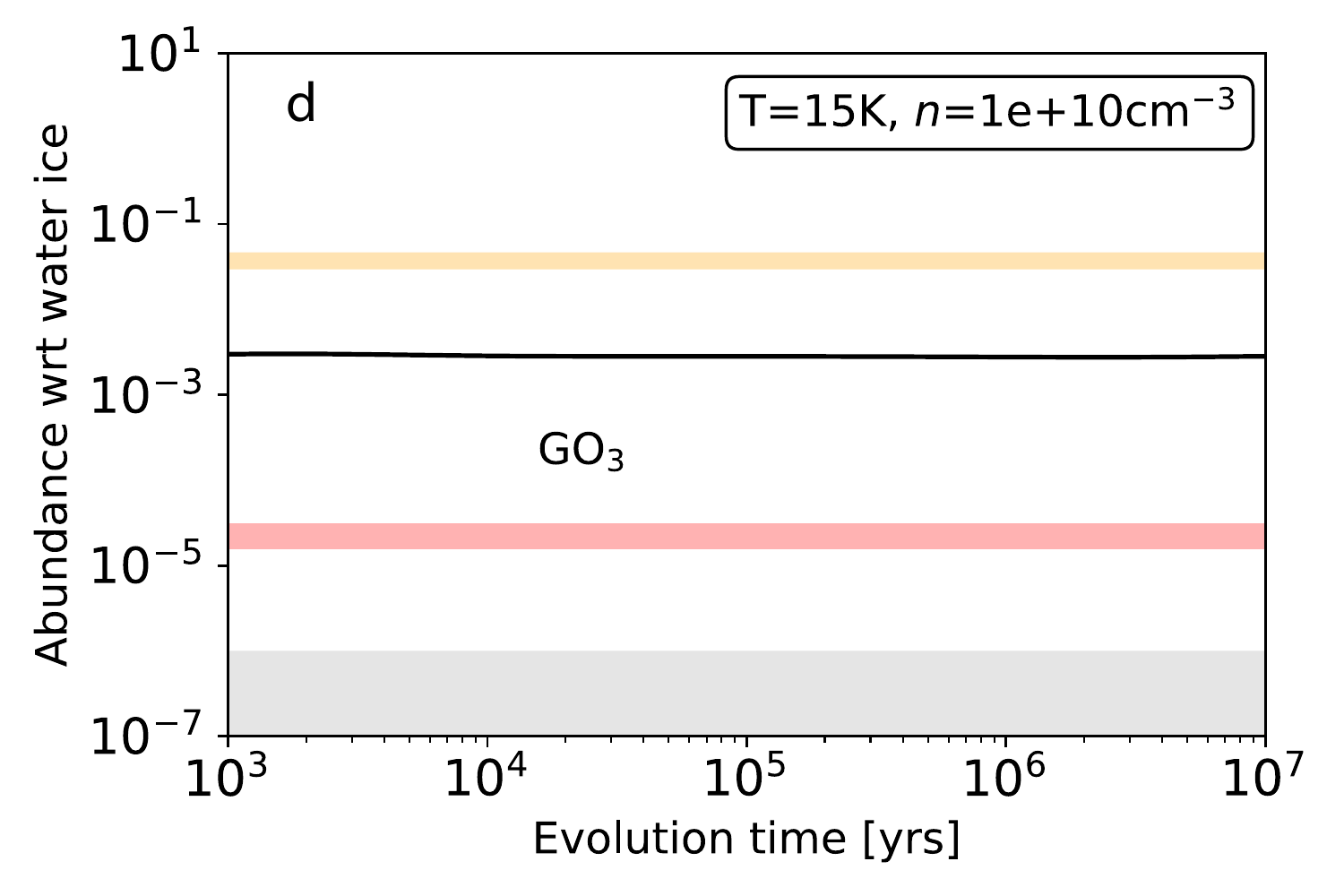}}
\subfigure{\includegraphics[width=0.33\textwidth]{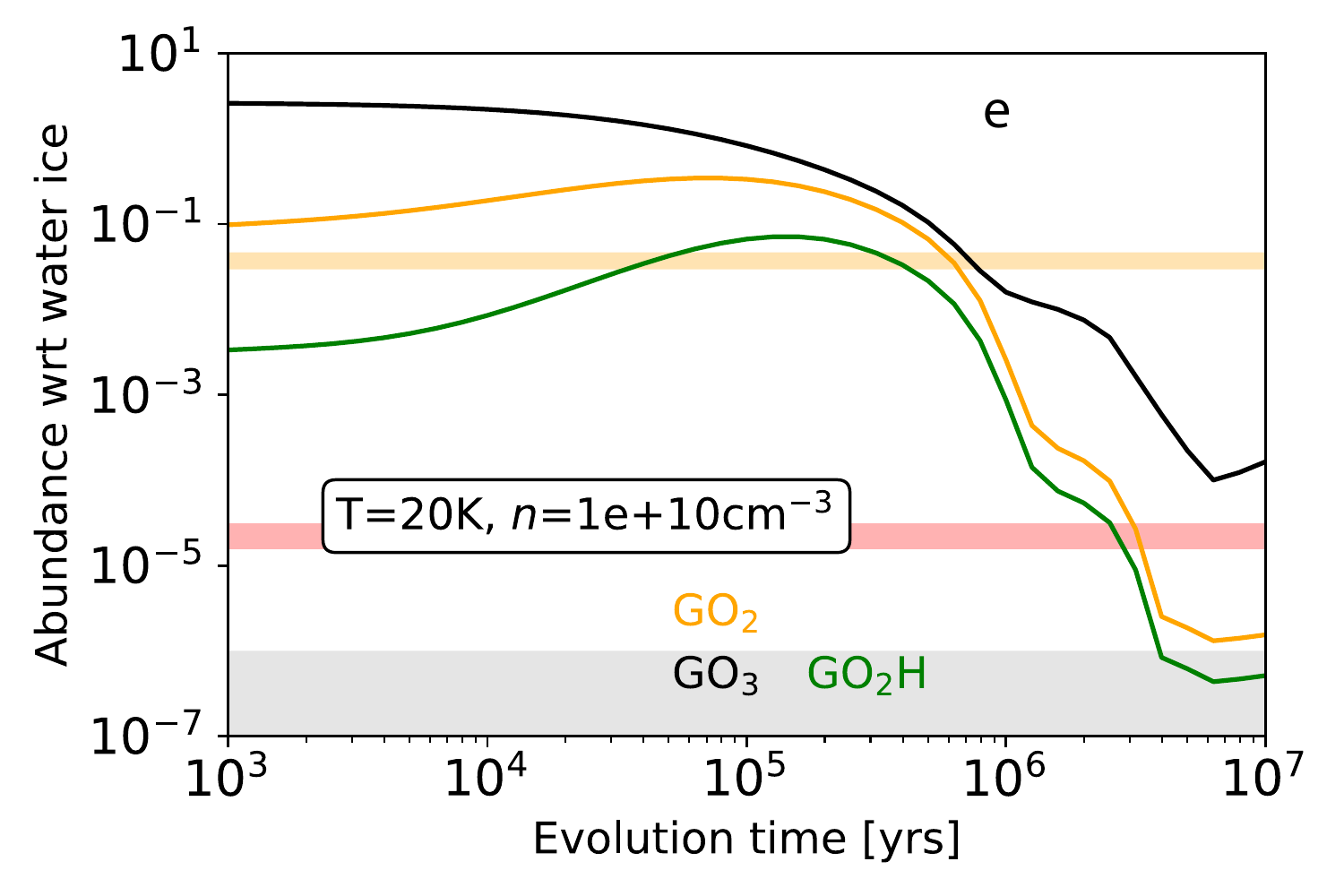}}
\subfigure{\includegraphics[width=0.33\textwidth]{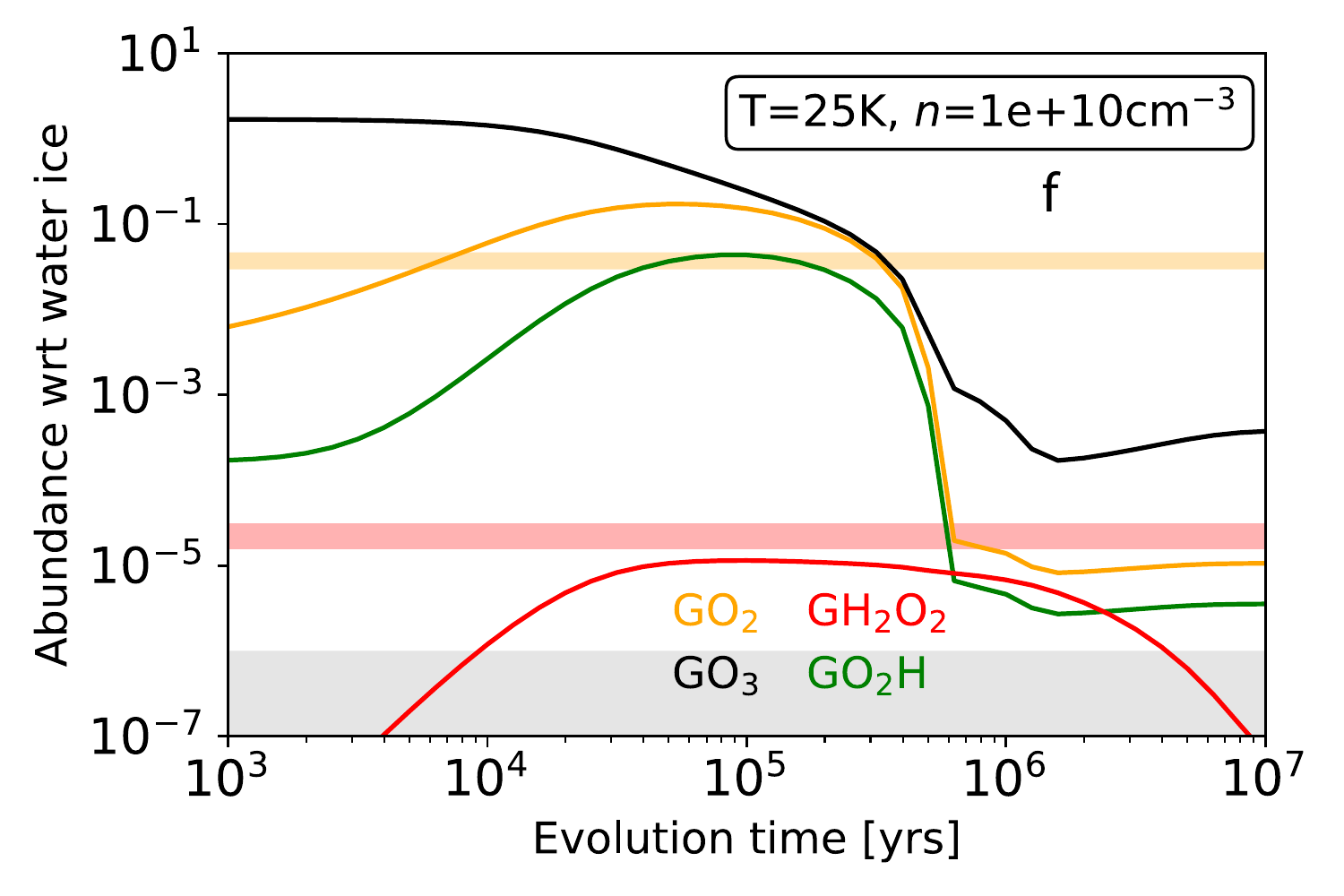}}\\
\caption{Evolving abundances as function of time for the reset scenario for six different combinations of temperature (15, 20 or 25 K), density ($10^{9}$ or $10^{10}$ cm$^{-3}$) and molecular diffusion-to-binding energy ratio of 0.5. The barrier width for quantum tunnelling is $b_{qt}=2$ {\AA}. The orange shaded regions indicates the limits to the measured abundance of \ce{O2} ice in the coma of comet 67P. The red shaded regions indicates the limits to the measured abundance of \ce{H2O2}, and the top of the grey shaded region marks the upper limit to the measured abundance of \ce{O3} ice in comet 67P. The chemical network utilised includes \ce{O3} chemistry.}
\label{red_evolutions_0_5}
\end{figure*}

In Section \ref{abun_mosaic} the color mosaics revealed a promising set of physical and chemical parameters that supported the production of \ce{O2} ice in situ under conditions suitable for the PSN: $b_{\rm{qt}}=2$ {\AA} (rather than the fiducial value of 1 {\AA}), $T$=15-25 K, $n=10^{9}-10^{10}$ cm$^{-3}$ and evolution times up to 0.5 Myr. The ratio of diffusion-to-binding energy for ice species appeared to play a minor role, with both values of 0.3 and 0.5 facilitating \ce{O2} ice production.

In Figure \ref{red_evolutions_0_3} evolving abundances are plotted for \ce{O2}, \ce{O3}, \ce{H2O2} and \ce{O2H} ices with respect to \ce{H2O} ice, for the three different temperatures and two densities outlined above, and with $E_{\rm{diff}}/E_{\rm{bin}}$=0.3. Details about each plot can be found in the labels. Each panel includes shaded regions indicating the observed abundances for \ce{O2} ice (yellow), \ce{H2O2} ice (red) and upper limit for \ce{O3} ice (grey).

For all six panels, it is seen that \ce{O2} ice is within the measured abundance values, at least for a period during evolution. In all cases an initial production is seen, with the abundances of \ce{O2} ice across model setups peaking at 30-40\% with respect to \ce{H2O} ice between $\sim 5\times 10^{4}-5\times 10^{5}$ yrs, with the highest abundances reached at 20 K. For temperatures at 15-20 K, the \ce{O2} ice abundances are generally higher than for $T$=25K. By $\sim 3\times 10^{5}$ to $10^{6}$ yrs, depending on model setup, the \ce{O2} ice abundance drops at least two orders of magnitude below maximum. Hence, the observed abundance of \ce{O2} ice can be reproduced in all six setups, but only at early times.

Regarding ice species besides \ce{O2}, Fig. \ref{red_evolutions_0_3} shows that \ce{O3} is the most abundant of the plotted species, at least until $10^{5}$ yrs. At early evolutionary stages, \ce{O3} ice is even more abundant than \ce{H2O} ice. This is because the calculations are started with free oxygen atoms (reset scenario). This high level is not seen in the measurements of comet 67P, in which an upper limit of $10^{-6}$ with respect to \ce{H2O} ice was reported.

Likewise for \ce{H2O2} ice, it is mostly produced at a high abundance (>$10^{-3}$ with respect to \ce{H2O} ice) in these models. At 25 K after $\sim 0.5$ Myr of evolution, \ce{H2O2} ice is the most abundant of the plotted species. At no point in time in any of the plots does the abundance of \ce{H2O2} ice match the levels observed (red shaded region): for 15 K it is at least an order of magnitude too low, and between 20-25 K it is overproduced by between a factor of three and two orders of magnitude.

For a ratio of diffusion-to-binding energy of 0.5, similar plots are shown in Fig. \ref{red_evolutions_0_5}. Between 20-25 K there are evolutionary times when \ce{O2} ice is reproduced to match the observations. However the highest abundance reached is at 20 K. For $T$=15 K, no reproduction of the observed abundances of neither \ce{O2} ice nor \ce{H2O2} ice is seen, and \ce{O3} ice is (next to \ce{H2O} ice) the dominant carrier of elemental oxygen. At $T$=25K for $n=10^{9}$ cm$^{-3}$, the abundance of \ce{H2O2} ice matches that measured in the comet both by $\sim$0.03 Myr and by $\sim$ 3 Myr, and by 0.03 Myr the \ce{O2} ice abundance is only $\sim2$ times lower than the observed abundance. Along the same lines, for $T$=25 K and $n=10^{9}$ cm$^{-3}$ by $\sim 0.4$ Myr \ce{O2} ice matches the observed abundance, and \ce{H2O2} ice is only $\sim2$ times higher than the observed level. This indicates that something close to a match with the observed abundances between both \ce{O2} and \ce{H2O2} ice abundances with respect to \ce{H2O} ice is reached at 25 K. However, all the models are still vastly overproducing the \ce{O3} ice abundance compared with the measurements.

This parameter space investigation has shown that there are sweet spots, both for the physical and chemical setup, where \ce{O2} ice can be produced to match the measurements on comet 67P, even after expanding the grain-surface chemical network to include \ce{O3} ice chemistry. However, the low measured abundance of \ce{O3} ice remains unexplained by the models.

A possible way to adjust the models to lower the \ce{O3} ice production could be to increase the reaction barrier for the \ce{iO2 ->[\mathrm{iO}]iO3} reaction, where the default barrier is at $E_{\rm{act}}$=500 K. Because the abundance levels of \ce{O2} ice and \ce{H2O2} ice in Fig. \ref{red_evolutions_0_5} panel c ($T$=25K, $E_{\rm{diff}}/E_{\rm{bin}}$=0.5, $n=10^{-9}$ cm$^{-3}$, $b_{\rm{qt}}=$ 2{\AA} and $\zeta=10^{-17}$ s$^{-1}$) featured evolutionary stages when they were simultaneously in proximity to the observed levels, this model setup is now tested with three different values for the activation energy for the \ce{iO2 ->[\mathrm{iO}]iO3} reaction. In addition to the fiducial 500 K, also 1000 K and 2000 K activation energies are tested. It is noted that these are likely too high relative to what is known from laboratory work where \ce{O3} is seen to form readily at low temperatures \citep{lamberts2013}. However, given that \ce{O3} ice is not seen to be efficiently produced in space including comets, the hypothetical situation with a higher barrier is explored, as a means to mitigate as of yet unknown chemical pathways away from \ce{O3} ice. To compare the effects of these changes, the same tests of activation energies are performed for the setup in panel b in the same figure ($T=20$ K, keeping all other parameters the same).

\begin{figure*}[h]
\subfigure{\includegraphics[width=0.5\textwidth]{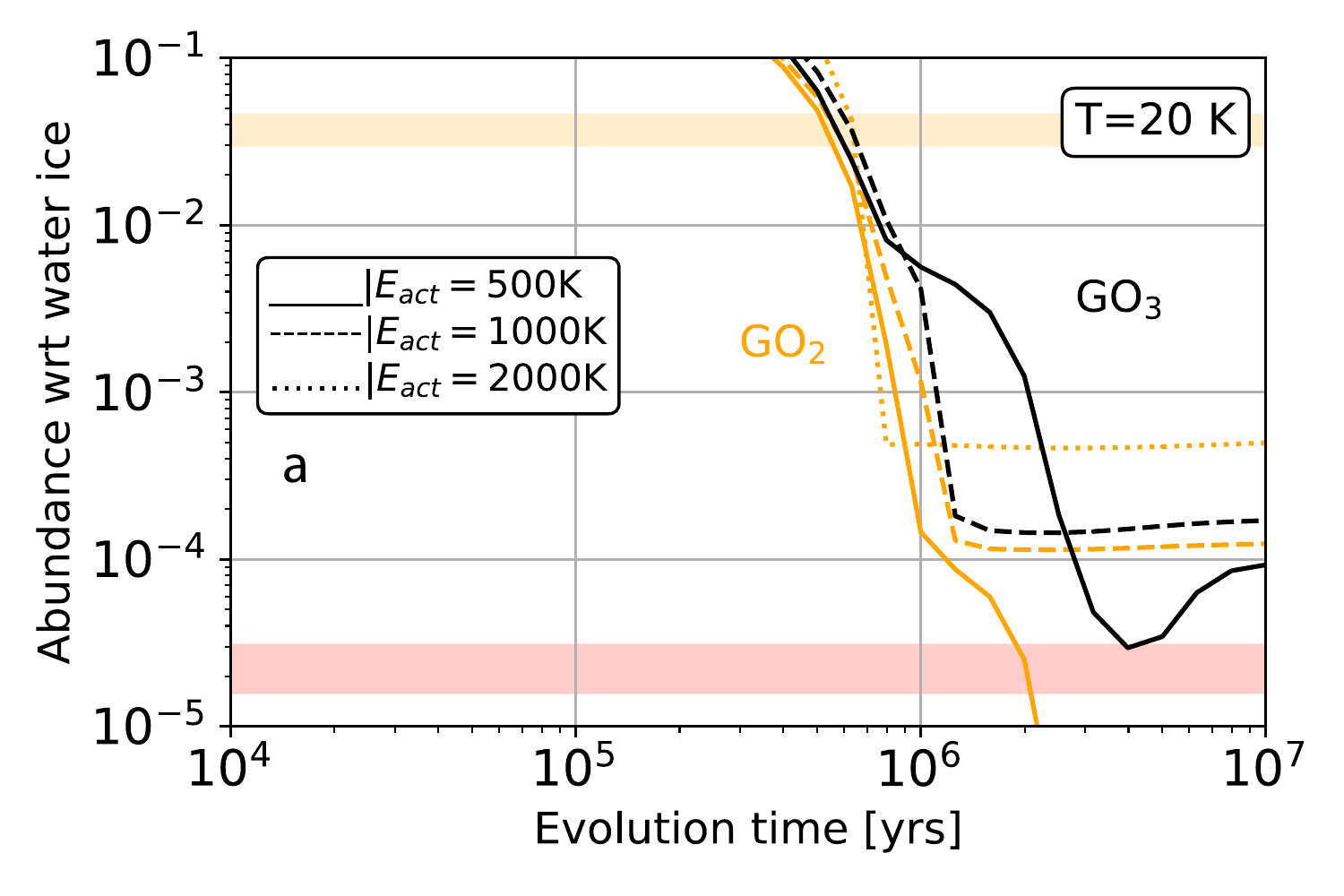}}
\subfigure{\includegraphics[width=0.5\textwidth]{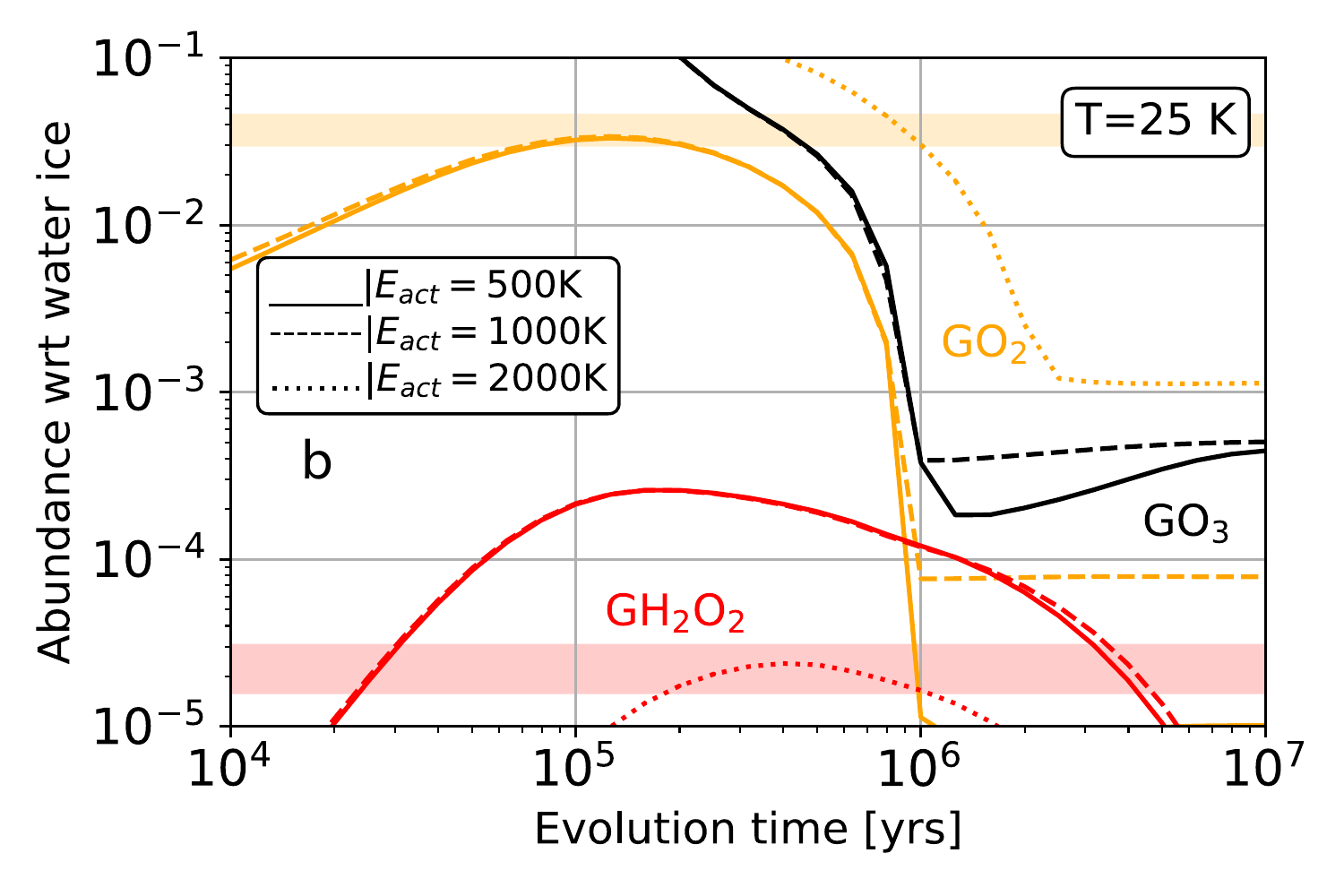}}\
\caption{Evolving abundances for \ce{O2} ice, \ce{H2O2} ice and \ce{O3} ice for three different activation energies for the reaction iO + i\ce{O2}. a) is for $T$=20 K. b) is for $T$= 25 K. The barrier width for quantum tunnelling is $b_{qt}=2$ {\AA}, and the molecular diffusion-to-binding energy is 0.5.}
\label{e_act}
\end{figure*}

In Fig. \ref{e_act} panels a and b, abundances for \ce{O2}, \ce{O3} and \ce{H2O2} ices are plotted as a function of time, with solid, dashed and dotted profiles representing activation energies for the \ce{O3} ice production reaction of 500 K, 1000 K and 2000 K, respectively. In panel a only the abundance of \ce{O2} ice is found to match the observed level. However, in panel b, at 25 K, it is seen that for at $E_{\rm{act}}$= 2000 K, \ce{O2} ice and \ce{H2O2} ice are both reproduced to within the observed values between 0.8-1 Myr evolution. For the same evolutionary timescale, \ce{O3} ice is much lower in abundance ($\sim10^{-16}$ with respect to \ce{H2O} ice), thus also agreeing with the upper limit for the cometary abundance. This is therefore a sweet spot in the physical and chemical parameter space, in which the observed abundances of all three species are reproduced. It is noteworthy that the ionisation level for this sweet spot includes the contributions from both SLRs and CRs, whereas it was found in Paper 1 that \ce{O2} ice be reproduced only without CRs. However, the framework here is different from that in Paper 1, as \ce{O3} chemistry and the updated binding energy for atomic oxygen have been included. Note that the anomalously high activation required for O + \ce{O2} is likely masking as of yet unknown routes in the \ce{O3} chemistry.

\subsection{Including a primordial source of \ce{O2} ice}

\begin{figure*}[h]
\subfigure{\includegraphics[width=0.33\textwidth]{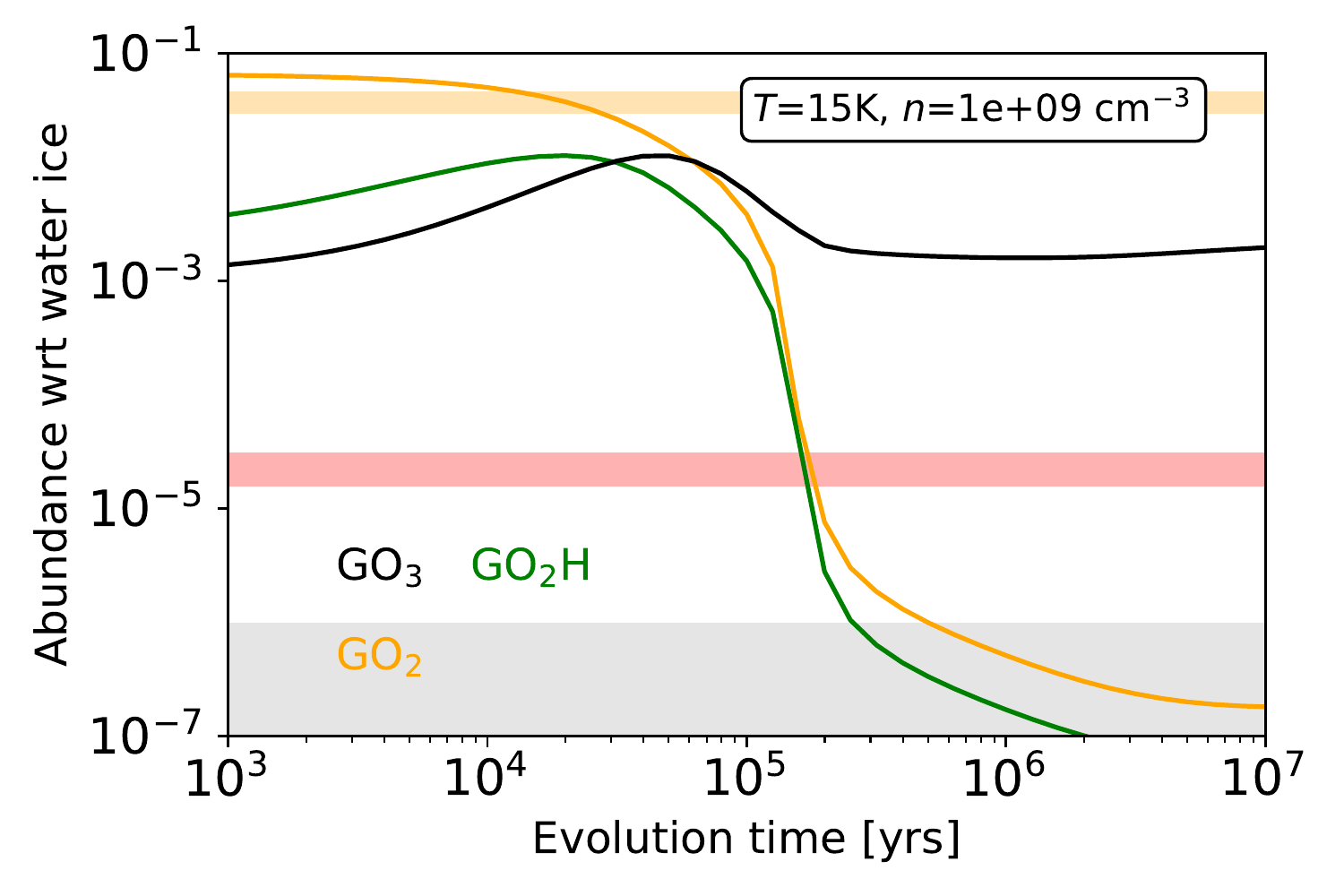}}
\subfigure{\includegraphics[width=0.33\textwidth]{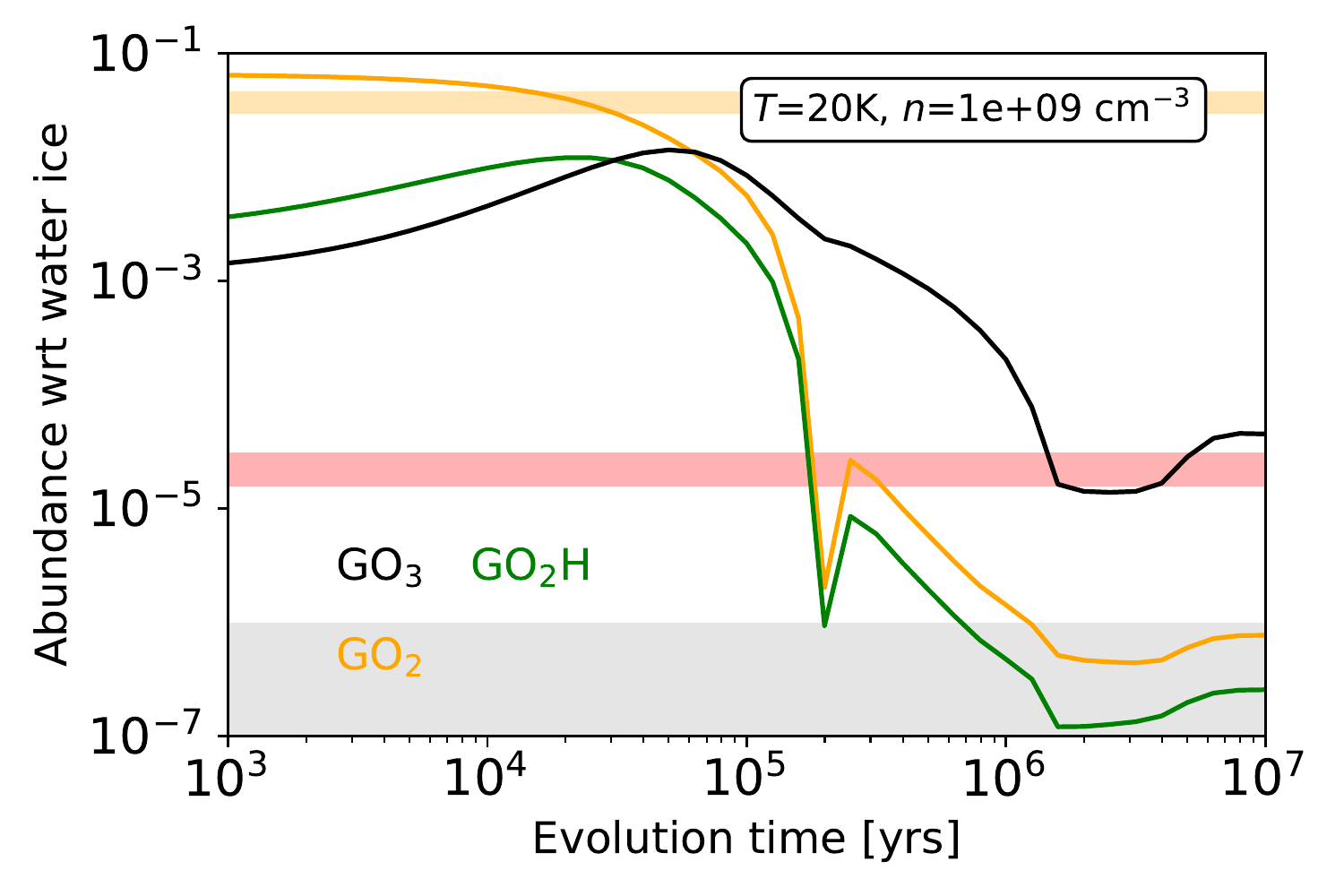}}
\subfigure{\includegraphics[width=0.33\textwidth]{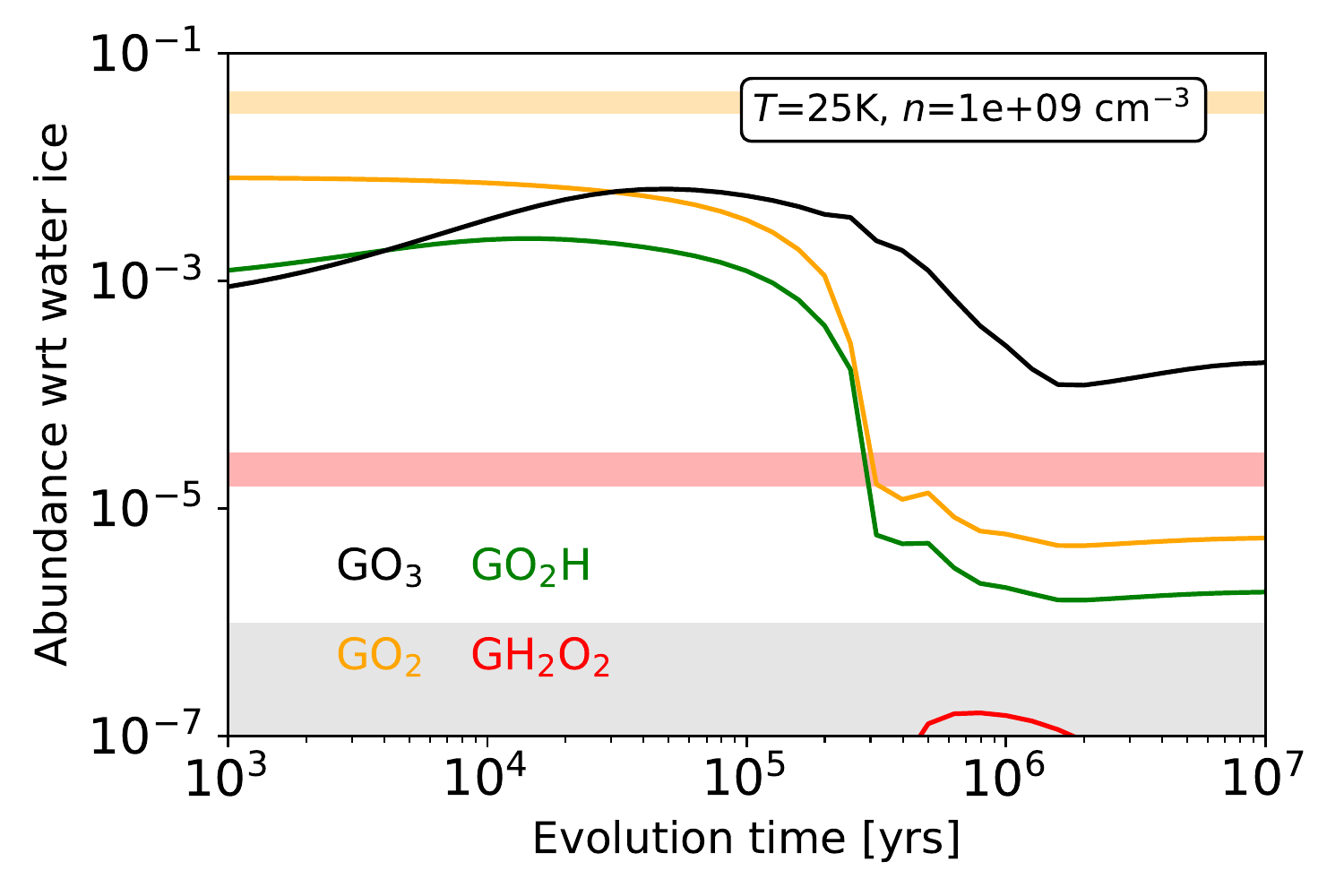}}\\
\subfigure{\includegraphics[width=0.33\textwidth]{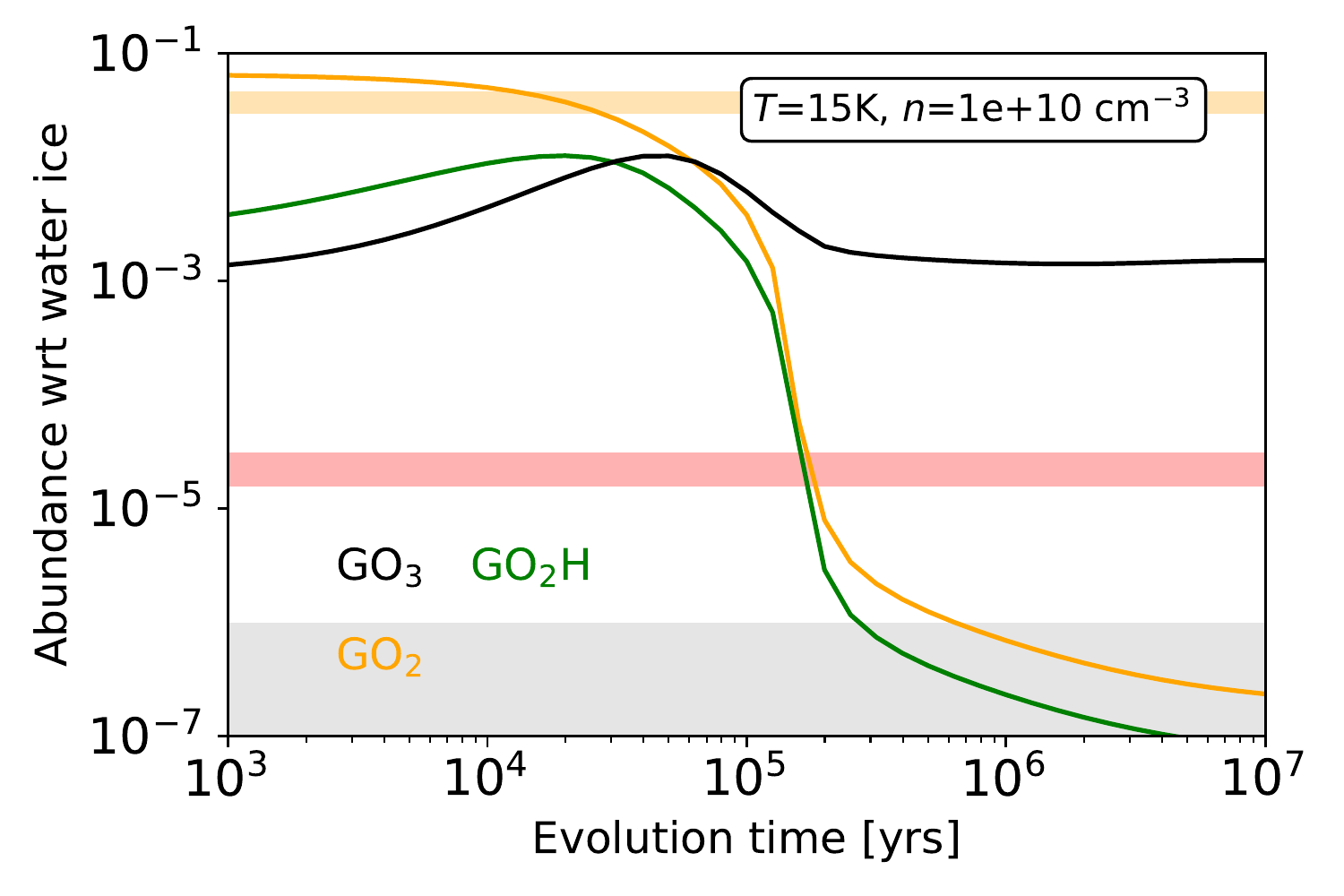}}
\subfigure{\includegraphics[width=0.33\textwidth]{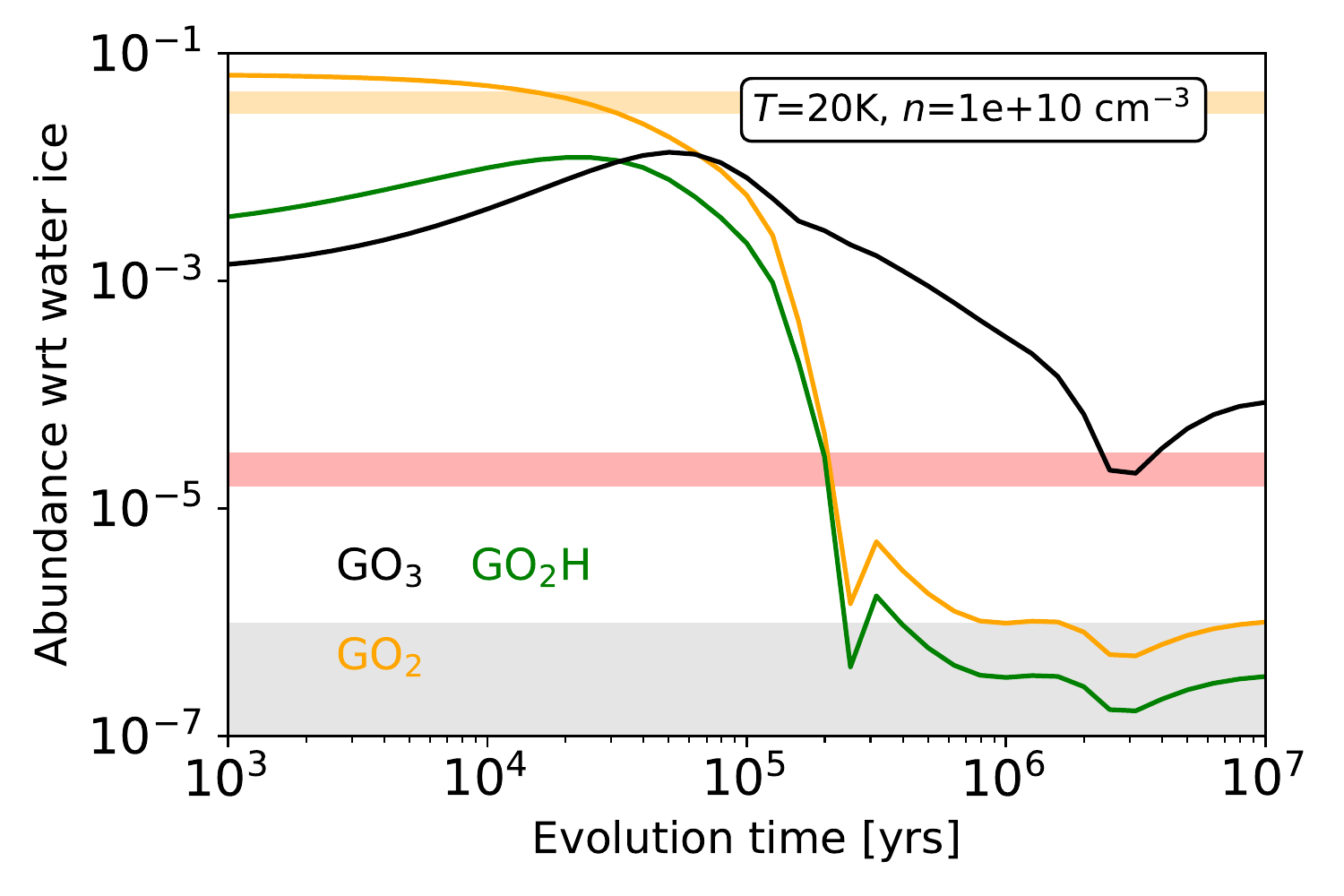}}
\subfigure{\includegraphics[width=0.33\textwidth]{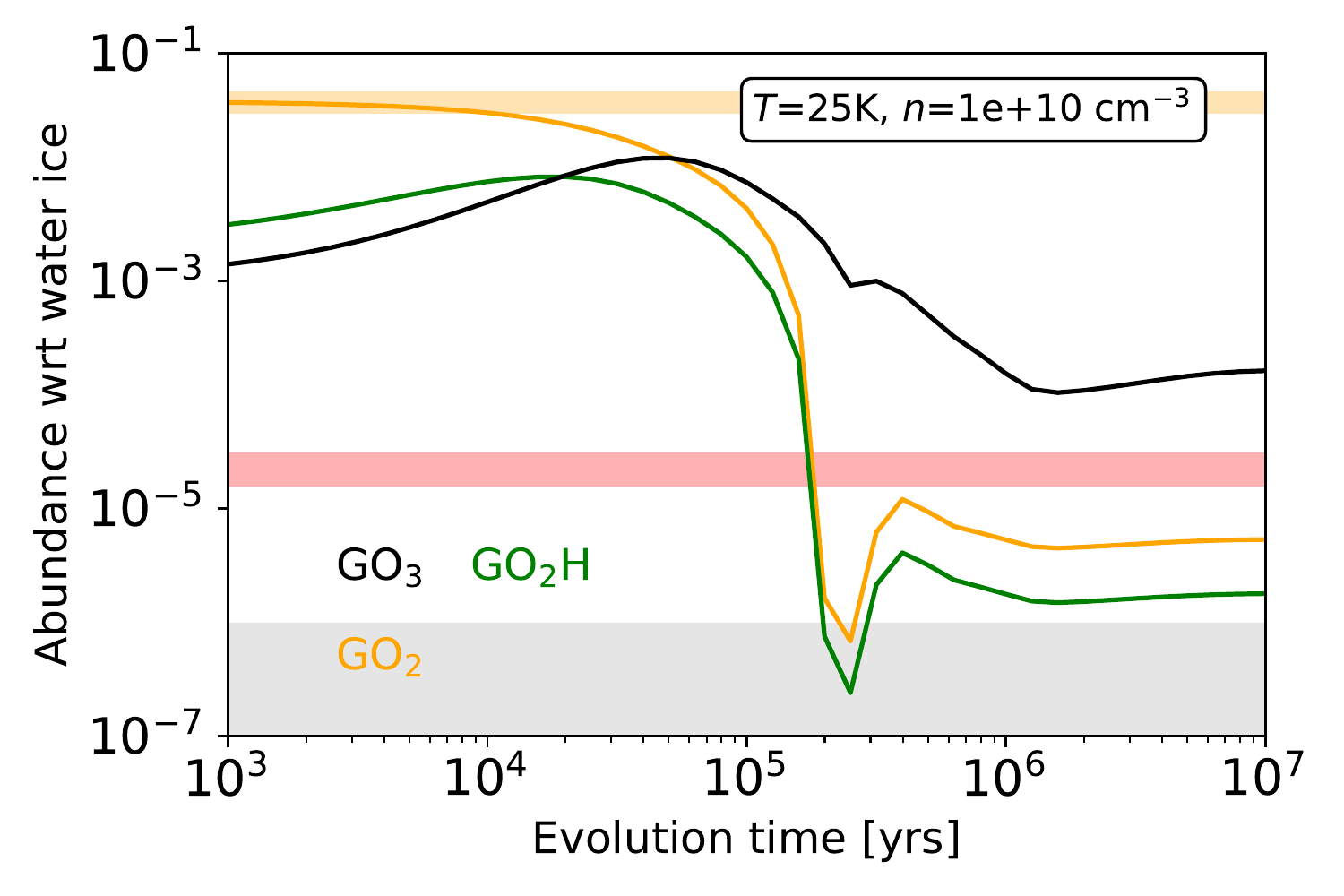}}\\
\caption{Evolving abundances as function of time for the inheritance scenario, plus 5\% extra elemental oxygen as \ce{O2}, for six different combinations of temperature (15, 20 or 25 K), density ($10^{9}$ or $10^{10}$ cm$^{-3}$) and molecular diffusion-to-binding energy ratio of 0.5. The barrier width for quantum tunnelling is $b_{qt}=2$ {\AA}. The orange shaded regions indicates the limits to the measured abundance of \ce{O2} ice in the coma of comet 67P. The red shaded regions indicates the limits to the measured abundance of \ce{H2O2}, and the top of the grey shaded region marks the upper limit to the measured abundance of \ce{O3} ice in comet 67P.}
\label{mols_moxy}
\end{figure*}

Here, the results from the scenario starting with 5\% of atomic oxygen locked in \ce{O2} is tested, exploring the theory for the primordial origin of \ce{O2} ice, as suggested by \citet{taquet2016} and \citet{mousis2016}. This is a modification of the inheritance scenario so that remaining oxygen is already locked in molecules (mostly \ce{H2O}, CO and \ce{CO2}, see the inheritance scenario abundances in Table \ref{old_abund}). In Fig. \ref{mols_moxy} evolving abundance profiles for this scenario are shown, for temperatures between 15-25 K from left to right, and densities between $10^{9}-10^{10}$ cm$^{-3}$ from top to bottom. The ionisation rate is at $\zeta=10^{-17}$ s$^{-1}$, the ratio of diffusion-to-binding energy is 0.5, and the barrier width for quantum tunnelling is 2{\AA}. The activation energy for the O + \ce{O2} reaction is the fiducial value of 500 K. 

It is seen in Fig. \ref{mols_moxy}, panels a, b, d and e, that an early abundance level of \ce{O2} ice at or above the observed mean is maintained until a few times $10^{4}$ yrs for temperatures between 15-20 K. At 25 K, the abundance of \ce{O2} ice is lower early on, and in particular for $n$=$10^{9}$ cm$^{-3}$ the abundance is $<10^{-2}$ with respect to \ce{H2O} ice, which is outside the observed range. Simultaneously with \ce{O2} ice matching the observed abundance in some cases, the abundance of \ce{O3} ice is 1-2 orders of magnitude lower than \ce{O2} ice by $3\times10^{3}$ yrs for all cases except in panel c. Between a few times $10^{4}$ yrs and 10 Myr the \ce{O2} ice abundance decreases by >2 orders of magnitude compared with the initial abundance, and from $\sim10^{5}$, yrs \ce{O3} ice is the dominant oxygen carrier of the plotted species. Hence, this scenario can reproduce the observed abundance of \ce{O2} at early stages, and with lower abundances of \ce{H2O2} and \ce{O3} ices, although abundances for the latter two species are still higher than measured.


\section{Discussion}
\label{discuss}
The abundance of \ce{O2} ice in comet 67P has motivated much work as to its origin. \citet{rubin2015}, \citet{mousis2016,mousis2018}, \citet{taquet2016} and \citet{dulieu2017} have all attempted to explain the \ce{O2} level from a point of view of possible chemical origins. \citet{mousis2016} and \citet{rubin2015} approached the problem by investigating chemical processing of either \ce{H2O} ice or \ce{H2O2} ice into \ce{O2} ice. \citet{mousis2018} attempted an explanation with a scenario in which ice-covered grains were transported from the midplane to the upper layers of the disk. Here, the grains undergo photochemical processing producing \ce{O2} ice, which is then cycled back to the disk midplane. It is noted that \citet{mousis2018} only consider the \ce{O2} ice abundances, not \ce{H2O2} or \ce{O3} ices. On the other hand, \citet{taquet2016} approached it from a broader point-of-view considering primordial production pathways in the parent cloud, and utilising extensive chemical networks and subsequently tracking the primordially-produced \ce{O2} ice to the disk midplane. While those works all concluded that a primordial origin and subsequent retention of \ce{O2} ice was the most likely, Paper 1 concluded that starting from a fully atomised disk midplane, there is an evolutionary phase during which a range of radii in the outer icy midplane (outside the \ce{O2} iceline) will reasonably reproduce the measured \ce{O2} / \ce{H2O} ice level.

The investigation here has taken the step further from Paper 1: exploring the physical and chemical parameter space that may facilitate the production of \ce{O2} ice in-situ in the PSN, and simultaneously tracing the abundances of the chemically related species \ce{H2O2}, \ce{O3} and \ce{O2H}, as well as using a more massive disk model more appropriate for the PSN.

\subsection{Chemical starting conditions}
\label{discuss_chem_start}

The chemical starting conditions' effects on the \ce{O2} ice production were explored. It is clear that only an atomised start can facilitate the production, although models starting with a percentage of elemental oxygen in \ce{O2} did retain this \ce{O2} ice for a short period of evolution. The presence of atoms other than oxygen (such as sulphur), however, does not have a large impact on the production of \ce{O2} ice. That means that oxygen-only and hydrogenated oxygen species lock up the majority of the available elemental oxygen.

\subsection{Dependence on ionisation levels}

The ionisation levels have been shown to facilitate different chemical evolution of abundances. The chemical timescales are shorter for higher ionisation levels. Ionisation Level 1 only reproduced the observed \ce{O2} abundance for models without \ce{O3}, and ionisation Levels 2 and 3 cause similar evolutionary trends, meaning that Level 3 does not facilitate any more \ce{O2} ice production than Level 2. It is noteworthy that the sweet spot for reproduction of \ce{O2}, \ce{H2O2} and \ce{O3} ices to within the observed abundances was found for a ionisation level of $\zeta=10^{-17}$ s$^{-1}$, which is the local ISM value for dense clouds. This ionisation resembles ionisation Level 2 at 100-130 AU in the PSN disk midplane (see Fig. \ref{high_mass_phys} panel b), which is also the radial range covering the temperature and density of the sweet spot, which is just outside the \ce{O2} iceline.

\subsection{Changing $E_{\rm{bin}}$ for atomic oxygen, and inclusion of \ce{O3}}
\label{discuss_ozone}

\ce{O3} was introduced and included in the chemical network with production through \ce{\mathrm{i}O2 ->[\mathrm{iO}] \mathrm{i}O3} (activation energy of 500 K), and destruction through barrierless hydrogenation. After this change, the binding energy $E_{\rm{bin}}$ for oxygen atoms was increased from 800 K to 1660 K, as measured by \citet{he2015}. This change was expected to make oxygen atoms less volatile, thus having them reside on the grain surfaces at higher temperatures than before. This was expected to aid the production of \ce{O2} and \ce{O3} ices through the pathway \ce{\mathrm{i}O ->[\mathrm{iO}] \mathrm{i}O2 ->[\mathrm{iO}] \mathrm{i}O3}.

The expansion to include \ce{O3} chemistry induces very different behaviours of \ce{O2}, \ce{O3}, and the chemically related \ce{H2O2}. Still assuming the PSN midplane, and ionisation Level 1, \ce{O2} ice is not reproduced to match the observed cometary abundances.

The updated binding energy for atomic oxygen also had an effect, as the parameter space investigation revealed \ce{O2} ice production matching the observed value, even when including \ce{O3} chemistry. This change made the oxygen atoms less volatile, thus adsorbing to the grain at higher temperatures. 


\subsection{Narrowing down on \ce{O2} ice production}
\label{discuss_sweet}

With the lower produced level of \ce{O2} ice after inclusion of \ce{O3} chemistry the question remained if \ce{O2} ice, in any case, can be produced to within the measured levels. Besides updating $E_{\rm{bin}}$ for atomic oxygen, two parameters for grain-surface chemistry, which had been kept constant thus far were now varied: the barrier width for quantum tunnelling, and the ratio of diffusion-to-binding energy of icy molecules. Adjusting these parameters, in particular the barrier width for quantum tunnelling, appears to have important impacts on the production level of \ce{O2} ice. Increasing the barrier for quantum tunnelling from 1 to 2 {\AA}, for temperatures ranging between 15-25 K, and densities 10$^{9}$-10$^{10}$ cm$^{-3}$, results in \ce{O2} ice produced to within the mean measured level in comet 67P. Increasing the barrier width from 1 to 2 {\AA} lowers the mobility of H and \ce{H2} on the grain-surfaces. This means that O+O reactions proceed more efficiently than hydrogenation of O, \ce{O2} and \ce{O3} ice, which in turn leads to higher abundances of \ce{O2} and \ce{O3} ices for a $b_{\rm{qt}}$ of 2 {\AA}. Generally, with this wider tunnelling barrier, the measured mean abundance of \ce{O2} ice can be reproduced relatively early in the evolution, from 0.05-0.5 Myr. At 25 K the corresponding level of \ce{H2O2} ice was close to its observed value. \ce{O3} ice, however, remains more abundant than \ce{O2} ice. Early in the evolution, \ce{O3} ice is even more abundant than \ce{H2O} ice, but this is explained by \ce{O3} ice being a precursor to \ce{H2O} ice in the reaction pathway
\newline \ce{\mathrm{i}O ->[\mathrm{iO}] \mathrm{i}O2 ->[\mathrm{iO}] \mathrm{i}O3 ->[\mathrm{iH}] \mathrm{i}OH ->[\mathrm{iH}] \mathrm{i}H2O},
\newline
due to the availability of free oxygen. Indeed, in Fig. \ref{red_evolutions_0_3}, showing the time evolution of species, \ce{O3} is the dominant oxygen carrier at early times, and \ce{H2O} at later times. The exception to this are the cases of a low ratio of diffusion-to-binding energy of 0.3, and temperatures of 25 K (panels c and f in Fig. \ref{red_evolutions_0_3}), where \ce{H2O2} dominates at later times. Here, a lower diffusion energy and higher temperature means a higher mobility of icy molecules on the grain surface. This leads to OH (the product of \ce{O3} + H) reacting with \ce{O2} ice (precursor to \ce{O3} ice and product of the hydrogenation of \ce{O3} ice), thereby lowering the abundance of \ce{O3} ice, and increasing the abundance of first \ce{O2H} ice, and subsequently \ce{H2O2} ice via the reaction pathway \ce{\mathrm{i}O2 ->[\mathrm{iH}]\mathrm{i}O2H ->[\mathrm{iH}]\mathrm{i}H2O2}.

\subsection{Activation energy for \ce{O3} ice production pathway}
\label{issues}

A crucial element in modelling chemical evolution is the assumed activation energies for reactions. Especially for grain-surface chemistry, these energies remain somewhat uncertain, due to the difficulty of estimating them through production rates in the lab. In this work, the fiducial activation energy for the reaction \ce{\mathrm{i}O + \mathrm{i}O2 -> \mathrm{i}O3} was $E_{\rm{act}}$=500 K \citep{lamberts2013}, resulting in model \ce{O3} ice at a higher abundance than its parent species, \ce{O2} ice. This means that the modelled abundance of \ce{O3} ice is several orders-of-magnitude higher than the upper limit determined in comet 67P (see Figs. \ref{red_evolutions_0_3} and \ref{red_evolutions_0_5}).

Testing out higher activation energies for the production of \ce{O3} ice (thereby impeding the \ce{O + O2} reaction) revealed that for $E_{\rm{act}}$=2000 K, the production of \ce{O3} ice was sufficiently impeded to bring the abundance below the limit observed in the comet. This test simultaneously showed a match between both the abundances of \ce{O2} ice and \ce{H2O2} ice and the cometary abundances at 25 K between 0.8-1 Myr evolution for an ionisation level of $\zeta=10^{-17}$ s$^{-1}$, starting from the reset scenario. Thus, a sweet spot in the parameter space, where all three species match observed abundances, was found. However, it should be emphasised that such a high barrier is not supported in previous analyses of laboratory experiments. An upper limit for the activation energy of O + \ce{O2} of $E_{\rm{act}}$ = 150 K on amorphous silicate surfaces was derived from experimental results by \citet{minnisale2014}. \citet{lamberts2013} required an activation energy of $E_{\rm{act}}$ = 500 K in order to reproduce experimental results from \citet{ioppolo2010} of \ce{O3} production on thick ices, and in agreement with this result, \citet{taquet2016} were able to reproduce the low abundances \ce{O3} and \ce{O2H} on ice mantles in dark clouds using an activation energy of $E_{\rm{act}}$ = 300 K.

In this work \ce{O3} has remained overproduced for these levels of activation energies, even when increasing it to $E_{\rm{act}}$ = 1000 K. Only using an artificially high energy of $E_{\rm{act}}$ = 2000 K for the O + \ce{O2} reaction lowered the abundance of \ce{O3} ice to the observed level. This could suggest that as-of-yet unknown reactions may be missing from the network: reactions involving either \ce{O2} or O that do not lead to \ce{O3} (so O and \ce{O2} get locked up in other molecules than \ce{O3}) and/or new routes for destruction of \ce{O3} ice.

\subsection{Location of \ce{O2} ice production sweet spot in PSN disk midplane}

Taking the sweet spot for matching abundances of \ce{O2} ice, \ce{H2O2} ice and \ce{O3} ice, it is interesting to look at where in the PSN disk midplane the physical conditions for this match may be satisfied.

A close look at Fig. \ref{high_mass_phys}a shows that indeed in the radial range $\sim$100-130 AU is found a temperature of $\sim$25 K and is found a density of $10^{9}-10^{10}$ cm$^{-3}$. Thus, for the sweet-spot \ce{O2} ice production scenario, this would be the predicted radial range for formation of both comets 1P and 67P, and a formation time scale by 0.8-1 Myr, given the model assumptions. This radial range is 2-3 times larger than the orbit of the present day Kuiper belt (30-50 AU).

\subsection{Primordial origin of \ce{O2} ice}

A scenario that did predict early presence of \ce{O2} ice to within the observed level, and \ce{H2O2} ice and \ce{O3} ice orders of magnitude lower than that of \ce{O2} ice, is the scenario starting with an initial abundance of \ce{O2} of $\sim$5\% with respect to the total elemental oxygen abundance, which was inspired by the results of \citet{taquet2016} assuming that the \ce{O2} is inherited from the parent cloud. Given early formation of the comets (by thousands to tens of thousands of years after formation of the PSN midplane), this scenario remains likely for the formation of the ices, and the fiducial values of $b_{\rm{qt}}=$ 1 {\AA} and $E_{\rm{act}}$= 500 K for the \ce{O + O2} reaction in the ice. This scenario agrees with the findings of \citet{taquet2016}.

\section{Conclusion}
\label{concl}

Since the somewhat unexpected detection of abundant \ce{O2} ice in the coma of comet 67P, several studies have attempted to explain the origin of the \ce{O2}. Building on the results from Paper 1, this work has investigated the possibility of in-situ formation of \ce{O2} ice on grains in the midplane of the PSN disk midplane. 

While a high abundance of \ce{O2} ice, matching that observed in comet 67P, was reproduced outside the \ce{O2} ice in the PSN disk midplane at intermediate evolutionary stages when assuming the initial chemistry to be reset, the same production was not seen after including \ce{O3} ice chemistry into the chemical network. For the fiducial choice of parameters for grain-surface chemistry, and an activation energy of 500 K for the \ce{O + O2 ->O3} reaction on the grains, \ce{O3} ice was in most cases found to be the dominant oxygen-carrier next to \ce{H2O} ice, and \ce{O2} ice was orders of magnitude too low in abundance compared to the abundance observed in comet 67P.

In order to test the sensitivity of the production of \ce{O2}, \ce{O3}, and \ce{H2O2}  ices to the assumed parameters for grain-surface chemistry, in particular the barrier width for quantum tunnelling $b_{\rm{qt}}$ and the ratio of diffusion-to-binding energy of ice molecules $E_{\rm{diff}}/E_{\rm{bin}}$, a parameter space investigation was conducted. Several temperatures and densities were also tested, and the reset scenario assumed for initial abundances. This led to a sweet-spot set of parameters being revealed: $b_{\rm{qt}}$= 2{\AA}, $E_{\rm{diff}}/E_{\rm{bin}}=0.3-0.5$ $T$=15-25 K and $n=10^{9}-10^{10}$ cm$^{-3}$ which facilitated \ce{O2} reproduction matching the observed level. However, the abundances of \ce{O3} and \ce{H2O2} ices were still in disagreement with the observed values by orders of magnitude.

As a last adjustment of the chemistry intended to lower the \ce{O3} ice abundance, the activation energy for production of \ce{O3} ice from the association of O and \ce{O2} ices was increased in order to mitigate possible unknown chemical pathways away from \ce{O3}. For an activation energy of $E_{\rm{act}}= 2000$ K, $E_{\rm{diff}}/E_{\rm{bin}}=0.5$, and the remaining physical and chemical conditions as given above, the abundances of \ce{O2} ice, \ce{H2O2} ice, and the upper limit for \ce{O3} ice in the comet were all reproduced. This matches a formation location in the PSN disk midplane between 120-150 AU, just outside the \ce{O2} iceline. However, this high activation energy for \ce{O3} ice production is not supported by laboratory estimates, and thus more laboratory work is needed to determine potential missing chemical pathways for \ce{O3} ice chemistry.

A model starting out with a percentage of elemental oxygen locked in \ce{O2} and thus assuming a primordial origin of \ce{O2}, also reproduced the observed abundance of \ce{O2} ice at early stages of evolution, without increasing $E_{\rm{act}}$ for the O + \ce{O2} reaction. Here, the \ce{O3} and \ce{H2O2} ice abundances were below the \ce{O2} ice abundance, but not matching the observed abundance levels. However, since the observed abundances of all three ices species are only reproduced in this work in the case for a set of rather extreme choices for the chemical parameters, the most plausible explanation for the origin of the cometary \ce{O2} ice remains the primordial one, as originally proposed by \citet{taquet2016}.

\begin{acknowledgements}

The authors thank Ewine van Dishoeck and Arthur Bosman for many useful discussions and comments that helped the investigation and improved the quality of the manuscript. The authors also thank an anonymous referee for their review of this work.
Astrochemistry in Leiden is supported by the European Union A-ERC grant 291141 CHEMPLAN and the Netherlands Research School for Astronomy (NOVA). CW also acknowledges the Netherlands Organisation for Scientific Research (NWO, grant 639.041.335) and the University of Leeds for financial support.
\end{acknowledgements}

\begin{appendix}
\section{Appendix: additional figures}
\begin{table}
\captionsetup{justification=centering}
\caption{Initial abundances (with respect to H$_{\rm nuc}$) for atomic (reset scenario) and molecular (inheritance scenario) setups.}             
\centering                          
\begin{tabular}{l l c}              
\hline\hline                        
Species 	& Atomic 		      & Molecular     \\              
\hline                              
\\
   H		& 9.1$\times 10^{-5}$ & 5.0$\times 10^{-5}$    \\      
   He	 	& 9.8$\times 10^{-2}$ & 9.8$\times 10^{-2}$    \\
   \ce{H2} 	& 5.0$\times 10^{-1}$ & 5.0$\times 10^{-1}$    \\
   N 		& 6.2$\times 10^{-5}$ &                                \\
   O 		& 5.2$\times 10^{-4}$ &                                   \\ 
   C		& 1.8$\times 10^{-4}$ &                                    \\
   S		& 6.0$\times 10^{-6}$ &                                    \\
   \ce{H2O}	&		      &3.0$\times 10^{-4}$         \\
   CO		        &		      &6.0$\times 10^{-5}$        \\
   \ce{CO2}	&		      &6.0$\times 10^{-5}$         \\
   \ce{CH4}	&		      &1.8$\times 10^{-5}$         \\
   \ce{N2}	&		      	      &2.1$\times 10^{-5}$         \\
   \ce{NH3}	&		      &2.1$\times 10^{-5}$         \\
   \ce{CH3OH}	&		      &4.5$\times 10^{-5}$         \\
   \ce{H2S}	&		      &6.0$\times 10^{-6}$         \\ 
   \ce{O2}		&			&0\\                 
   \\
   \hline                                   
\label{old_abund}
\end{tabular}
\end{table}

\begin{figure*}
\subfigure{\includegraphics[width=0.33\textwidth]{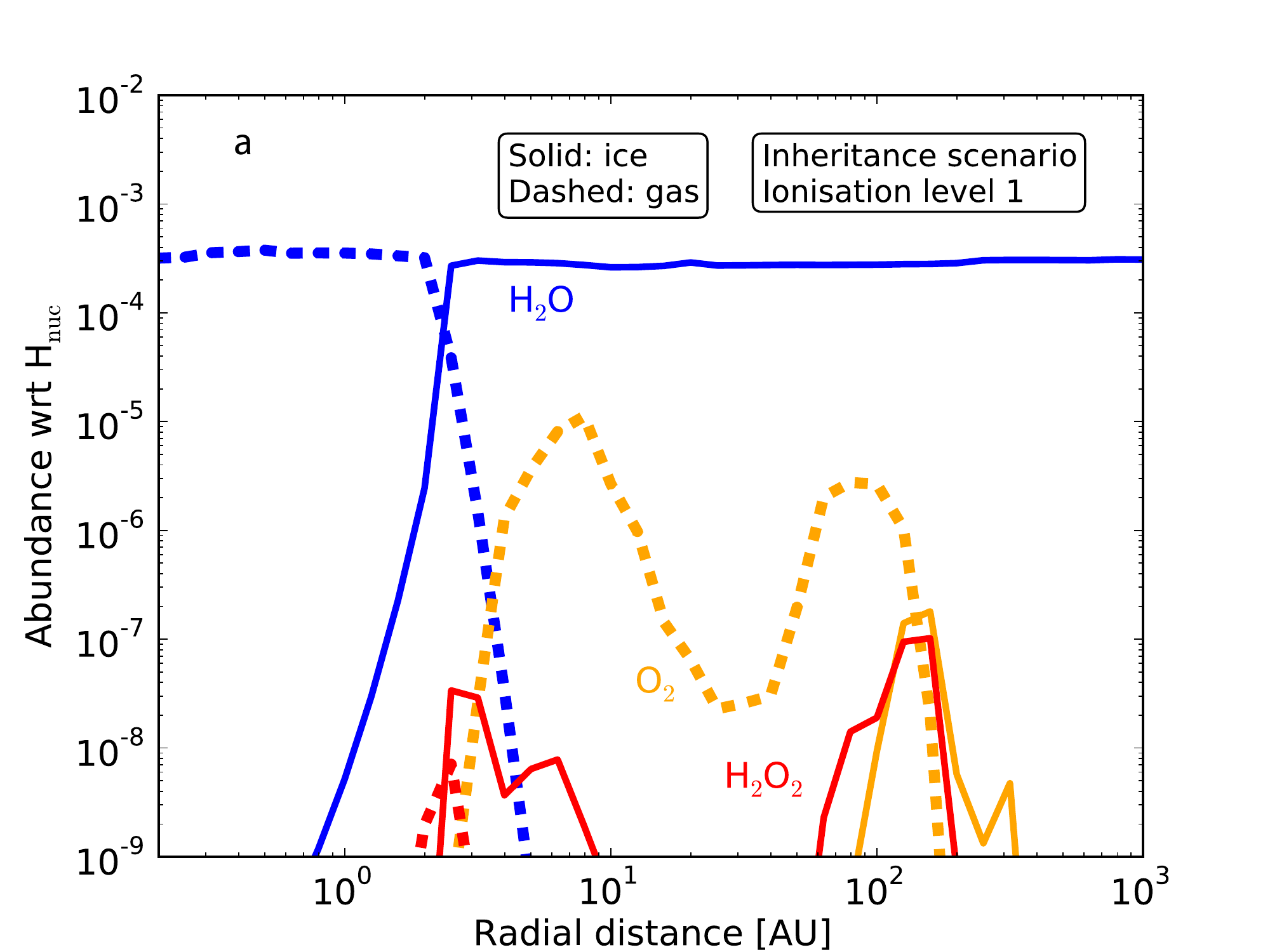}}
\subfigure{\includegraphics[width=0.33\textwidth]{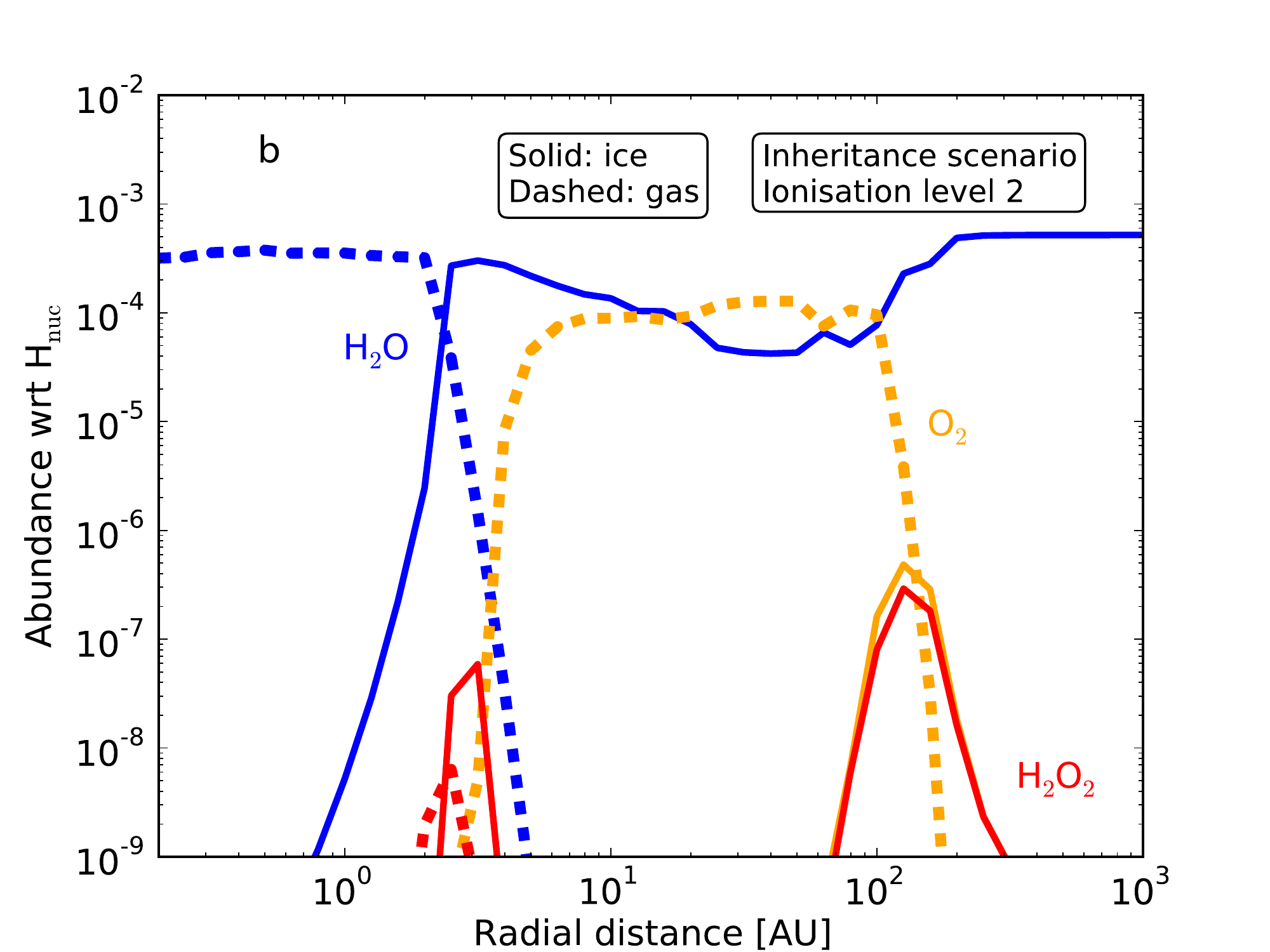}}
\subfigure{\includegraphics[width=0.33\textwidth]{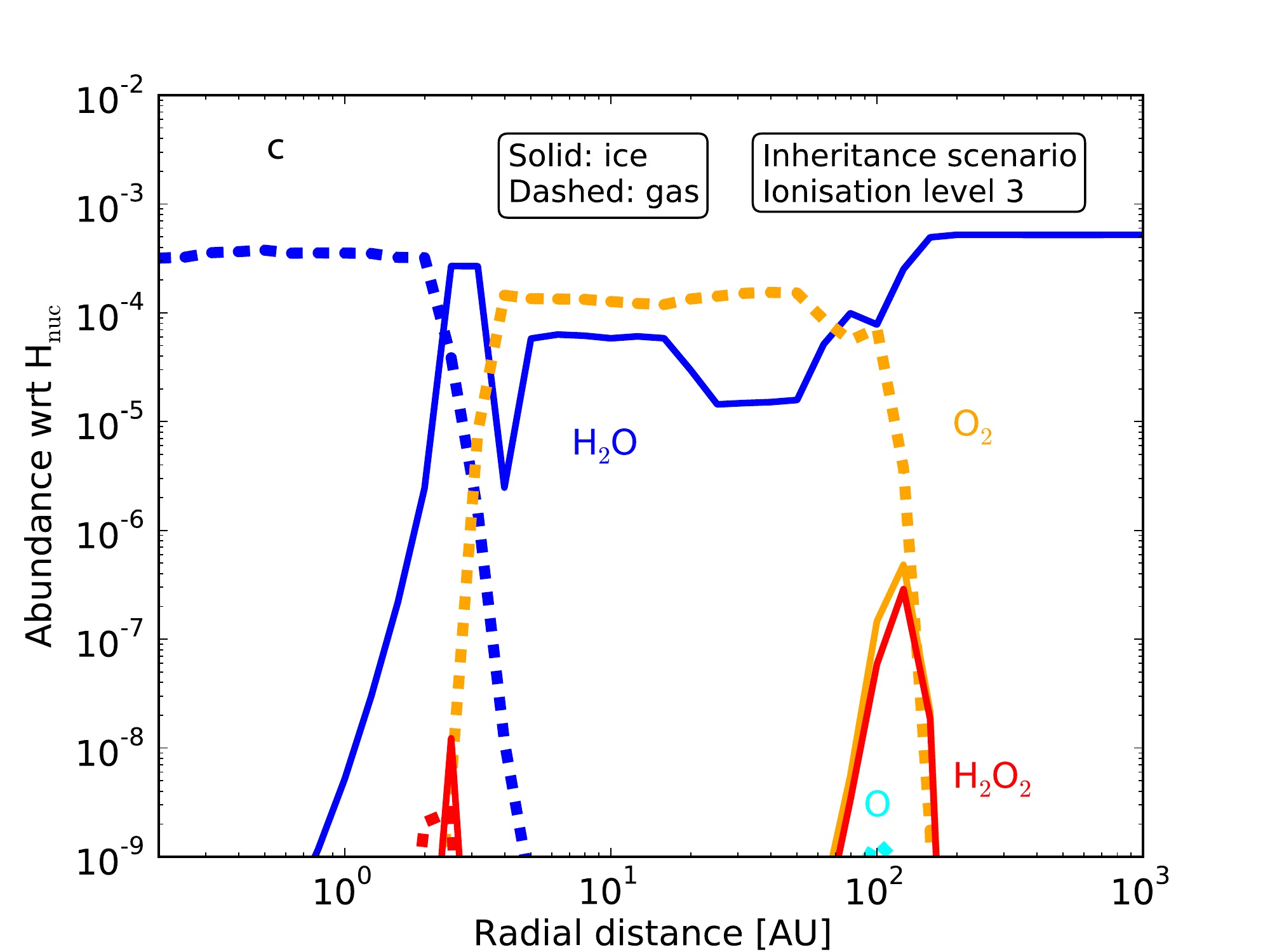}}\\
\subfigure{\includegraphics[width=0.33\textwidth]{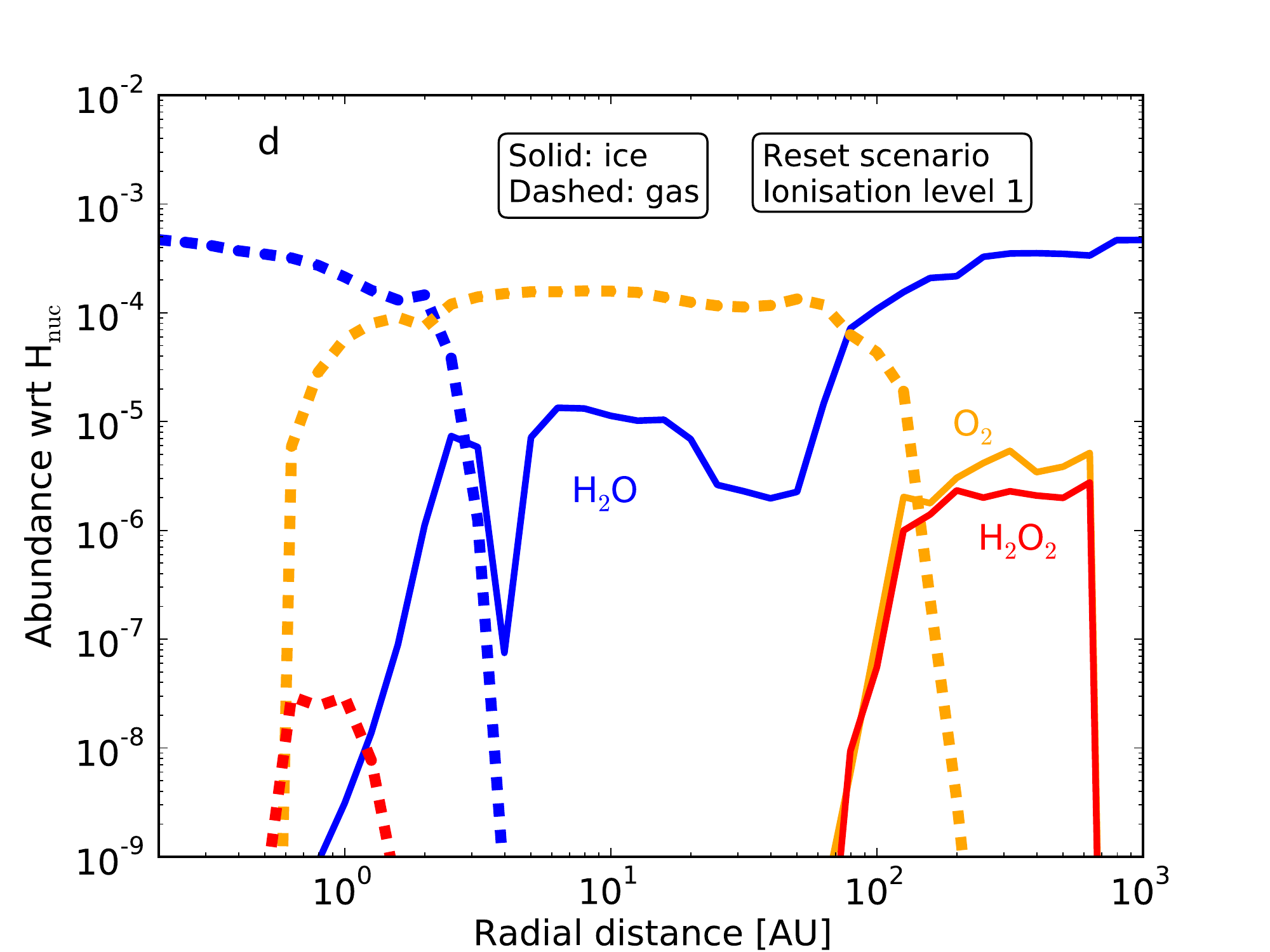}}
\subfigure{\includegraphics[width=0.33\textwidth]{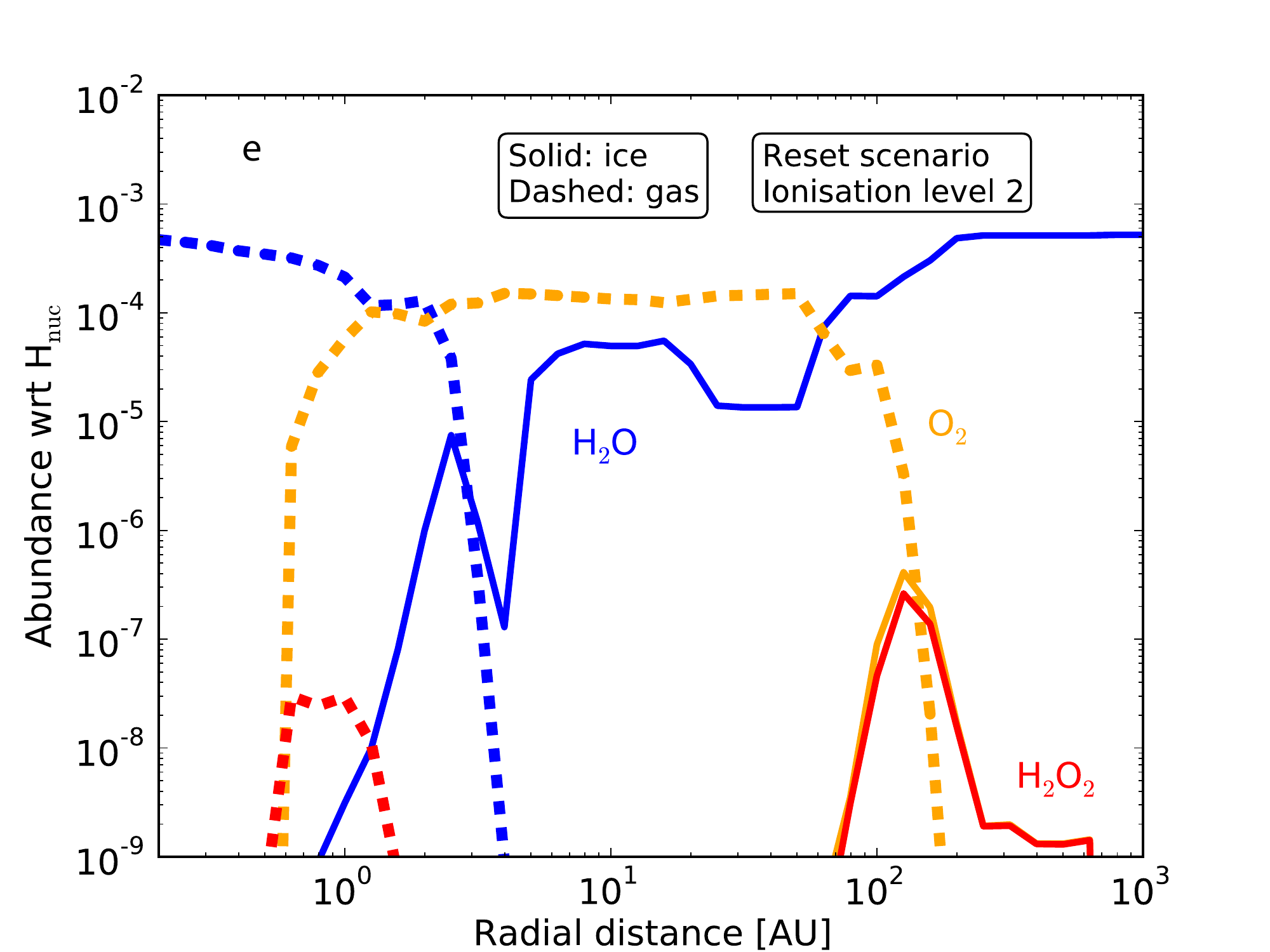}}
\subfigure{\includegraphics[width=0.33\textwidth]{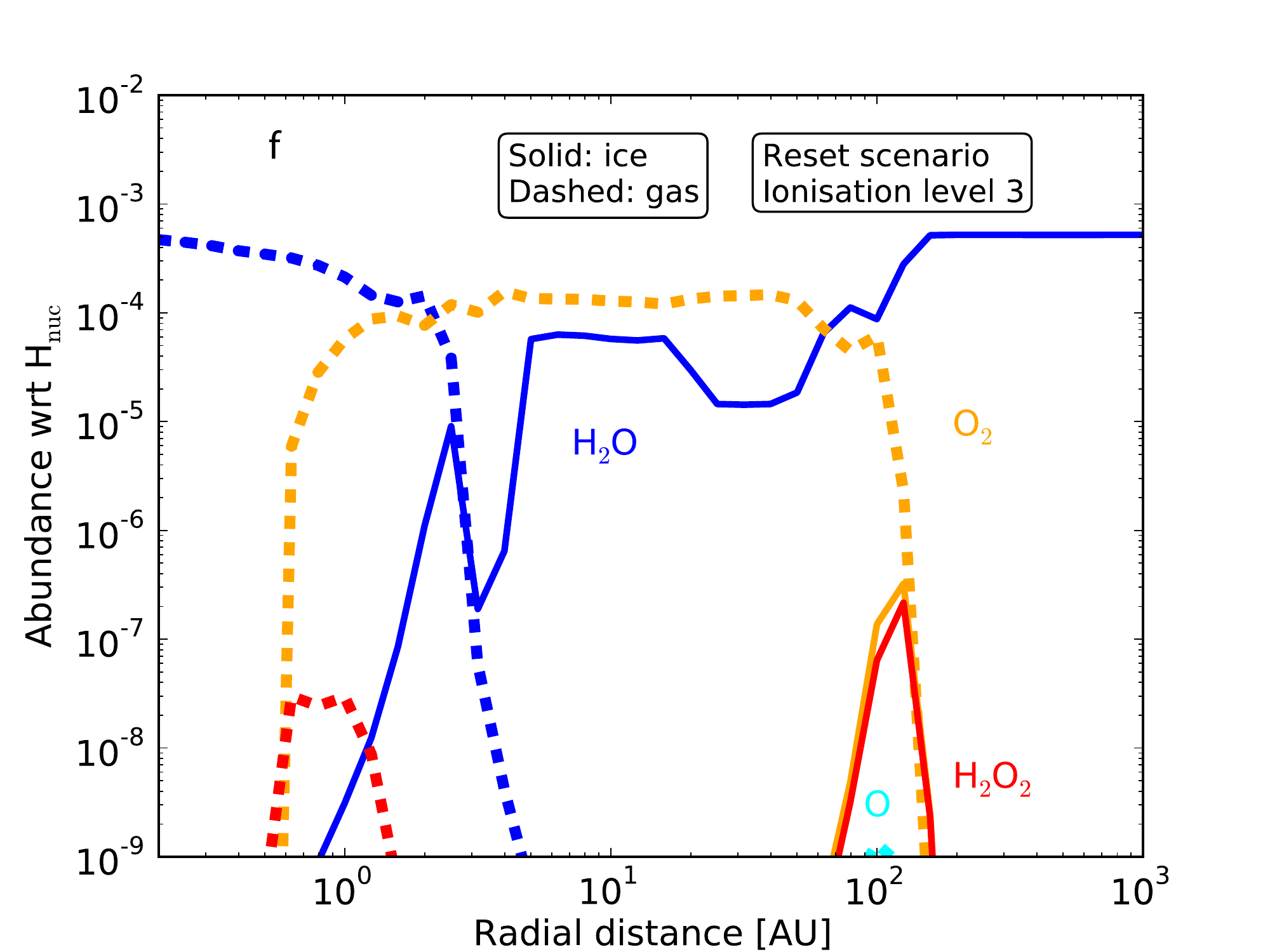}}\\
\subfigure{\includegraphics[width=0.33\textwidth]{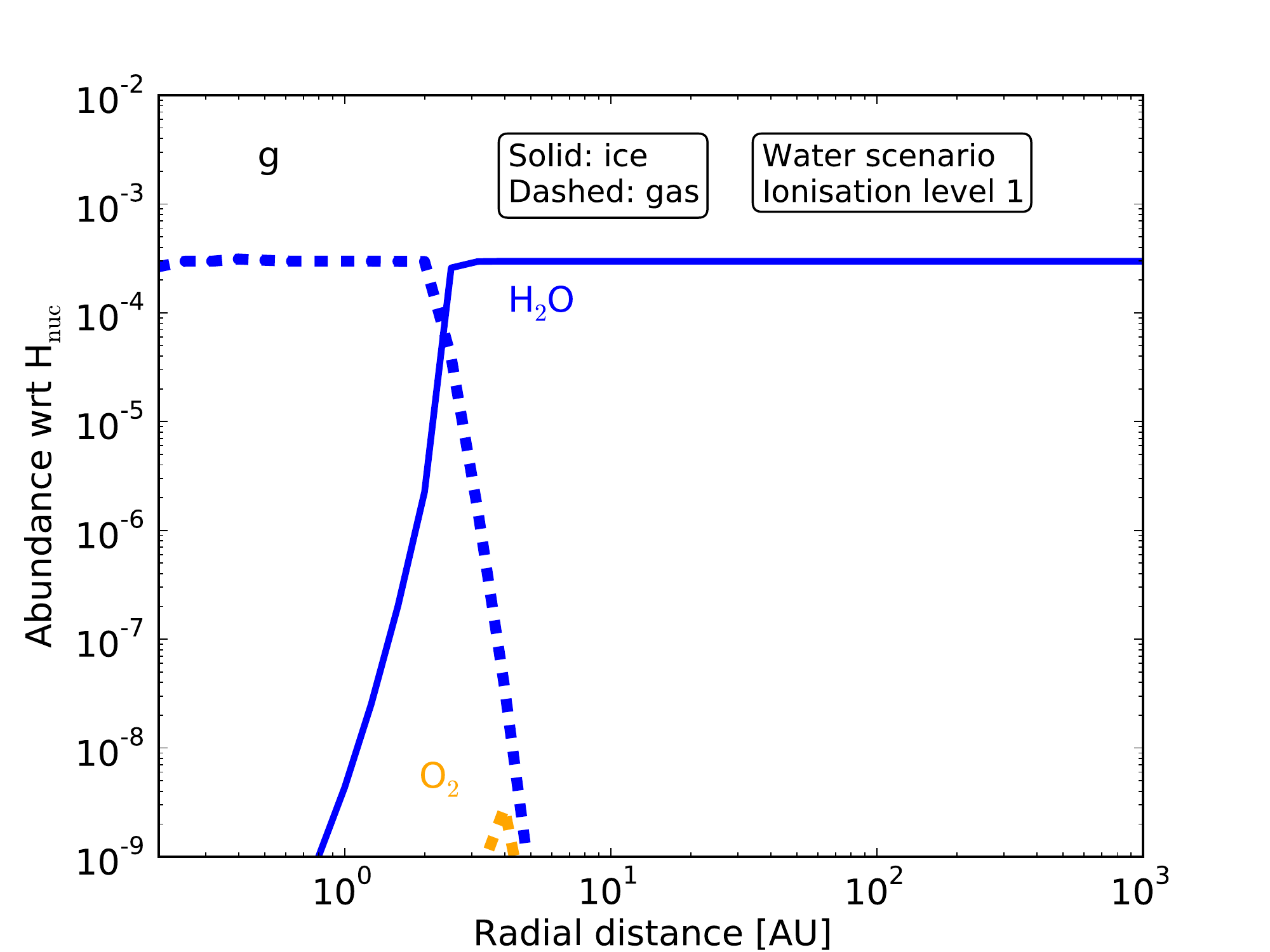}}
\subfigure{\includegraphics[width=0.33\textwidth]{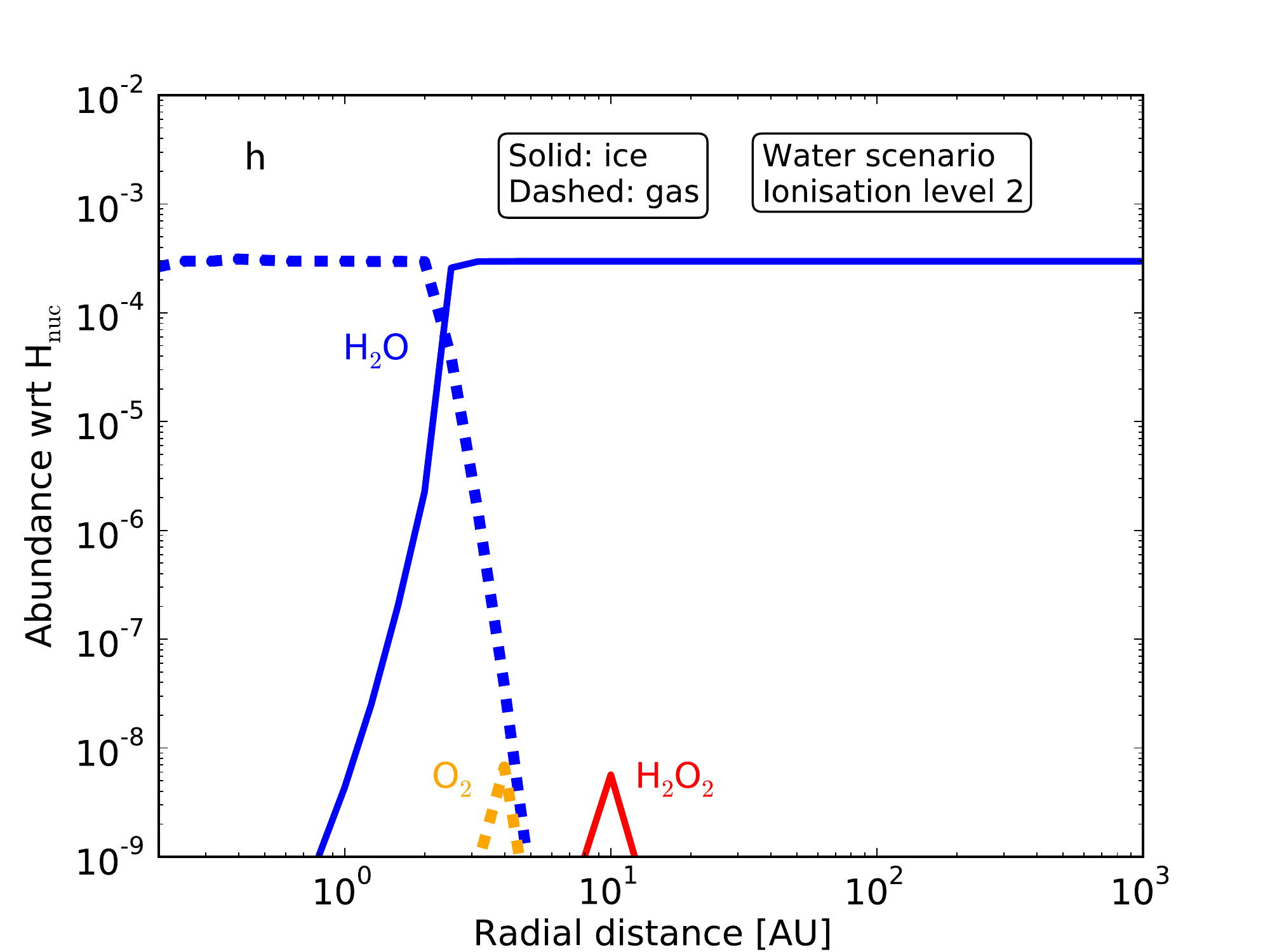}}
\subfigure{\includegraphics[width=0.33\textwidth]{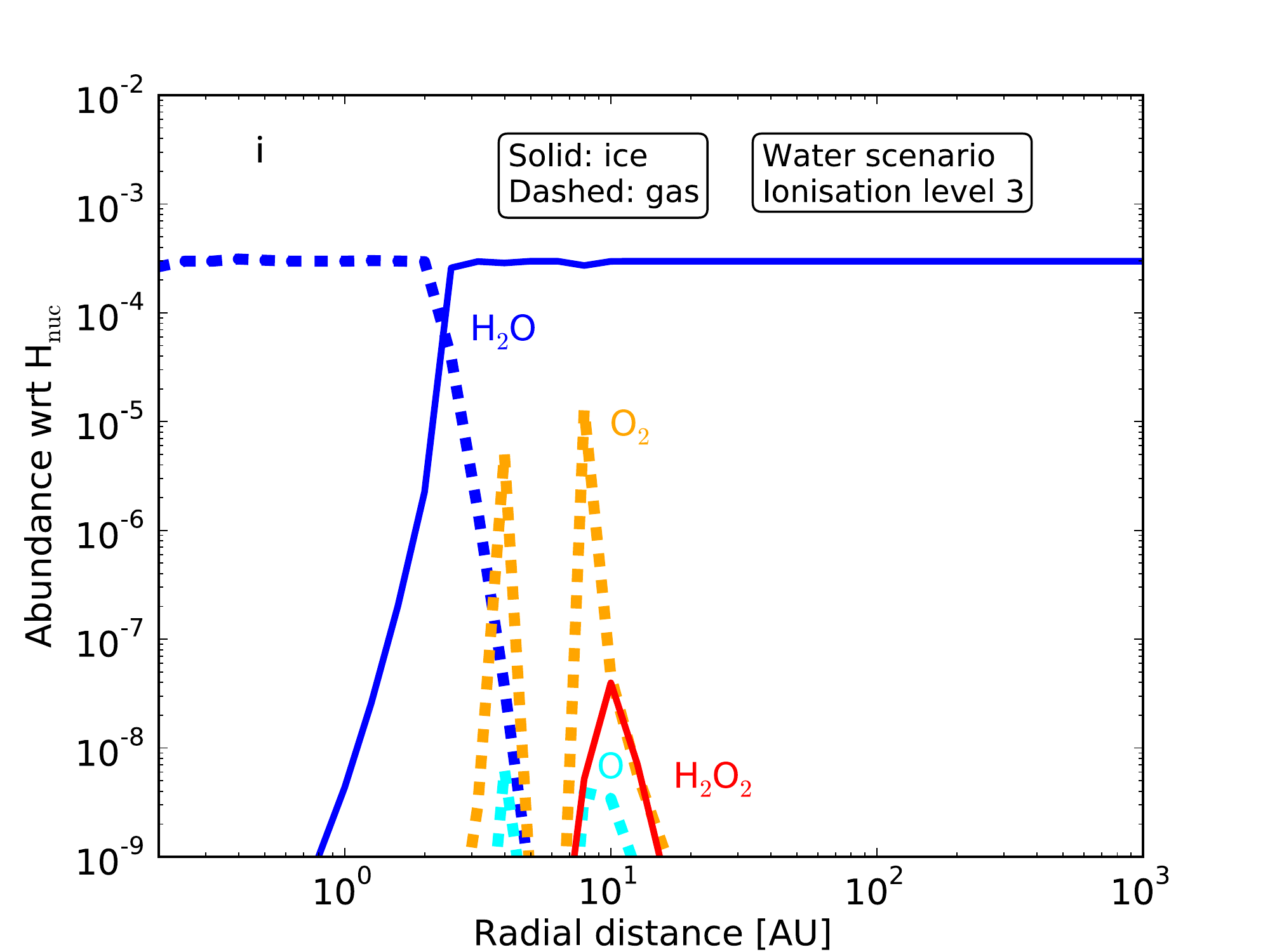}}\\
\subfigure{\includegraphics[width=0.33\textwidth]{mmsn_oxygen_ion1.pdf}}
\subfigure{\includegraphics[width=0.33\textwidth]{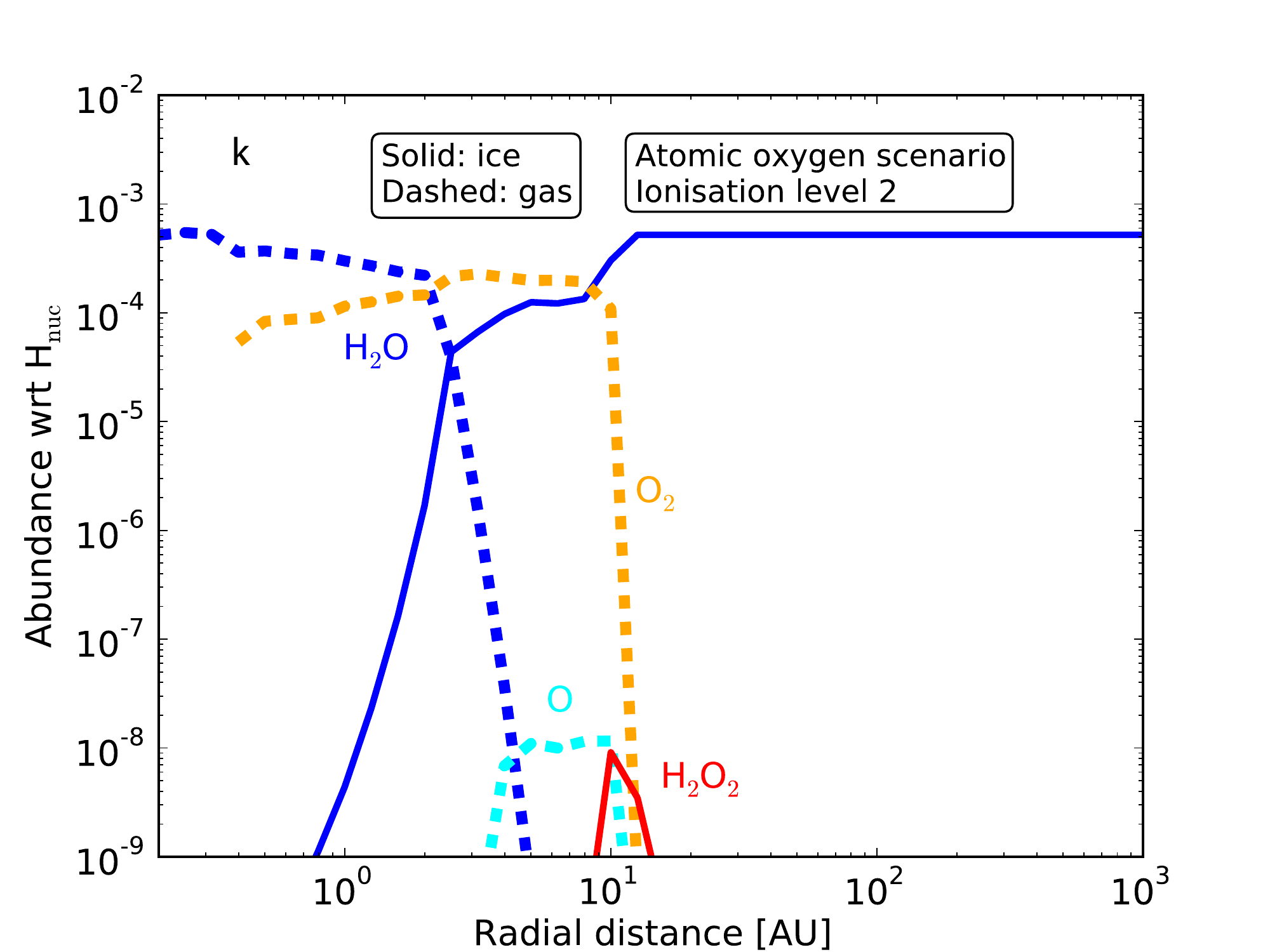}}
\subfigure{\includegraphics[width=0.33\textwidth]{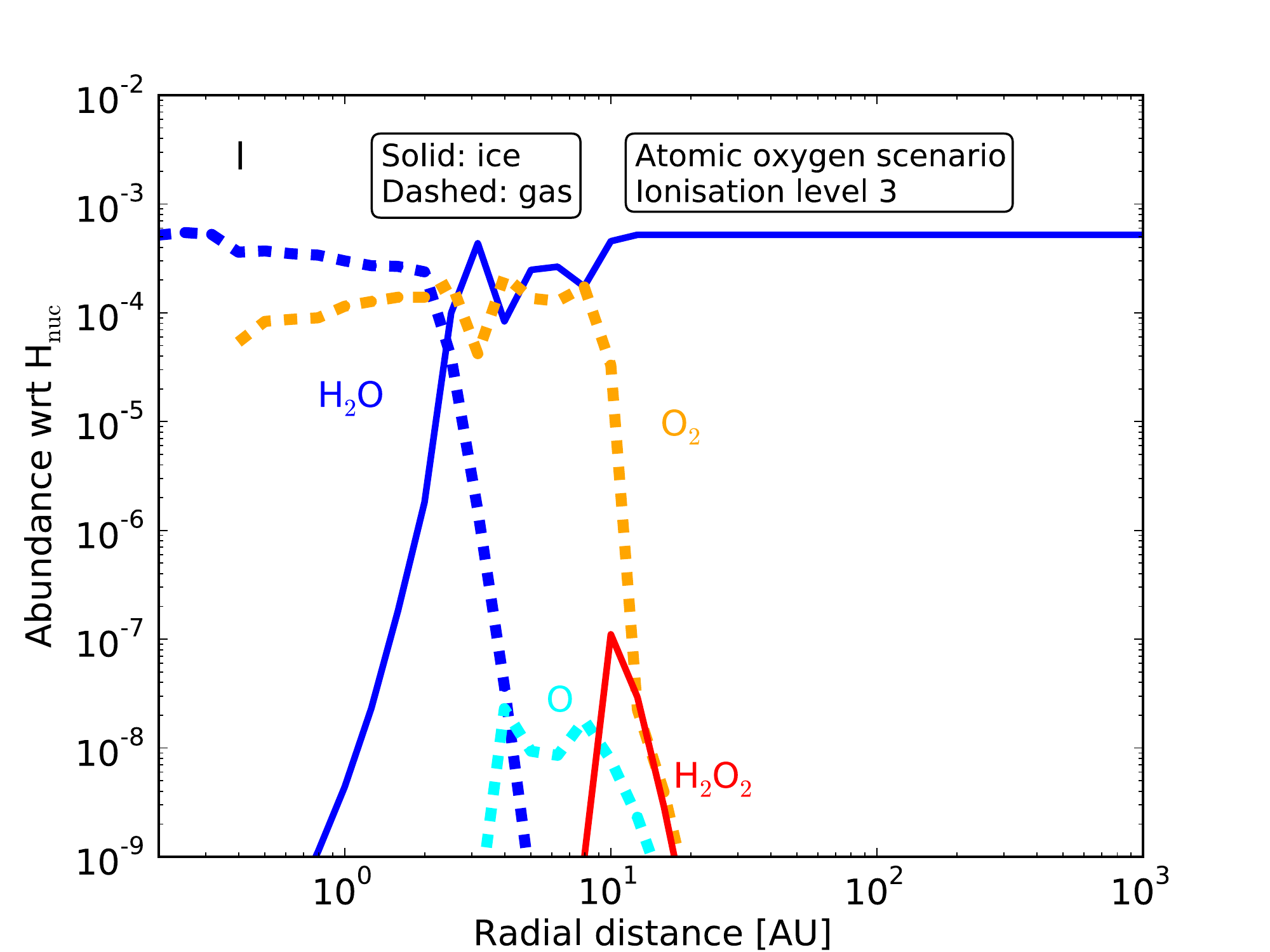}}\\
\caption{Abundances by 10 Myr evolution for. Top to bottom are the inheritance scenario, the reset scenario, the water scenario, and the oxygen scenario, see the panels. Left to right are changing ionisation levels. The chemical network utilised does not include \ce{O3} chemistry.}
\label{abun_10}
\end{figure*}

\begin{figure*}
\subfigure{\includegraphics[width=0.33\textwidth]{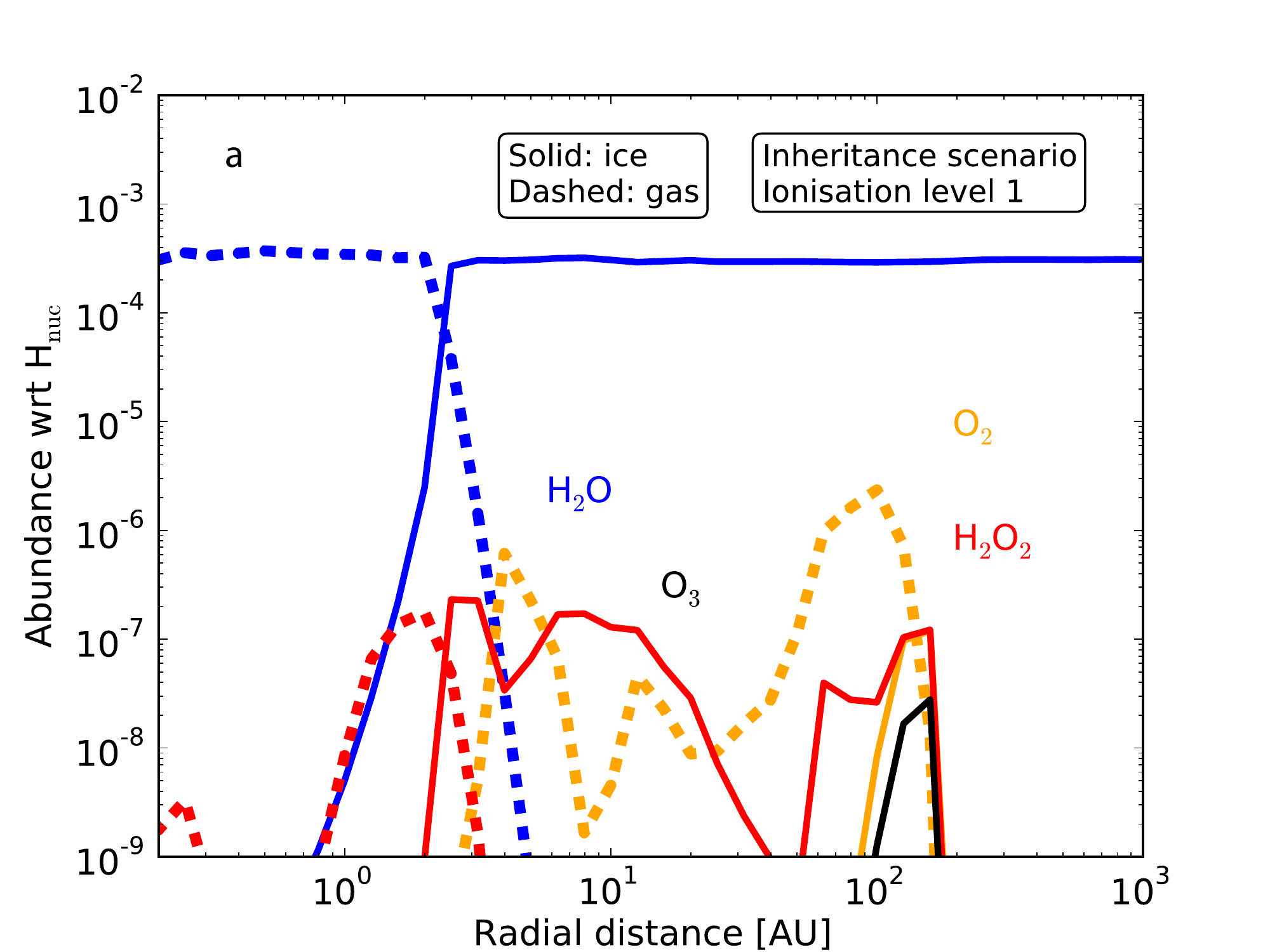}}
\subfigure{\includegraphics[width=0.33\textwidth]{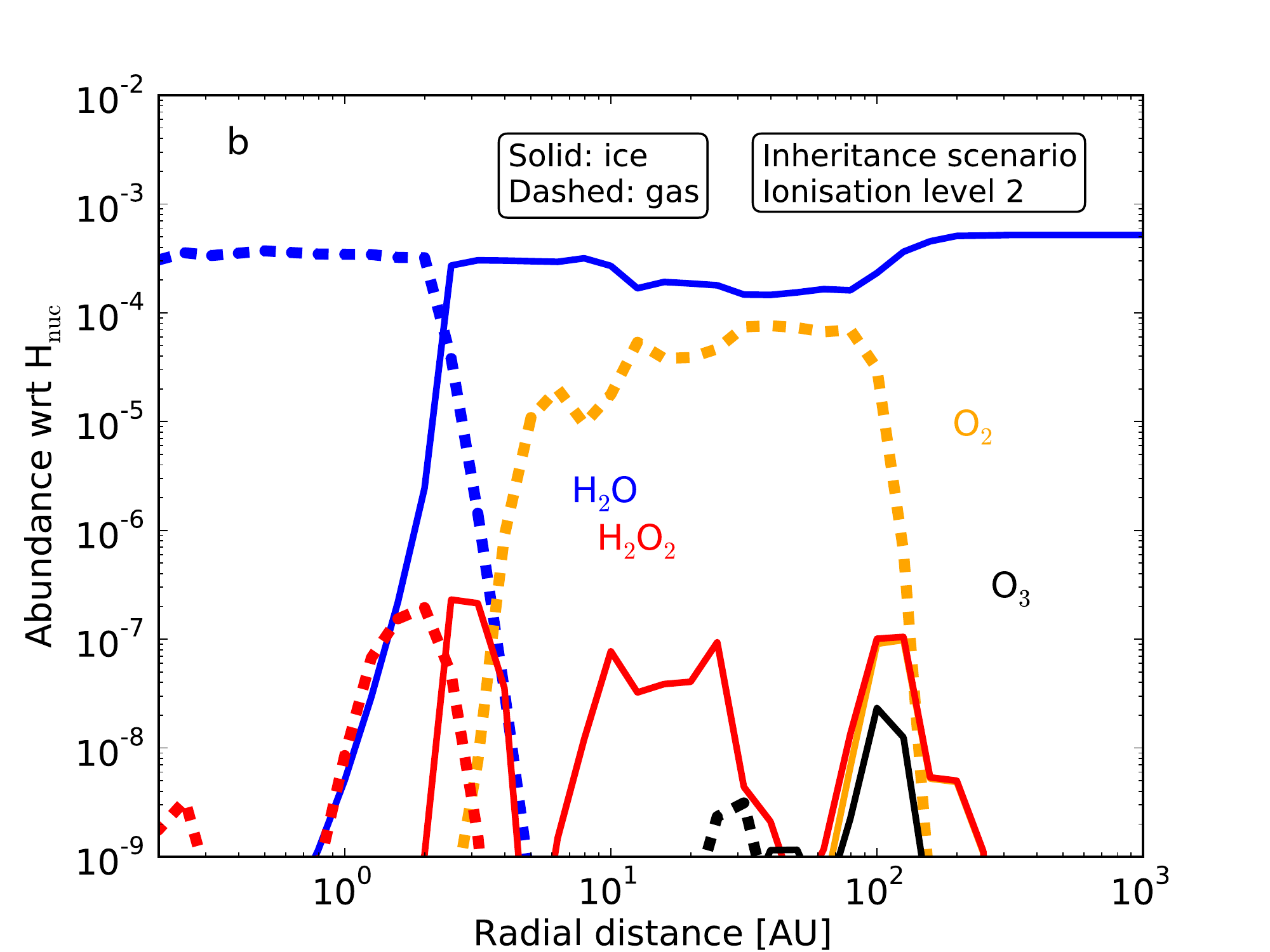}}
\subfigure{\includegraphics[width=0.33\textwidth]{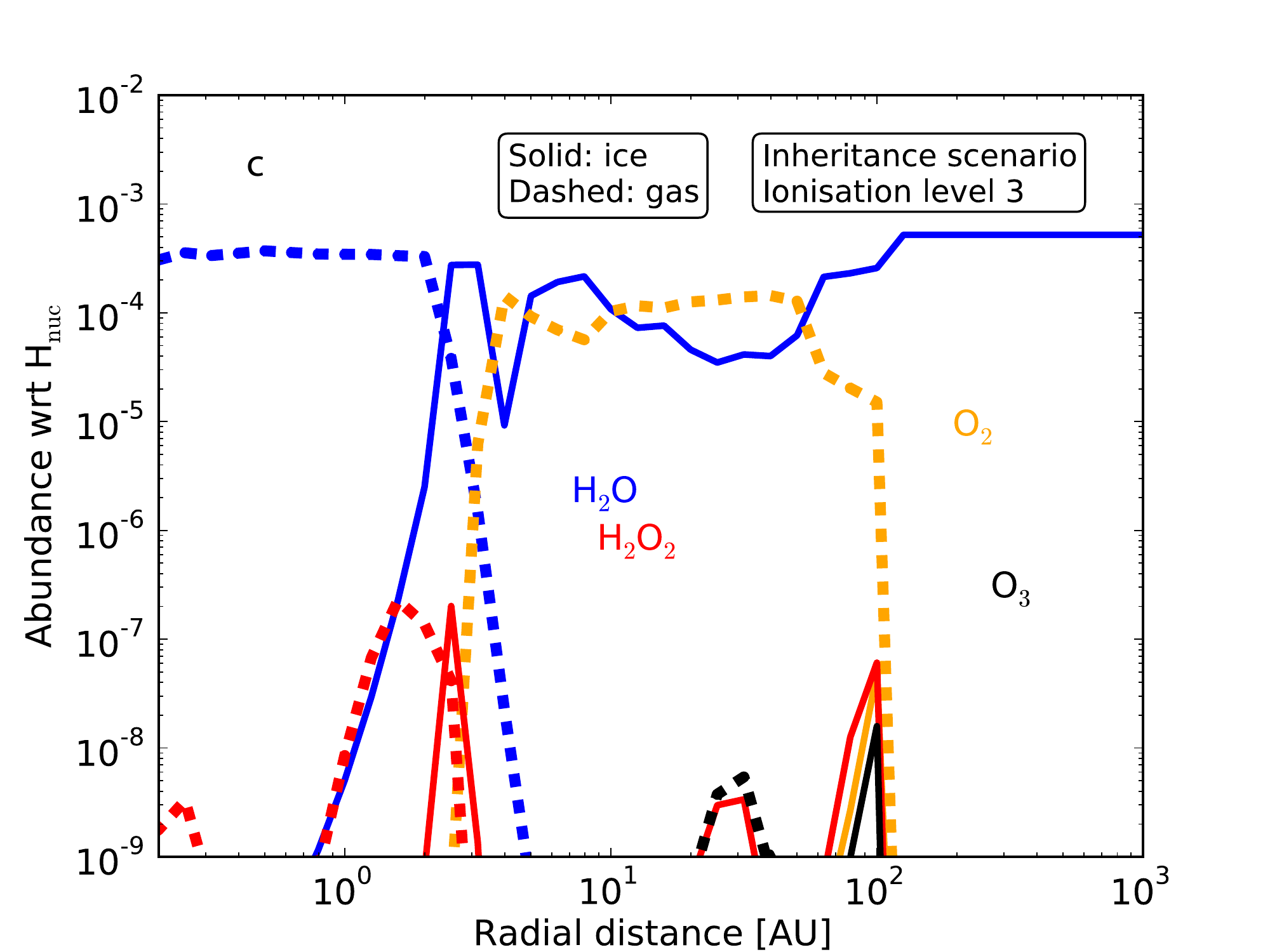}}\\
\subfigure{\includegraphics[width=0.33\textwidth]{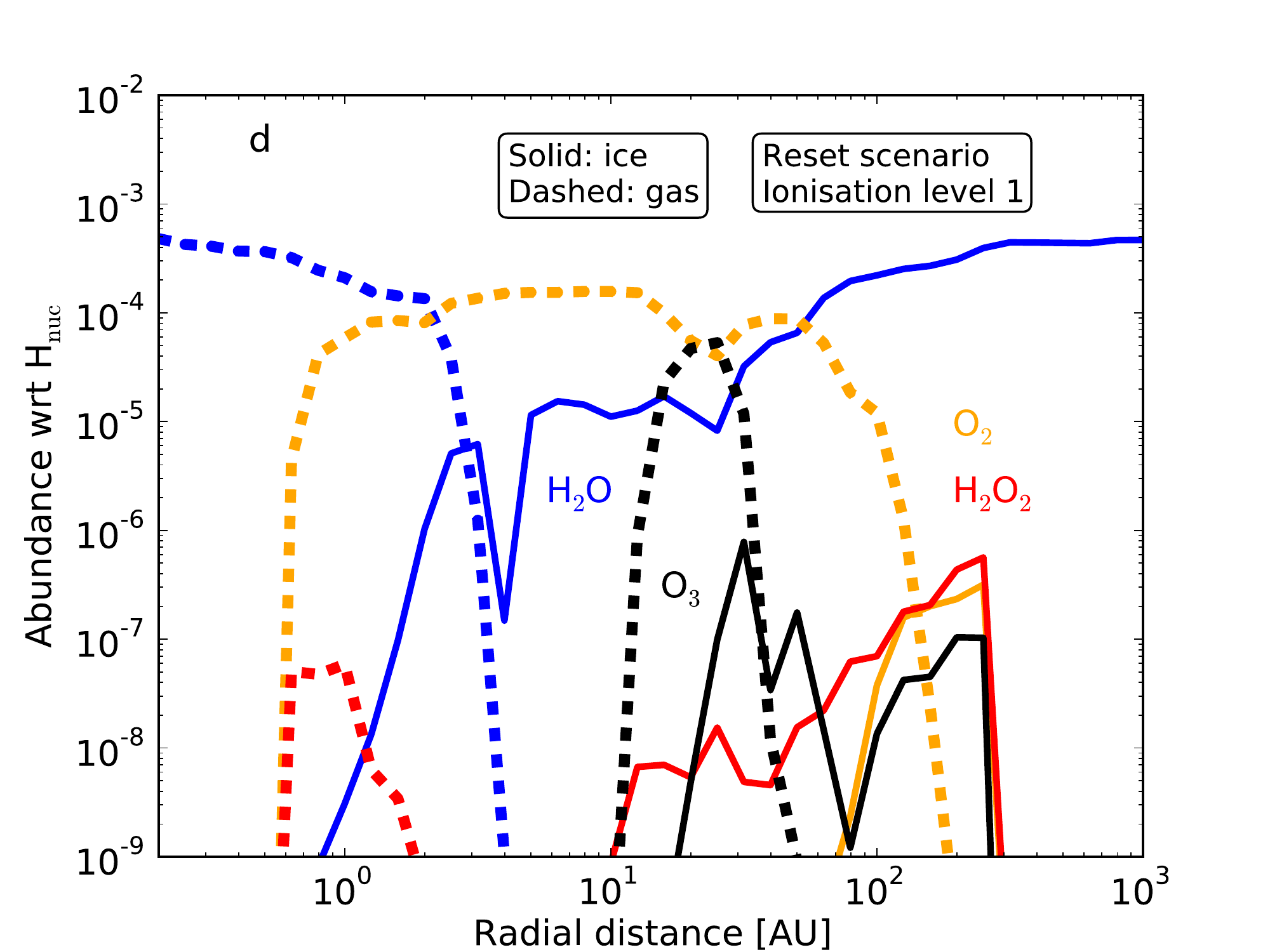}}
\subfigure{\includegraphics[width=0.33\textwidth]{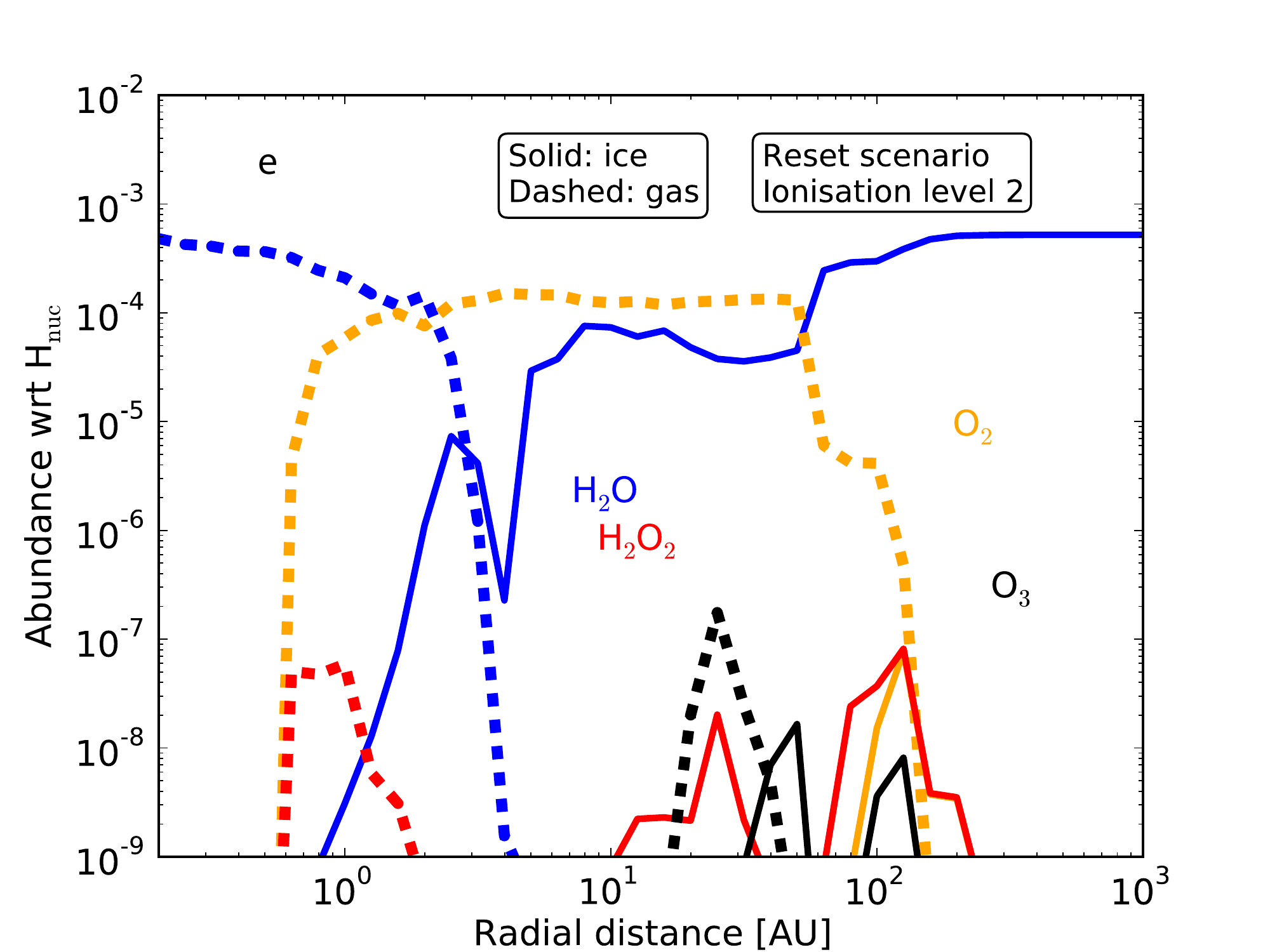}}
\subfigure{\includegraphics[width=0.33\textwidth]{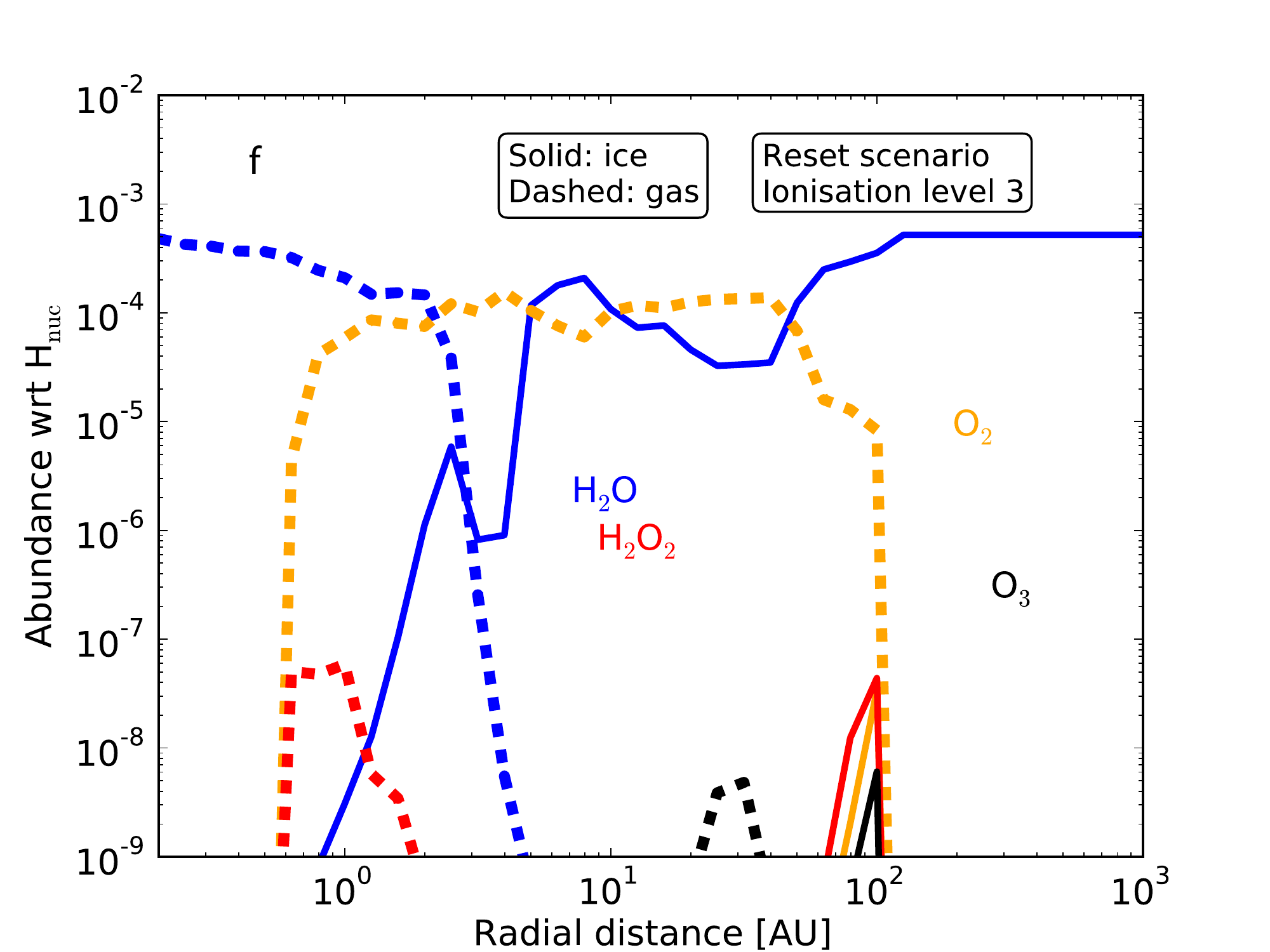}}\\ 
\subfigure{\includegraphics[width=0.33\textwidth]{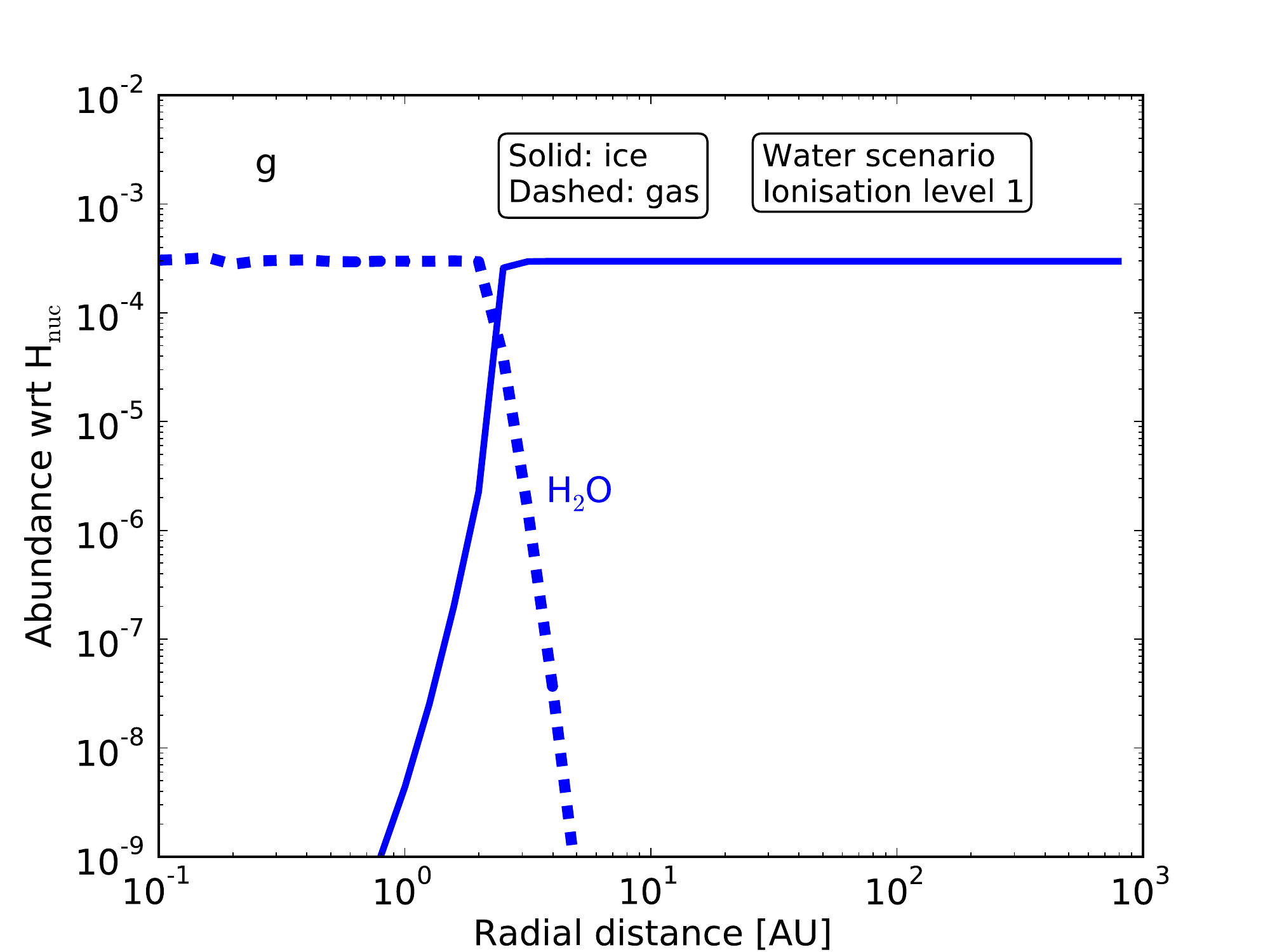}}
\subfigure{\includegraphics[width=0.33\textwidth]{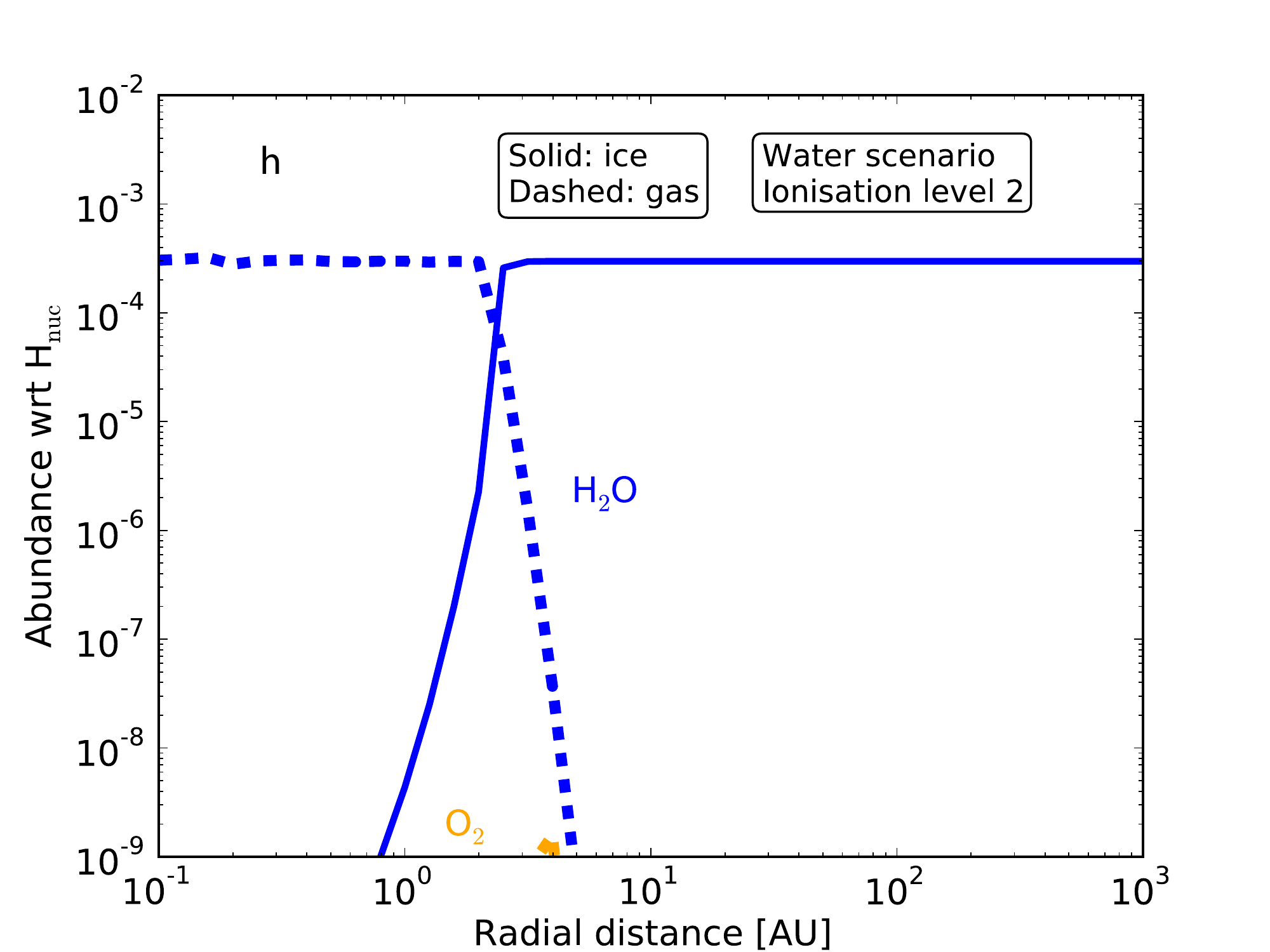}}
\subfigure{\includegraphics[width=0.33\textwidth]{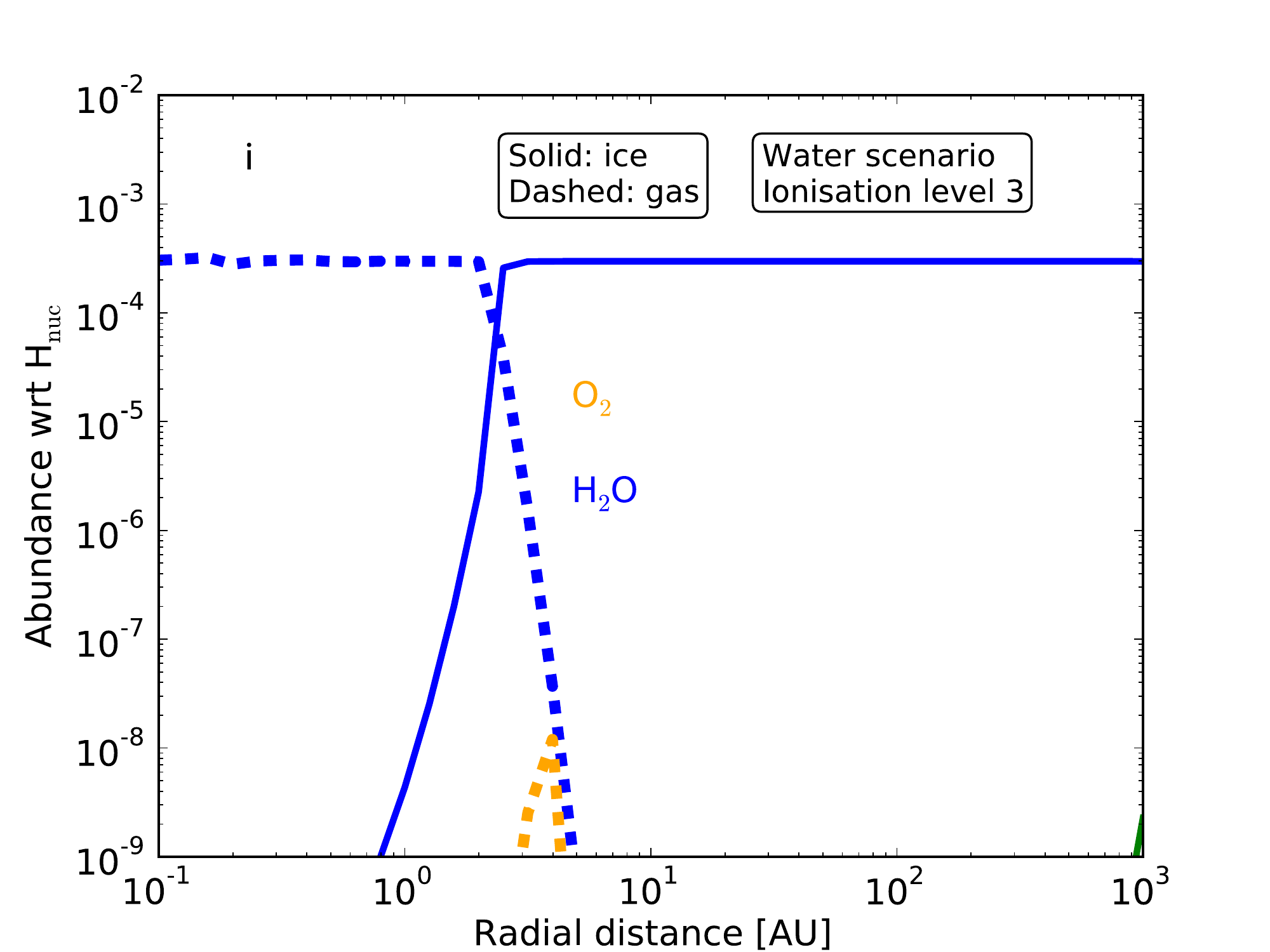}}\\
\subfigure{\includegraphics[width=0.33\textwidth]{mmsn_ozone_oxygen_ion1.pdf}}
\subfigure{\includegraphics[width=0.33\textwidth]{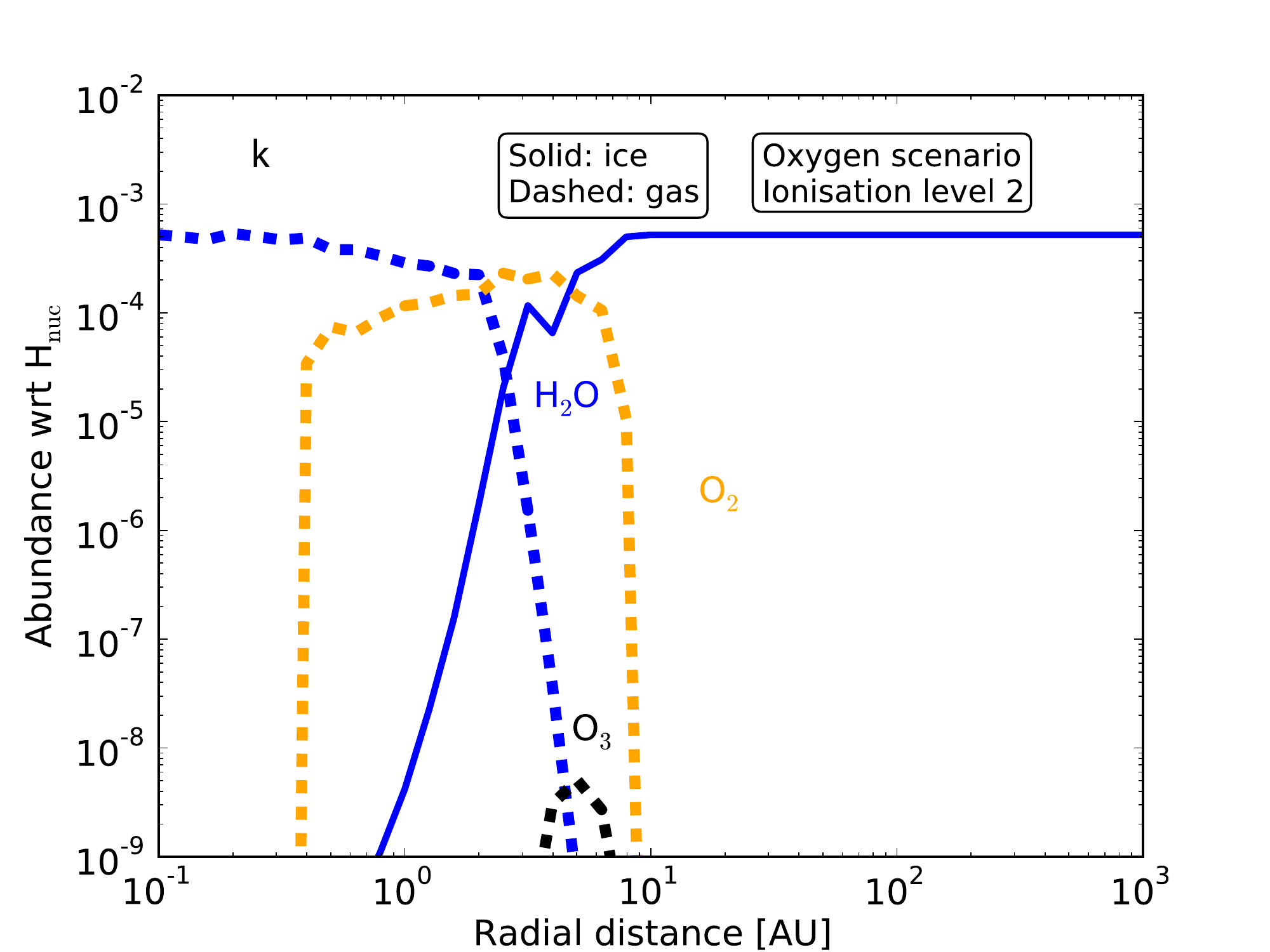}}
\subfigure{\includegraphics[width=0.33\textwidth]{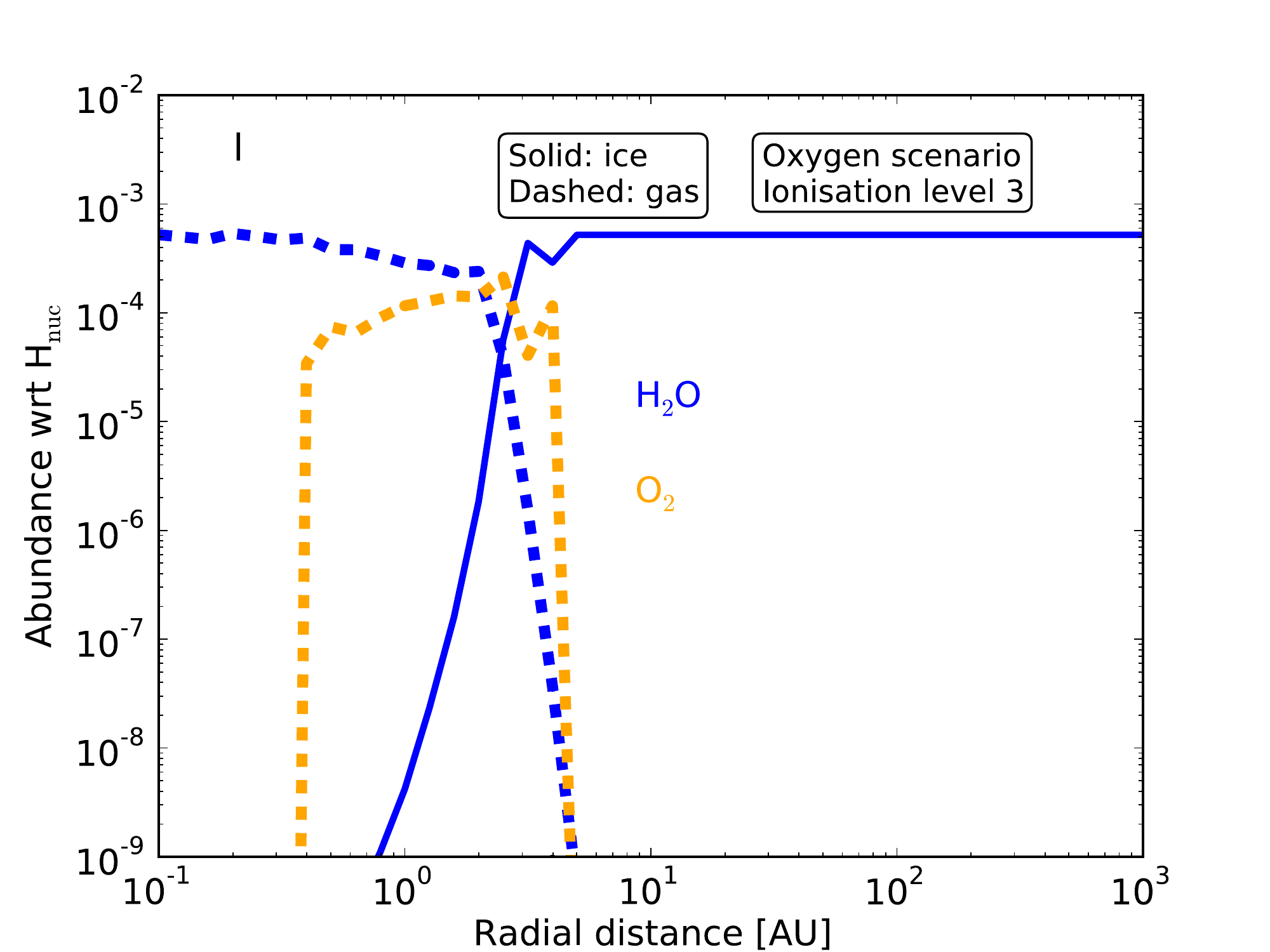}}\\
\caption{Abundances by 10 Myr evolution for. Top to bottom are the inheritance scenario, the reset scenario, the water scenario, and the oxygen scenario, see the panels. Left to right are changing ionisation levels. The chemical network utilised includes \ce{O3} chemistry.}
\label{ozone_10}

\end{figure*}

Fig. \ref{abun_10} shows the gas and ice abundances of O, \ce{O2}, \ce{H2O} and \ce{H2O2} as a function of radius by 10 Myr of evolution, for the chemical network without \ce{O3} chemistry. Fig. \ref{ozone_10} shows the gas and ice abundances of \ce{O2}, \ce{H2O} and \ce{H2O2} and \ce{O3} as a function of radius by 10 Myr of evolution, for the chemical network including \ce{O3}.
\begin{figure*}
\subfigure{\includegraphics[width=0.22\textwidth]{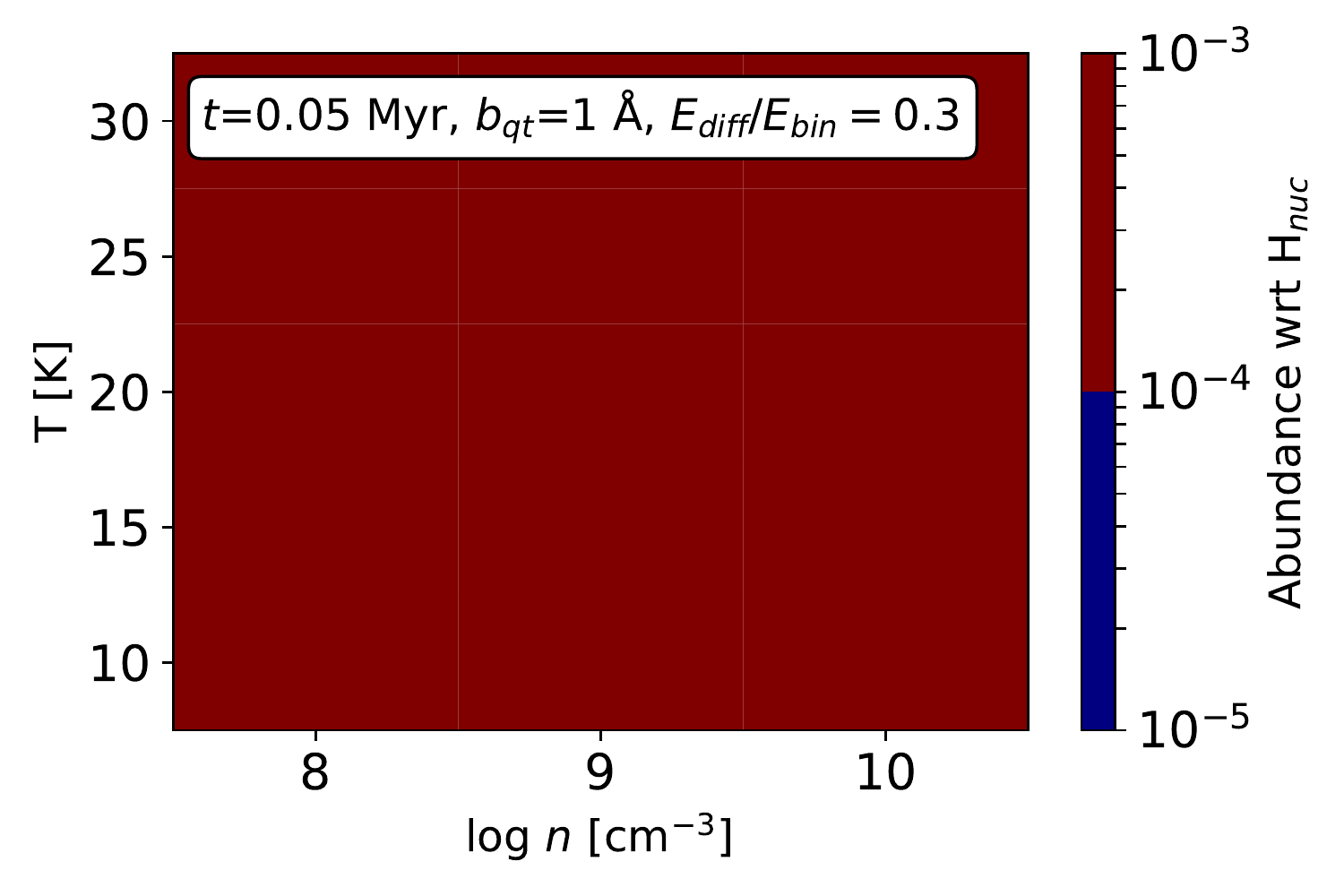}}
\subfigure{\includegraphics[width=0.22\textwidth]{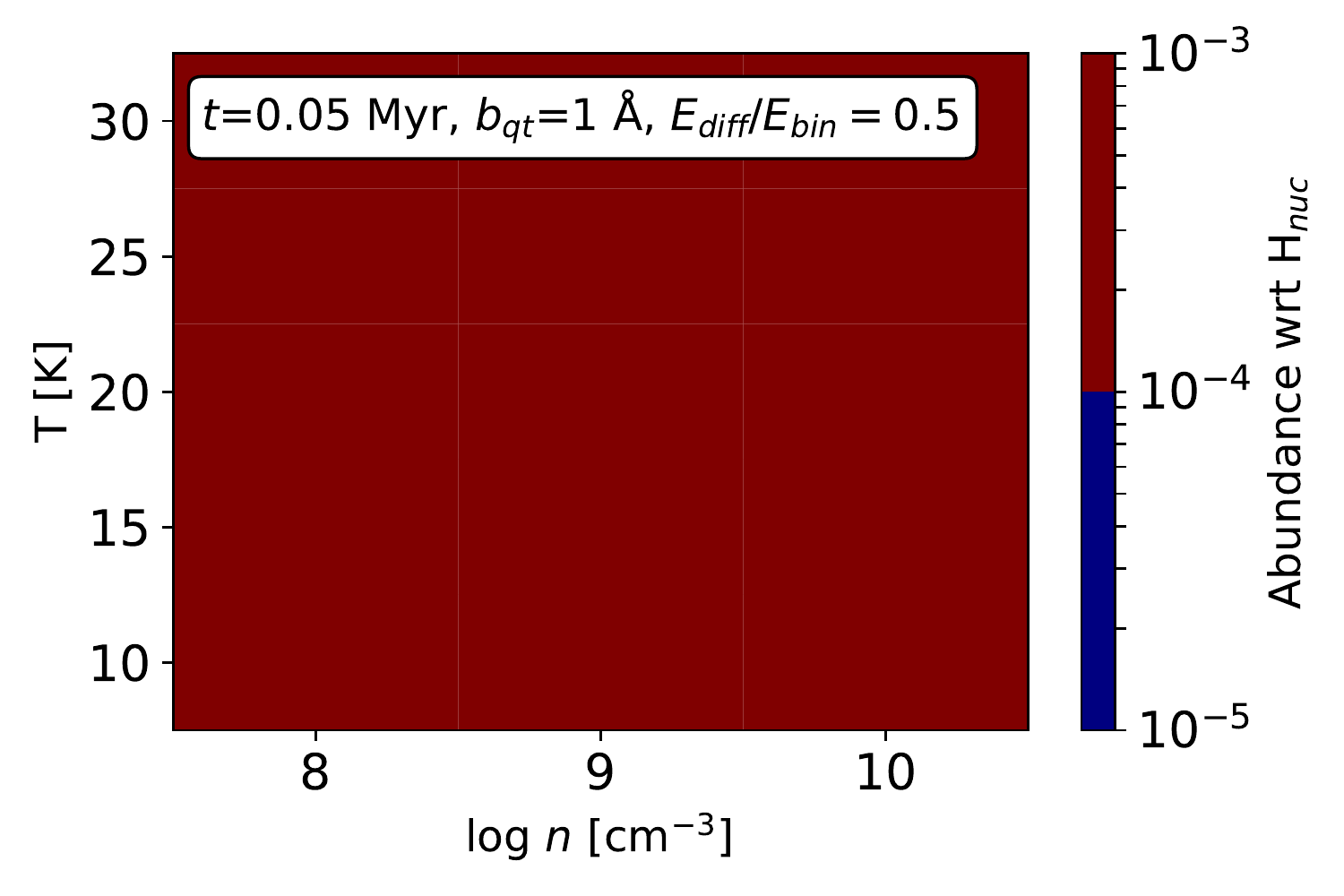}}
\subfigure{\includegraphics[width=0.22\textwidth]{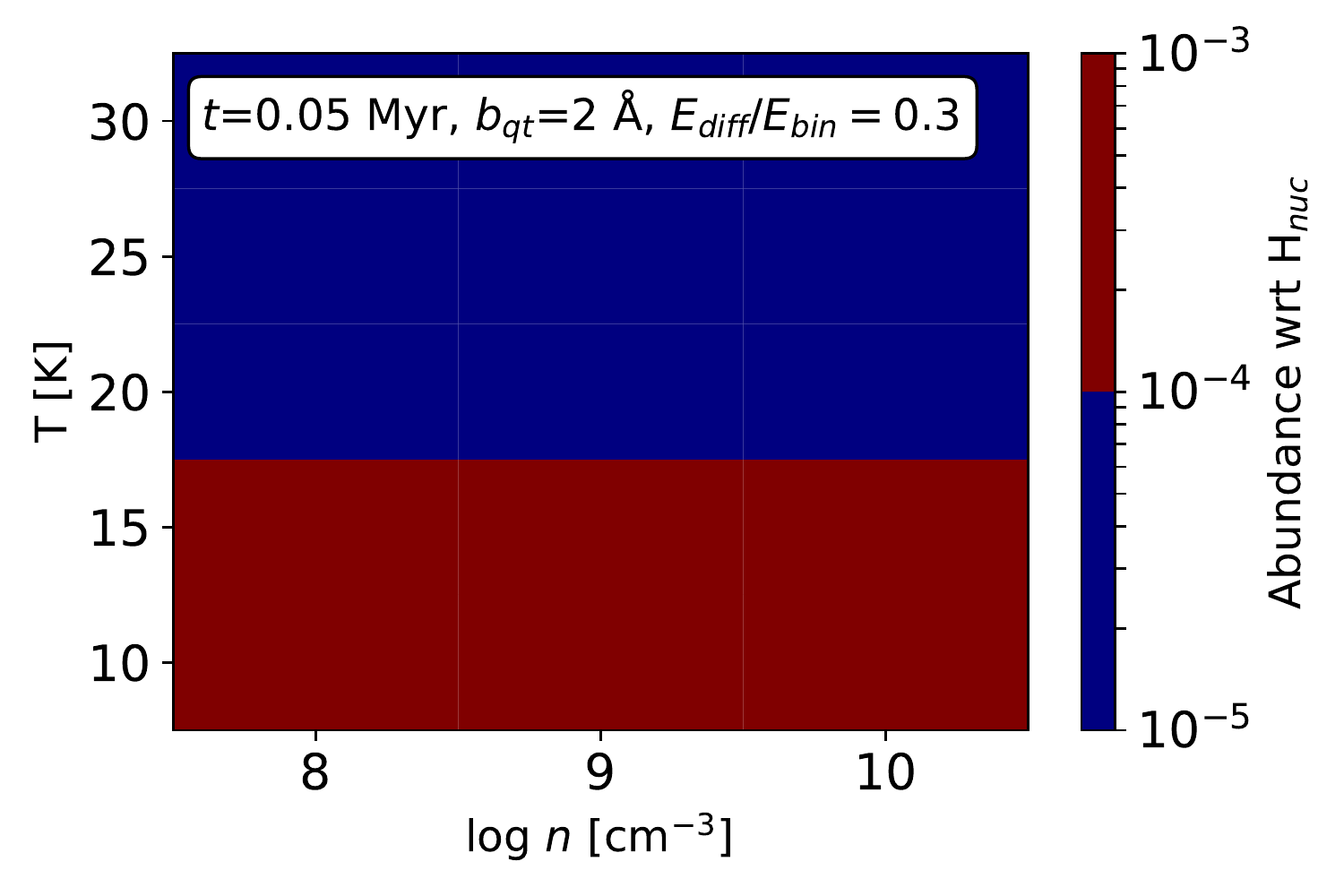}}
\subfigure{\includegraphics[width=0.22\textwidth]{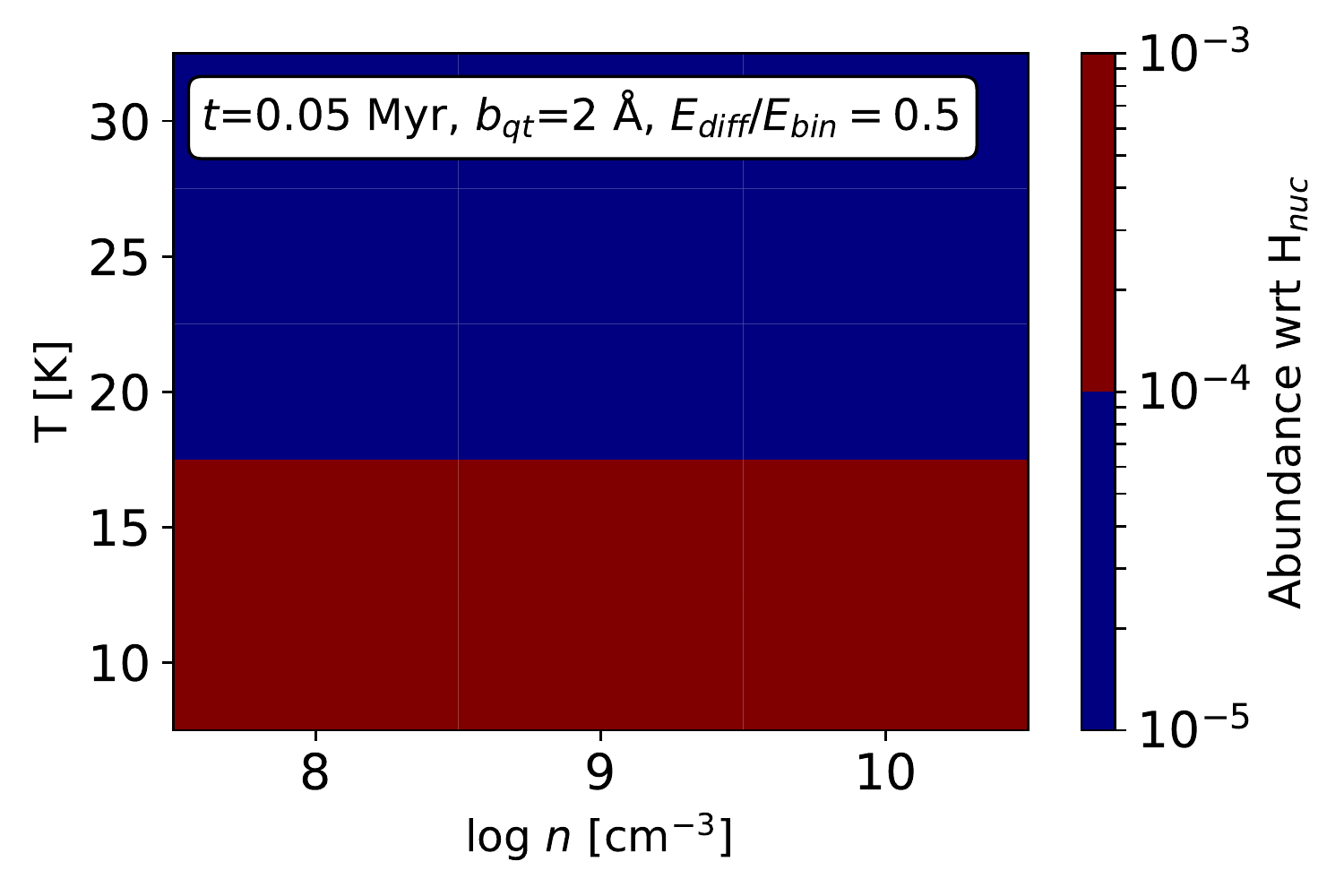}}\\
\subfigure{\includegraphics[width=0.22\textwidth]{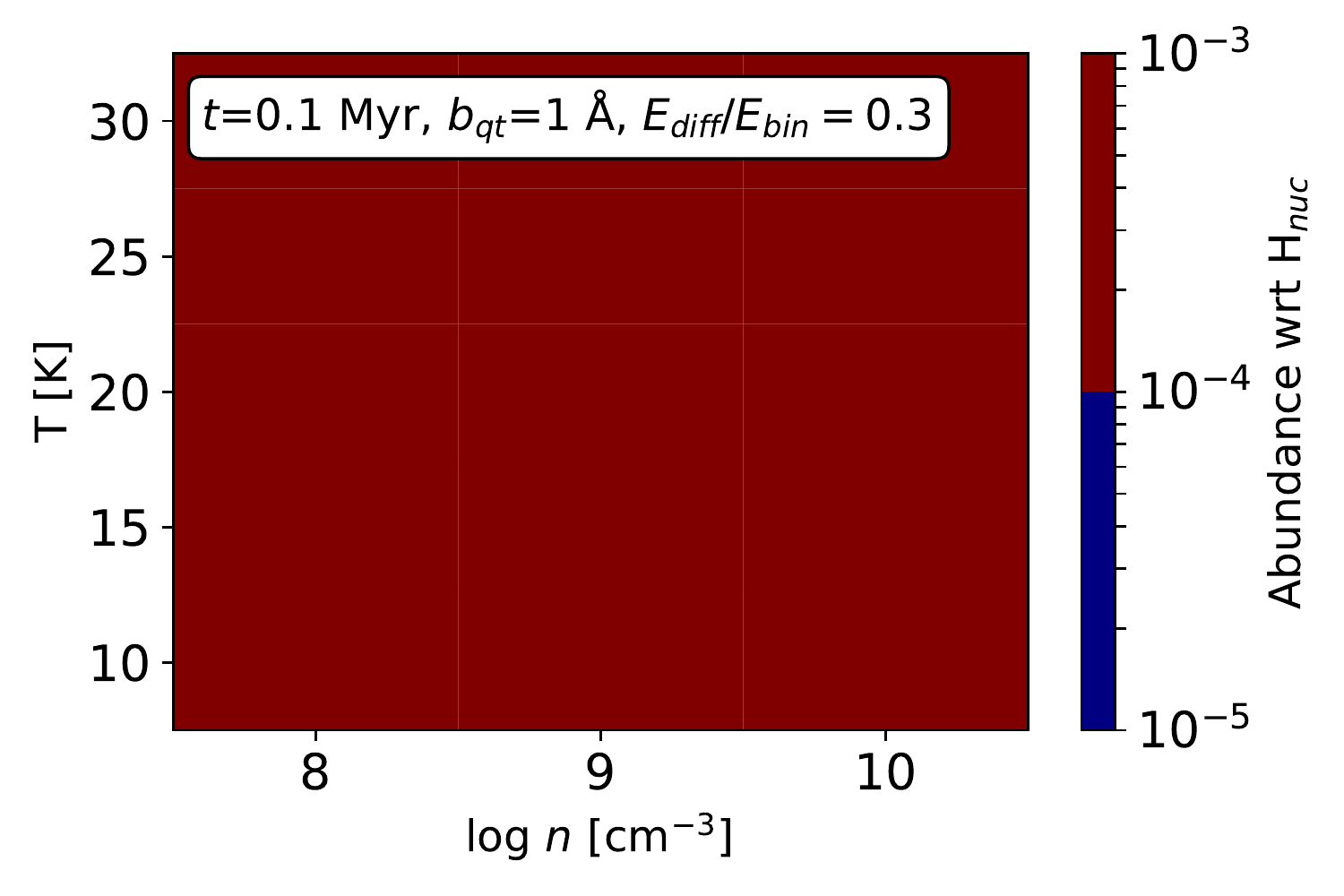}}
\subfigure{\includegraphics[width=0.22\textwidth]{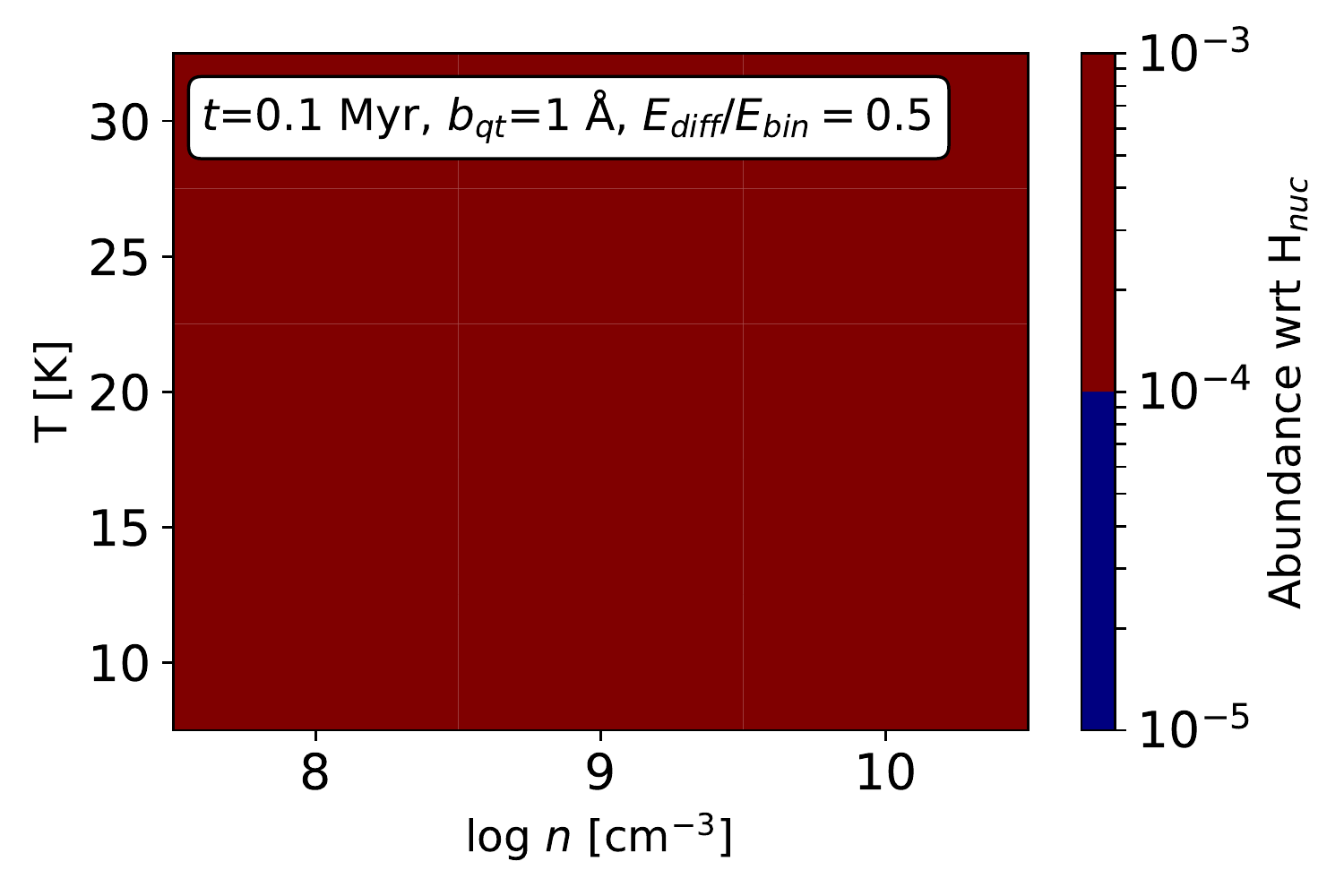}}
\subfigure{\includegraphics[width=0.22\textwidth]{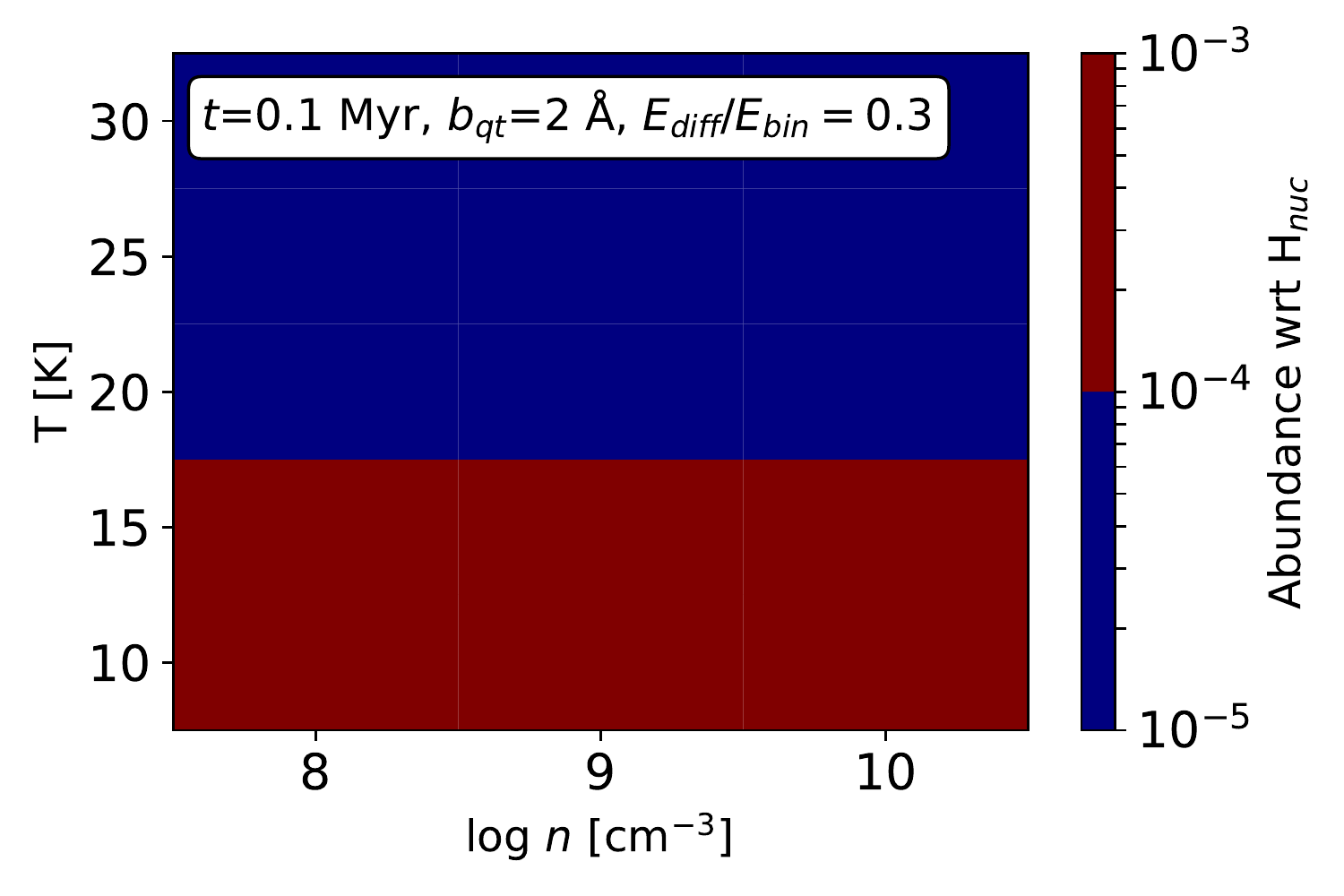}}
\subfigure{\includegraphics[width=0.22\textwidth]{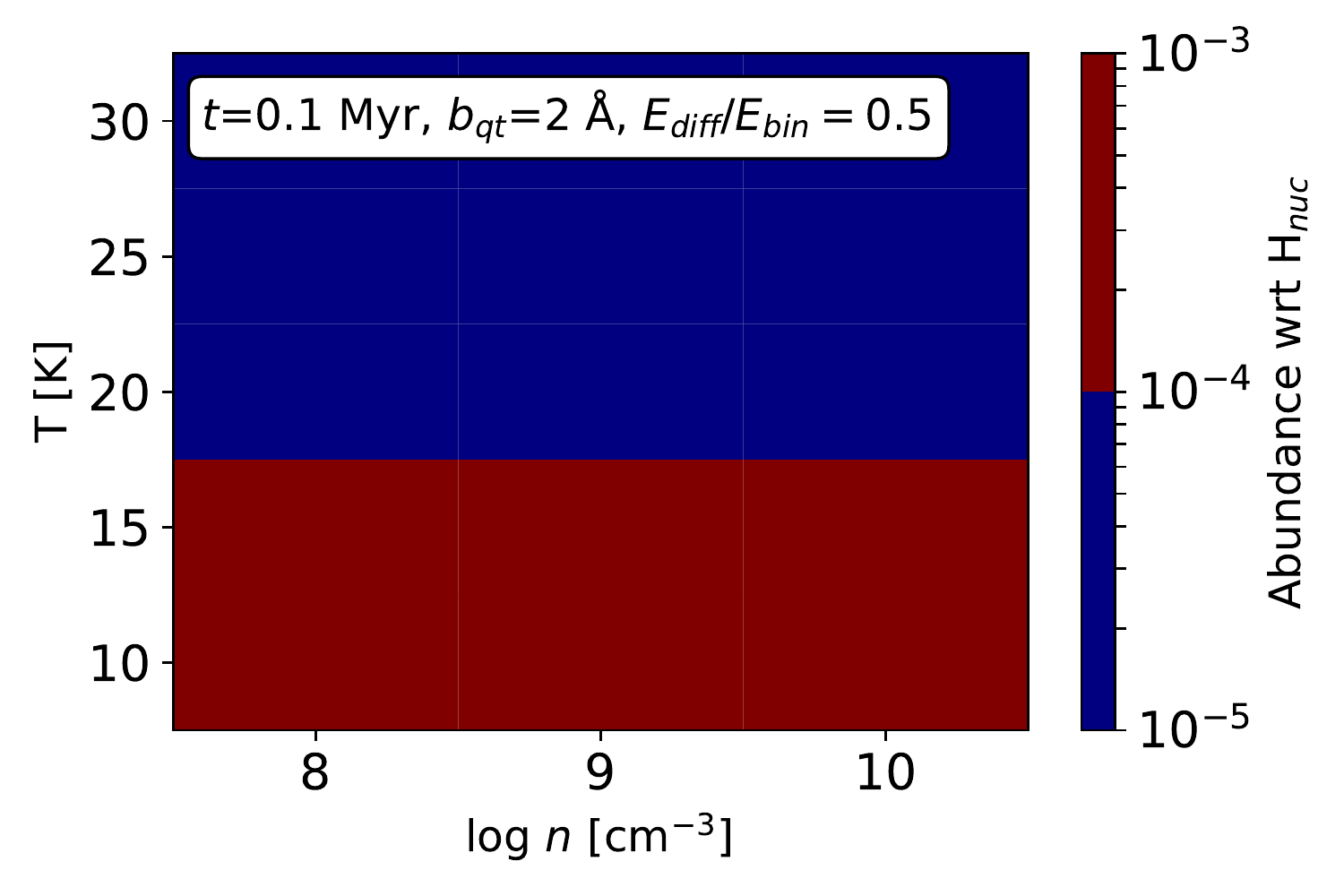}}\\
\subfigure{\includegraphics[width=0.22\textwidth]{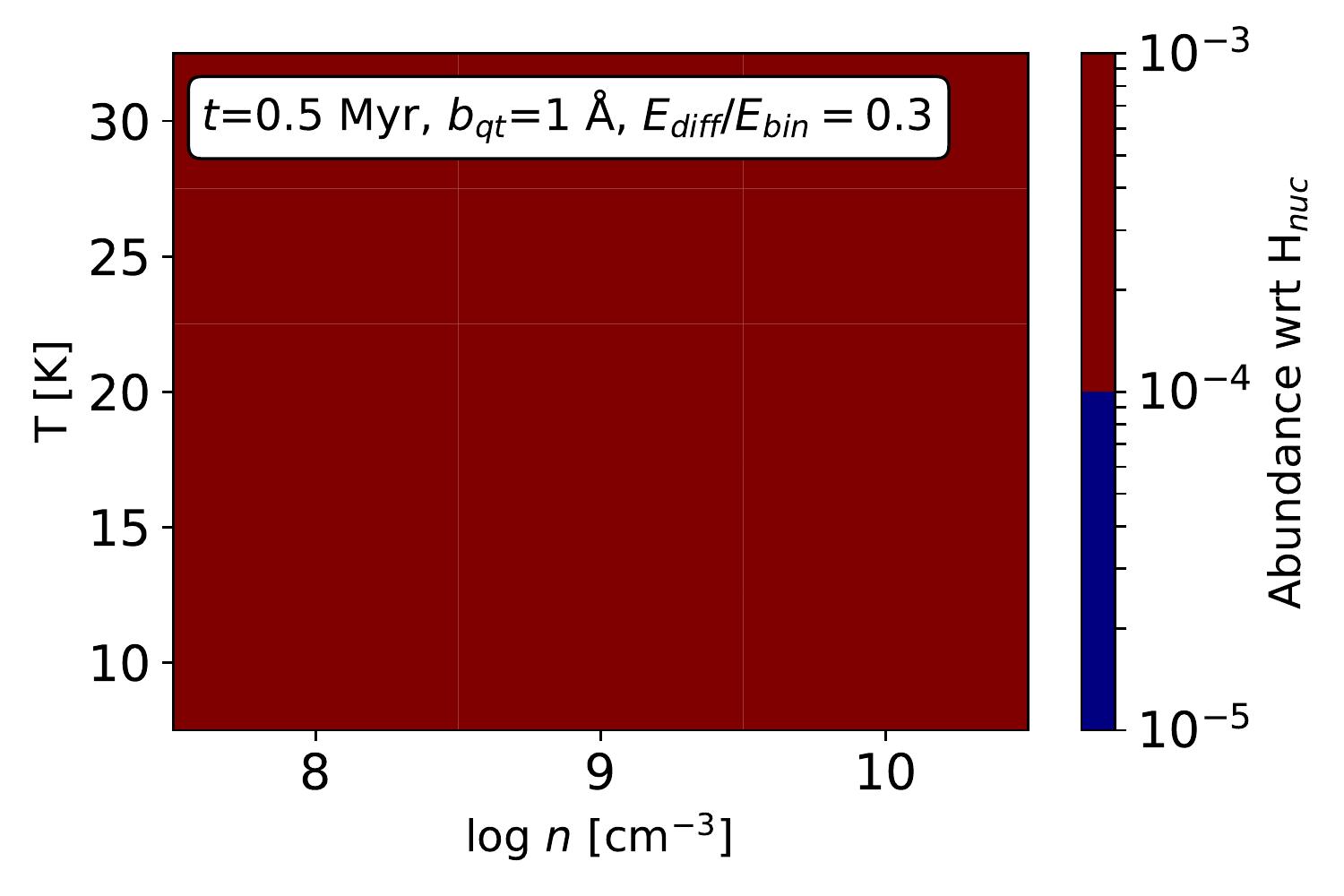}}
\subfigure{\includegraphics[width=0.22\textwidth]{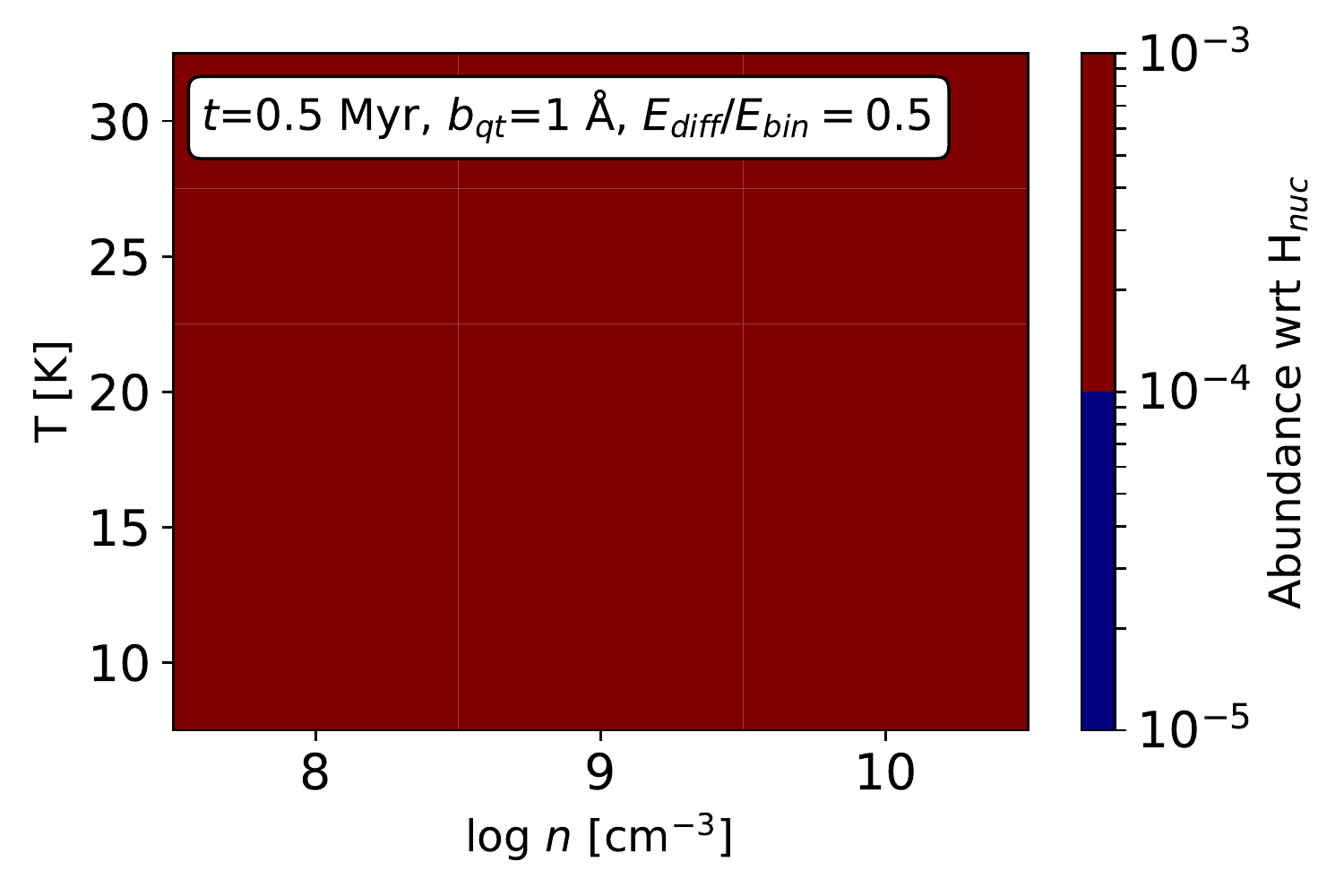}}
\subfigure{\includegraphics[width=0.22\textwidth]{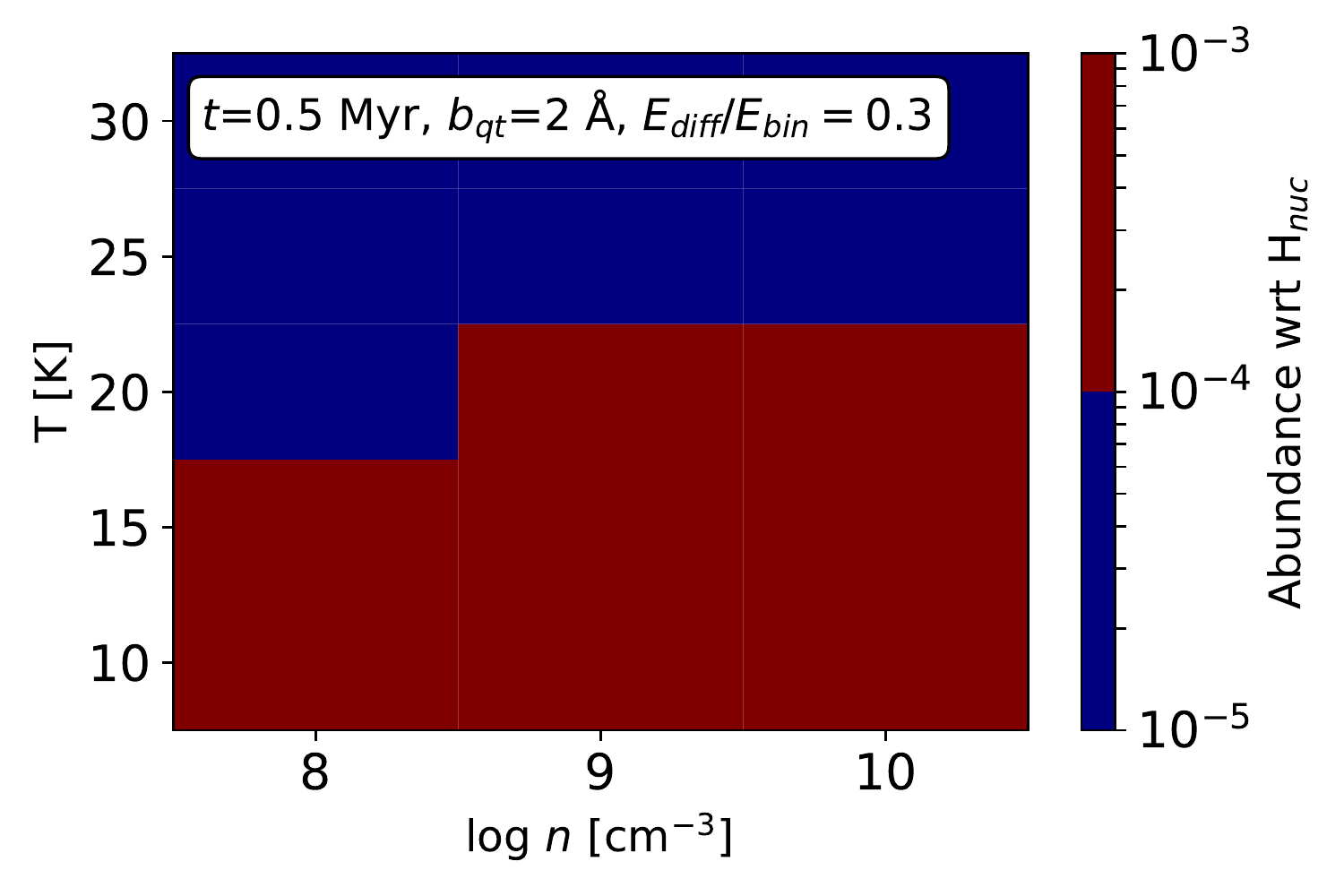}}
\subfigure{\includegraphics[width=0.22\textwidth]{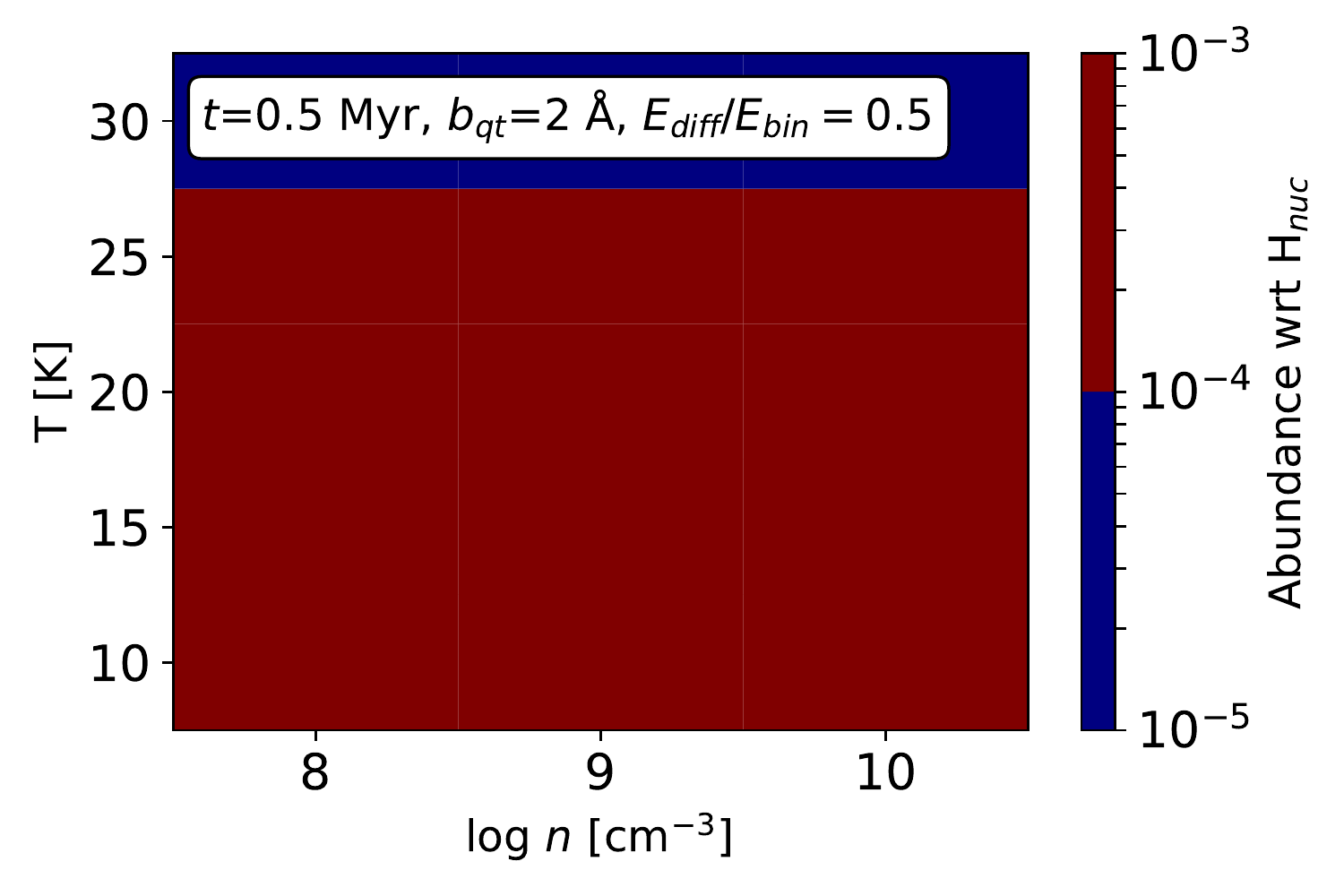}}\\
\subfigure{\includegraphics[width=0.22\textwidth]{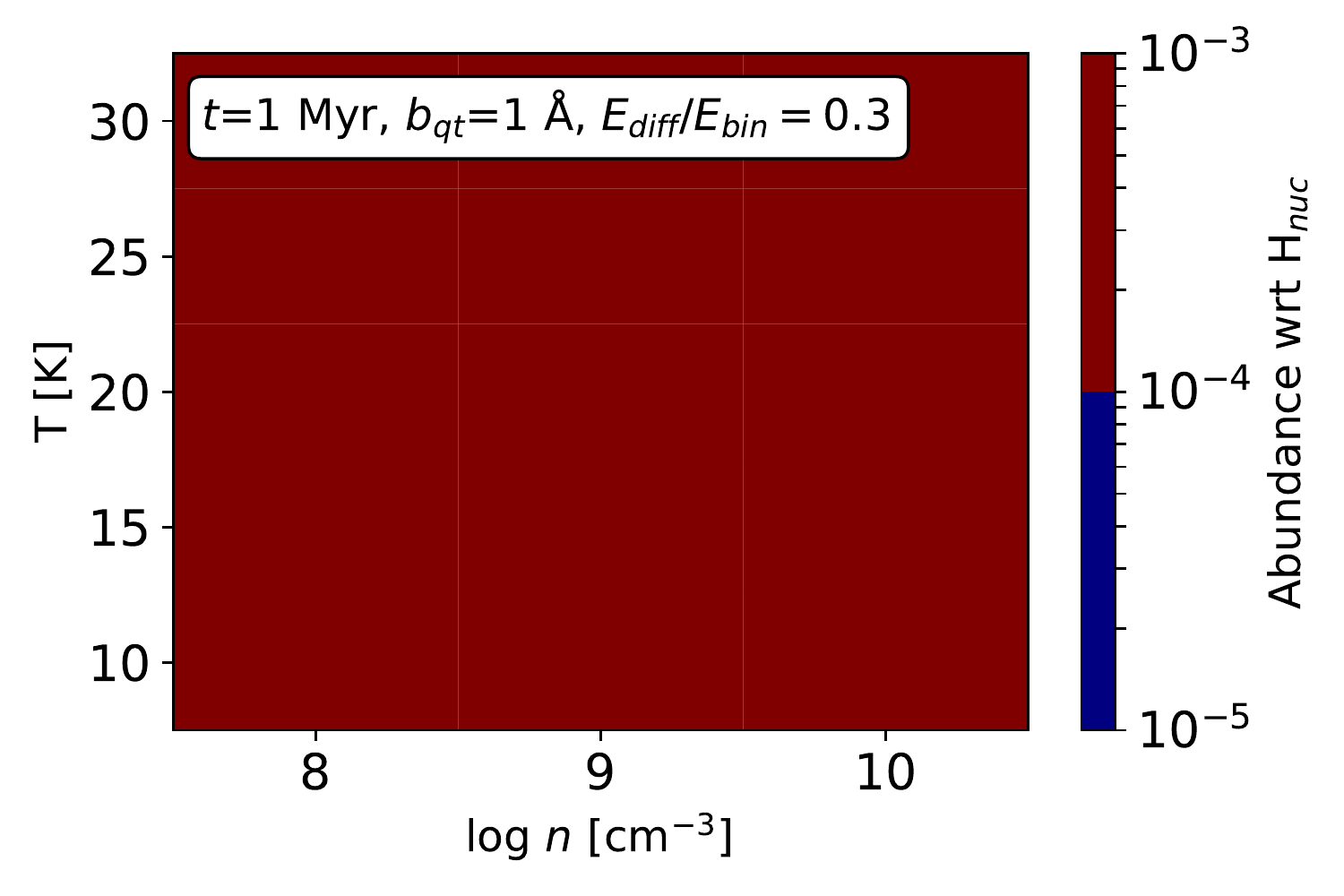}}
\subfigure{\includegraphics[width=0.22\textwidth]{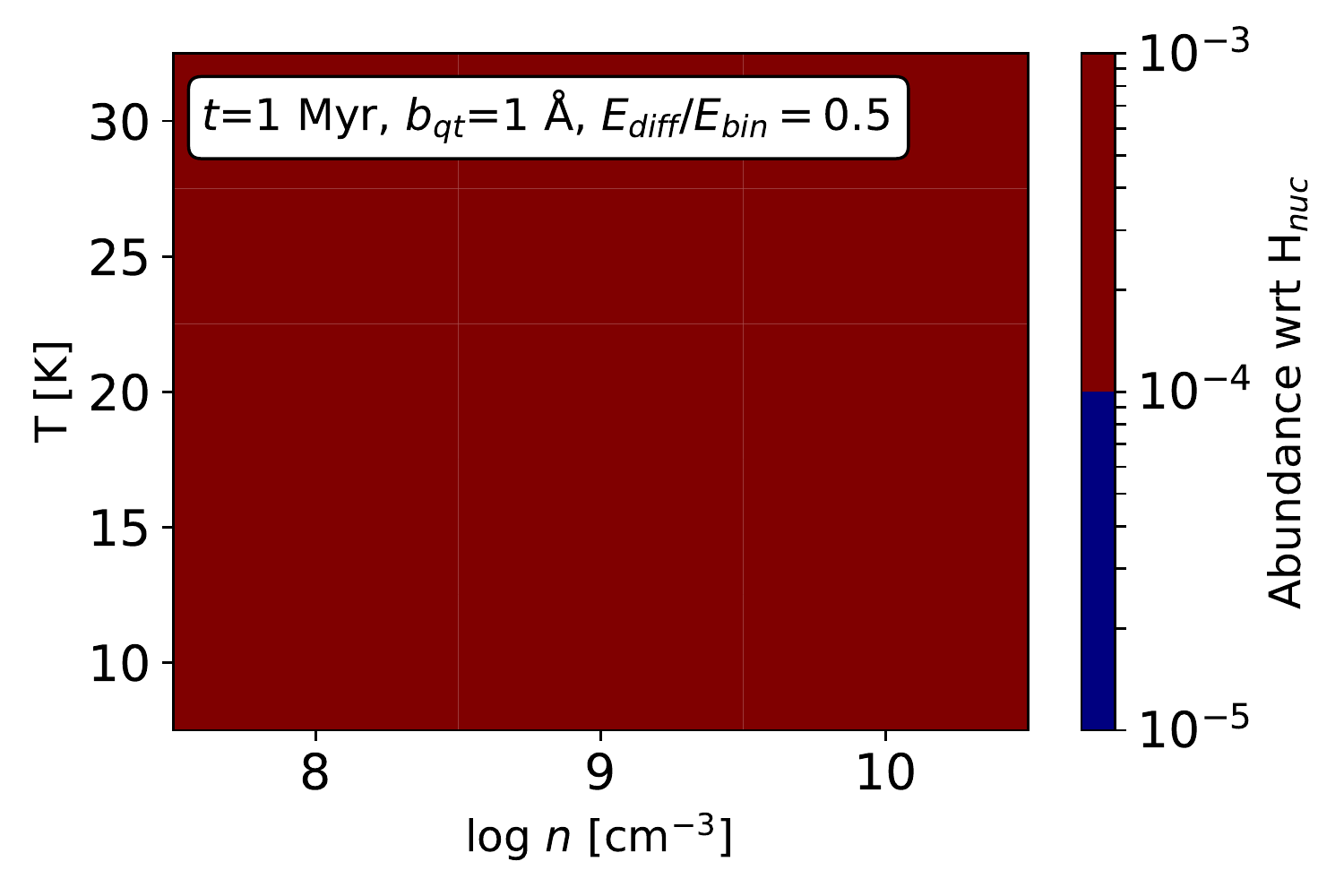}}
\subfigure{\includegraphics[width=0.22\textwidth]{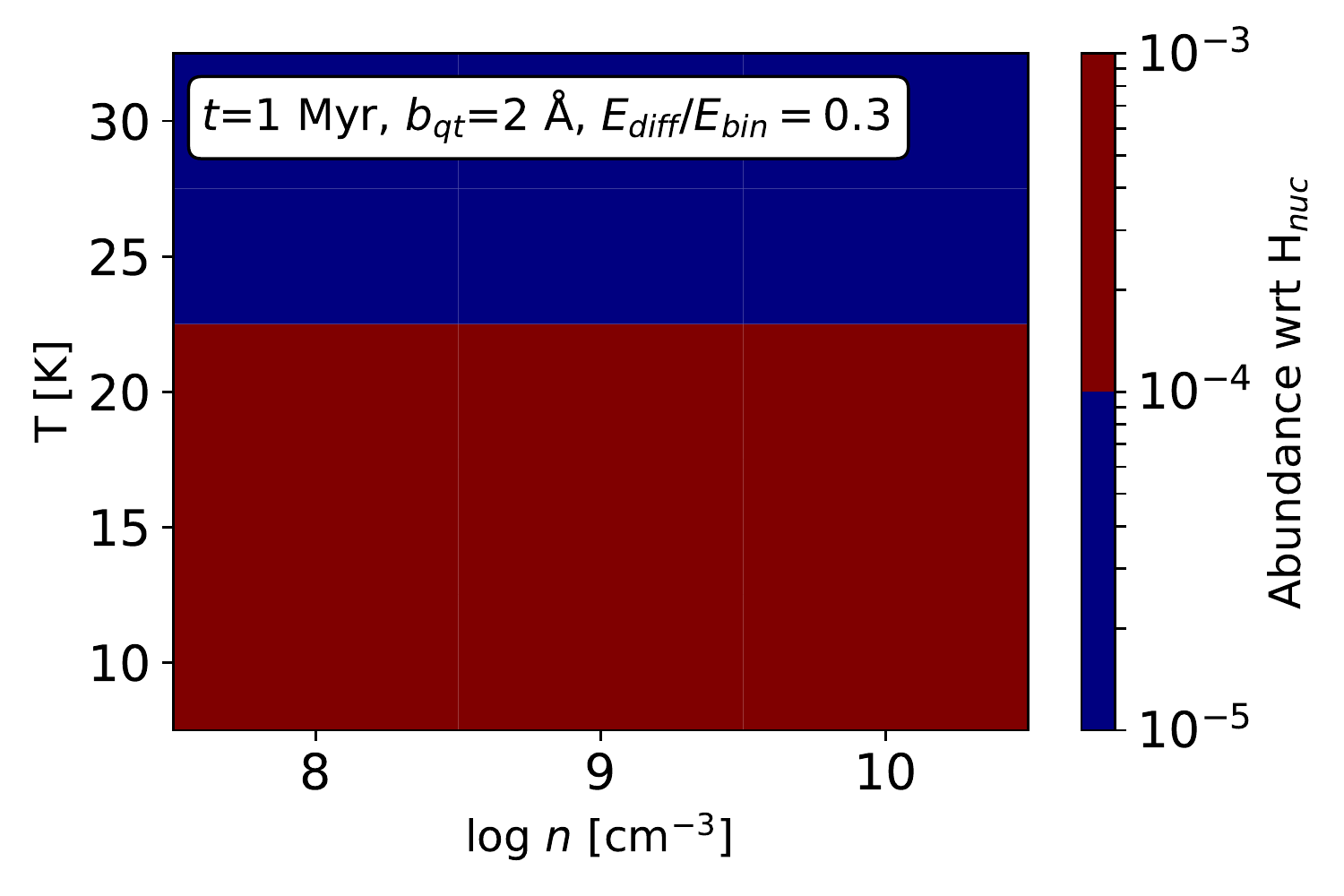}}
\subfigure{\includegraphics[width=0.22\textwidth]{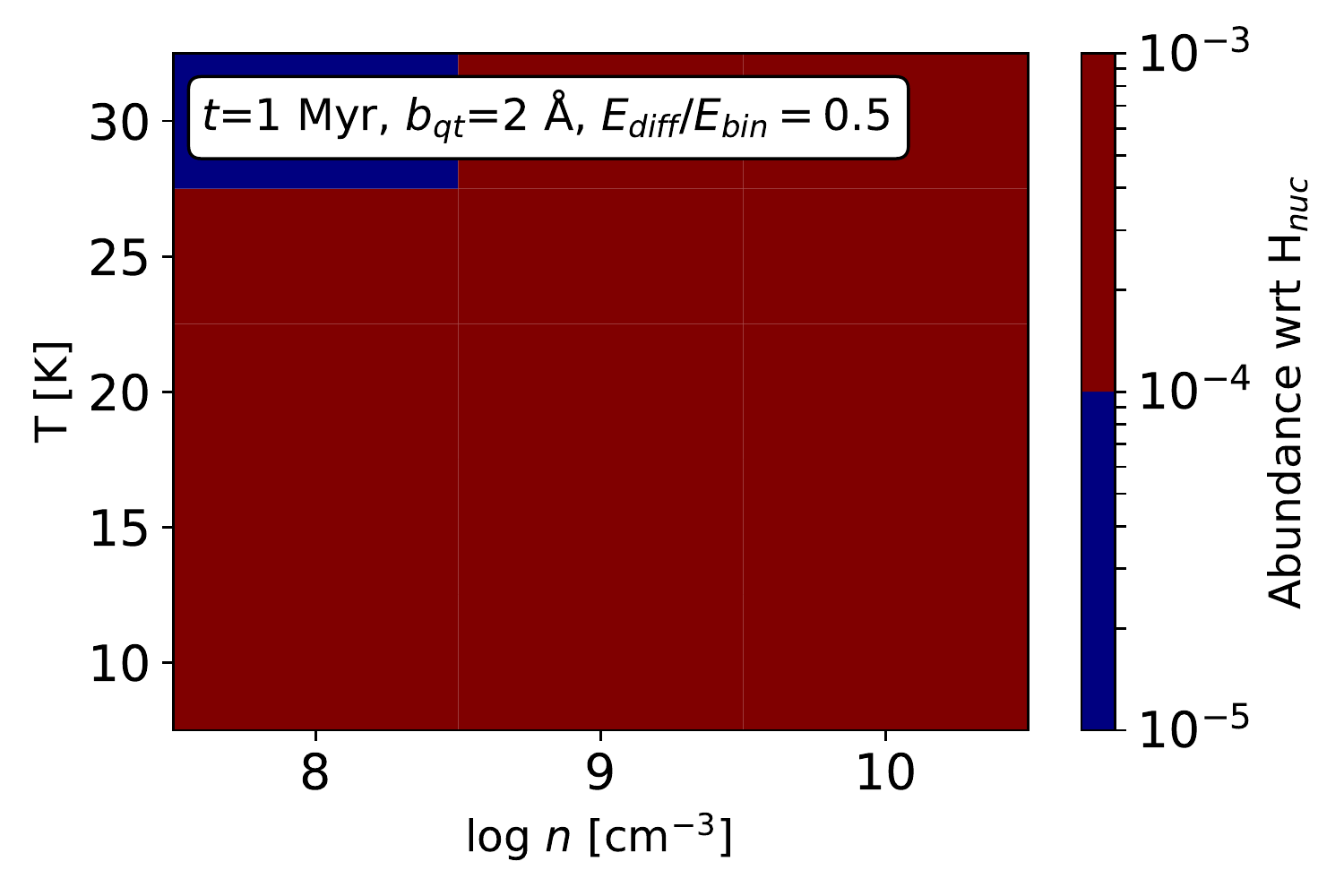}}\\
\subfigure{\includegraphics[width=0.22\textwidth]{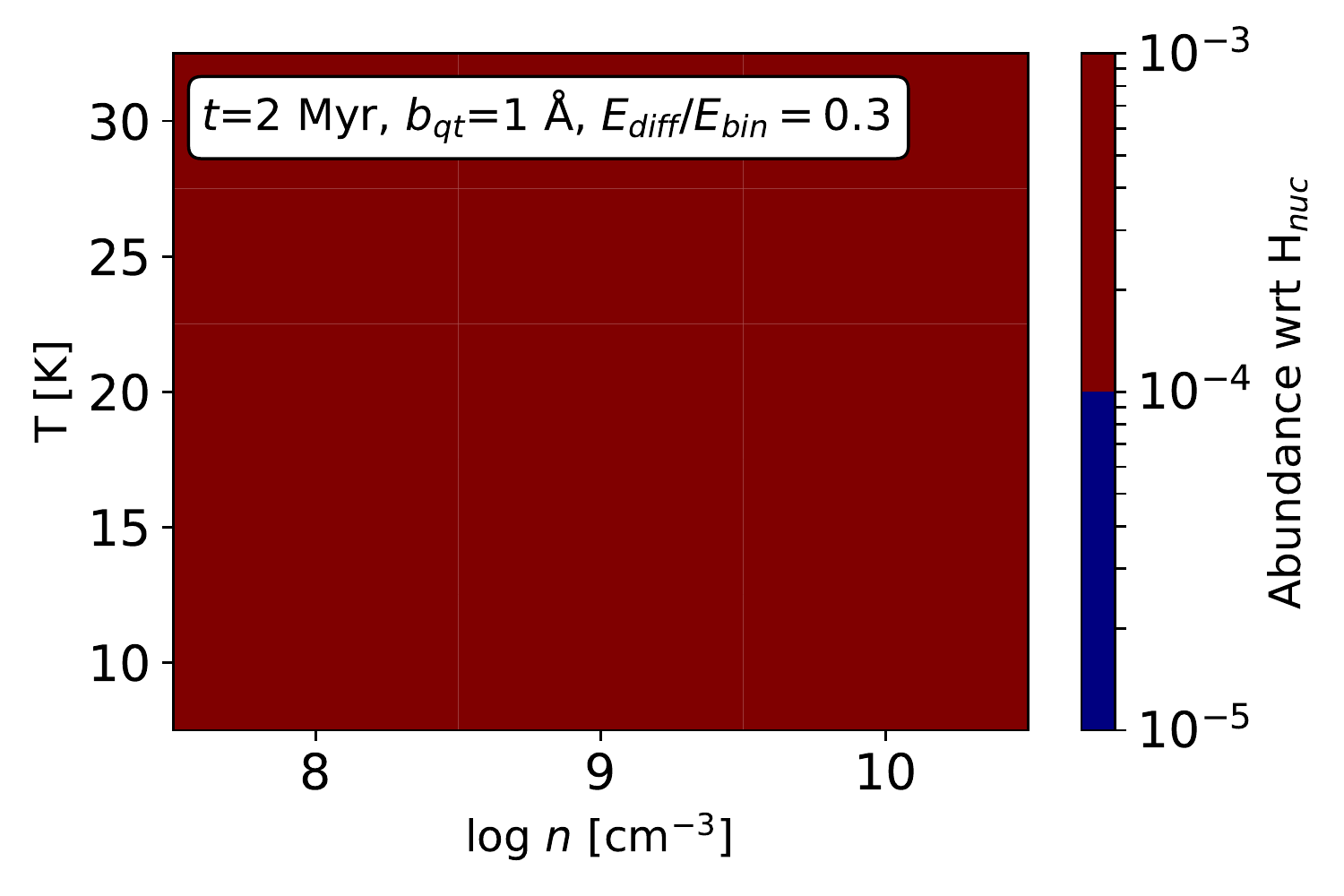}}
\subfigure{\includegraphics[width=0.22\textwidth]{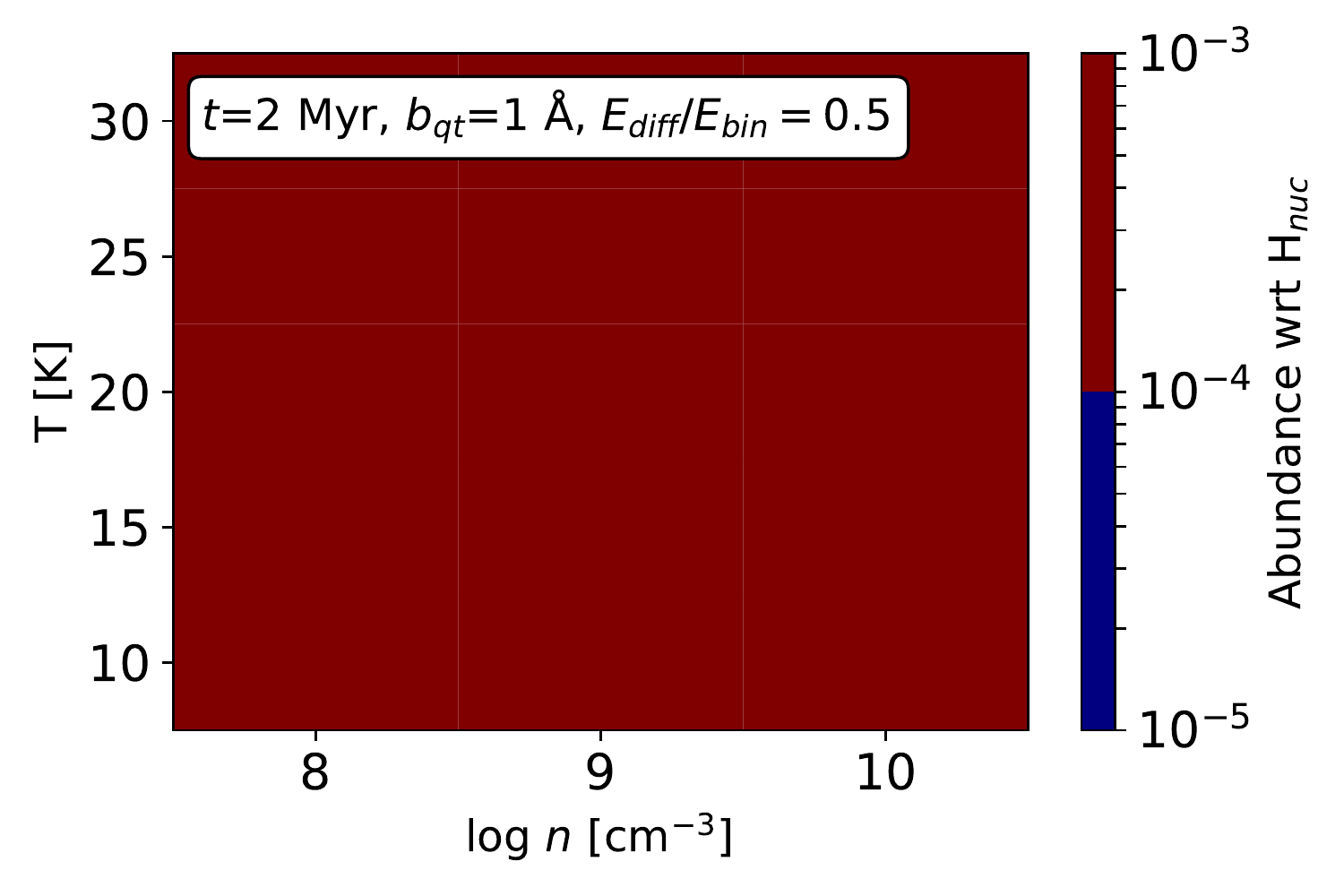}}
\subfigure{\includegraphics[width=0.22\textwidth]{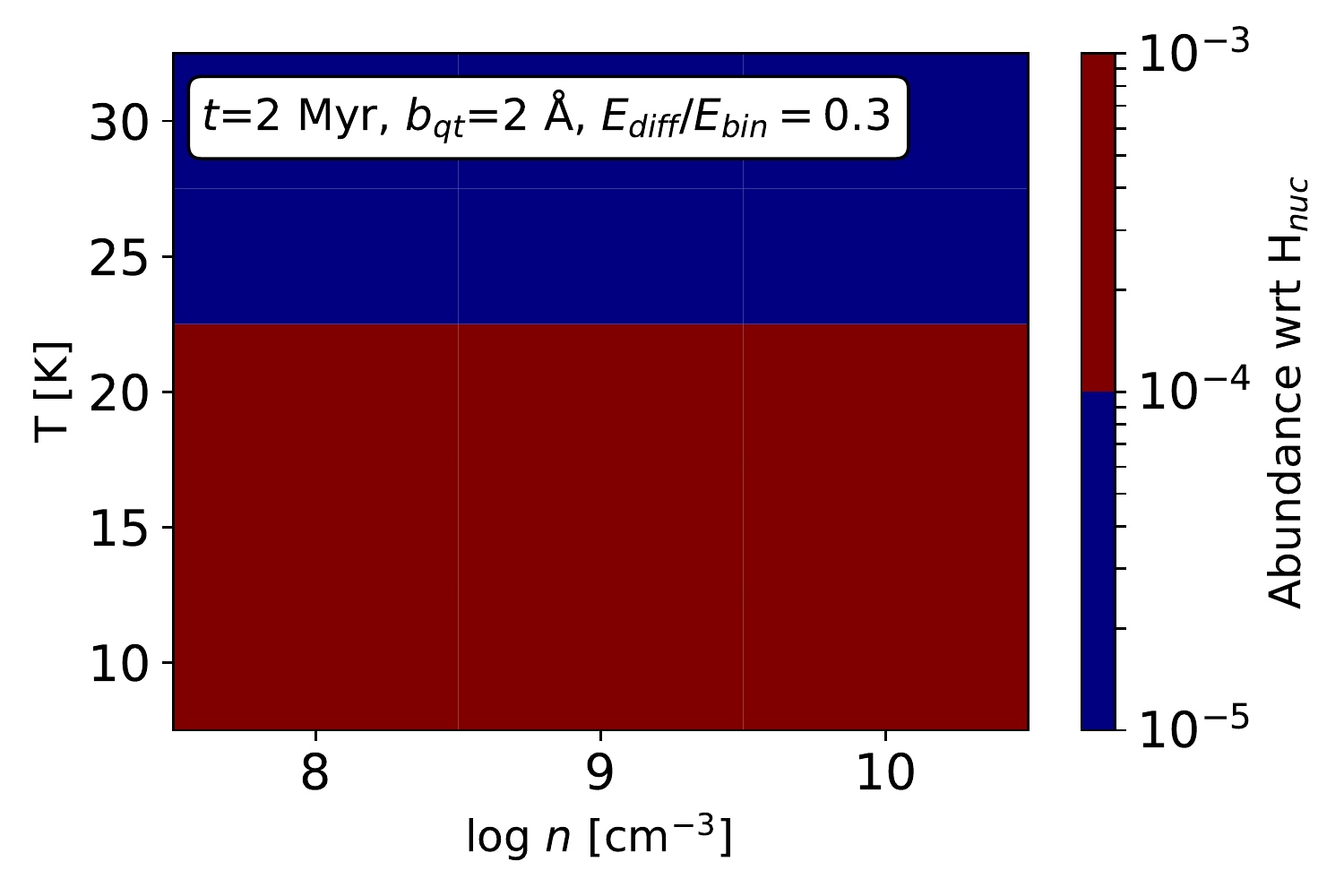}}
\subfigure{\includegraphics[width=0.22\textwidth]{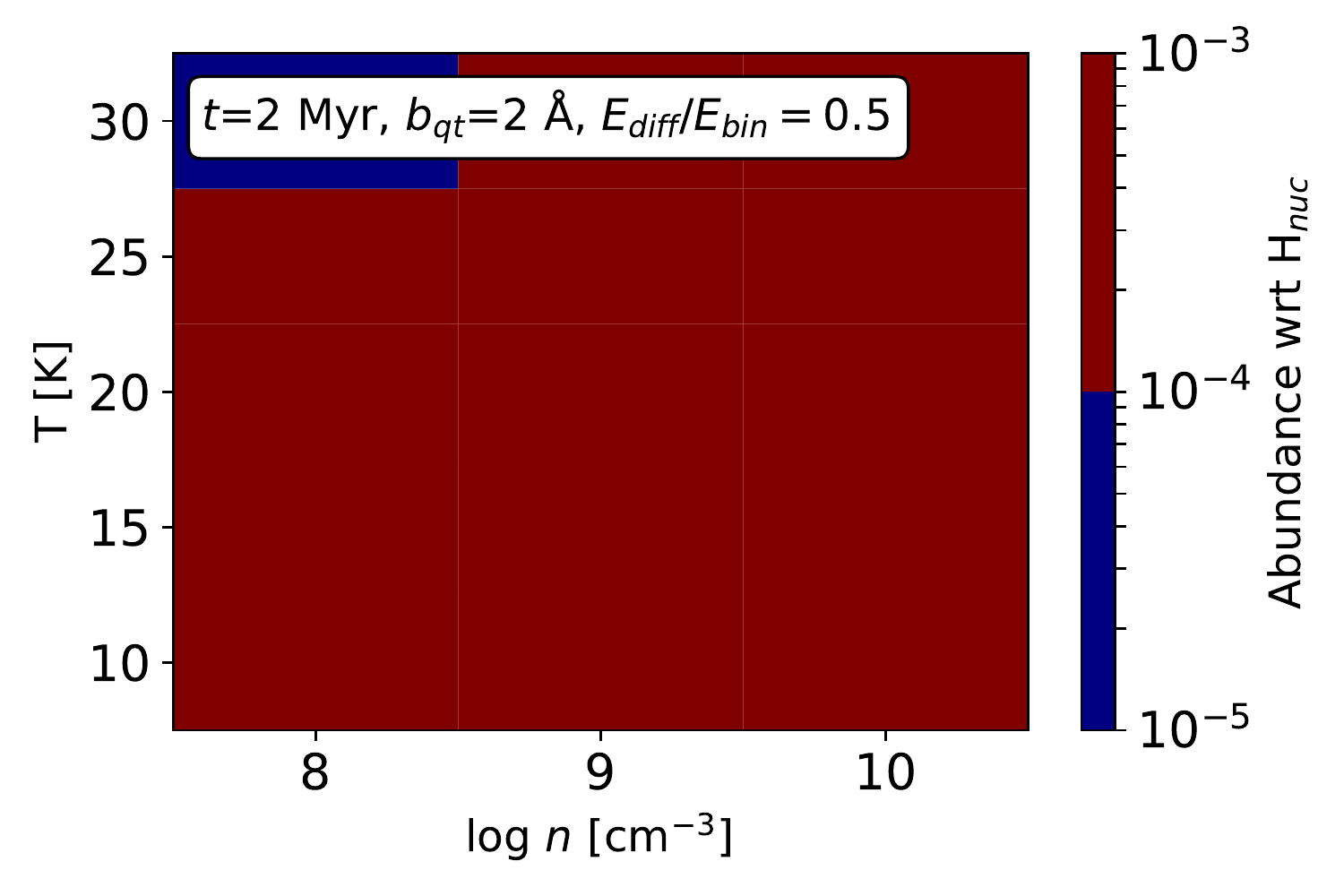}}\\
\subfigure{\includegraphics[width=0.22\textwidth]{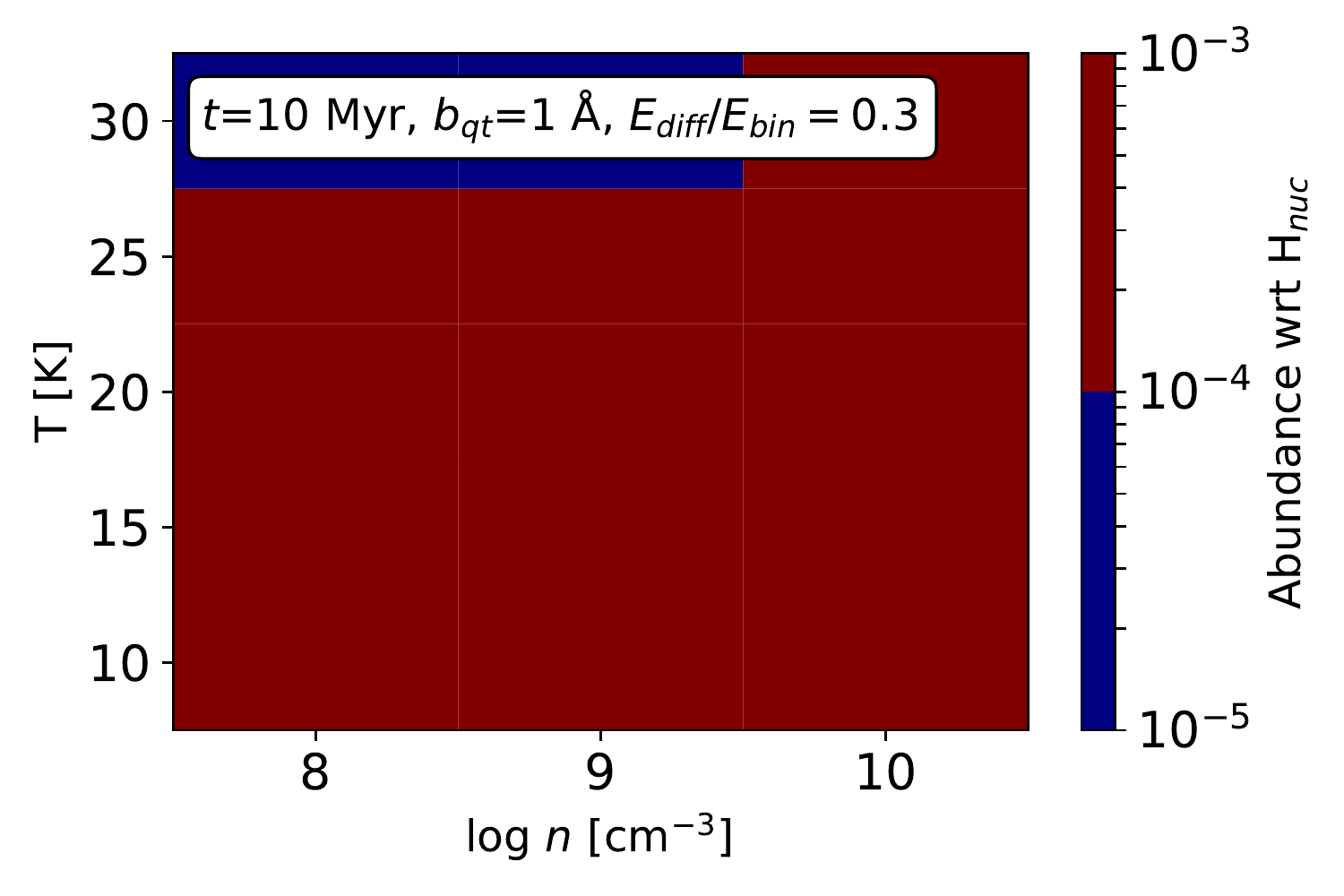}}
\subfigure{\includegraphics[width=0.22\textwidth]{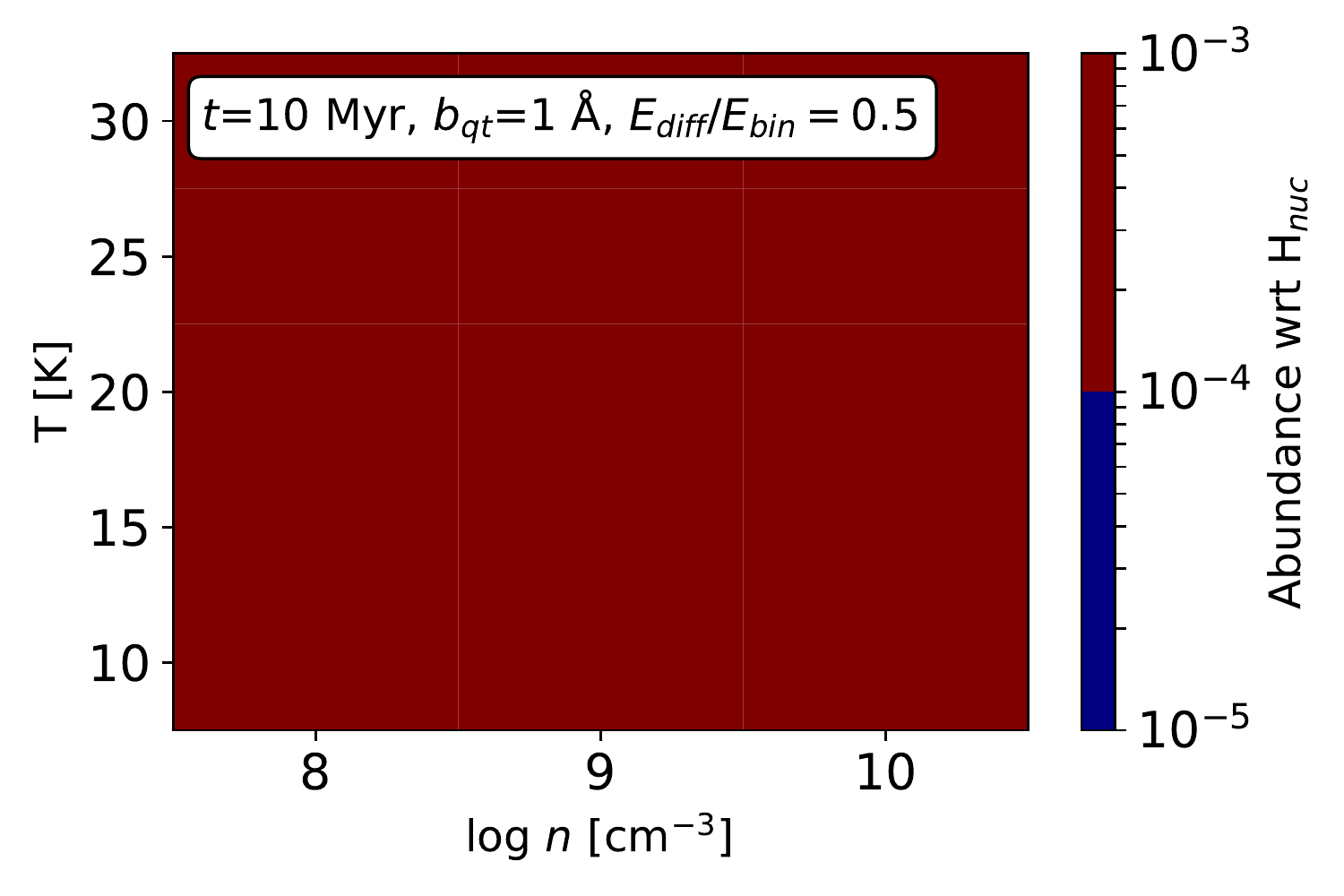}}
\subfigure{\includegraphics[width=0.22\textwidth]{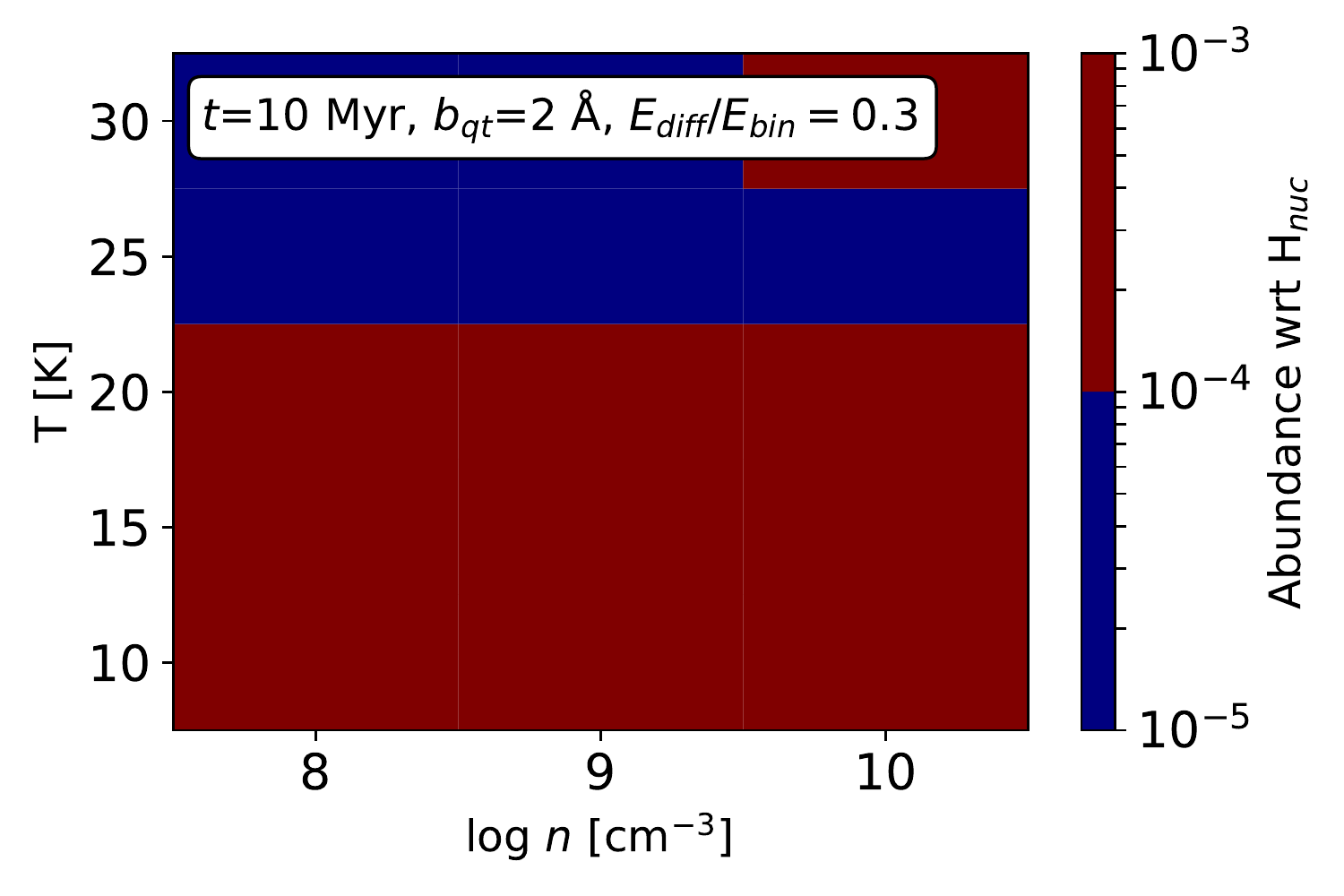}}
\subfigure{\includegraphics[width=0.22\textwidth]{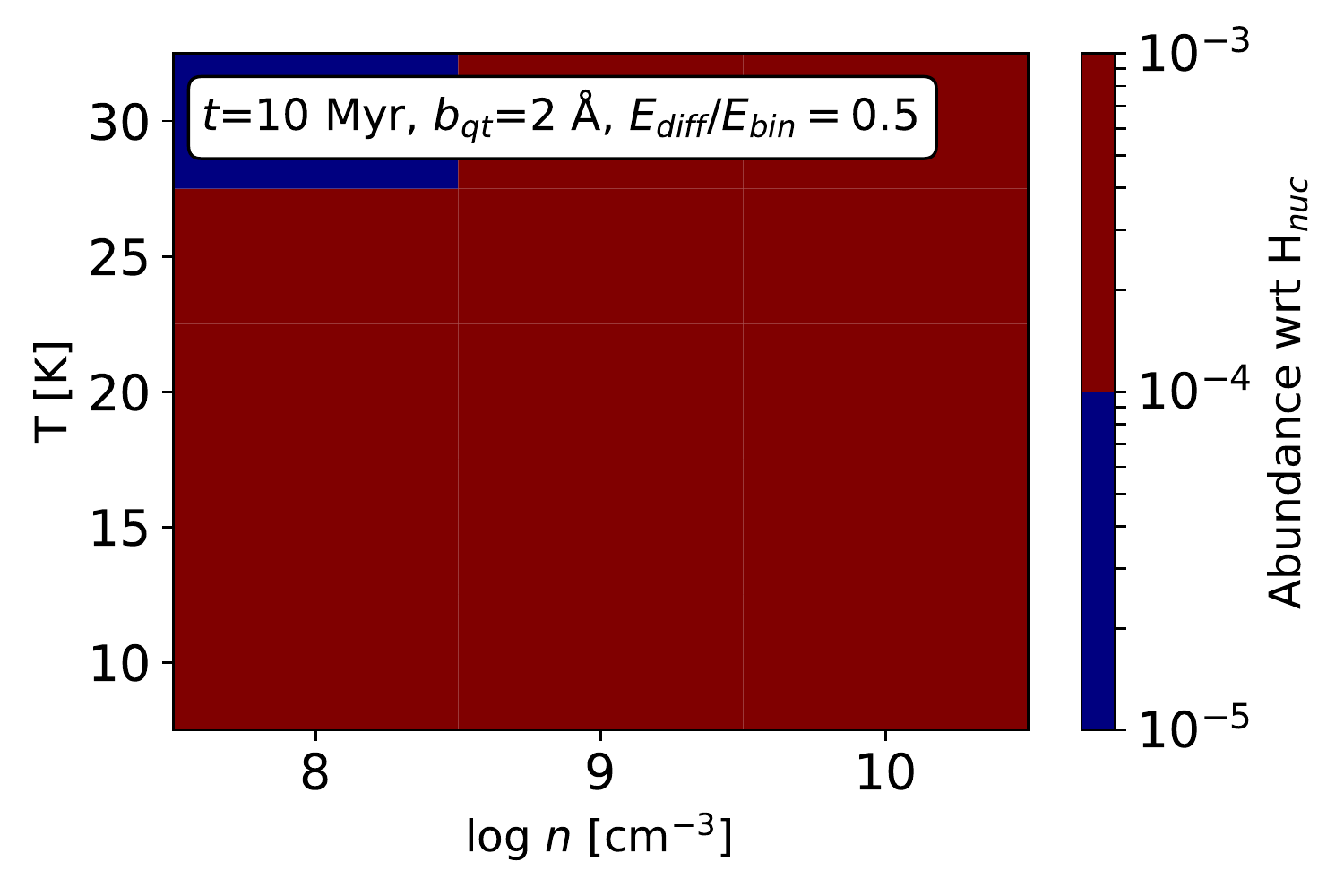}}\\
\caption{Abundances (given as colors) for \ce{H2O} ice as function of midplane density ($x$-axes), and temperature ($y$-axes) at different evolutionary steps for the reset scenario. From left to right are abundances from model runs with different parameters for grain-surface reactions: columns one and two feature $b_{qt}$= 1 {\AA}, columns three and four feature $b_{qt}$= 1 {\AA}, columns one and three are with $E_{\rm{diff}}/E_{\rm{bin}}$= 0.3, and columns two and four are with $E_{\rm{diff}}/E_{\rm{bin}}$= 0.5. Top to bottom are different evolutionary times, from 0.05 Myr (top) to 10 Myr (bottom). To the right of each plot is a colorbar, indicating the abundance level with respect to H$_{\rm{nuc}}$ for each color. The chemical network utilised includes \ce{O3} chemistry.}
\label{chem_params_h2o}
\end{figure*}

\begin{figure*}
\subfigure{\includegraphics[width=0.22\textwidth]{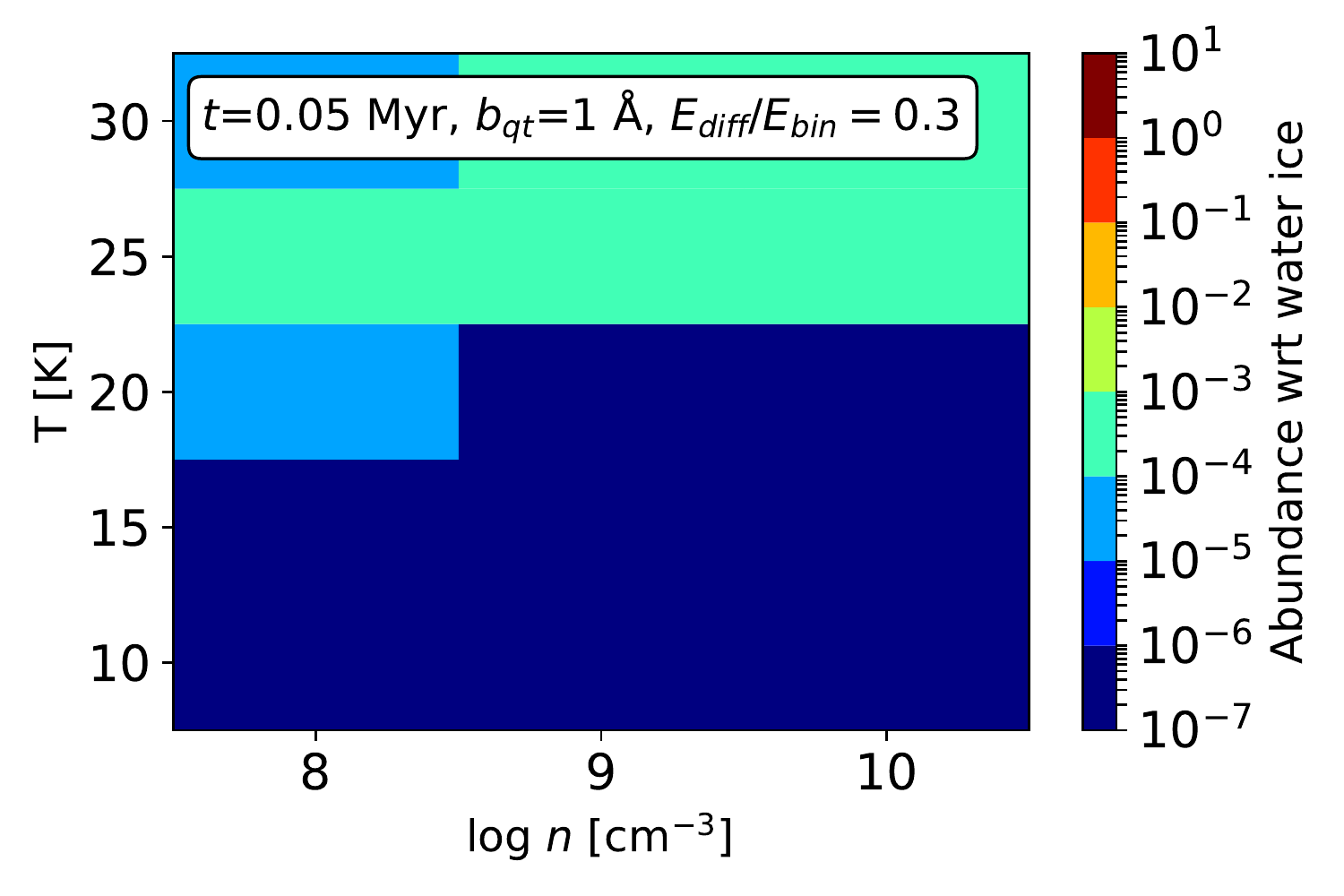}}
\subfigure{\includegraphics[width=0.22\textwidth]{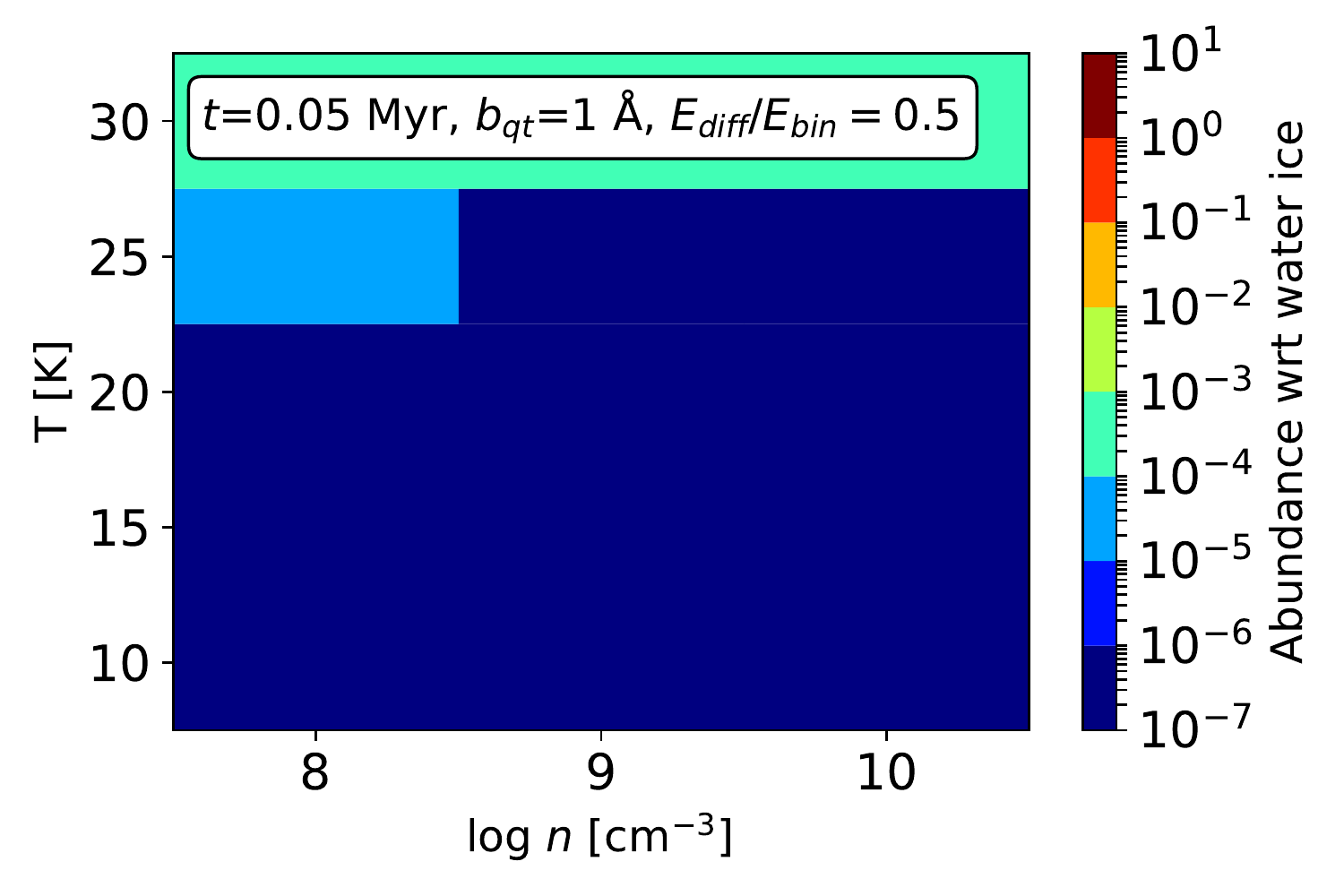}}
\subfigure{\includegraphics[width=0.22\textwidth]{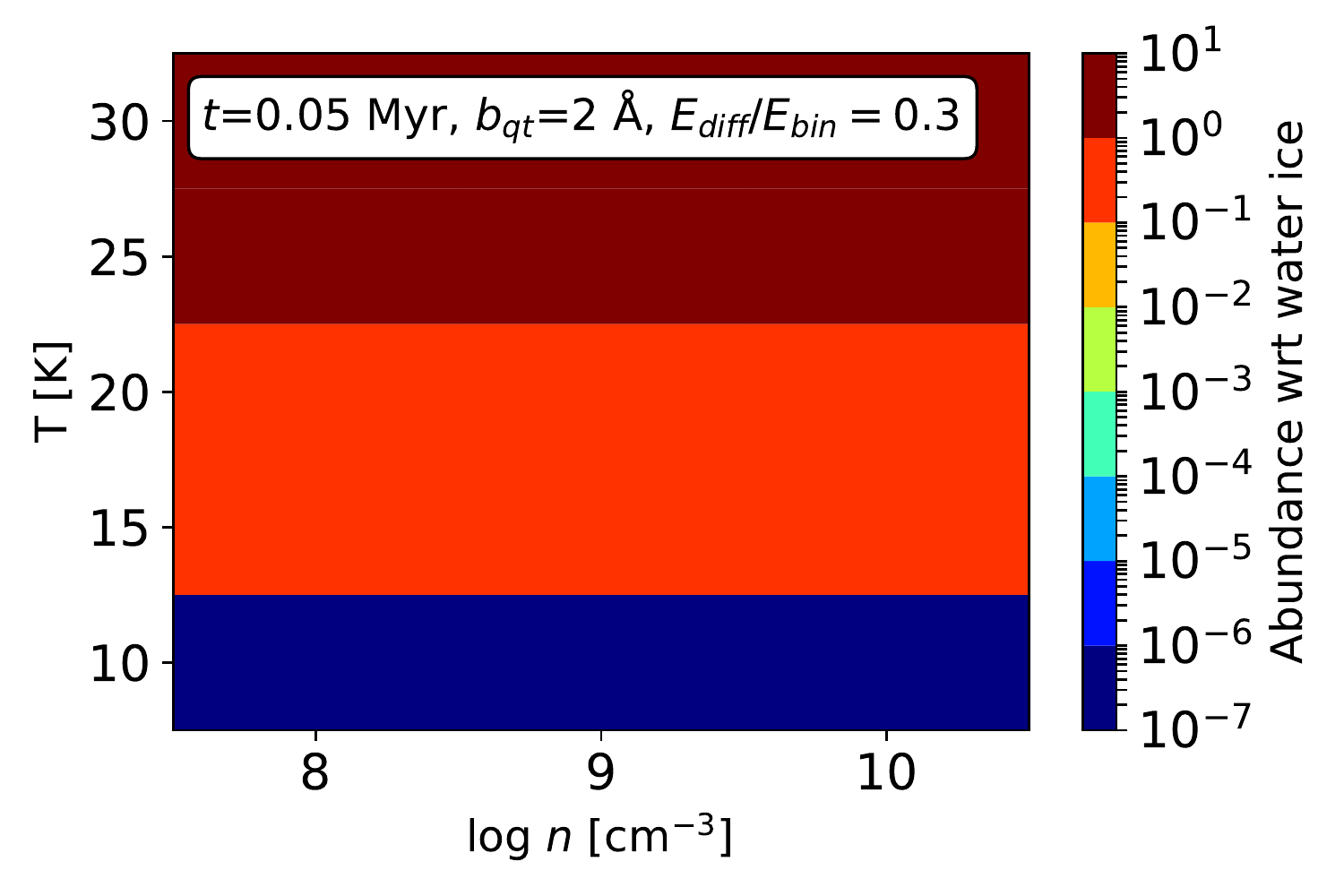}}
\subfigure{\includegraphics[width=0.22\textwidth]{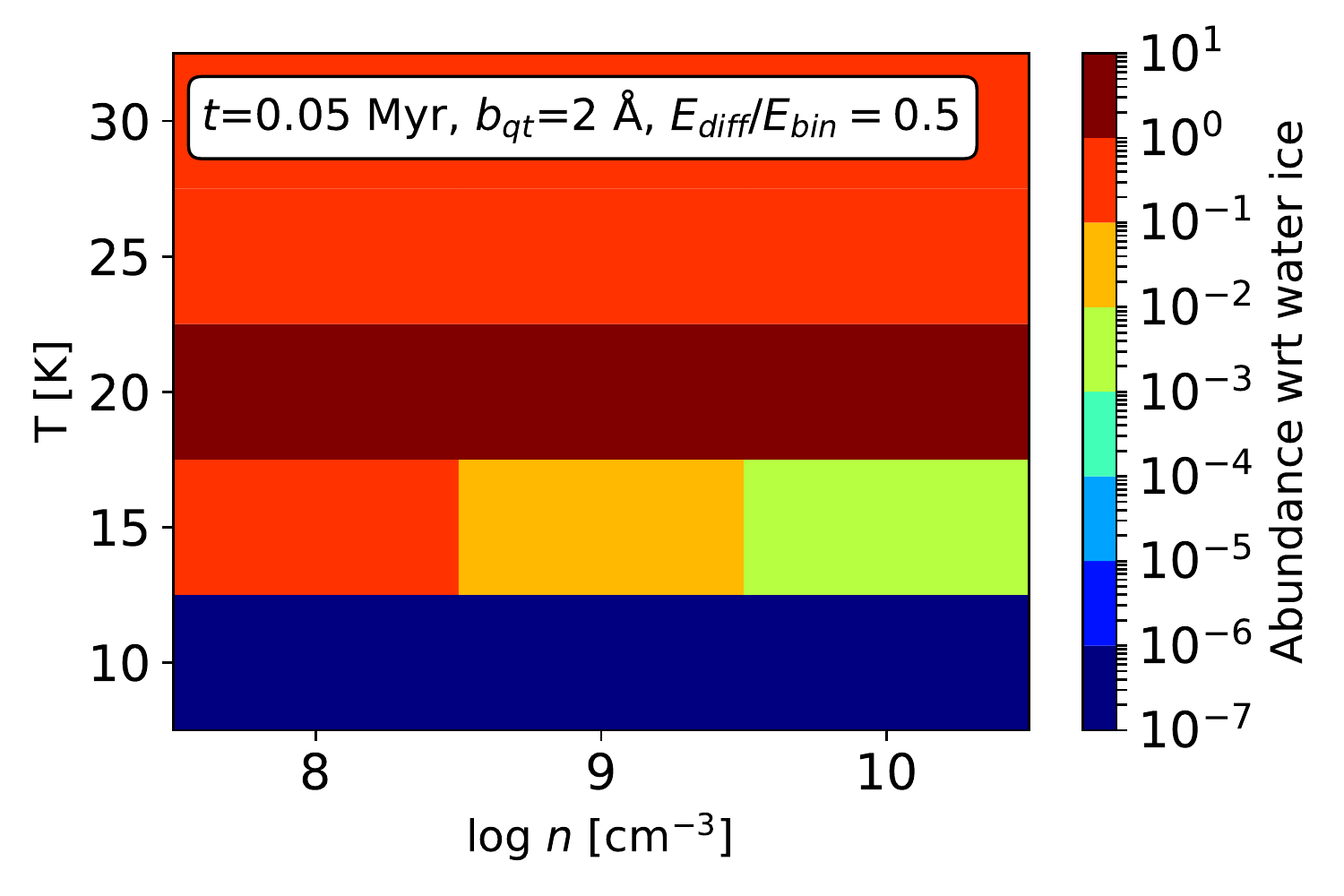}}\\
\subfigure{\includegraphics[width=0.22\textwidth]{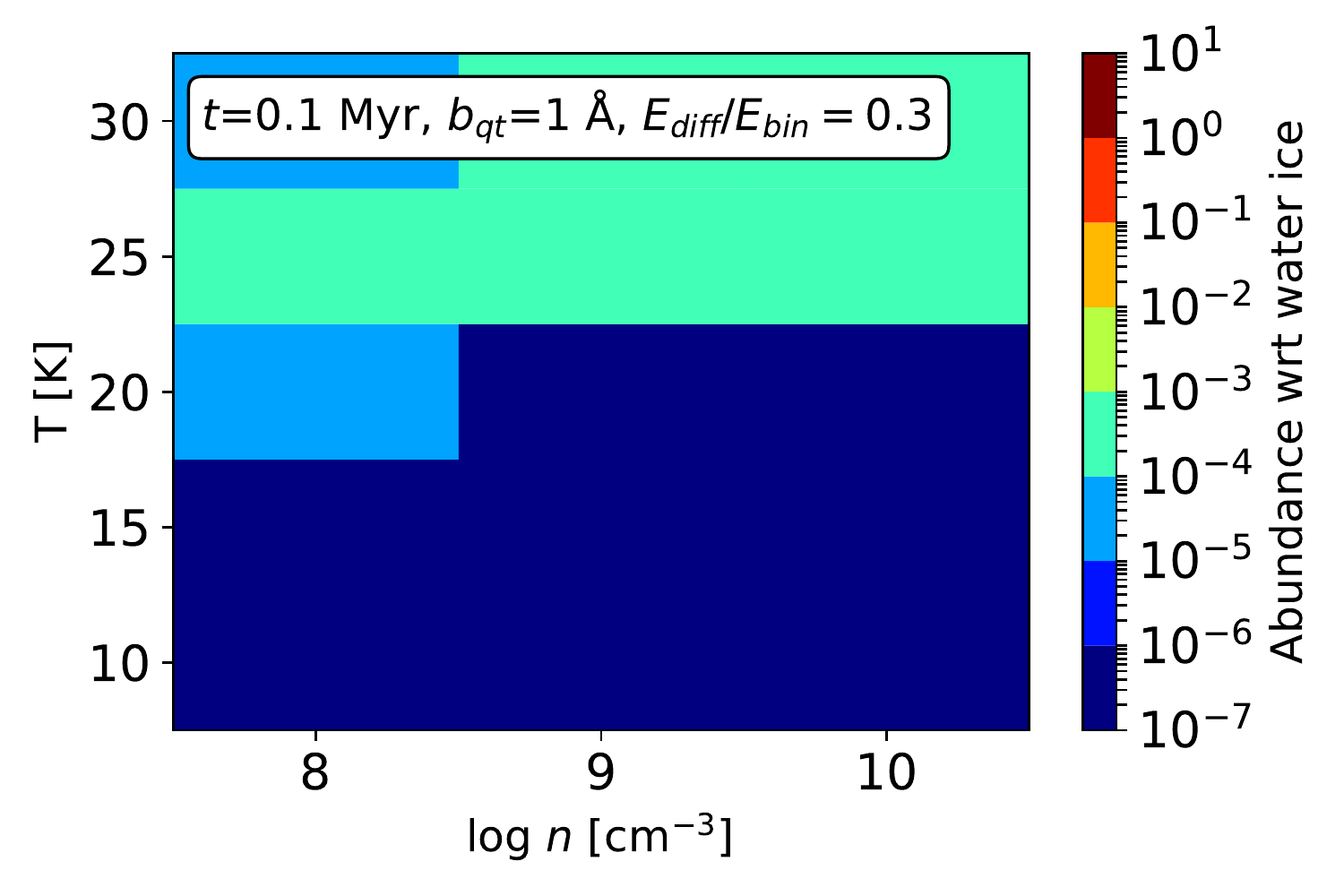}}
\subfigure{\includegraphics[width=0.22\textwidth]{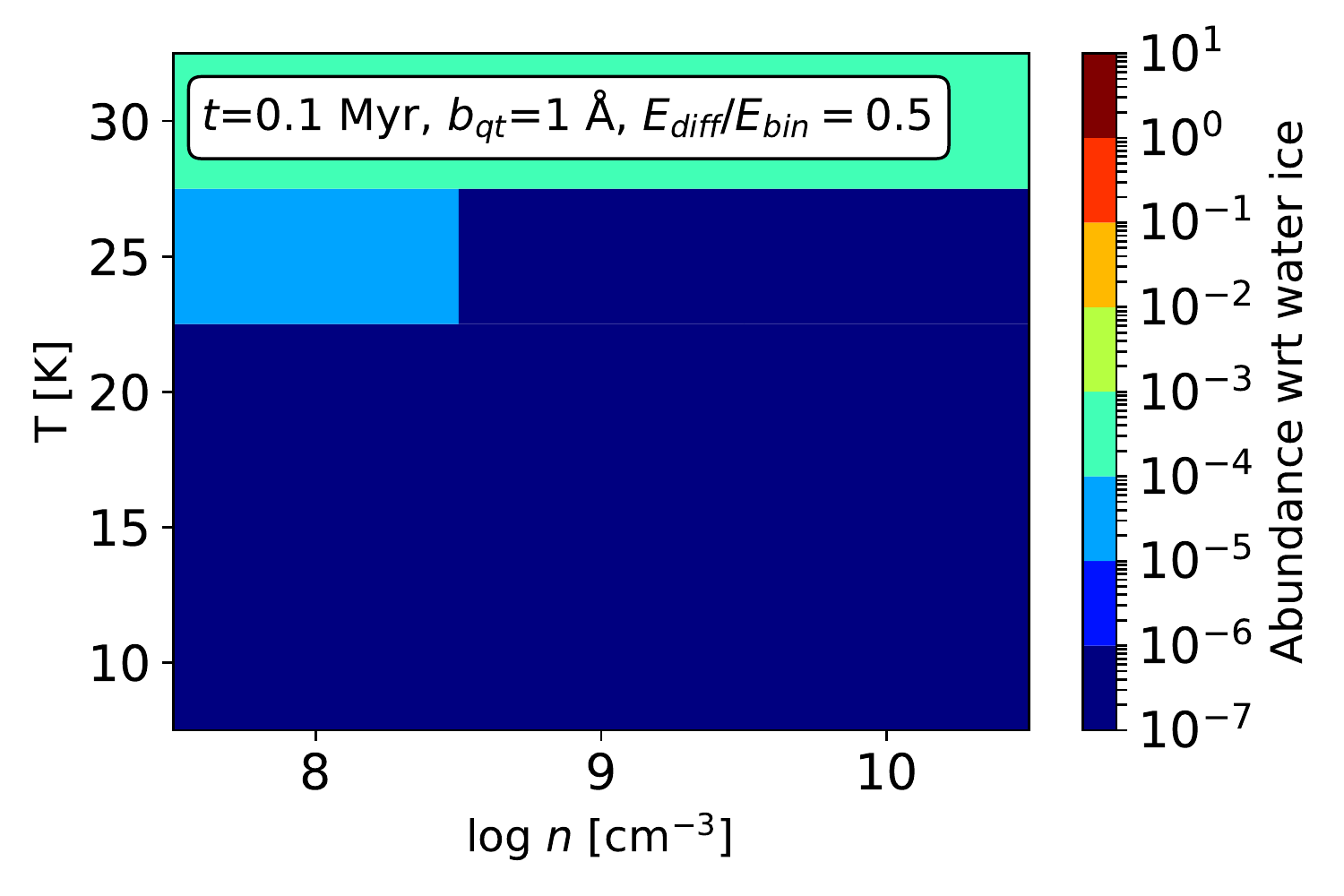}}
\subfigure{\includegraphics[width=0.22\textwidth]{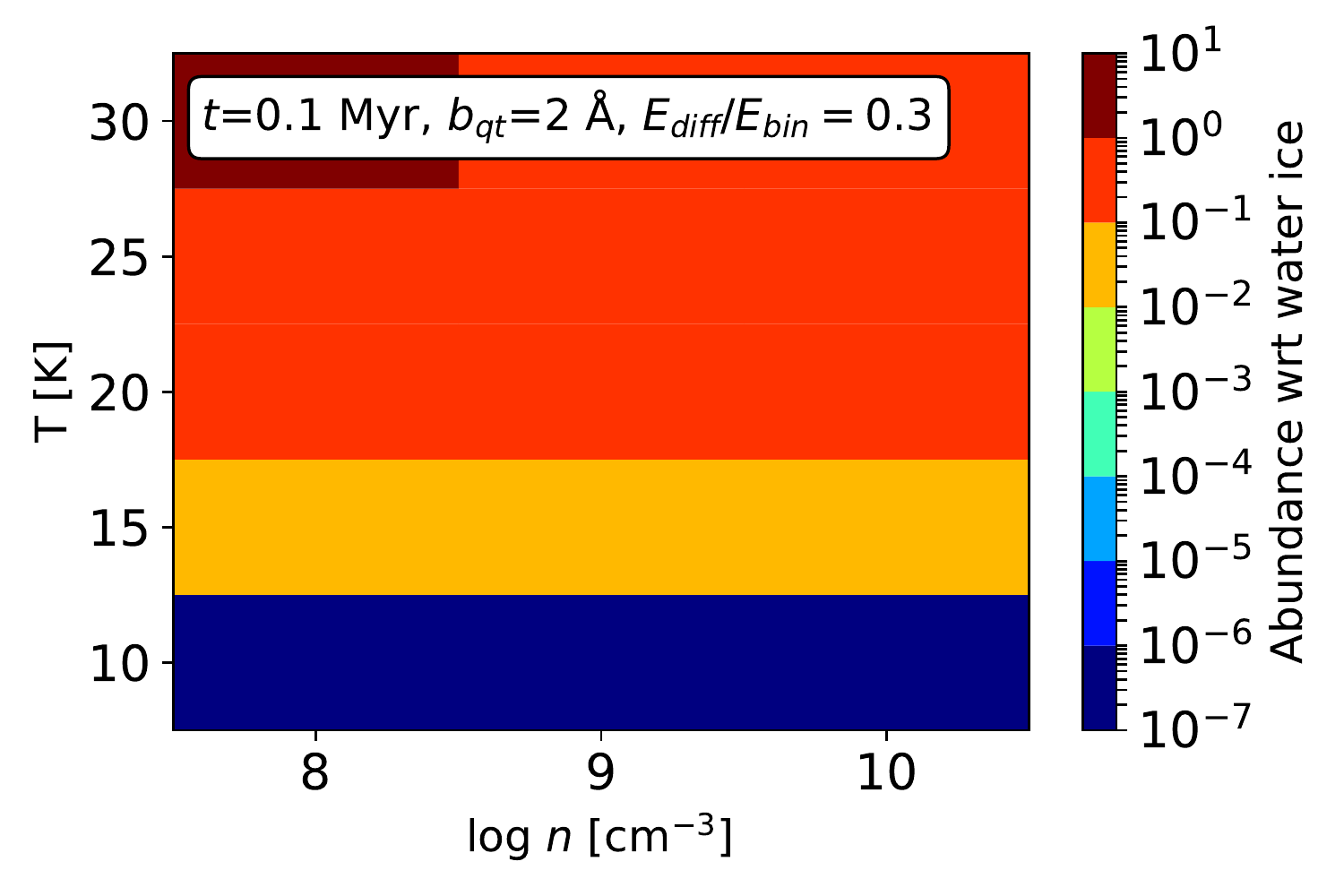}}
\subfigure{\includegraphics[width=0.22\textwidth]{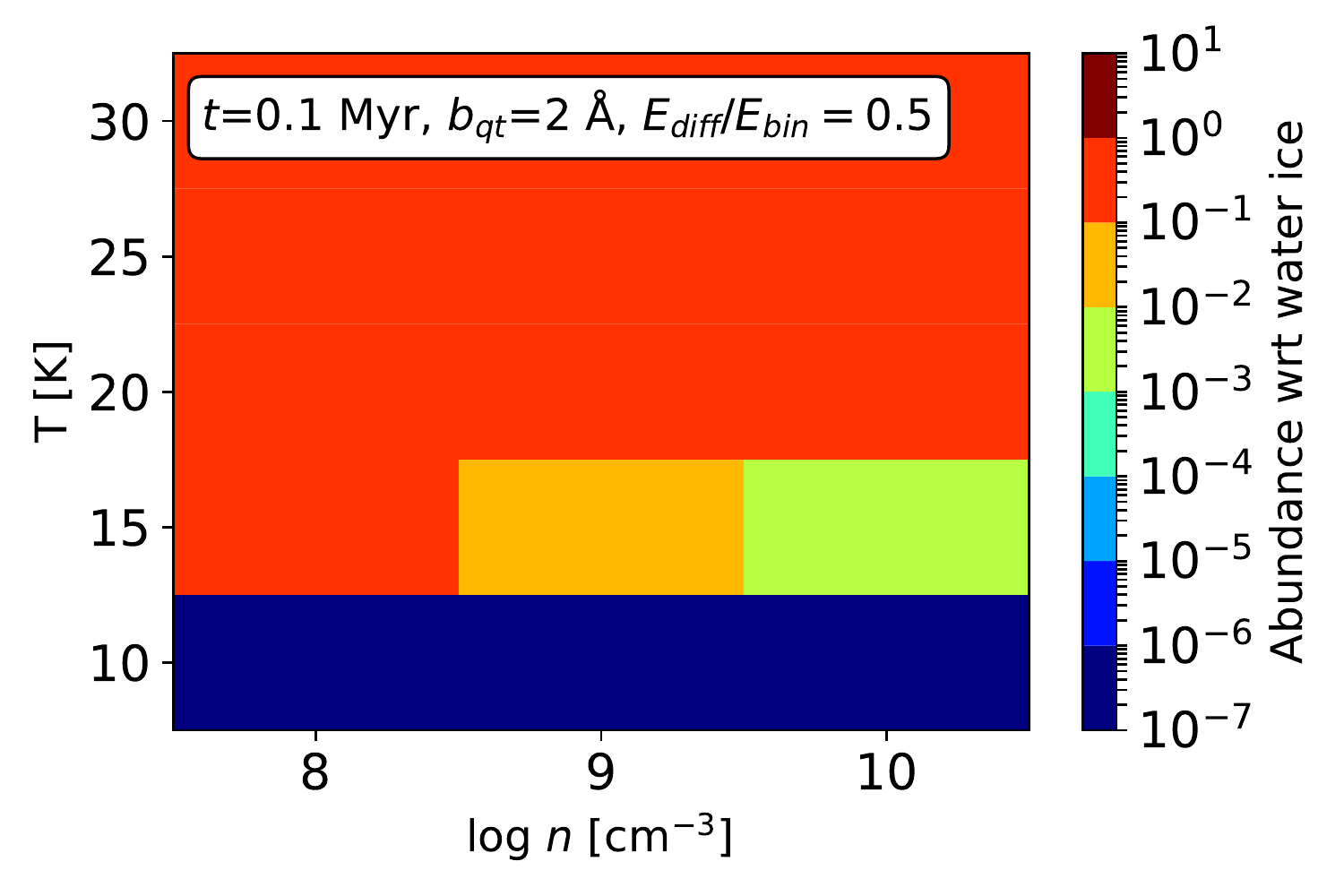}}\\
\subfigure{\includegraphics[width=0.22\textwidth]{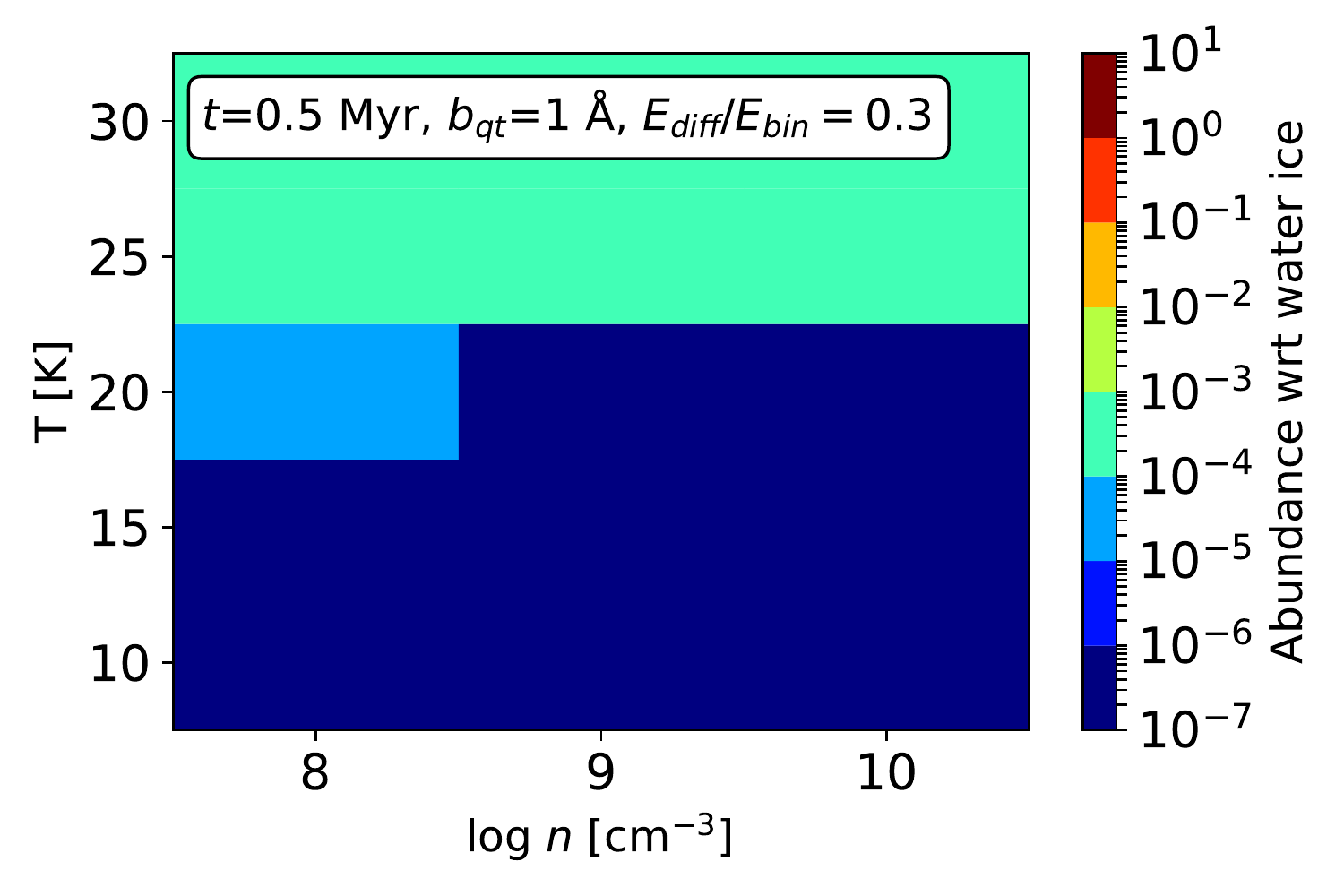}}
\subfigure{\includegraphics[width=0.22\textwidth]{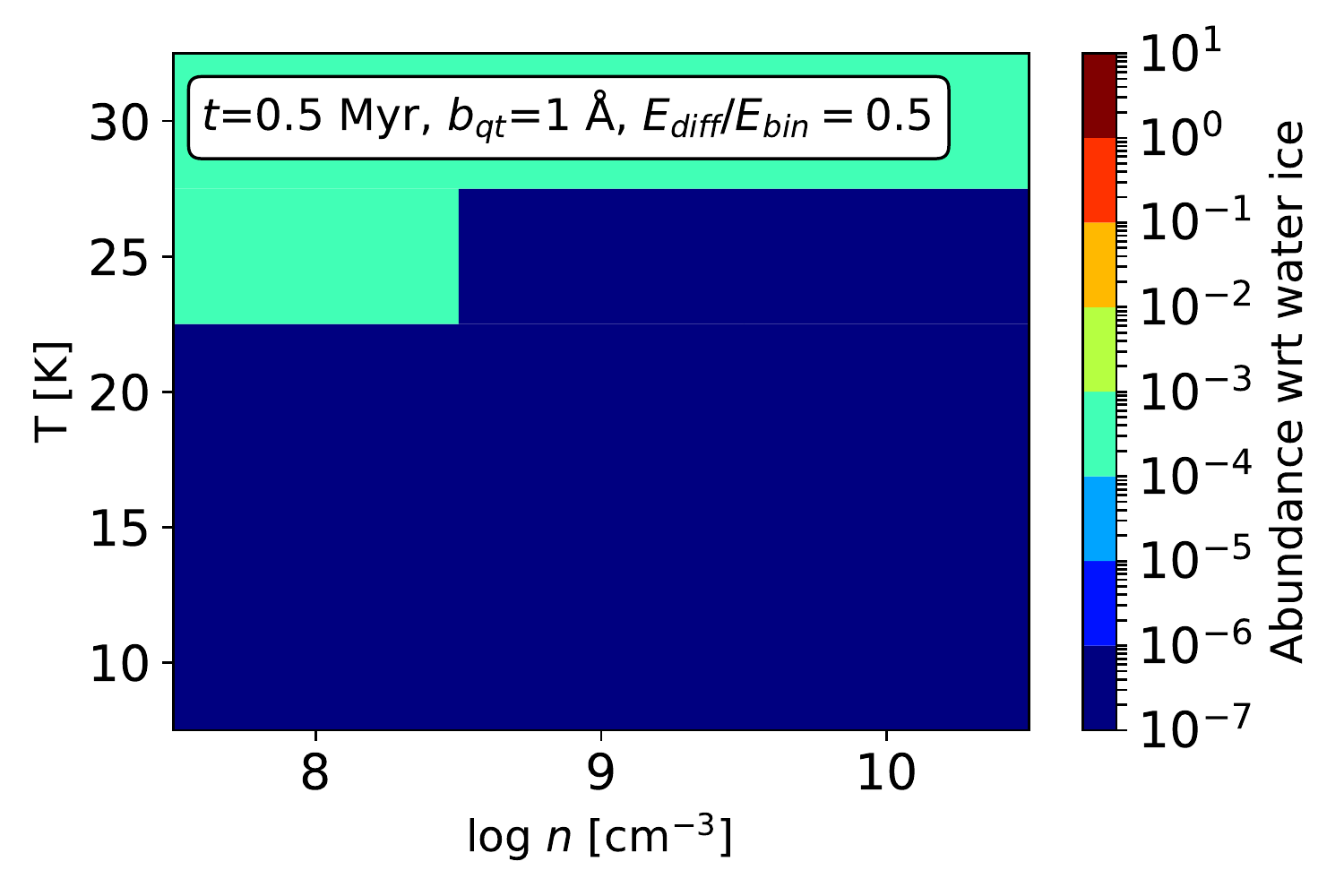}}
\subfigure{\includegraphics[width=0.22\textwidth]{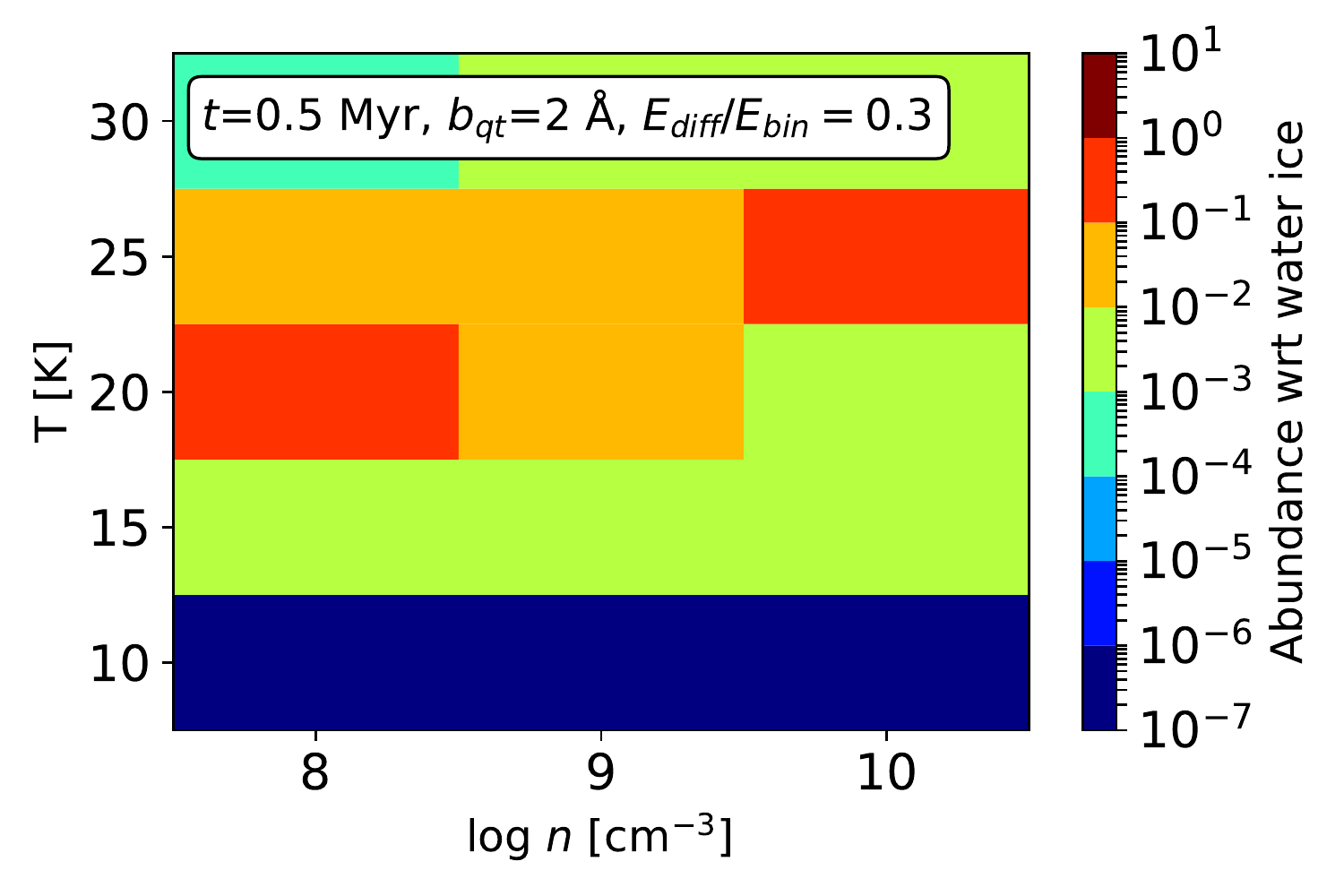}}
\subfigure{\includegraphics[width=0.22\textwidth]{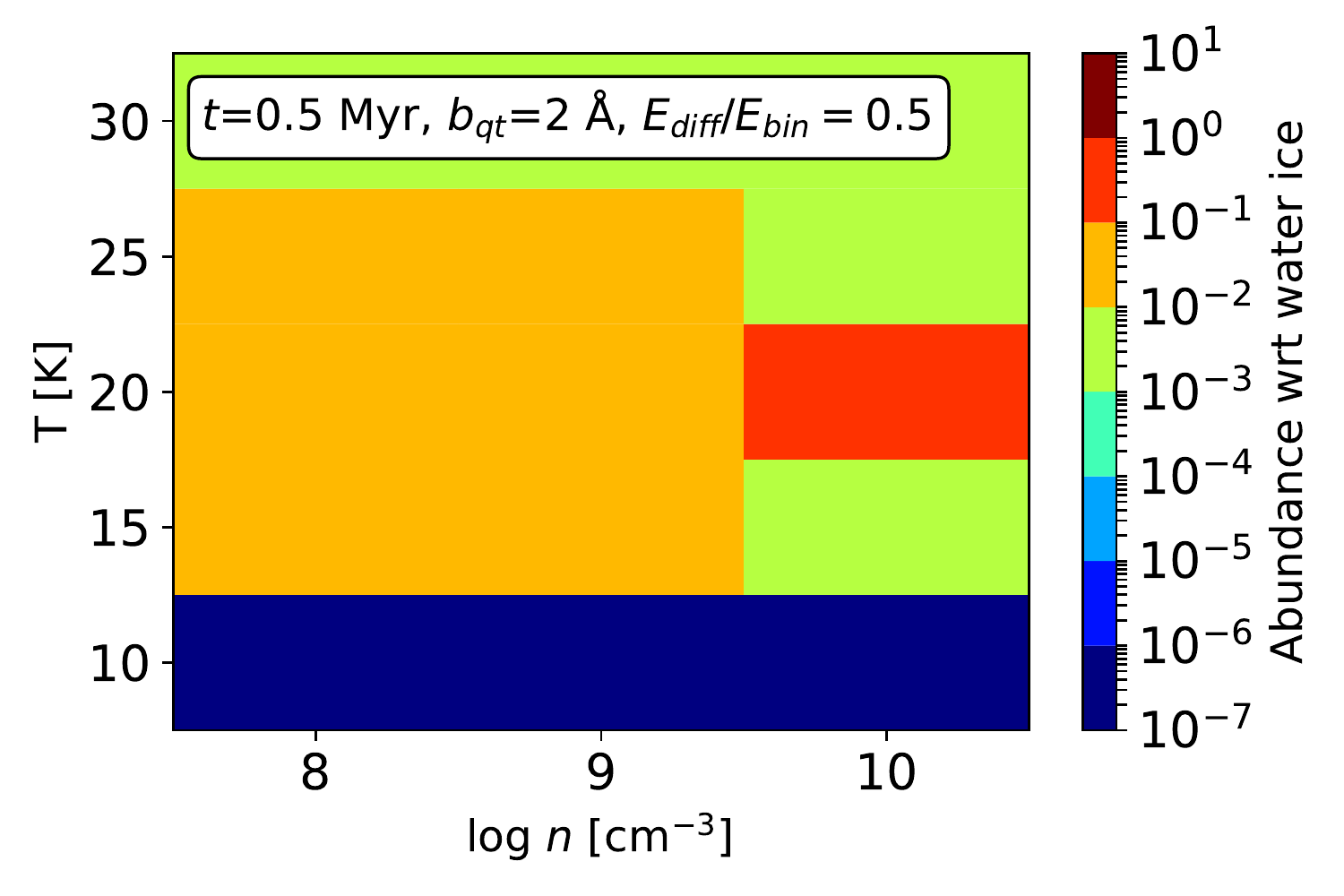}}\\
\subfigure{\includegraphics[width=0.22\textwidth]{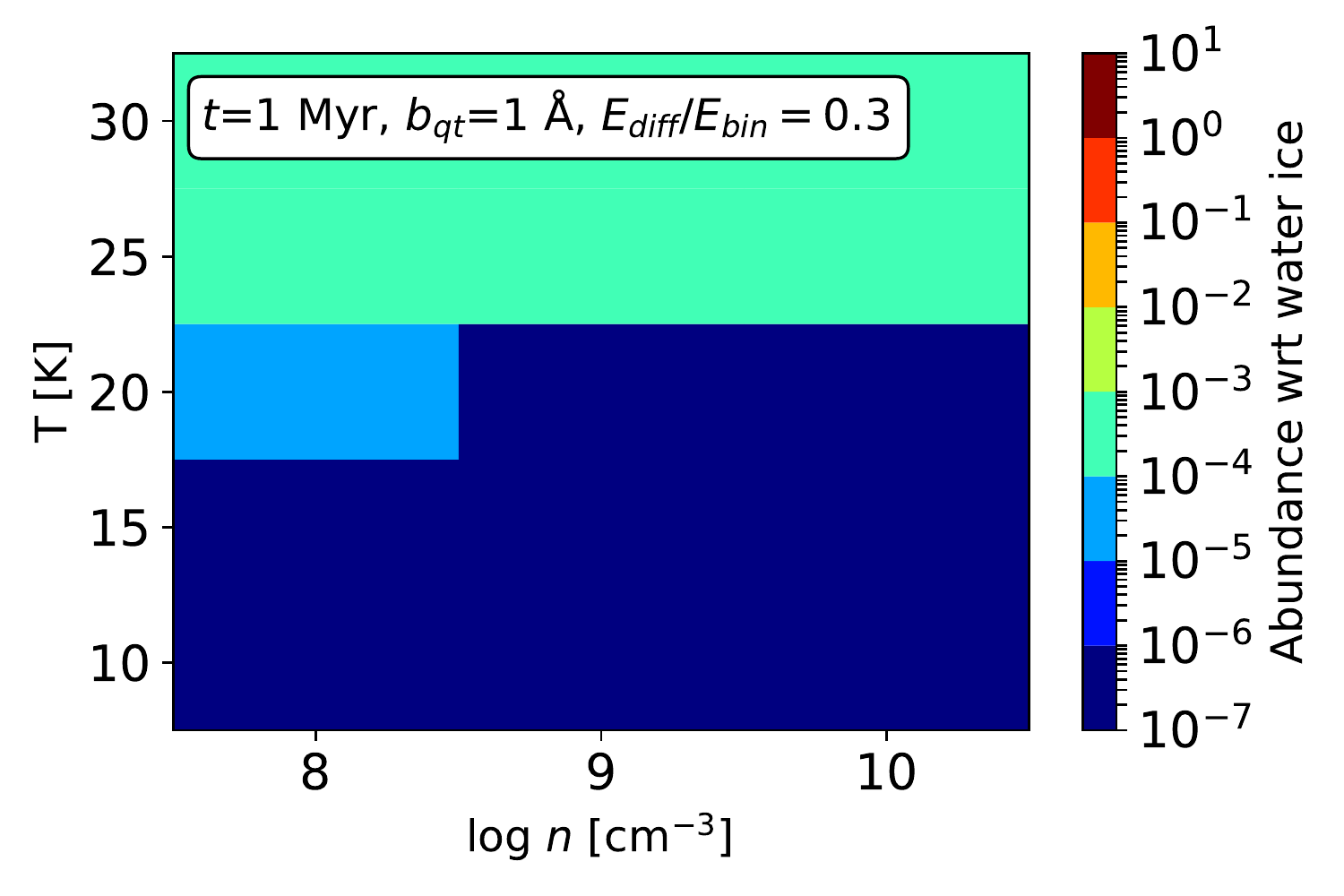}}
\subfigure{\includegraphics[width=0.22\textwidth]{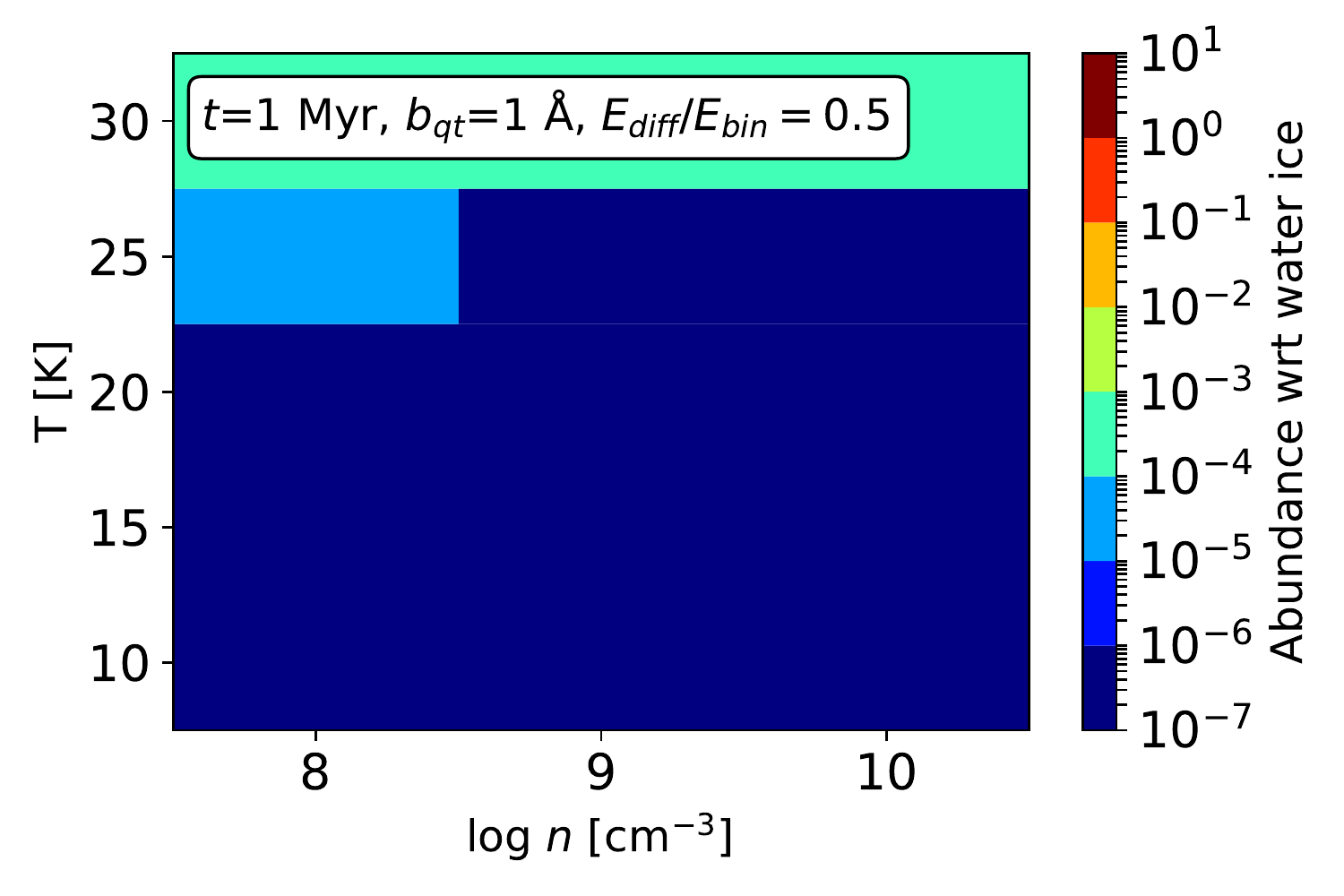}}
\subfigure{\includegraphics[width=0.22\textwidth]{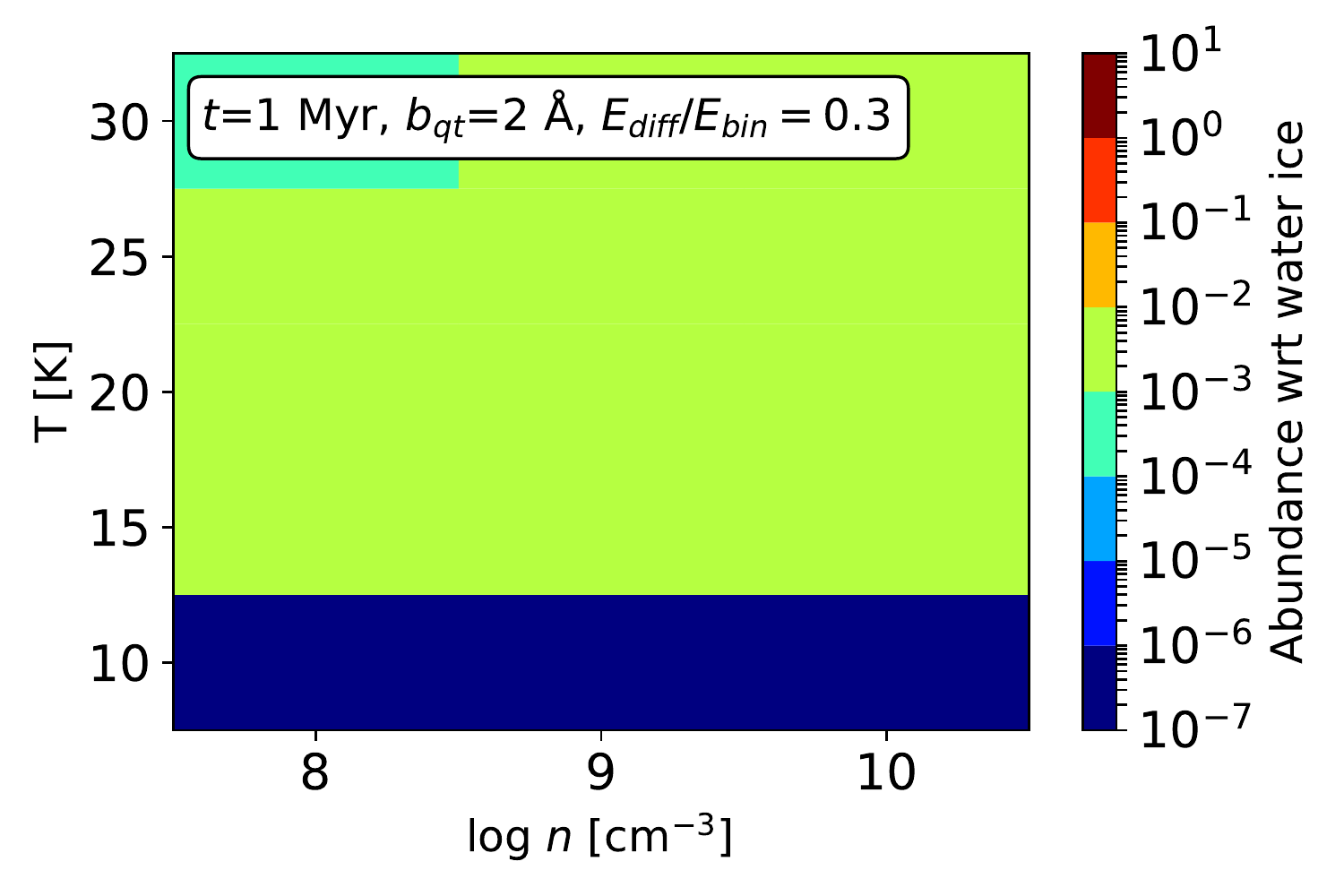}}
\subfigure{\includegraphics[width=0.22\textwidth]{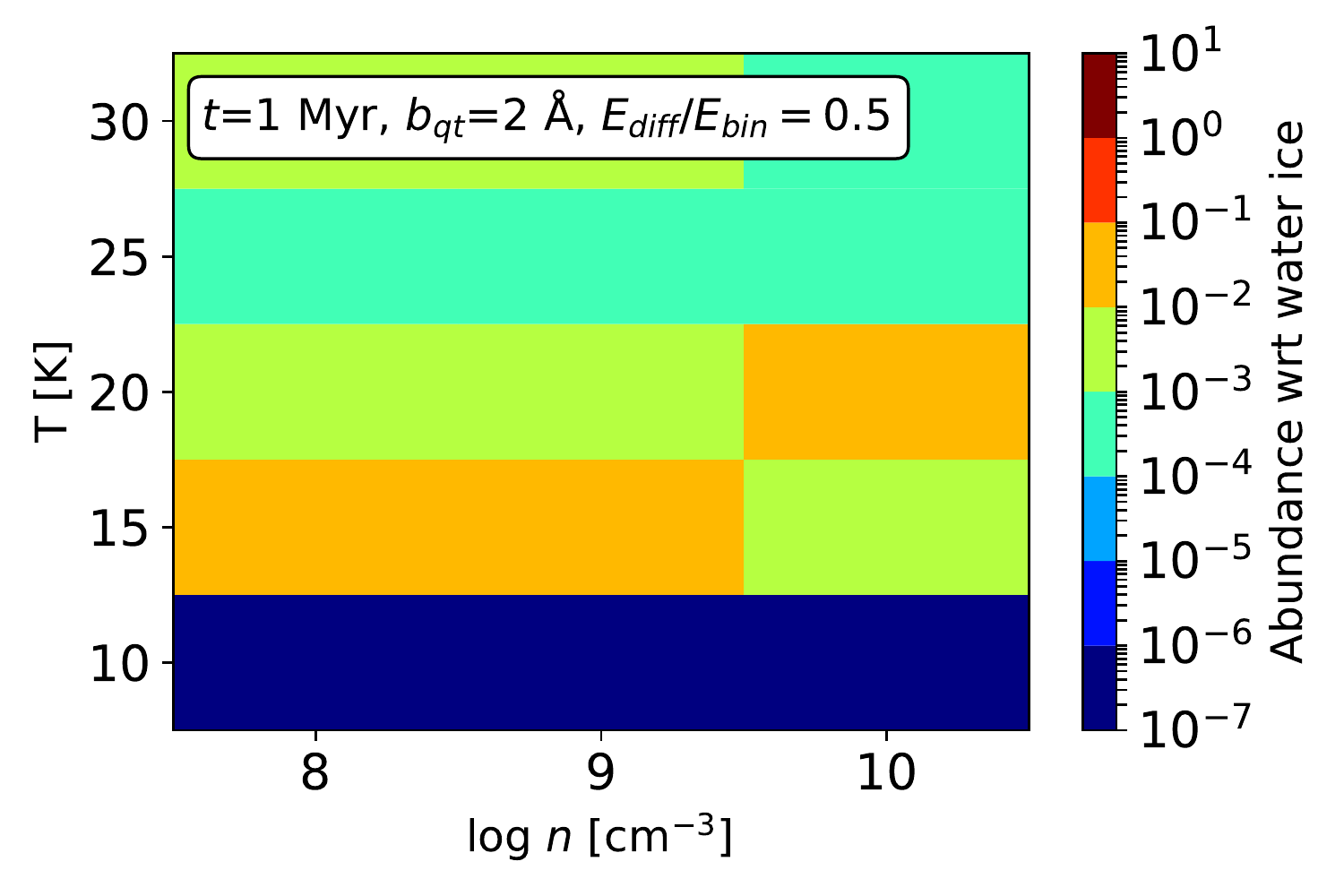}}\\
\subfigure{\includegraphics[width=0.22\textwidth]{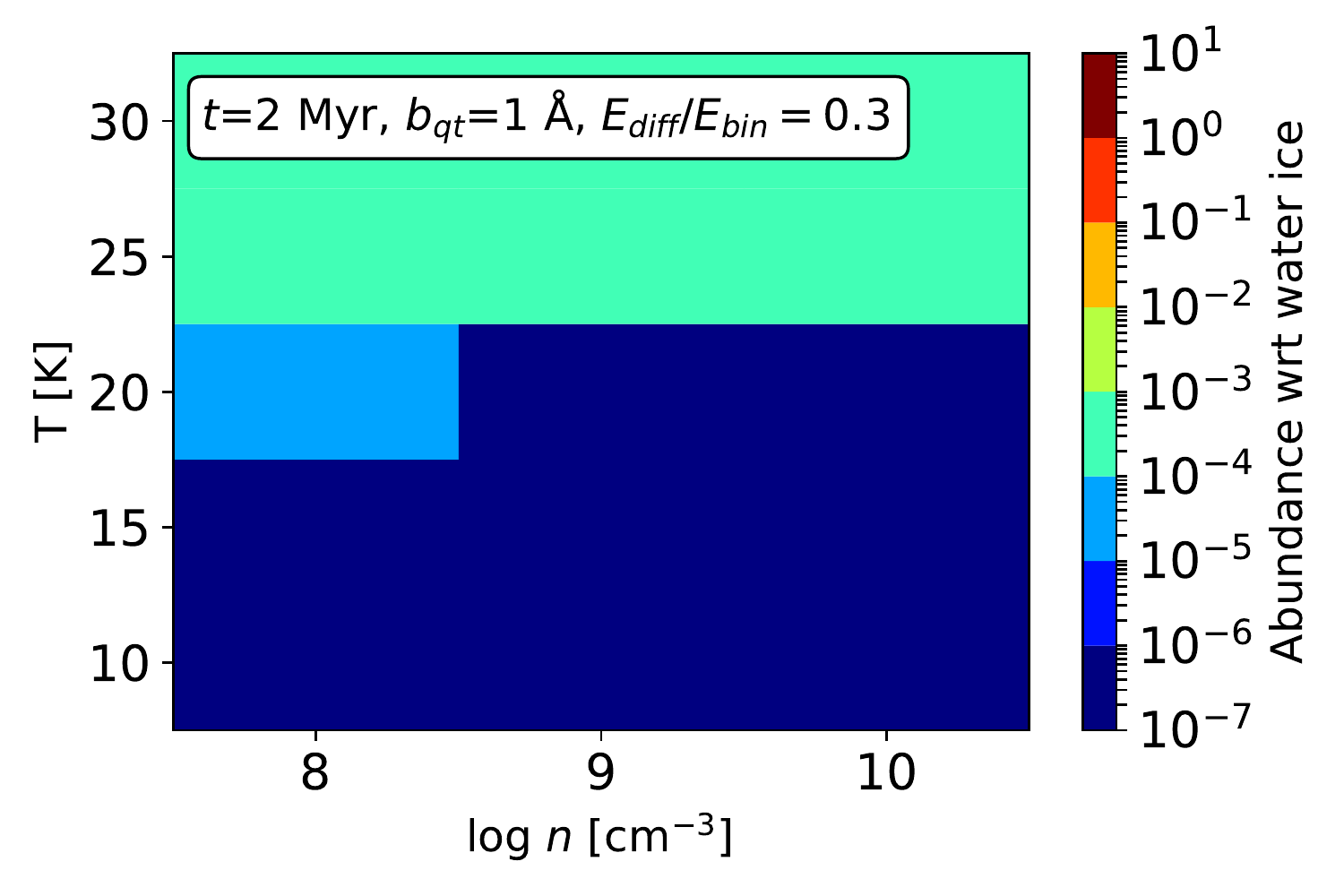}}
\subfigure{\includegraphics[width=0.22\textwidth]{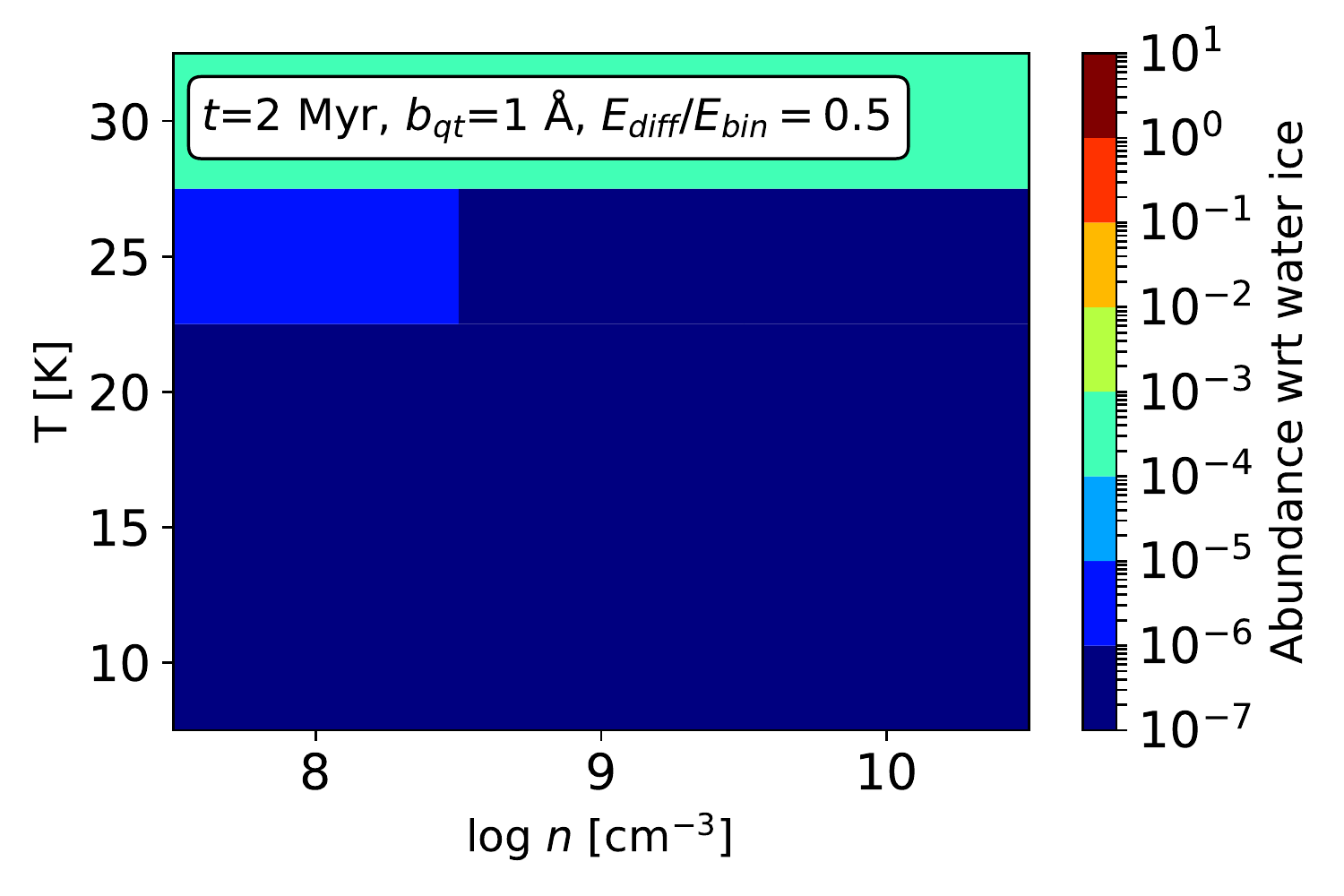}}
\subfigure{\includegraphics[width=0.22\textwidth]{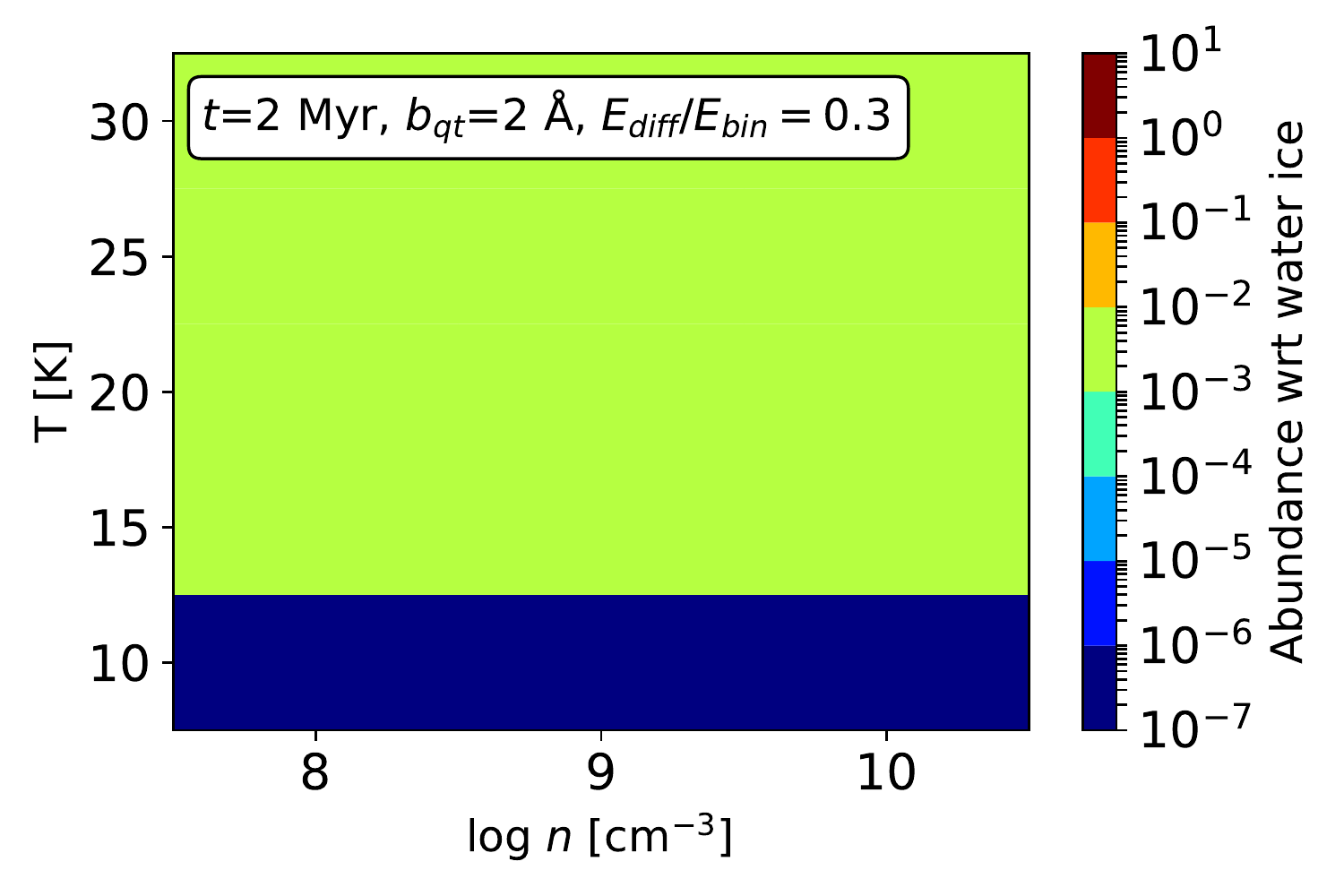}}
\subfigure{\includegraphics[width=0.22\textwidth]{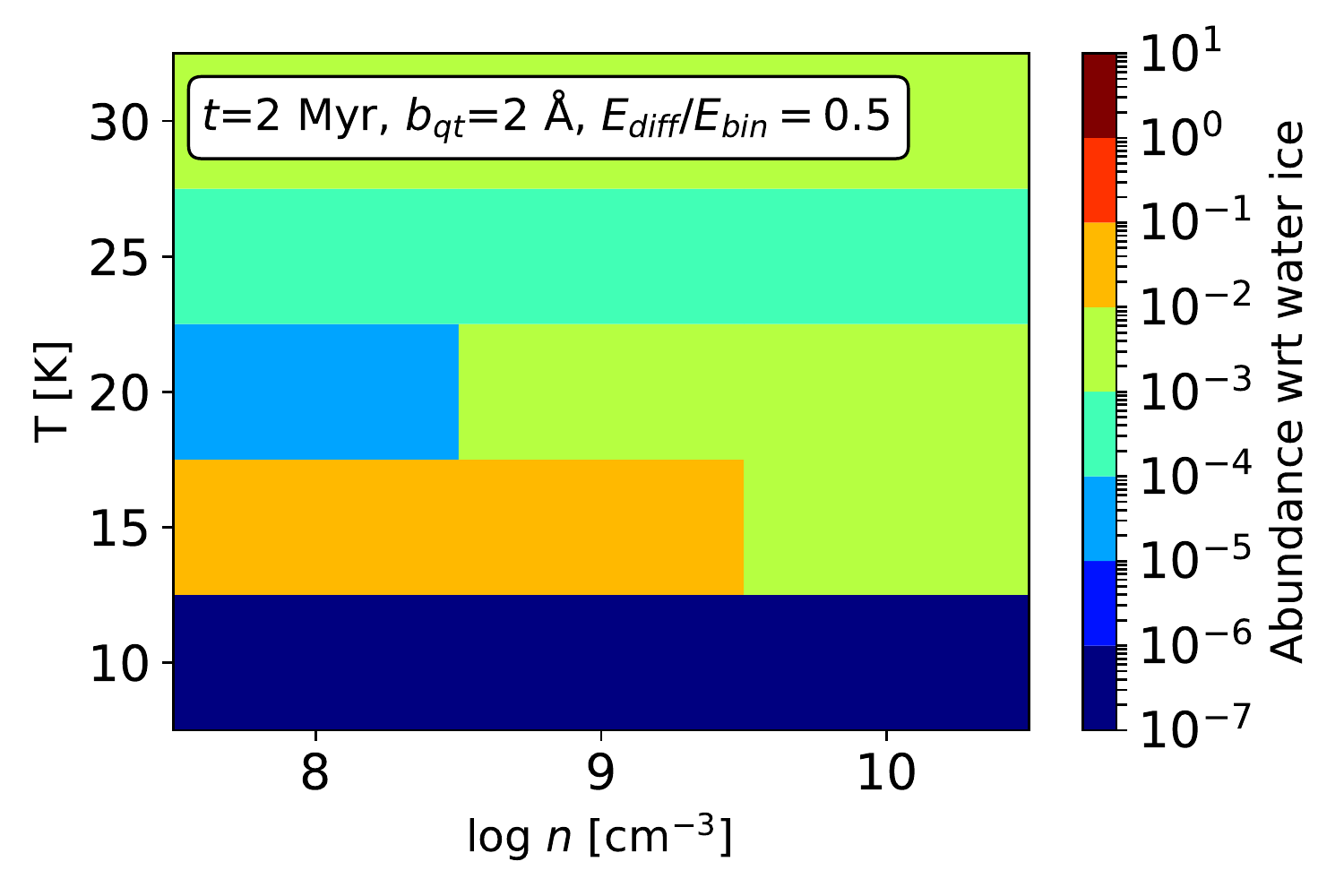}}\\
\subfigure{\includegraphics[width=0.22\textwidth]{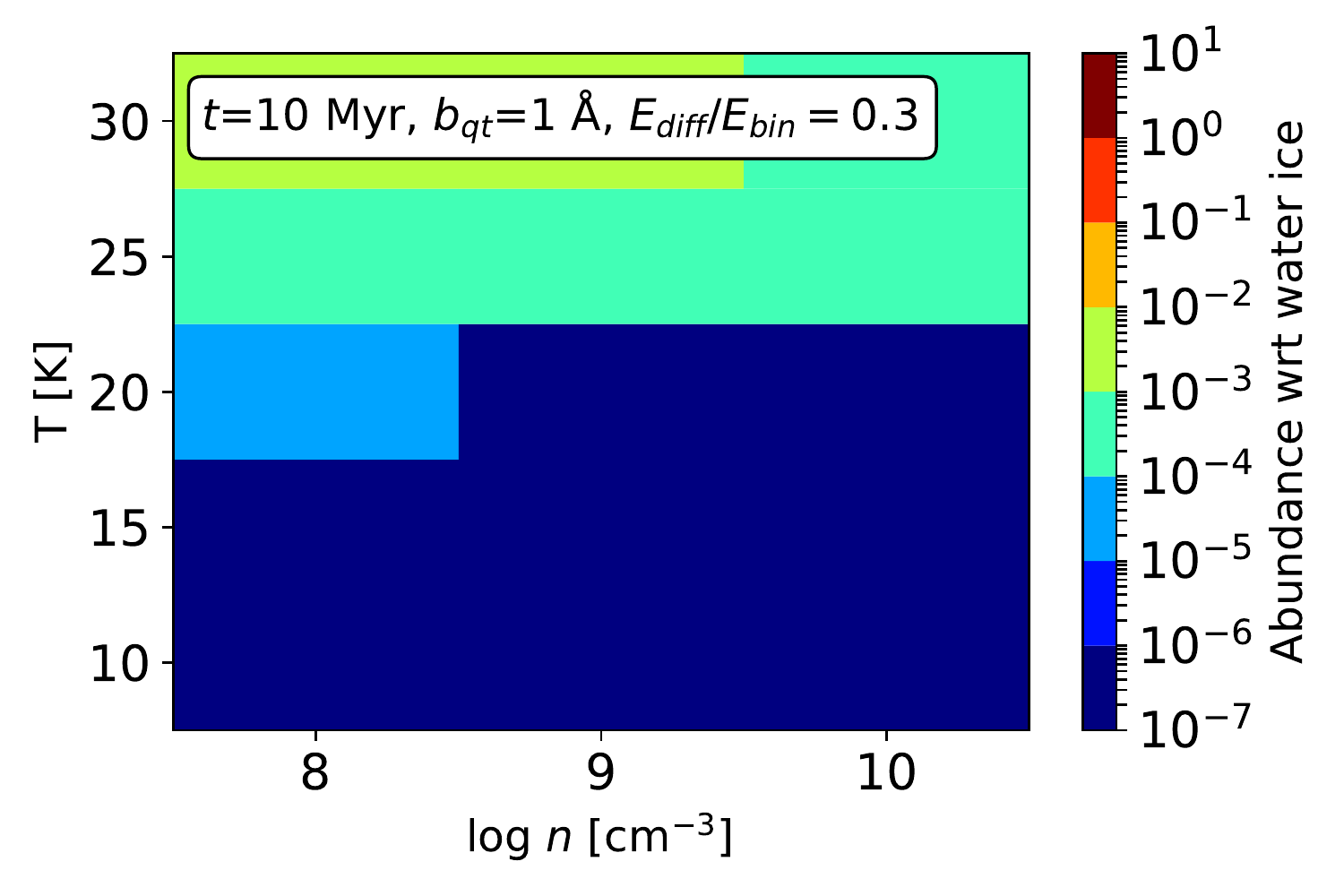}}
\subfigure{\includegraphics[width=0.22\textwidth]{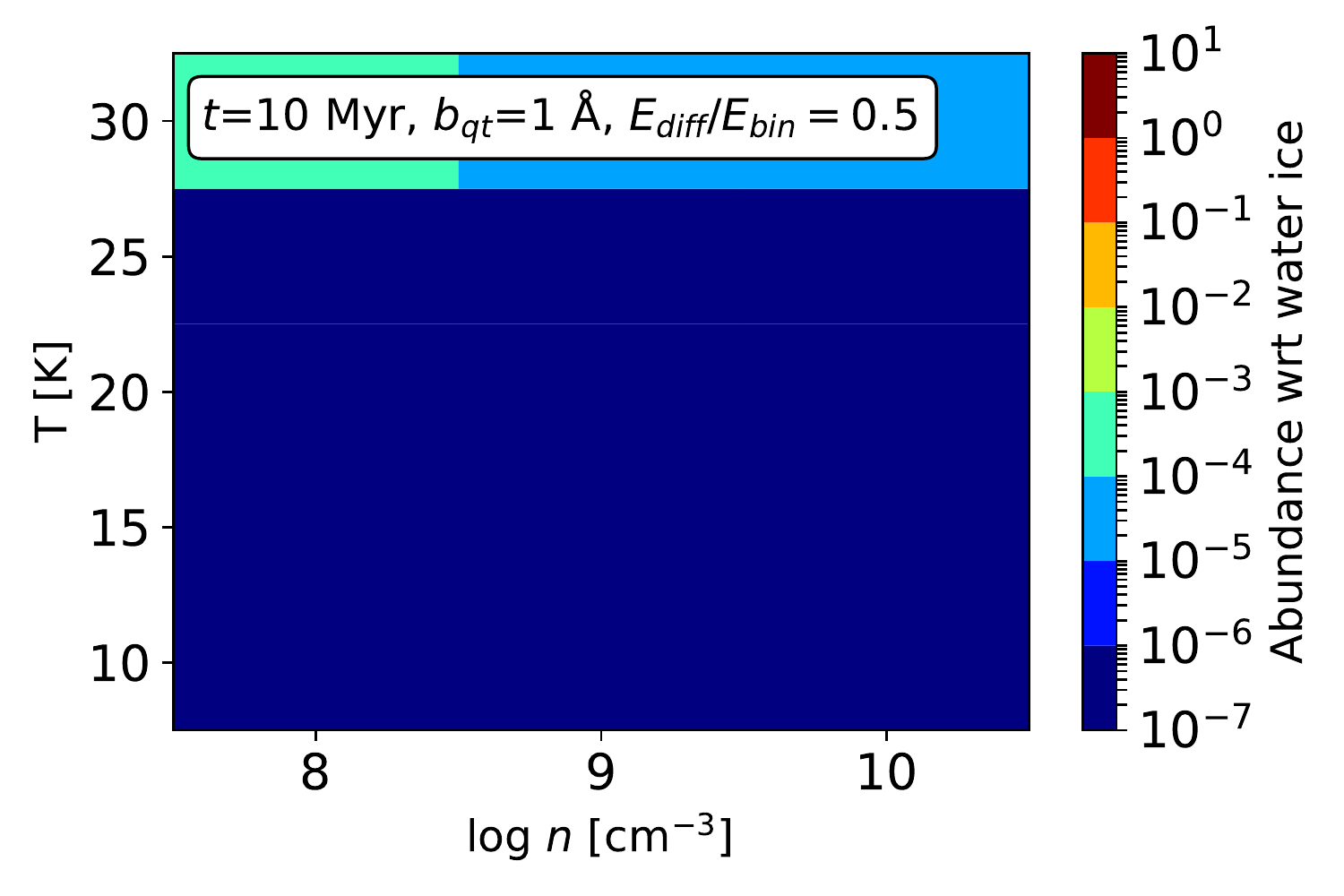}}
\subfigure{\includegraphics[width=0.22\textwidth]{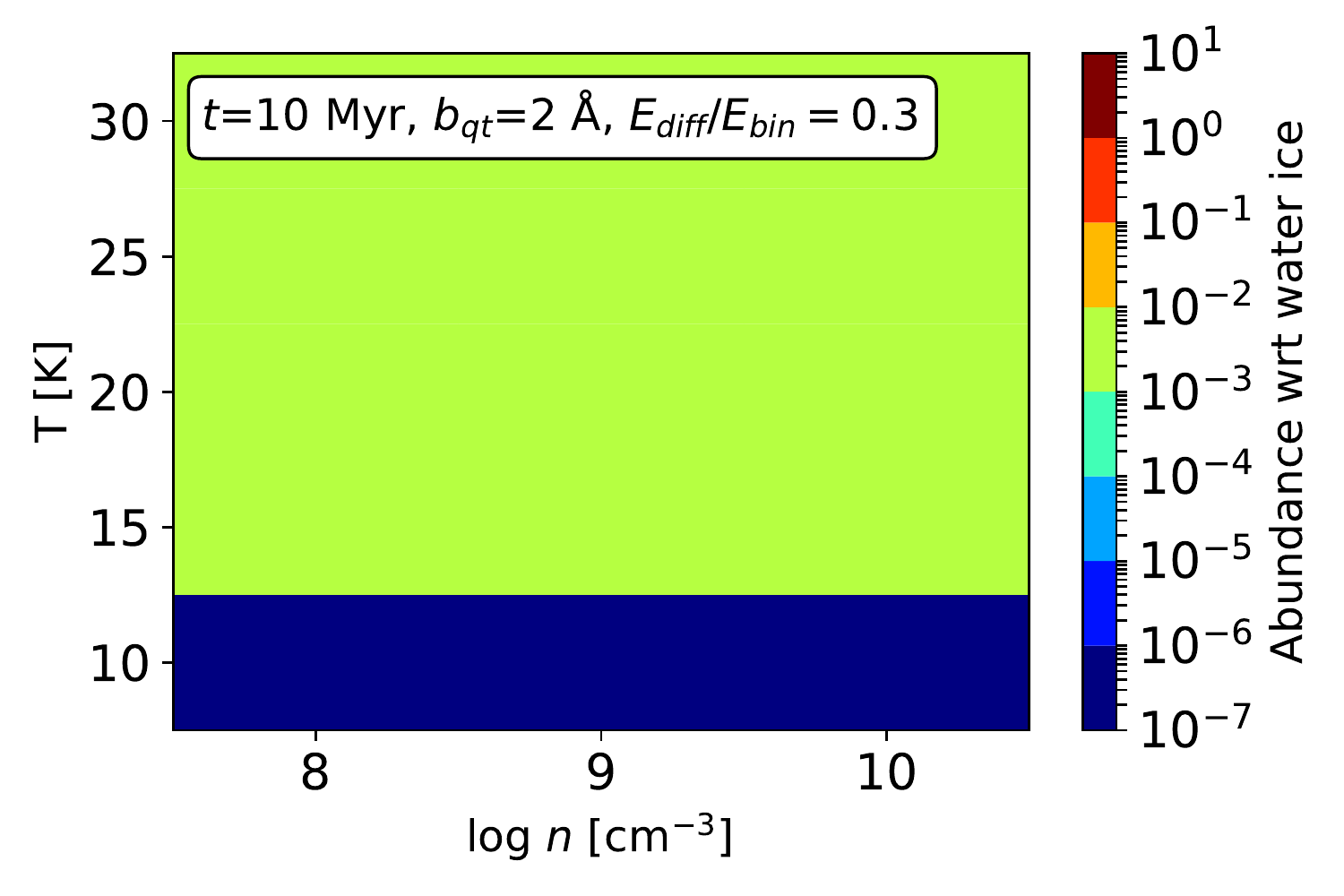}}
\subfigure{\includegraphics[width=0.22\textwidth]{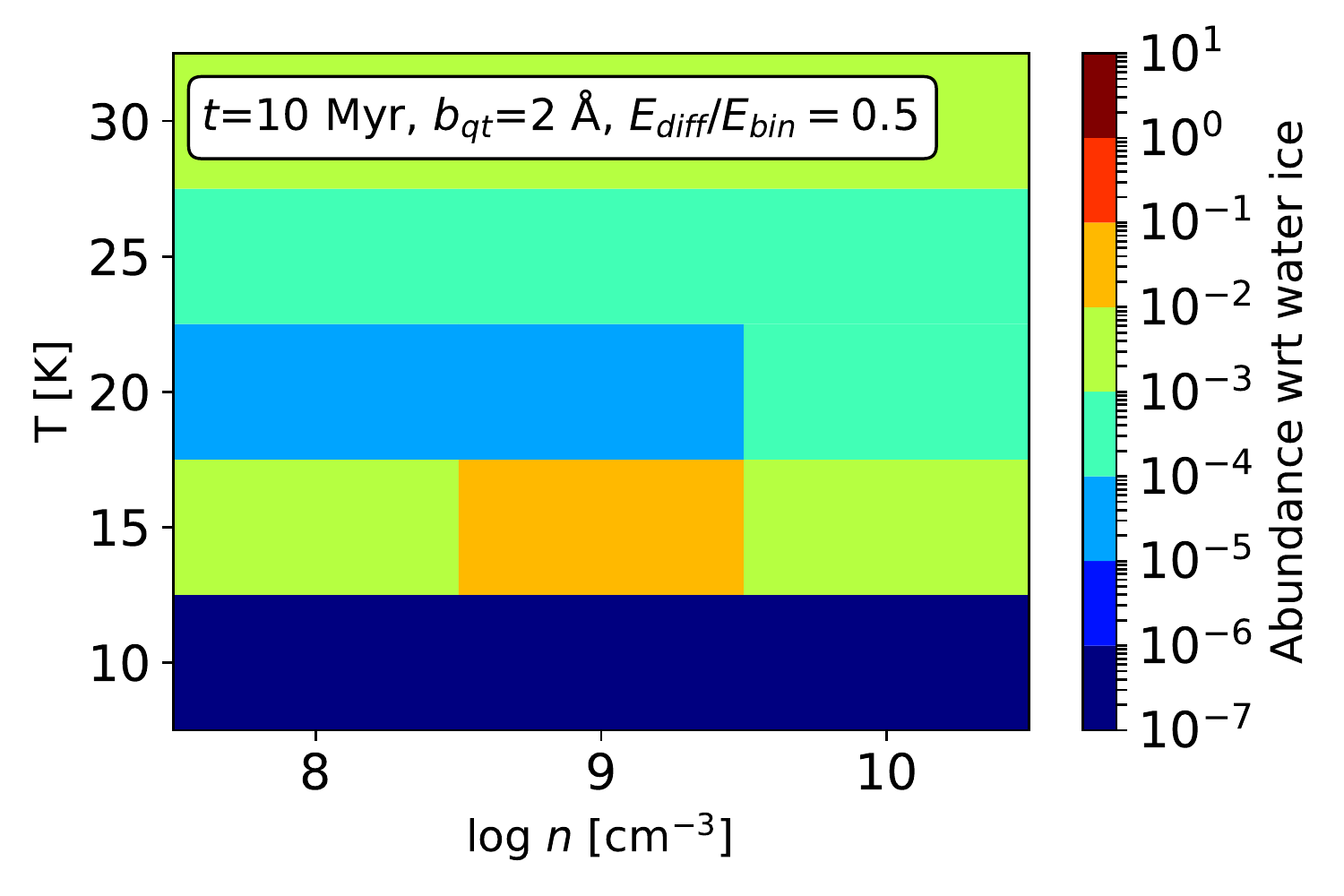}}\\
\caption{Abundances (given as colors) for \ce{O3} ice as function of midplane density ($x$-axes), and temperature ($y$-axes) at different evolutionary steps for the reset scenario. From left to right are abundances from model runs with different parameters for grain-surface reactions: columns one and two feature $b_{qt}$= 1 {\AA}, columns three and four feature $b_{qt}$= 1 {\AA}, columns one and three are with $E_{\rm{diff}}/E_{\rm{bin}}$= 0.3, and columns two and four are with $E_{\rm{diff}}/E_{\rm{bin}}$= 0.5. Top to bottom are different evolutionary times, from 0.05 Myr (top) to 10 Myr (bottom). To the right of each plot is a colorbar, indicating the abundance level with respect to \ce{H2O} ice for each color. The chemical network utilised includes \ce{O3} chemistry. Darkest hue of blue matches the upper limit for cometary \ce{O3} abundances.}
\label{chem_params_o3}
\end{figure*}

\begin{figure*}
\subfigure{\includegraphics[width=0.22\textwidth]{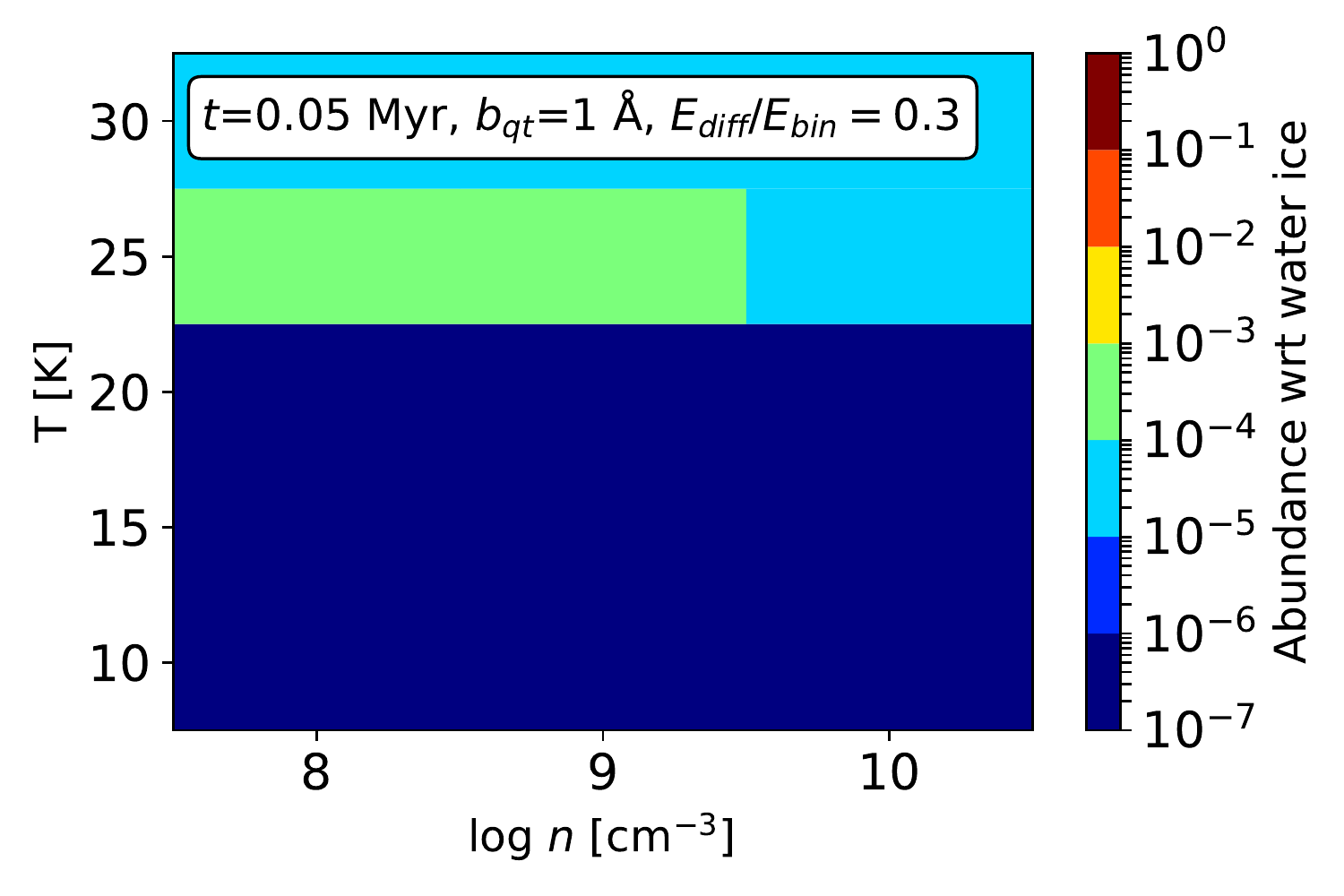}}
\subfigure{\includegraphics[width=0.22\textwidth]{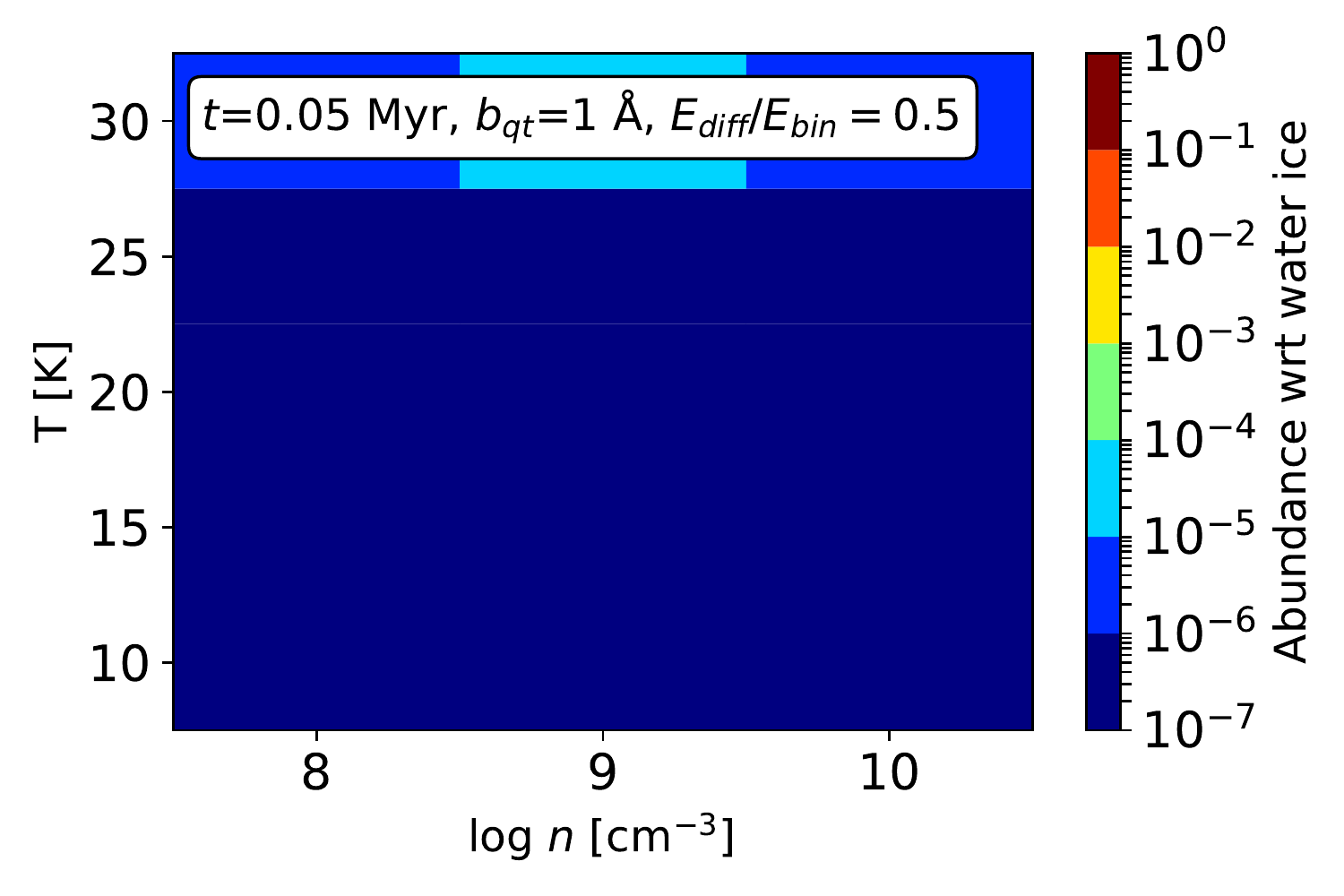}}
\subfigure{\includegraphics[width=0.22\textwidth]{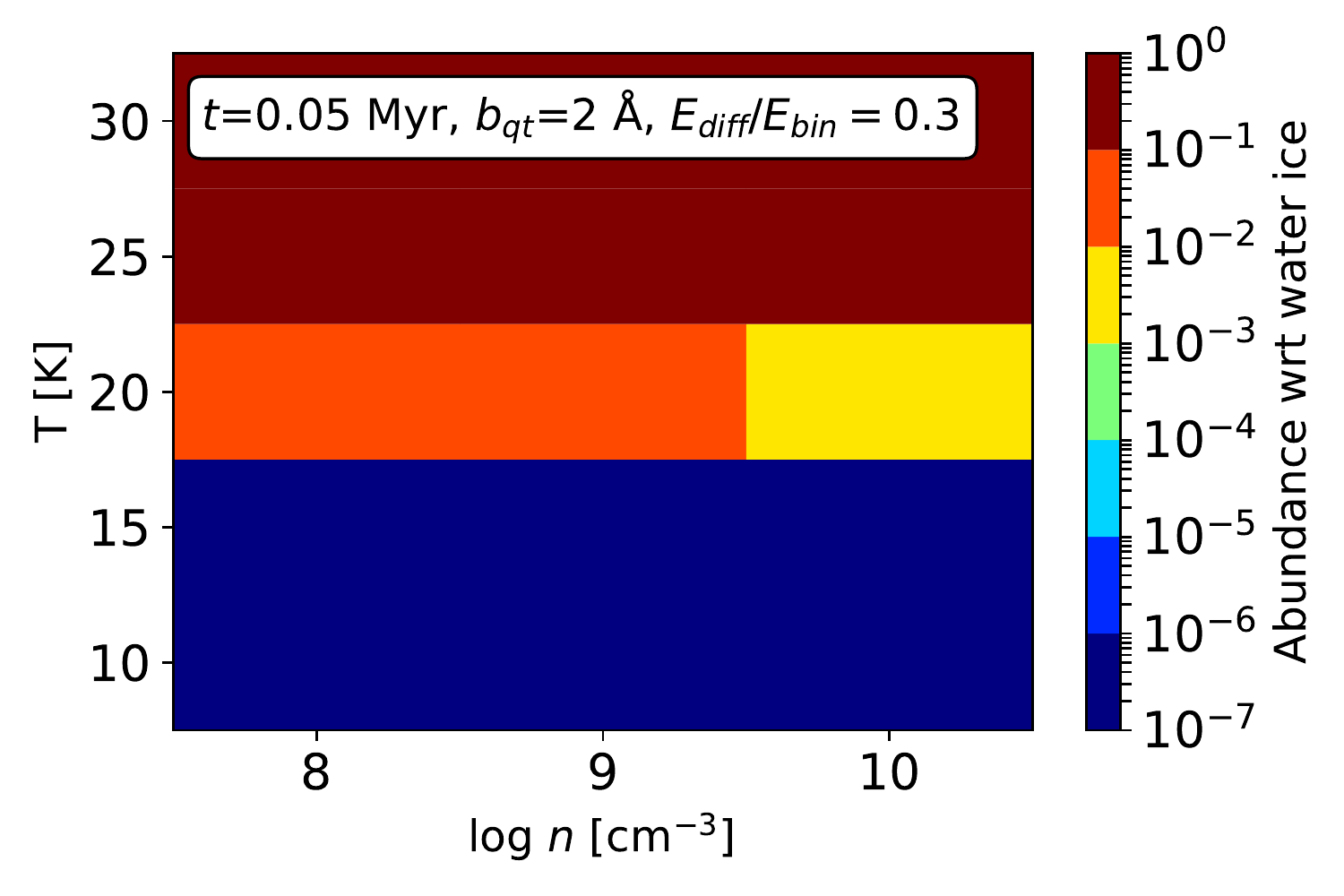}}
\subfigure{\includegraphics[width=0.22\textwidth]{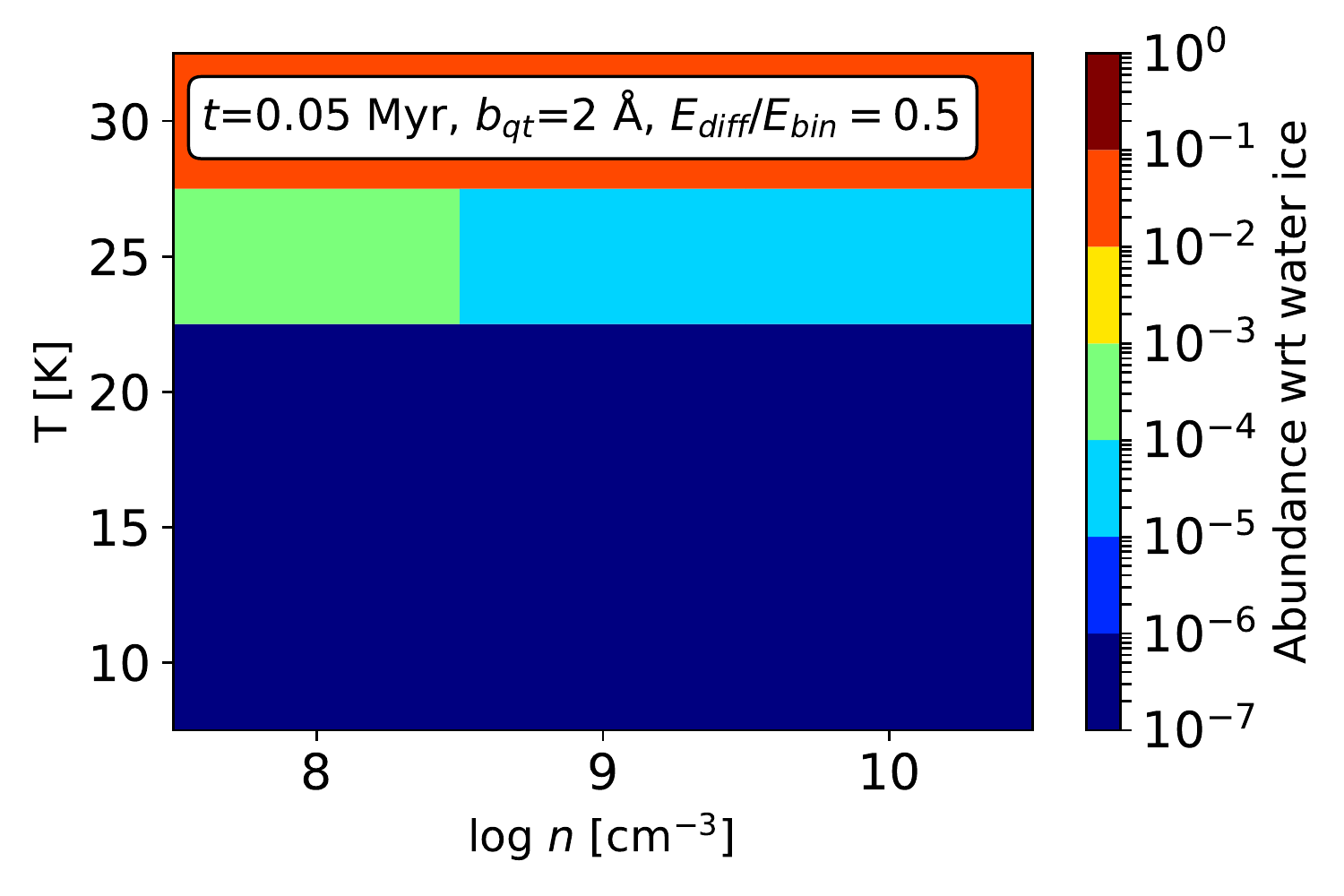}}\\
\subfigure{\includegraphics[width=0.22\textwidth]{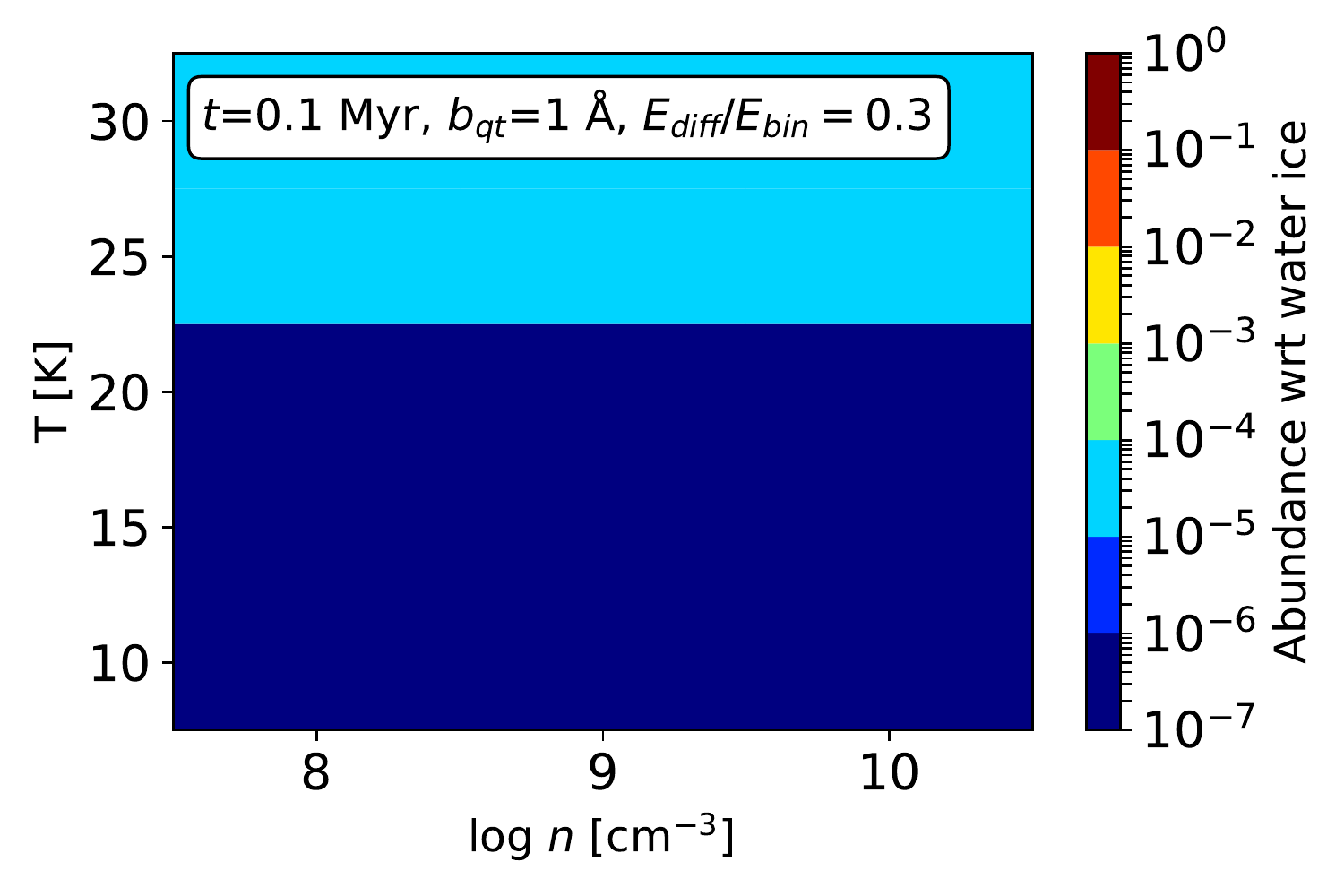}}
\subfigure{\includegraphics[width=0.22\textwidth]{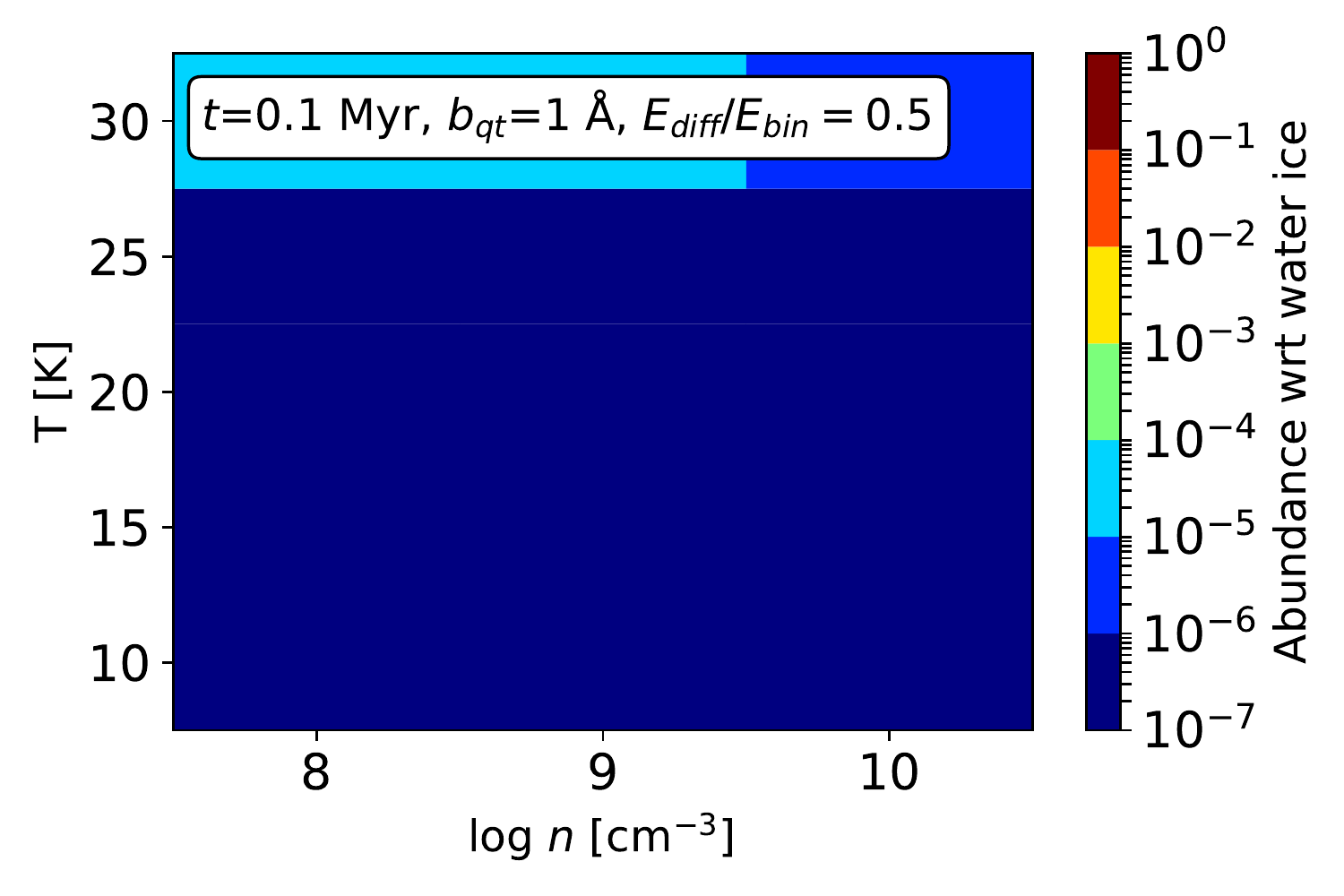}}
\subfigure{\includegraphics[width=0.22\textwidth]{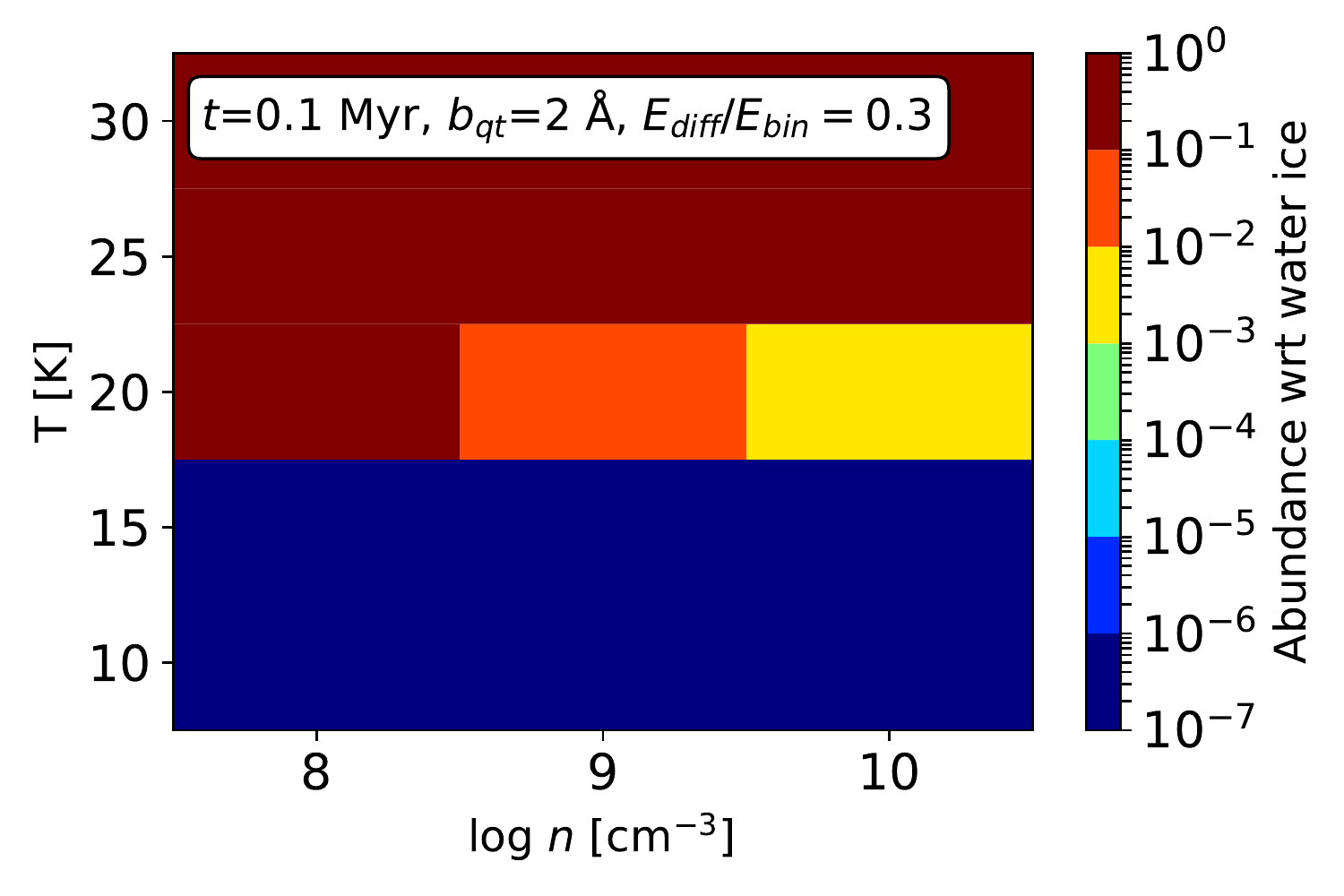}}
\subfigure{\includegraphics[width=0.22\textwidth]{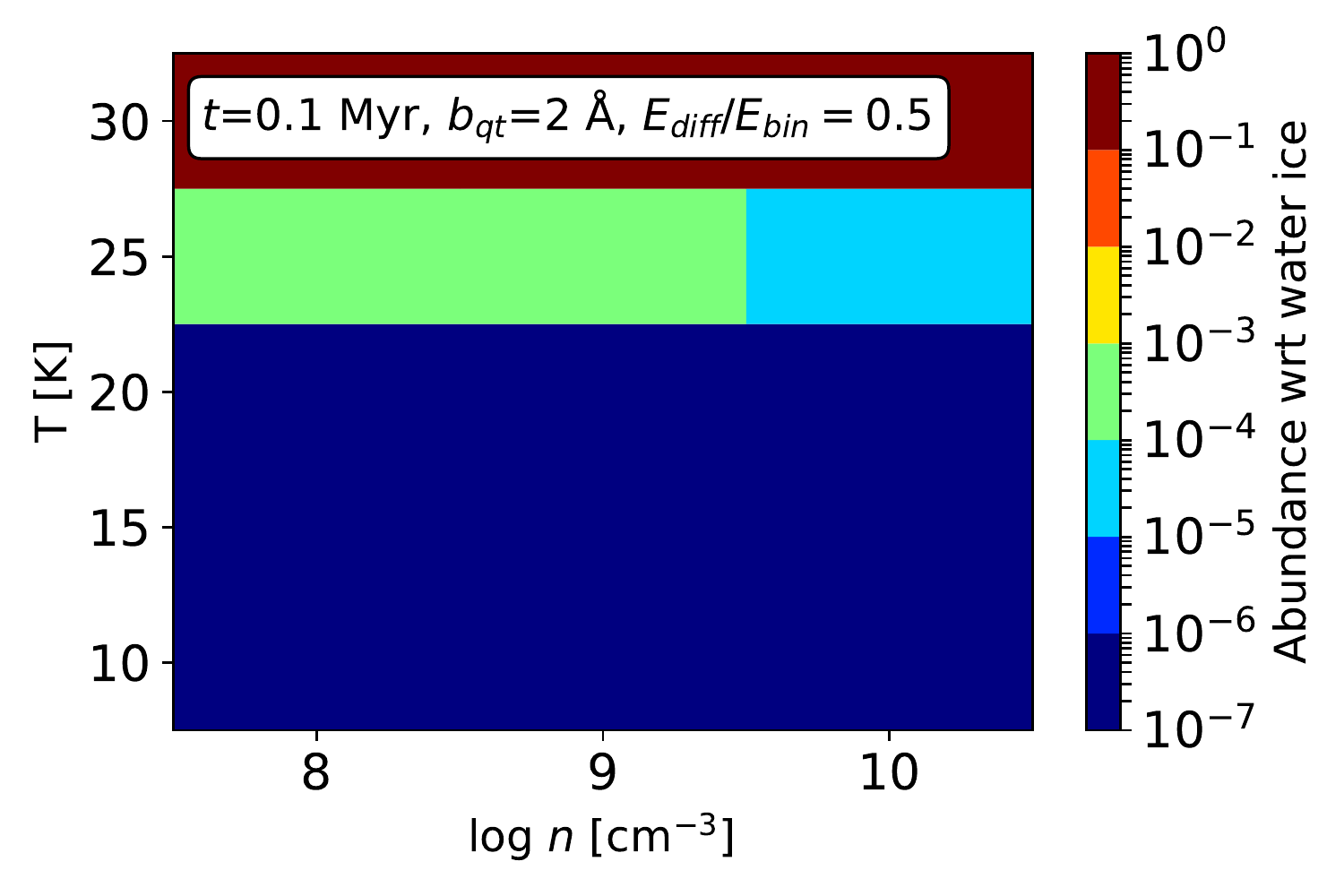}}\\
\subfigure{\includegraphics[width=0.22\textwidth]{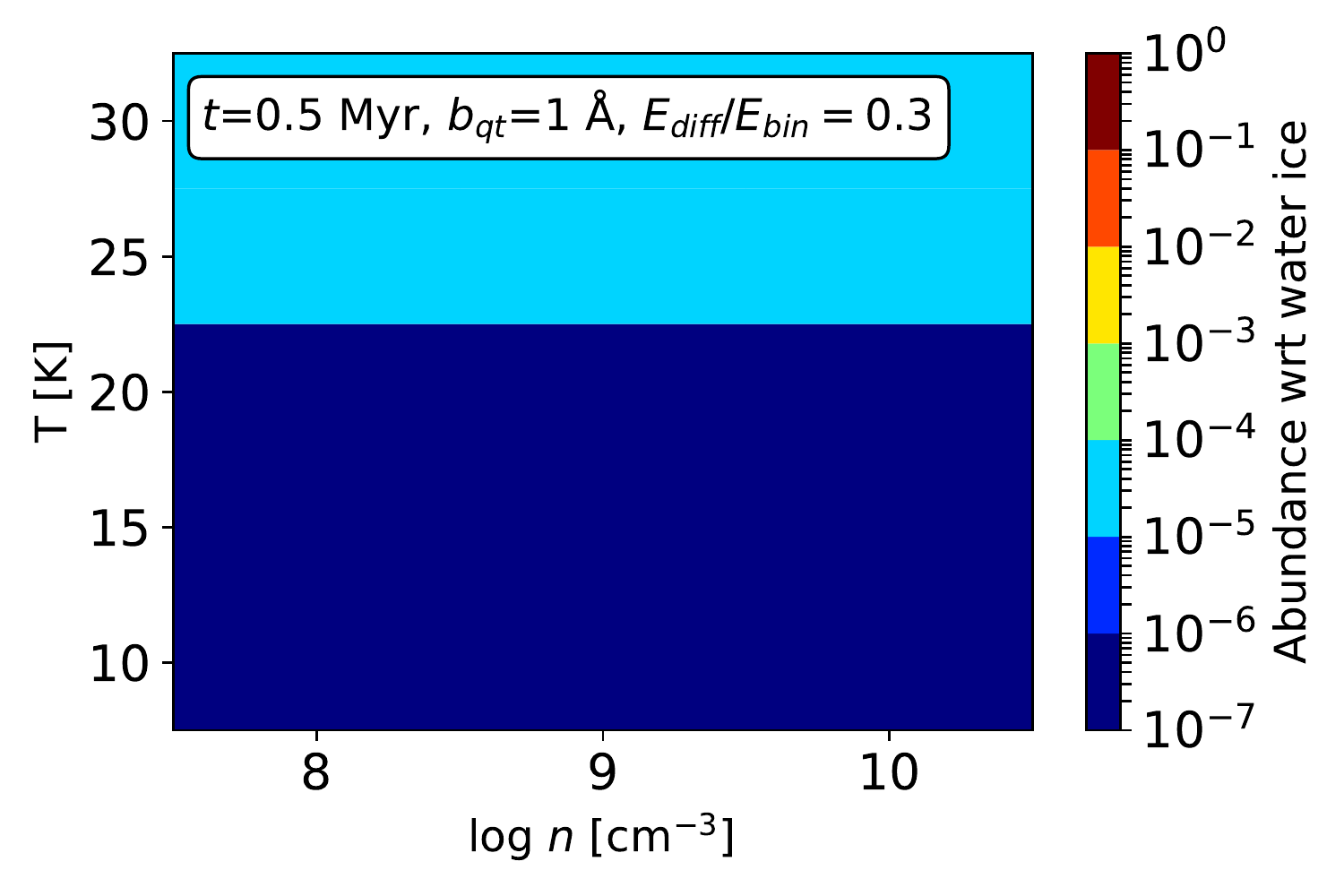}}
\subfigure{\includegraphics[width=0.22\textwidth]{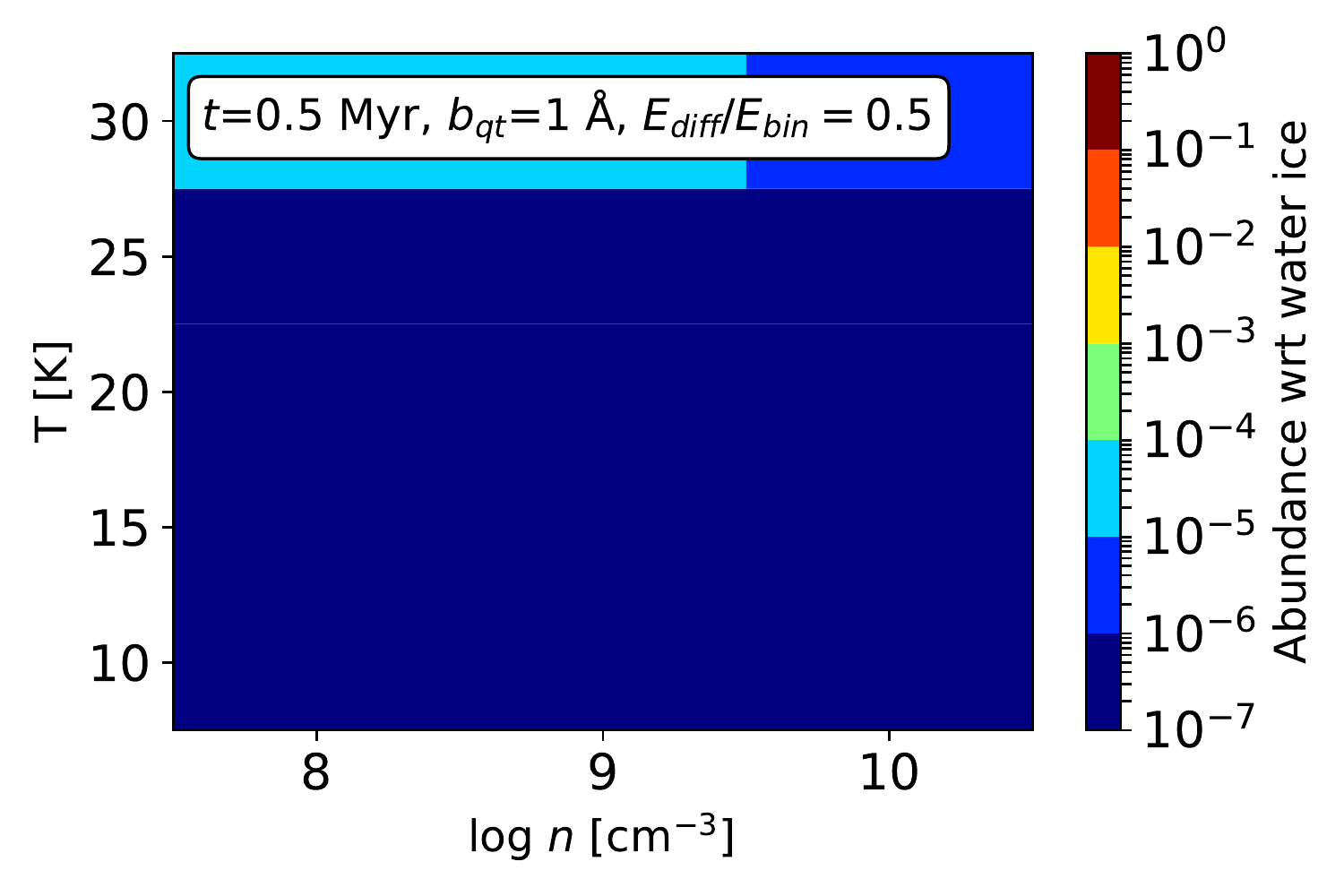}}
\subfigure{\includegraphics[width=0.22\textwidth]{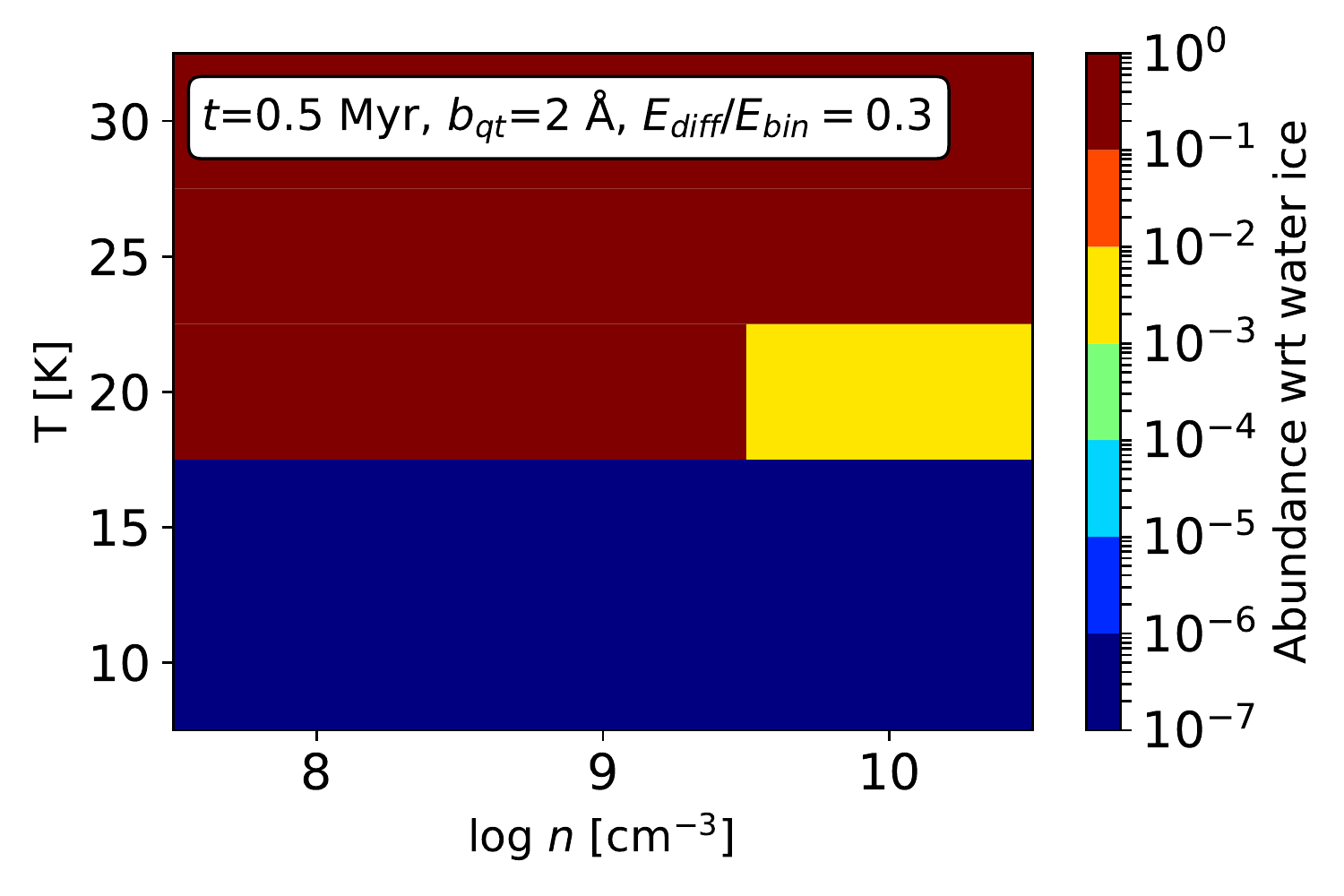}}
\subfigure{\includegraphics[width=0.22\textwidth]{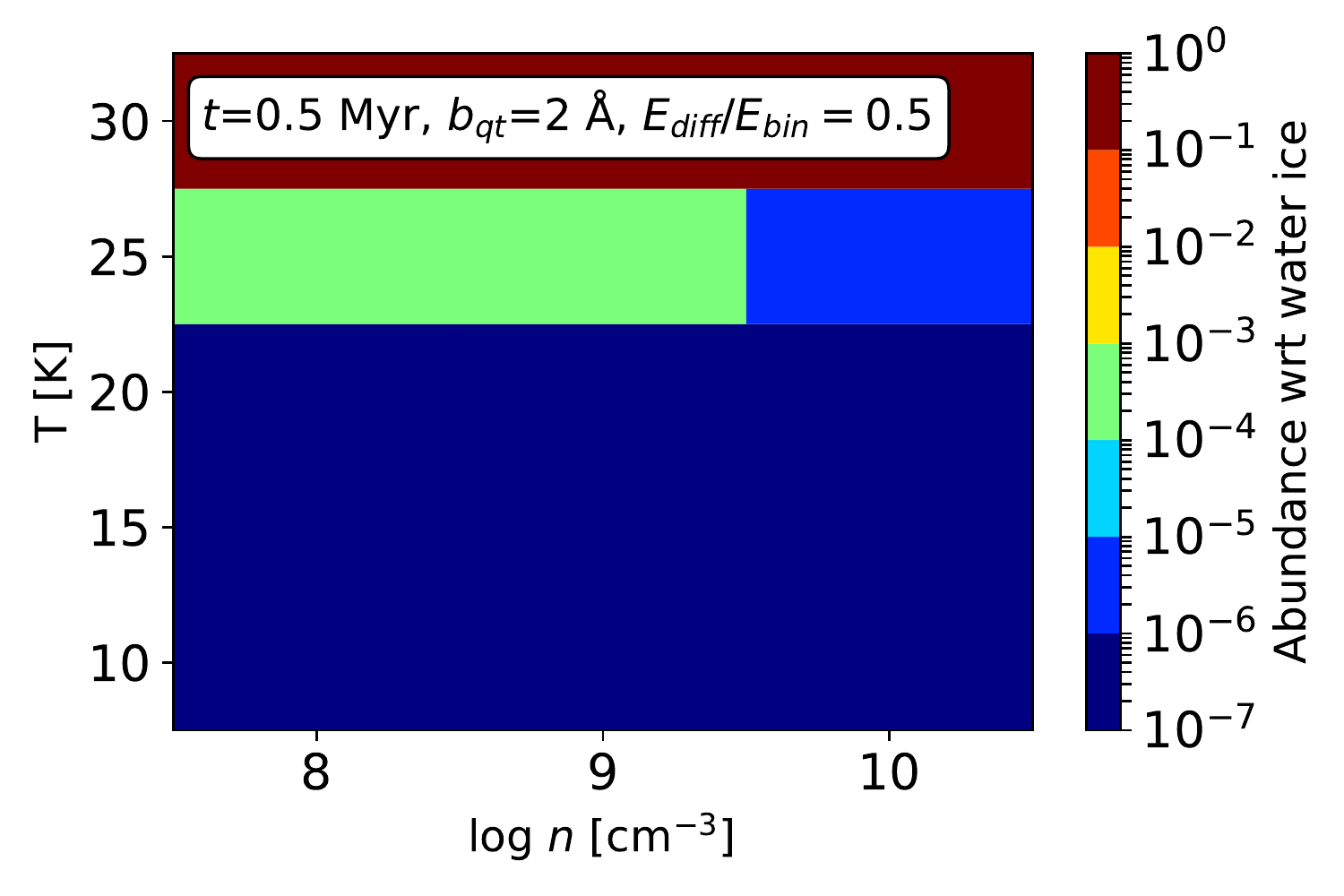}}\\
\subfigure{\includegraphics[width=0.22\textwidth]{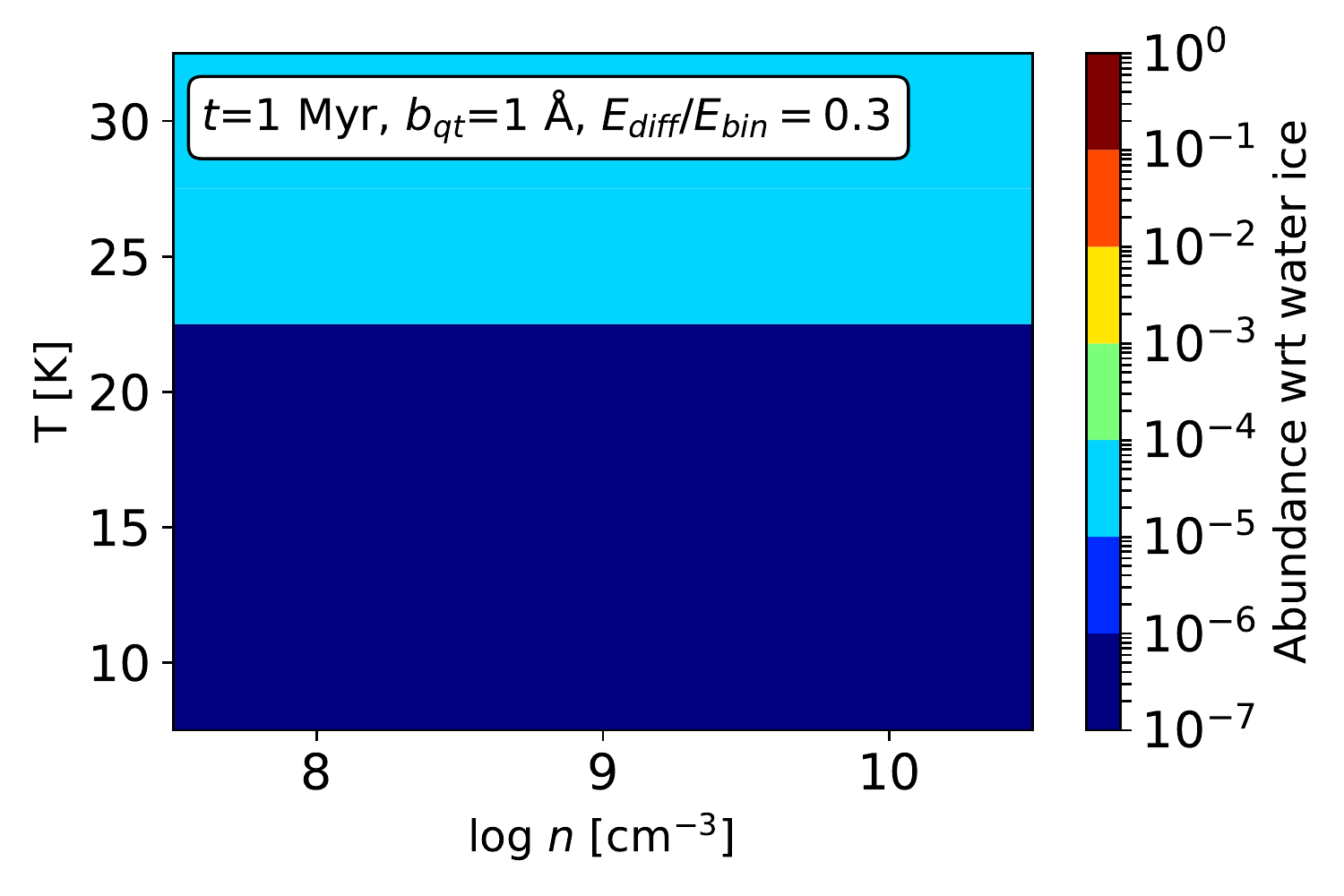}}
\subfigure{\includegraphics[width=0.22\textwidth]{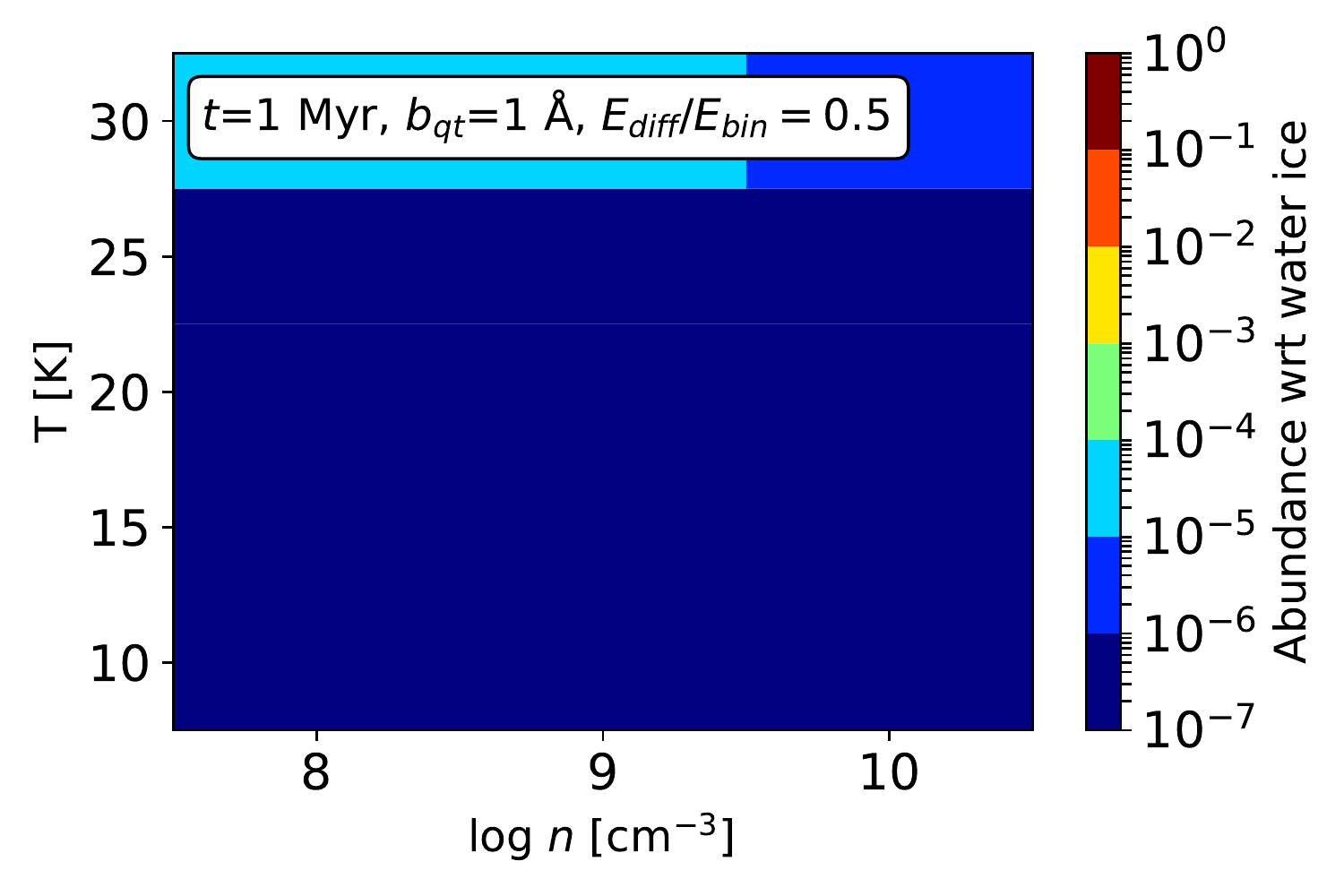}}
\subfigure{\includegraphics[width=0.22\textwidth]{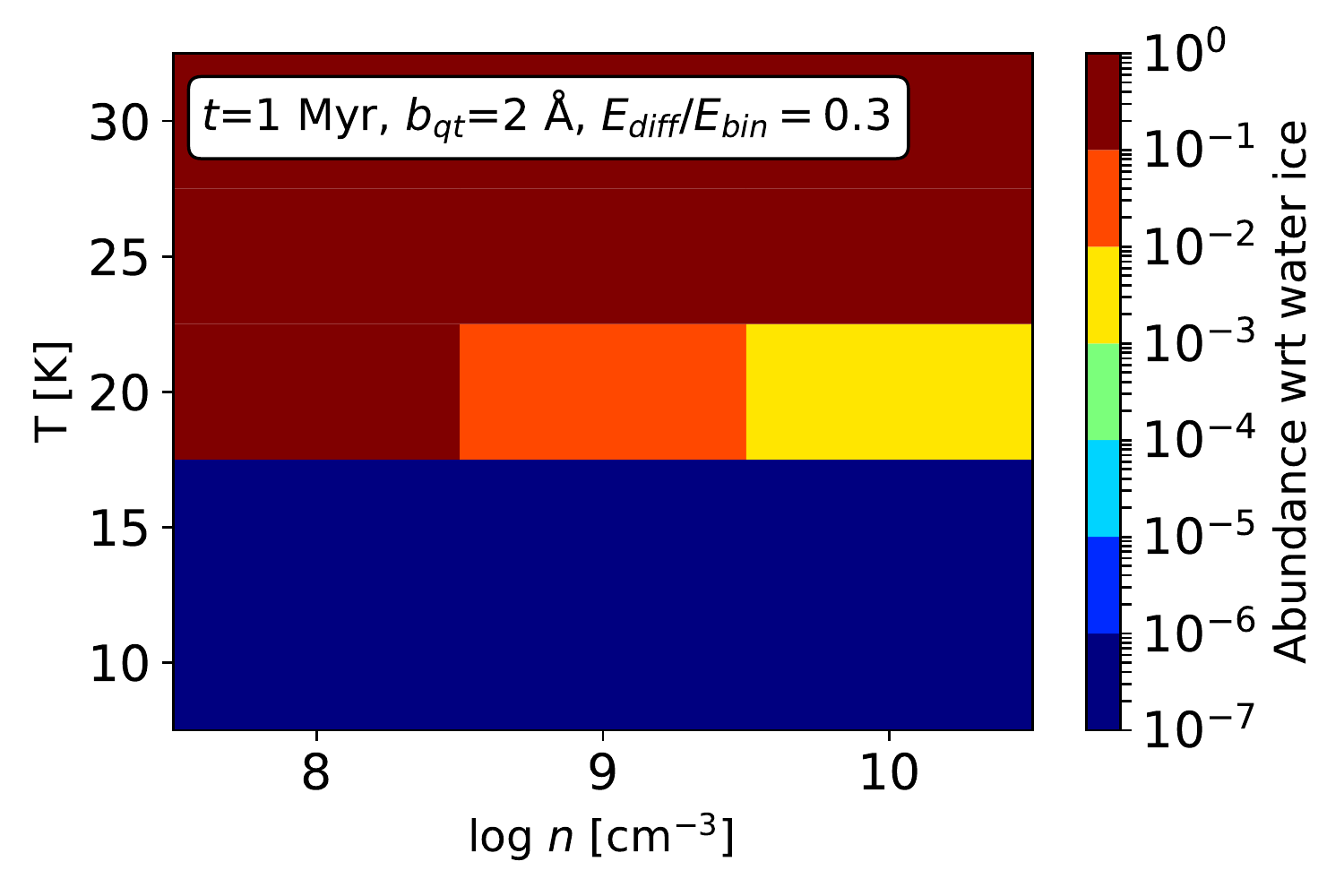}}
\subfigure{\includegraphics[width=0.22\textwidth]{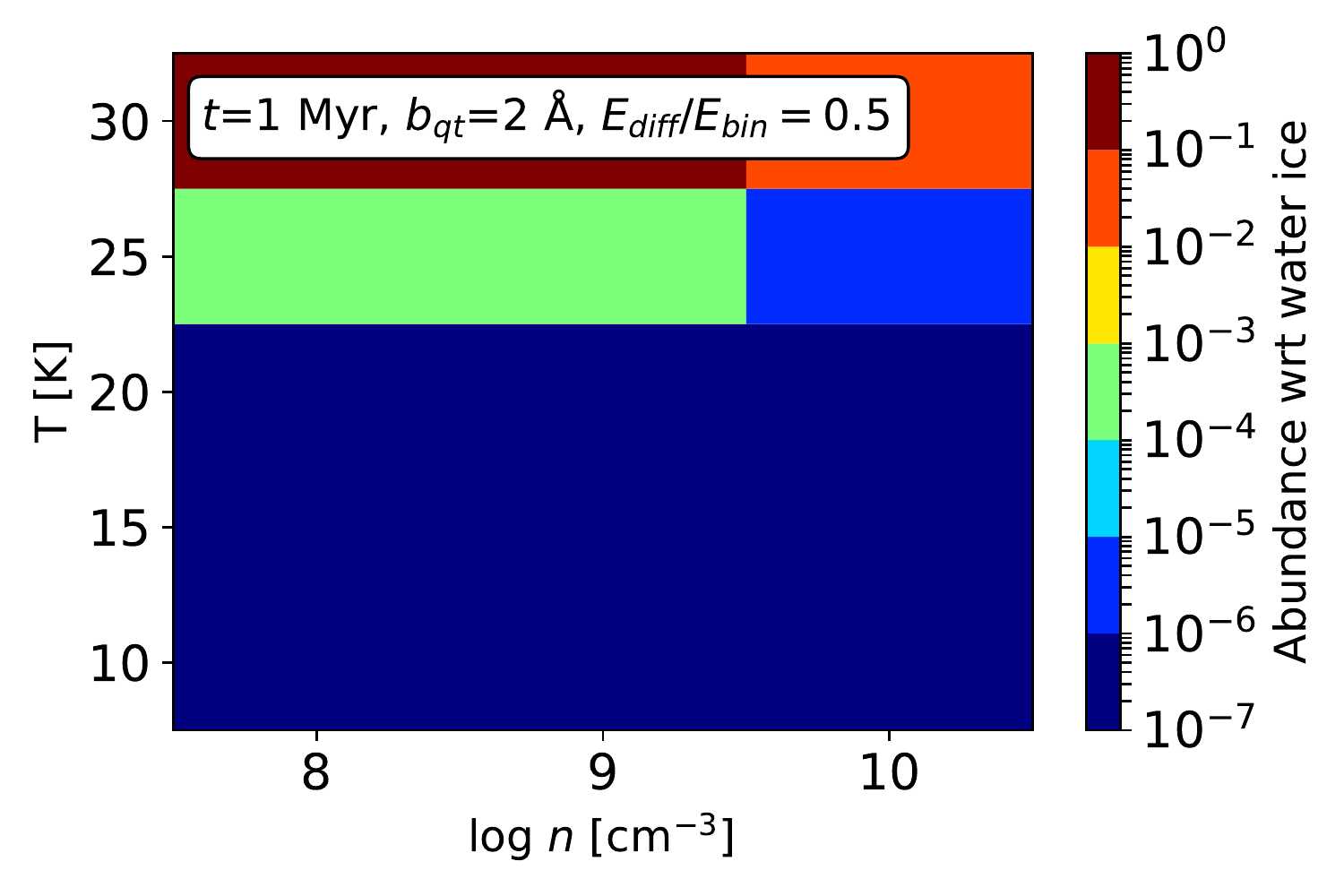}}\\
\subfigure{\includegraphics[width=0.22\textwidth]{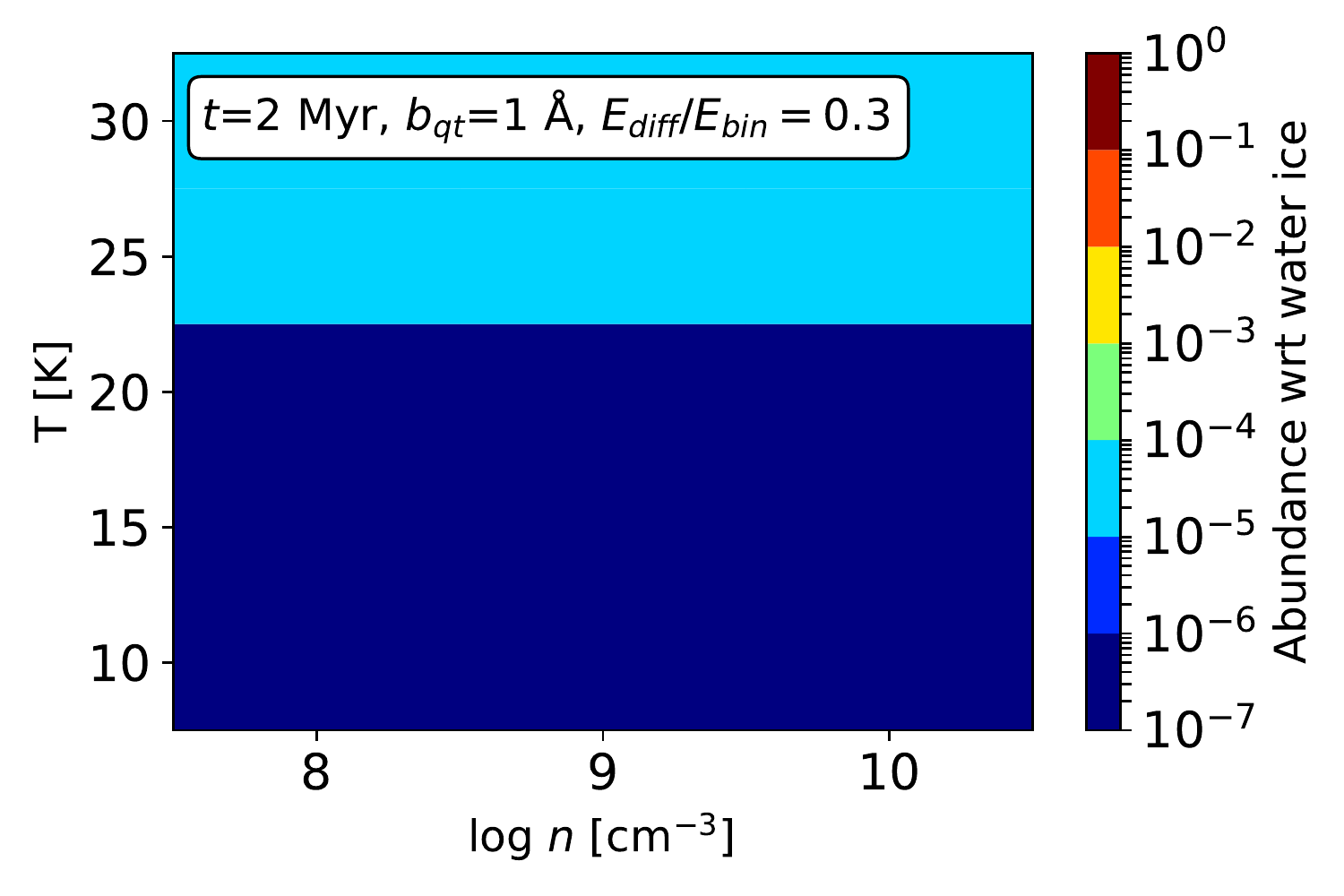}}
\subfigure{\includegraphics[width=0.22\textwidth]{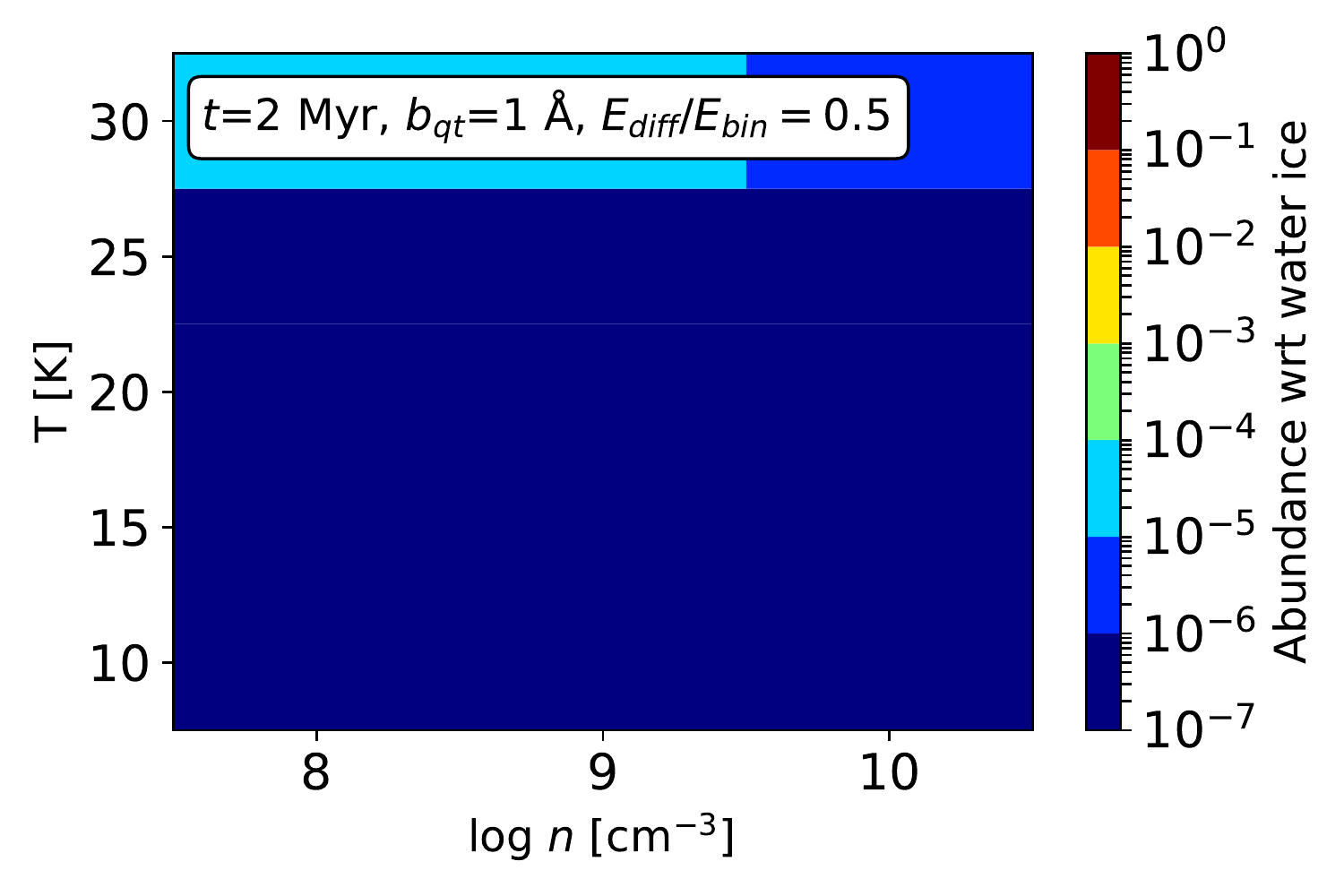}}
\subfigure{\includegraphics[width=0.22\textwidth]{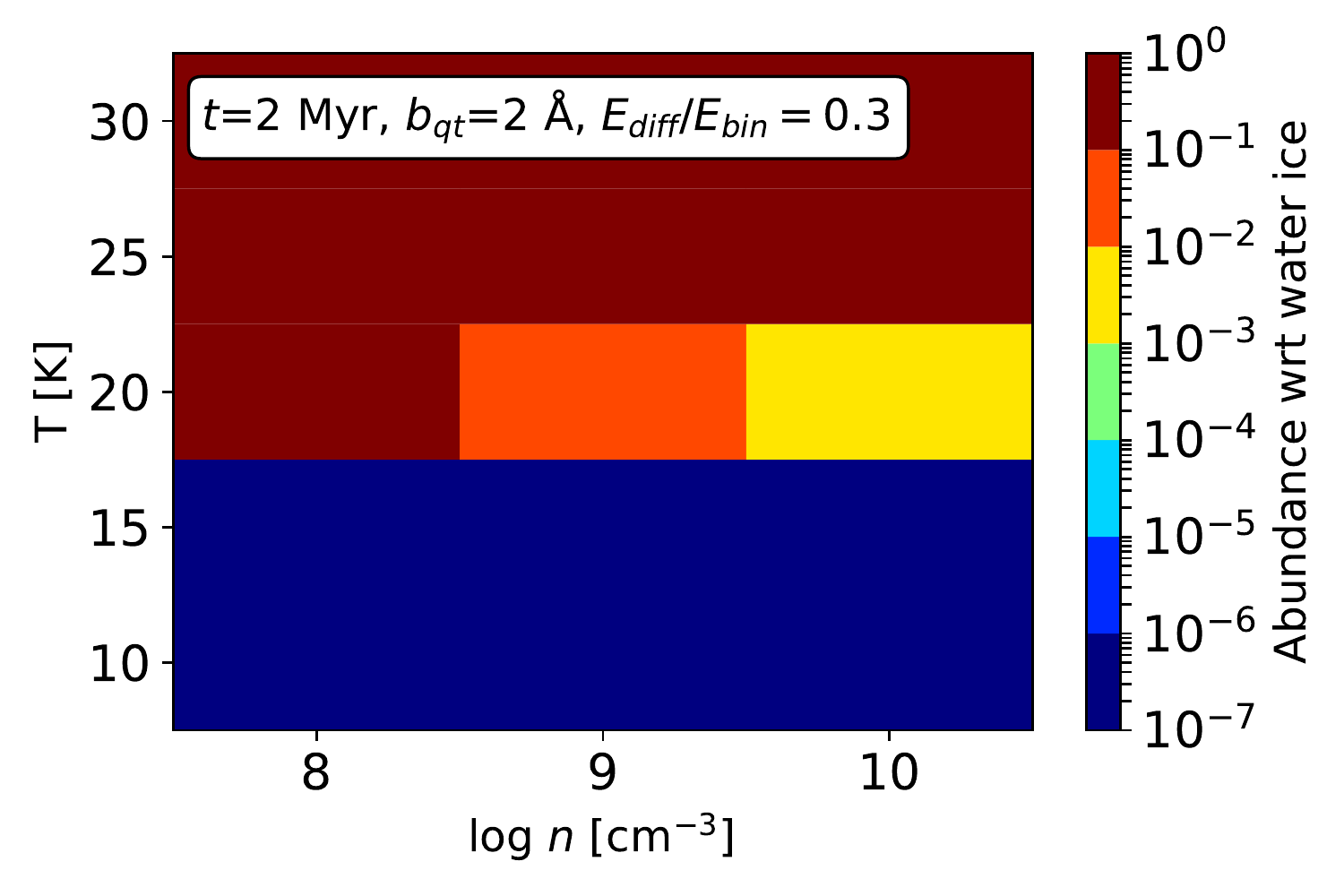}}
\subfigure{\includegraphics[width=0.22\textwidth]{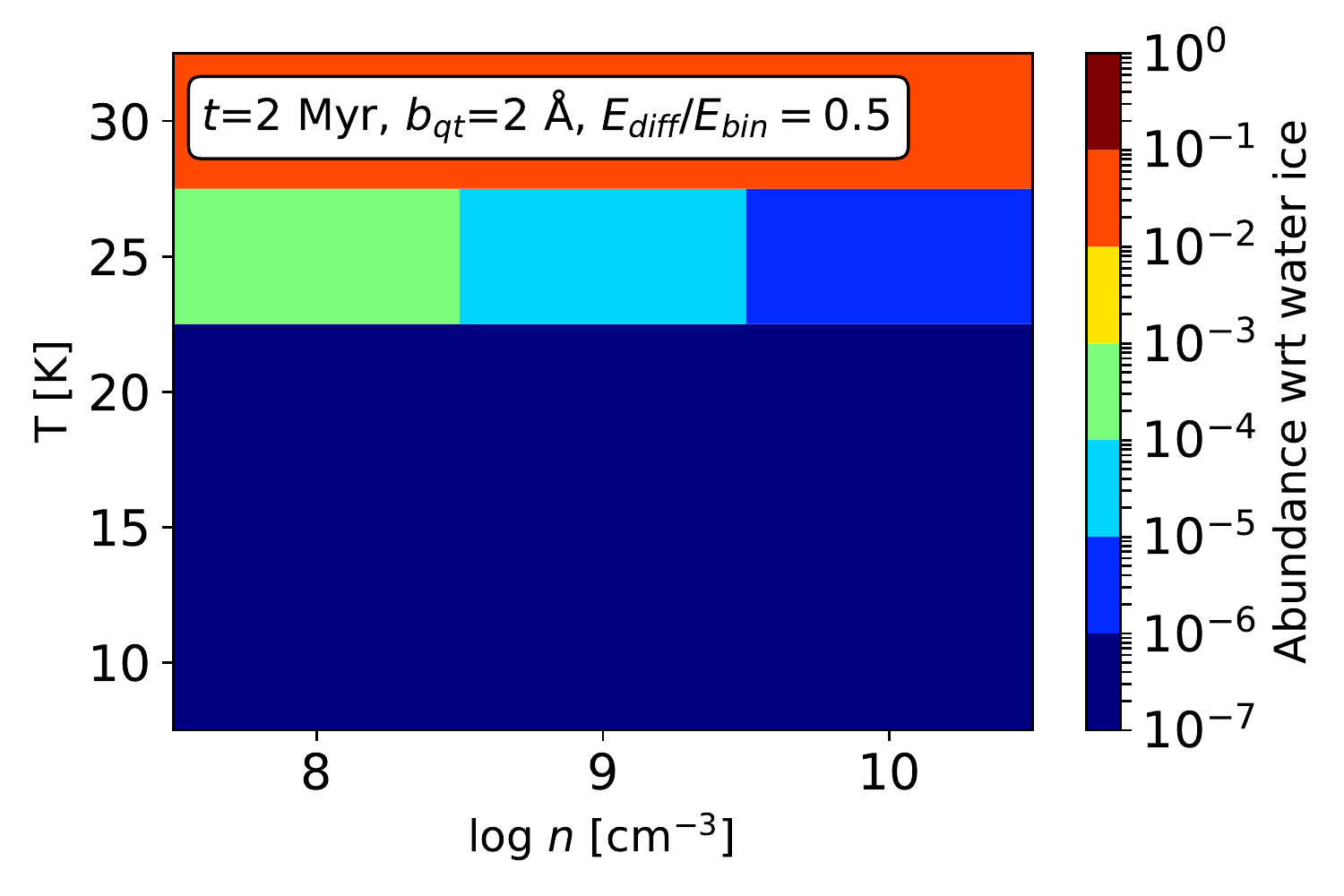}}\\
\subfigure{\includegraphics[width=0.22\textwidth]{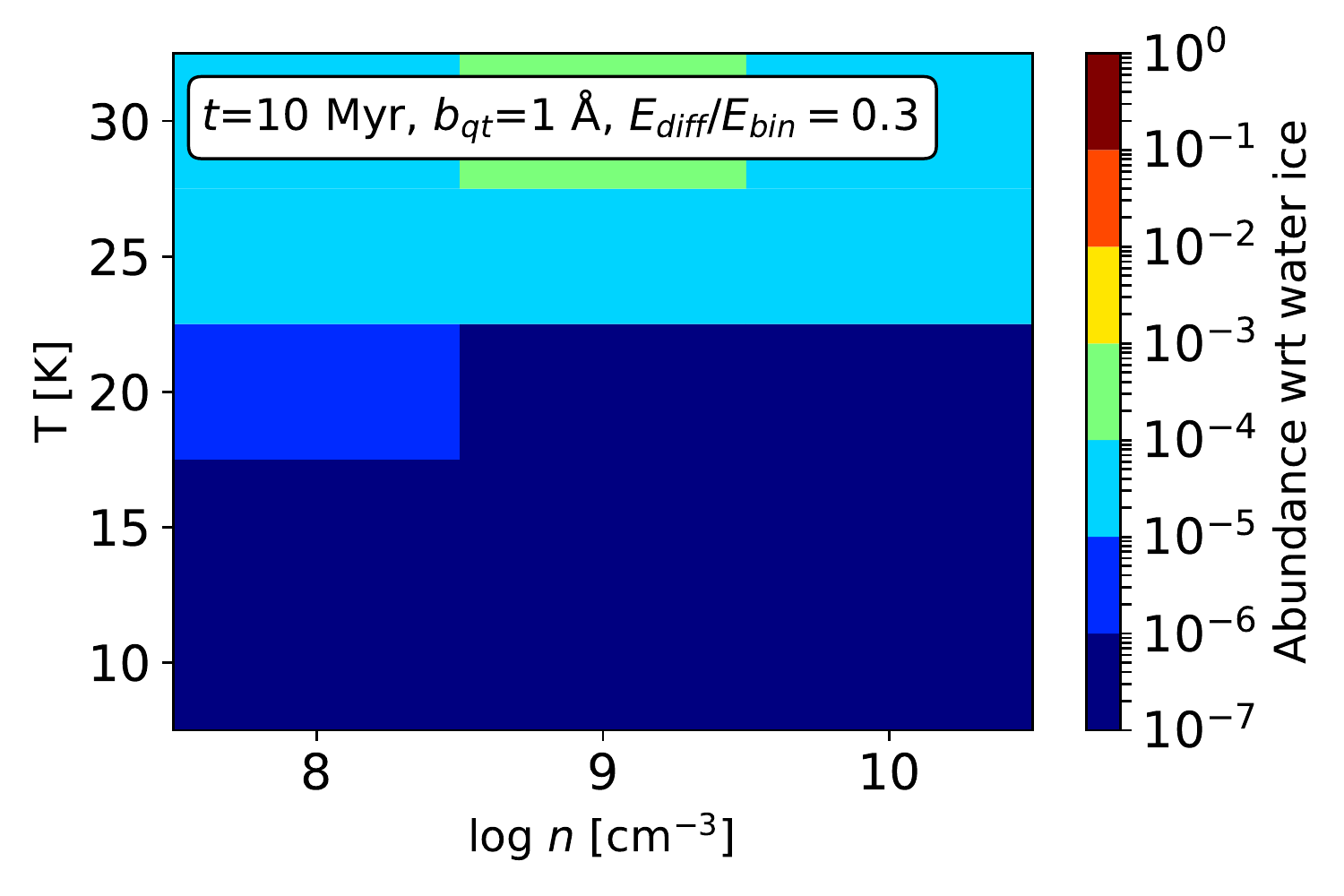}}
\subfigure{\includegraphics[width=0.22\textwidth]{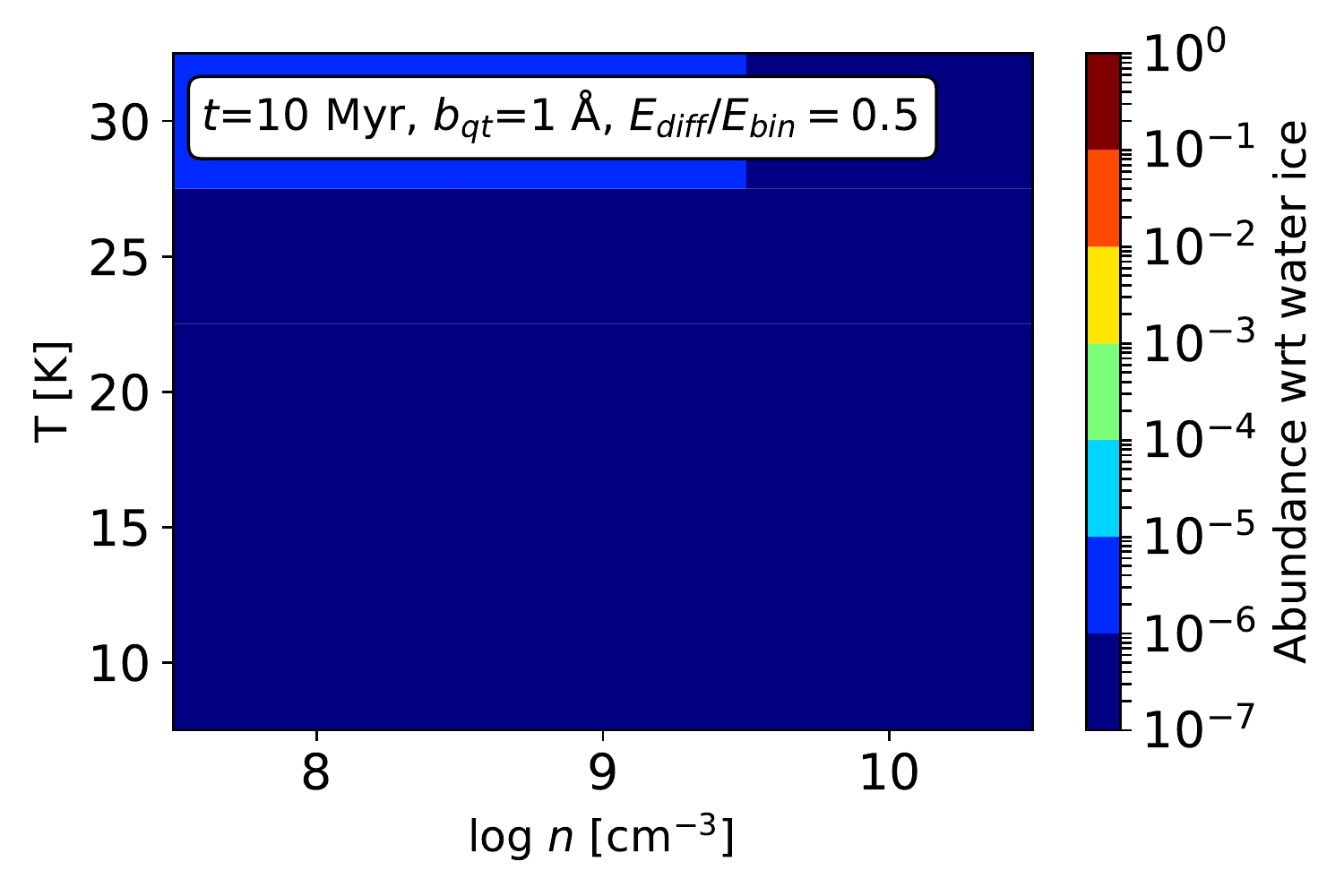}}
\subfigure{\includegraphics[width=0.22\textwidth]{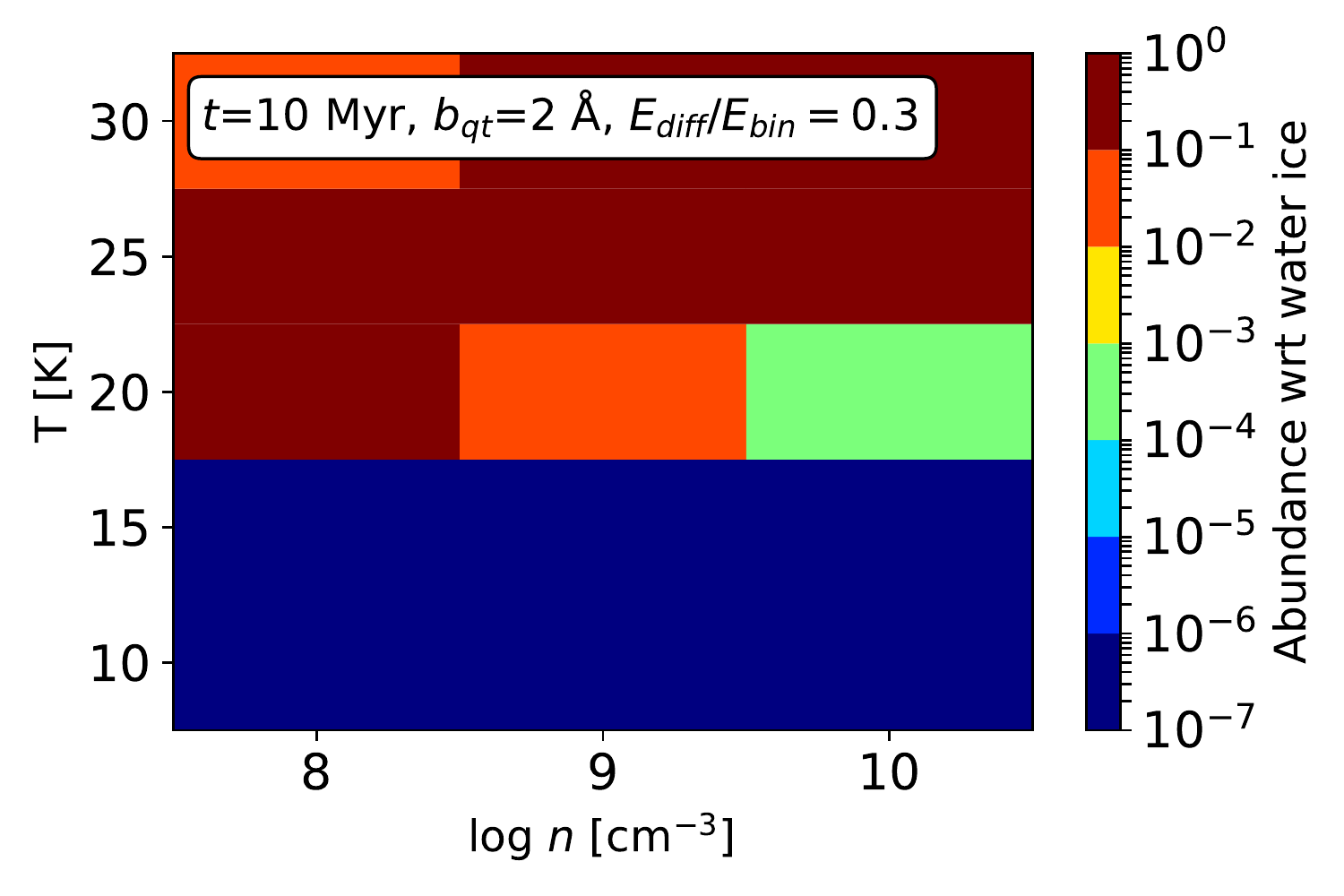}}
\subfigure{\includegraphics[width=0.22\textwidth]{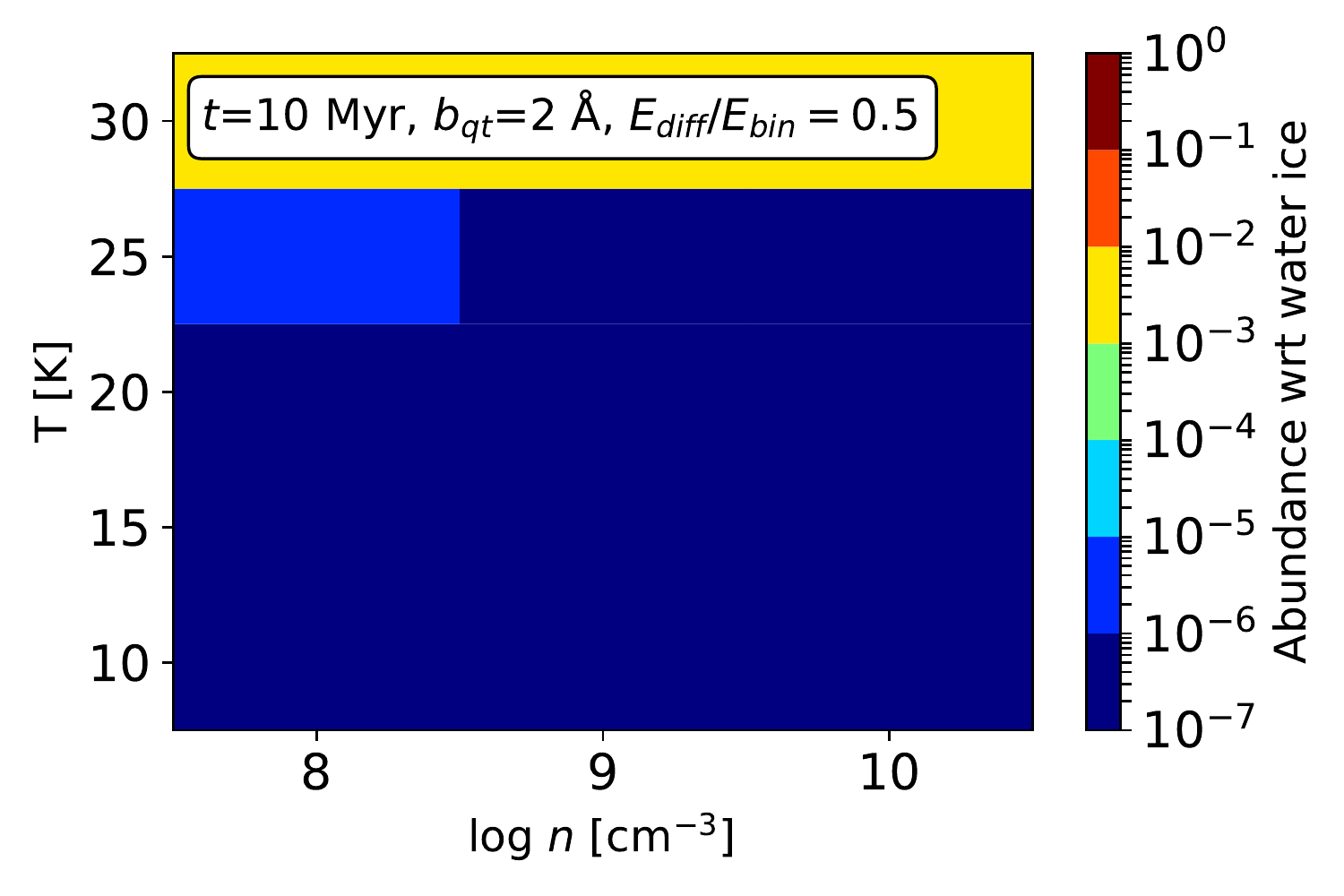}}\\
\caption{Abundances (given as colors) for \ce{H2O2} ice as function of midplane density ($x$-axes), and temperature ($y$-axes) at different evolutionary steps for the reset scenario. From left to right are abundances from model runs with different parameters for grain-surface reactions: columns one and two feature $b_{qt}$= 1 {\AA}, columns three and four feature $b_{qt}$= 1 {\AA}, columns one and three are with $E_{\rm{diff}}/E_{\rm{bin}}$= 0.3, and columns two and four are with $E_{\rm{diff}}/E_{\rm{bin}}$= 0.5. Top to bottom are different evolutionary times, from 0.05 Myr (top) to 10 Myr (bottom). To the right of each plot is a colorbar, indicating the abundance level with respect to \ce{H2O} ice for each color. The chemical network utilised includes \ce{O3} chemistry. Lightest hue of blue matches the cometary \ce{H2O2} abundances.}
\label{chem_params_h2o2}
\end{figure*}

\end{appendix}
\bibliographystyle{aa} 
\bibliography{bib_new}

\begin{thebibliography}{32}
\expandafter\ifx\csname natexlab\endcsname\relax\def\natexlab#1{#1}\fi

\bibitem[{{Aikawa} {et~al.}(1997){Aikawa}, {Umebayashi}, {Nakano}, \&
  {Miyama}}]{aikawa1997}
{Aikawa}, Y., {Umebayashi}, T., {Nakano}, T., \& {Miyama}, S.~M. 1997, \apjl,
  486, L51

\bibitem[{Bieler {et~al.}(2015)Bieler, Altwegg, Balsiger, Bar-Nun, Berthelier,
  Bochsler, Briois, Calmonte, Combi, De~Keyser, van Dishoeck, Fiethe, Fuselier,
  Gasc, Gombosi, Hansen, Hassig, Jackel, Kopp, Korth, Le~Roy, Mall, Maggiolo,
  Marty, Mousis, Owen, Reme, Rubin, Semon, Tzou, Waite, Walsh, \&
  Wurz}]{bieler15}
Bieler, A., Altwegg, K., Balsiger, H., {et~al.} 2015, Nature, 526, 678

\bibitem[{{Brown} \& {Bolina}(2007)}]{brown2007}
{Brown}, W.~A. \& {Bolina}, A.~S. 2007, \mnras, 374, 1006

\bibitem[{{Cazaux} \& {Tielens}(2002)}]{cazaux2002}
{Cazaux}, S. \& {Tielens}, A.~G.~G.~M. 2002, \apjl, 575, L29

\bibitem[{{Cleeves} {et~al.}(2013){Cleeves}, {Adams}, \&
  {Bergin}}]{cleeves13crex}
{Cleeves}, L.~I., {Adams}, F.~C., \& {Bergin}, E.~A. 2013, \apj, 772, 5

\bibitem[{{Cuppen} {et~al.}(2010){Cuppen}, {Ioppolo}, {Romanzin}, \&
  {Linnartz}}]{cuppen2010}
{Cuppen}, H.~M., {Ioppolo}, S., {Romanzin}, C., \& {Linnartz}, H. 2010,
  Physical Chemistry Chemical Physics (Incorporating Faraday Transactions), 12,
  12077

\bibitem[{{Cuppen} {et~al.}(2017){Cuppen}, {Walsh}, {Lamberts}, {Semenov},
  {Garrod}, {Penteado}, \& {Ioppolo}}]{cuppen2017}
{Cuppen}, H.~M., {Walsh}, C., {Lamberts}, T., {et~al.} 2017, \ssr, 212, 1

\bibitem[{{Dalgarno}(2006)}]{dalgarno2006}
{Dalgarno}, A. 2006, Proceedings of the National Academy of Science, 103, 12269

\bibitem[{{Dulieu} {et~al.}(2017){Dulieu}, {Minissale}, \&
  {Bockel{\'e}e-Morvan}}]{dulieu2017}
{Dulieu}, F., {Minissale}, M., \& {Bockel{\'e}e-Morvan}, D. 2017, \aap, 597,
  A56

\bibitem[{{Eistrup} {et~al.}(2016){Eistrup}, {Walsh}, \& {van
  Dishoeck}}]{eistrup2016}
{Eistrup}, C., {Walsh}, C., \& {van Dishoeck}, E.~F. 2016, \aap, 595, A83

\bibitem[{{Eistrup} {et~al.}(2018){Eistrup}, {Walsh}, \& {van
  Dishoeck}}]{eistrup2018}
{Eistrup}, C., {Walsh}, C., \& {van Dishoeck}, E.~F. 2018, \aap, 613, A14

\bibitem[{{Ennis} {et~al.}(2011){Ennis}, {Bennett}, \& {Kaiser}}]{ennis2011}
{Ennis}, C.~P., {Bennett}, C.~J., \& {Kaiser}, R.~I. 2011, Physical Chemistry
  Chemical Physics (Incorporating Faraday Transactions), 13, 9469

\bibitem[{{Garrod} \& {Herbst}(2006)}]{garrod2006}
{Garrod}, R.~T. \& {Herbst}, E. 2006, \aap, 457, 927

\bibitem[{{Hayashi}(1981)}]{hayashi1981}
{Hayashi}, C. 1981, Progress of Theoretical Physics Supplement, 70, 35

\bibitem[{{He} {et~al.}(2015){He}, {Shi}, {Hopkins}, {Vidali}, \&
  {Kaufman}}]{he2015}
{He}, J., {Shi}, J., {Hopkins}, T., {Vidali}, G., \& {Kaufman}, M.~J. 2015,
  \apj, 801, 120

\bibitem[{{Indriolo} {et~al.}(2015){Indriolo}, {Neufeld}, {Gerin}, {Schilke},
  {Benz}, {Winkel}, {Menten}, {Chambers}, {Black}, {Bruderer}, {Falgarone},
  {Godard}, {Goicoechea}, {Gupta}, {Lis}, {Ossenkopf}, {Persson},
  {Sonnentrucker}, {van der Tak}, {van Dishoeck}, {Wolfire}, \&
  {Wyrowski}}]{indriolo2015}
{Indriolo}, N., {Neufeld}, D.~A., {Gerin}, M., {et~al.} 2015, \apj, 800, 40

\bibitem[{{Ioppolo} {et~al.}(2008){Ioppolo}, {Cuppen}, {Romanzin}, {van
  Dishoeck}, \& {Linnartz}}]{ioppolo2008}
{Ioppolo}, S., {Cuppen}, H.~M., {Romanzin}, C., {van Dishoeck}, E.~F., \&
  {Linnartz}, H. 2008, \apj, 686, 1474

\bibitem[{{Ioppolo} {et~al.}(2010){Ioppolo}, {Cuppen}, {Romanzin}, {van
  Dishoeck}, \& {Linnartz}}]{ioppolo2010}
{Ioppolo}, S., {Cuppen}, H.~M., {Romanzin}, C., {van Dishoeck}, E.~F., \&
  {Linnartz}, H. 2010, Physical Chemistry Chemical Physics (Incorporating
  Faraday Transactions), 12, 12065

\bibitem[{{Jones} {et~al.}(2014){Jones}, {Kaiser}, \& {Strazzulla}}]{jones2014}
{Jones}, B.~M., {Kaiser}, R.~I., \& {Strazzulla}, G. 2014, \apj, 781, 85

\bibitem[{{Lamberts} {et~al.}(2013){Lamberts}, {Cuppen}, {Ioppolo}, \&
  {Linnartz}}]{lamberts2013}
{Lamberts}, T., {Cuppen}, H.~M., {Ioppolo}, S., \& {Linnartz}, H. 2013,
  Physical Chemistry Chemical Physics (Incorporating Faraday Transactions), 15,
  8287

\bibitem[{{Mart{\'{\i}}n-Dom{\'e}nech}
  {et~al.}(2015){Mart{\'{\i}}n-Dom{\'e}nech}, {Manzano-Santamar{\'{\i}}a},
  {Mu{\~n}oz Caro}, {Cruz-D{\'{\i}}az}, {Chen}, {Herrero}, \&
  {Tanarro}}]{martindomenech2015}
{Mart{\'{\i}}n-Dom{\'e}nech}, R., {Manzano-Santamar{\'{\i}}a}, J., {Mu{\~n}oz
  Caro}, G.~M., {et~al.} 2015, \aap, 584, A14

\bibitem[{{McElroy} {et~al.}(2013){McElroy}, {Walsh}, {Markwick}, {Cordiner},
  {Smith}, \& {Millar}}]{mcelroy13}
{McElroy}, D., {Walsh}, C., {Markwick}, A.~J., {et~al.} 2013, \aap, 550, A36

\bibitem[{{Minissale} {et~al.}(2014){Minissale}, {Congiu}, \&
  {Dulieu}}]{minnisale2014}
{Minissale}, M., {Congiu}, E., \& {Dulieu}, F. 2014, The Journal of Chemical
  Physics, 140, 074705

\bibitem[{{Mousis} {et~al.}(2016){Mousis}, {Ronnet}, {Brugger}, {Ozgurel},
  {Pauzat}, {Ellinger}, {Maggiolo}, {Wurz}, {Vernazza}, {Lunine},
  {Luspay-Kuti}, {Mandt}, {Altwegg}, {Bieler}, {Markovits}, \&
  {Rubin}}]{mousis2016}
{Mousis}, O., {Ronnet}, T., {Brugger}, B., {et~al.} 2016, \apjl, 823, L41

\bibitem[{{Mousis} {et~al.}(2018){Mousis}, {Ronnet}, {Lunine}, {Maggiolo},
  {Wurz}, {Danger}, \& {Bouquet}}]{mousis2018}
{Mousis}, O., {Ronnet}, T., {Lunine}, J.~I., {et~al.} 2018, ArXiv e-prints
  [\eprint[arXiv]{1804.03478}]

\bibitem[{{Rubin} {et~al.}(2015){Rubin}, {Altwegg}, {van Dishoeck}, \&
  {Schwehm}}]{rubin2015}
{Rubin}, M., {Altwegg}, K., {van Dishoeck}, E.~F., \& {Schwehm}, G. 2015,
  \apjl, 815, L11

\bibitem[{{Ruffle} \& {Herbst}(2000)}]{ruffle2000}
{Ruffle}, D.~P. \& {Herbst}, E. 2000, \mnras, 319, 837

\bibitem[{{Taquet} {et~al.}(2016){Taquet}, {Furuya}, {Walsh}, \& {van
  Dishoeck}}]{taquet2016}
{Taquet}, V., {Furuya}, K., {Walsh}, C., \& {van Dishoeck}, E.~F. 2016, \mnras,
  462, S99

\bibitem[{{Teolis} {et~al.}(2017){Teolis}, {Plainaki}, {Cassidy}, \&
  {Raut}}]{teolis2017}
{Teolis}, B.~D., {Plainaki}, C., {Cassidy}, T.~A., \& {Raut}, U. 2017, Journal
  of Geophysical Research (Planets), 122, 1996

\bibitem[{{Tielens} \& {Allamandola}(1987)}]{tielens1987}
{Tielens}, A.~G.~G.~M. \& {Allamandola}, L.~J. 1987, in Astrophysics and Space
  Science Library, Vol. 134, Interstellar Processes, ed. D.~J. {Hollenbach} \&
  H.~A. {Thronson}, Jr., 397--469

\bibitem[{{van Dishoeck} \& {Black}(1986)}]{dishoeck1986}
{van Dishoeck}, E.~F. \& {Black}, J.~H. 1986, in Interstellar Processes:
  Abstracts of Contributed Papers, ed. D.~J. {Hollenbach} \& H.~A. {Thronson},
  Jr.

\bibitem[{{Walsh} {et~al.}(2015){Walsh}, {Nomura}, \& {van Dishoeck}}]{walsh15}
{Walsh}, C., {Nomura}, H., \& {van Dishoeck}, E. 2015, \aap, 582, A88

\end{thebibliography}
\end{document}